\documentclass[notitlepage,superscriptaddress,aps,prl,twocolumn,noeprint]{revtex4-1}
\usepackage{amsmath}
\usepackage{amssymb}

\usepackage{upgreek}
\usepackage{mathtools}
\usepackage{graphicx}
\usepackage{afterpage}
\usepackage{amssymb}
\usepackage{epstopdf}
\usepackage{tikz}
\usepackage{verbatim}
\usepackage{lmodern}
\usepackage{multirow} 
\usepackage{bbm}
\usetikzlibrary{arrows,shapes,backgrounds} 
\usepackage{tikzpagenodes}

\tikzset{
    every picture/.style={remember picture},
    na/.style={baseline=-.5ex},
    background grid/.style={draw, black!50,step=.5cm}
}

\usepackage[colorlinks=true,citecolor=blue,linkcolor=magenta]{hyperref}
\usepackage{tikz}
\usepackage{color}
\usepackage{bm}
\usepackage[caption=false,position=top,singlelinecheck=off,justification=raggedright]{subfig}

\DeclarePairedDelimiter\bra{\langle}{\rvert}  
\DeclarePairedDelimiter\ket{\lvert}{\rangle}       
\DeclarePairedDelimiterX\braket[2]{\langle}{\rangle}{#1 \delimsize\vert #2}      
\def\lesssim{\ \raise.3ex\hbox{$<$}\kern-0.8em\lower.7ex\hbox{$\sim$}\ }
\def\gesim{\ \raise.3ex\hbox{$>$}\kern-0.8em\lower.7ex\hbox{$\sim$}\ }

\def\avg #1{\left< #1 \right>}

\begin{document}
\title{
Photoinduced non-reciprocal magnetism
}

    \author{Ryo Hanai}
    \email{hanai.r.7e4b@m.isct.ac.jp} 
    \affiliation{Department of Physics, Institute of Science Tokyo, 2-12-1 Ookayama, Meguro-ku, Tokyo, 152-8551, Japan}	
 
    \author{Daiki Ootsuki}
    \affiliation{Research Institute for Interdisciplinary Science, Okayama University, Okayama 700-8530, Japan}
    
    \author{Rina Tazai}
    \affiliation{Center for Gravitational Physics and Quantum Information, Yukawa Institute for Theoretical Physics, Kyoto University, Kyoto 606-8502, Japan}

\date{\today}

\begin{abstract}
Out of equilibrium, the action-reaction symmetry of interactions is often broken, leading to the emergence of various collective phenomena with no equilibrium counterparts. Although ubiquitous in classical active systems, implementing such non-reciprocal interactions in solid-state systems has remained challenging, as known quantum schemes require precise single-site control. Here, we propose a dissipation-engineering protocol that induces non-reciprocal interactions in solid-state platforms with light, which we expect to be achievable with state-of-the-art experimental techniques. Focusing on magnetic metals, we show microscopically that a light injection that introduces a decay channel to a virtually excited state gives rise to non-reciprocal interactions between localized spins, resulting in chase-and-runaway dynamics. Applying our scheme to layered ferromagnets, we show that a non-reciprocal phase transition to a many-body time-dependent chiral phase occurs. Our work paves the way to bring solid-state systems to the realm of non-reciprocal science, providing yet another possibility to control quantum matter with light.

\end{abstract}

\maketitle

\begin{figure*}[ht]
\centering
\includegraphics[width=0.99\linewidth,keepaspectratio]{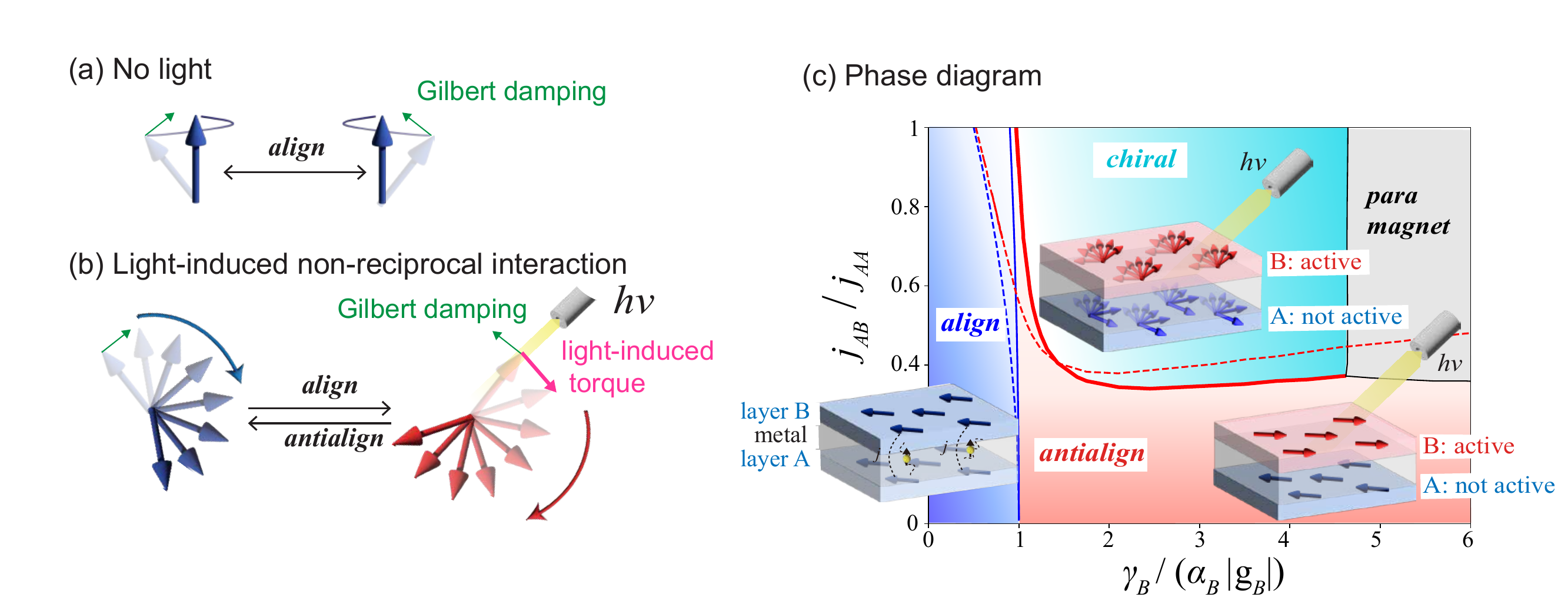}
\caption{
\textbf{Light control of magnetic interactions and magnetism via dissipation.}
(a) In the absence of light, the interaction between the local spins (thick blue arrow) is reciprocal. 
The spins are driven towards the equilibrium configuration [alignment in the ferromagnetic case illustrated here] through a magnetic friction called the Gilbert damping (green arrows).
(b) When the light is tuned to a frequency $h\nu$ that selectively activates the red spin, 
the light-induced torque (pink arrows) acts on the activated spin. 
As a result, the two spins effectively interact non-reciprocally, where the active (inactive) spin tries to anti-align (align) with the opponent's spin.
(c) Phase diagram of a layered ferromagnet under light injection, determined by our meanfield description (Eq.~\eqref{eq: non-reciprocal meanfield}).
Here, the two ferromagnetic layers (A and B) are separated by a non-magnetic metal and the laser is injected to introduce active dissipation to the B layer at rate $\gamma_{\rm B}$, making the interlayer magnetic interactions non-reciprocal.
When the light is off ($\gamma_{\rm B}=0$), the magnetization of the two layers aligns for ferromagnetic interlayer interaction $j_{\rm AB}(=j_{\rm BA})>0$ (blue region).
As the light-induced dissipation is turned on $\gamma_{\rm B}>0$, the system exhibits a phase transition to an antialigned configuration (red) 
at $\gamma_{\rm B}\simeq \alpha_{\rm B}|g_{\rm B}|$
and a non-reciprocal phase transition to a time-dependent chiral phase [where the two magnetizations exhibit many-body chase-and-runaway dynamics](cyan).
We set the intralayer interaction and Gilbert damping of the layer A (B) as $j_{\rm AA}=10{\rm meV}(j_{\rm BB}=5 {\rm meV}$) 
and
$\alpha_{\rm A}=0.1 (\alpha_{\rm B}=2\times 10^{-3})$,
respectively. 
The \textit{sd} coupling strength for B is 
$g_{\rm B}=-10{\rm meV}$, the filling $n=1$, and the temperature is
$k_{\rm B}T=9{\rm meV}$.
The dashed lines are the phase boundary at a lower temperature $k_{\rm B}T=5{\rm meV}$.
}
\label{fig: Light control of effective magnetic interactions}

\end{figure*}

\section{Introduction}

The free energy minimization principle for equilibrium systems states that all interactions between constituents must have action-reaction symmetry. 
However, this constraint is no longer present once the system is driven out of equilibrium \cite{Bowick2022, Shankar2022}. 
In fact, non-reciprocal interactions are ubiquitous in Nature: 
the brain is composed of inhibitory and exhibitory neurons that non-reciprocally interact~\cite{Wilson1972, Sompolinsky1986, Vreeswijk1996, Rieger1988}; 
the predator chases the prey and the prey runs away because their interaction is asymmetric;
colloids immersed in a chemically/optically active media exhibit non-reciprocal interactions \cite{Wimmer2013, Soto2014,  Yifat2018, Parker2020}.
Recent studies revealed that such non-reciprocal interactions fundamentally affect the collective properties of many-body systems 
\cite{Scheibner2020, Fruchart2023, Dadhichi2020, Loos2023, Fruchart2021, Hanai2019, Hanai2020, You2020, Saha2020, Zelle2023, Suchanek2023a, Suchanek2023b, Suchanek2023c, Brauns2024, 
Weis2022, Hanai2024}.
A prominent example
is the emergence of non-reciprocal phase transitions \cite{Fruchart2021, Hanai2019, Hanai2020, You2020, Saha2020, Zelle2023, Suchanek2023a, Suchanek2023b, Suchanek2023c, Weis2022, Brauns2024, Chiacchio2023, Nadolny2024}, where a time-dependent phase that exhibits a collective and persistent chase-and-runaway motion between macroscopic quantities arises.
Its critical point is characterized by the coalescence of a collective mode to the Nambu-Goldstone mode~\cite{Hanai2020, Fruchart2021, Zelle2023, Suchanek2023a, Suchanek2023b, Suchanek2023c, Weis2022} (instead of merely being degenerate as in the conventional cases),
which is strictly forbidden in equilibrium.

Non-reciprocal interactions are not restricted to the classical systems mentioned above. 
Quantum systems may also exhibit non-reciprocity \cite{Ashida2021, Zhang2022, Nadolny2024} by carefully tailoring dissipation microscopically \cite{Metelmann2015, Wang2023}.
Indeed, such 
non-reciprocity
has been implemented in synthetic quantum systems such as cold atoms \cite{Liang2022},  optomechanics \cite{Fang2017}, and circuit QED \cite{Wang2024}. 
However, these schemes require fine control of dissipation and gauge flux 
at a single-site/plaquette level, imposing a challenge to realize non-reciprocal interactions in solid-state systems.

\begin{figure}[ht]
\centering
\includegraphics[width=0.8\linewidth,keepaspectratio]{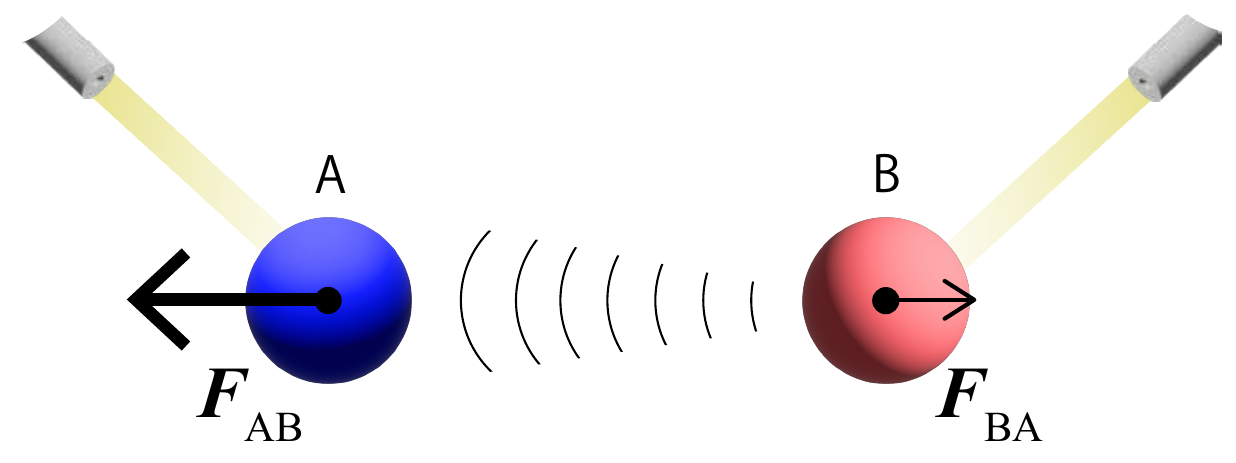}
\caption{
\textbf{Example of non-reciprocal interaction arising in soft active systems.}
Here, two metal colloids are immersed under light irradiation.
The light that reaches the metal colloid A(B) is partly absorbed, and the rest is reflected.
The reflected light reaches the metal colloid B(A), giving rise to an optical force $\bm F_{\rm BA(AB)}$. 
When the absorption/reflection properties of the two colloids are different, the optical forces need not obey Newton's third law $\bm F_{\rm AB}\ne -\bm F_{\rm BA}$.
}
\label{fig: colloid}
\end{figure}

In this work, we propose a novel dissipation-engineering scheme to realize non-reciprocal interactions in solid-state systems with light (Fig.~\ref{fig: Light control of effective magnetic interactions}(a),(b)).
Specifically, we show that the Ruderman–Kittel–Kasuya–Yosida (RKKY)~\cite{Ruderman1954, Kasuya1956, Yosida1957} interaction between localized spins in a magnetic metal can be made non-reciprocal by irradiating the sample at a frequency that opens a decay channel from the doubly occupied state of a selected subset of spins, while leaving the rest off-resonant.
The optical drive continuously removes these activated spins from the system—a loss that is compensated by the conduction band. Crucially, this process is most efficient when the activated spins reside in the energetically favored configuration, thereby generating a torque that pushes them away from the ground-state configuration, whereas the inactive spins experience no such torque
(Fig.~\ref{fig: Light control of effective magnetic interactions}(b)). 
The resulting exchange becomes intrinsically non-reciprocal: spin A attempts to align with spin B, but spin B attempts to anti-align with spin A, leading to a chase-and-run dynamics.
We estimate the injection power needed for their emergence is within reach of the current techniques.
Applying this scheme to a layered ferromagnet, we find that the light-induced interlayer non-reciprocal interaction triggers a non-reciprocal phase transition \cite{Fruchart2021} to a time-dependent chiral phase in which the magnetization of the two layers perpetually chase and flee one another (Fig. \ref{fig: Light control of effective magnetic interactions}(c)).
The origin of the non-reciprocity of our proposal is the imbalance in the amount of energy injected into each spin, conceptually akin to those arising in soft active matter ~\cite{Soto2014, Yifat2018, Parker2020, Tan2022}, yet distinct from earlier quantum schemes that use engineered gauge fluxes to control interference effects~\cite{Metelmann2015, Wang2023}.

We expect our scheme to be applicable to a wide class of quantum materials such as Mott insulating phases in strongly correlated electrons \cite{Mott1968, Imada1998},
multi-band superconductivity \cite{Kondo1963, Suh1959},
and
optical phonon-mediated superconductivity that arises e.g. in ${\rm SrTiO_3}$ \cite{Gorkov2016, Schooley1964}.
This broad applicability is anticipated because our proposal does not rely on any properties specific to magnetic metals: as long as the interaction is mediated via a virtual high-energy state, our scheme should equally work.

\section{Results}
\subsection{The dissipation engineering scheme}

To illustrate our idea of creating non-reciprocal interactions in solid-state platforms, 
we first briefly review how such non-reciprocal interactions arise in soft active systems, which inspired this work.
Consider a metal colloidal system under light irradiation as a paradigmatic example \cite{Yifat2018, Parker2020} (Fig. \ref{fig: colloid}).
When light reaches metal colloid B, part of it is absorbed while the rest is reflected. The reflected light then reaches metal colloid A, exerting an optical force $\bm F_{\rm AB}$ on it. 
A similar optical force $\bm F_{\rm BA}$ is induced from A to B; however, Newton’s third law is not necessarily satisfied ($\bm F_{\rm AB}\ne -\bm F_{\rm BA}$).
This can be seen clearly by considering the extreme case where one colloid, say A, is a perfect absorber, leading to $\bm F_{\rm BA}=0$ while $\bm F_{\rm AB}\ne 0$.
This has been verified experimentally, where the ``chase-and-runaway'' dynamics of the colloids is observed~\cite{Yifat2018, Parker2020}.

Fundamentally, the interaction is non-reciprocal because (a) the energy conservation law is violated for the degrees of freedom of interest, and (b) a mediating field facilitates the interaction. 
Condition (a) indicates that there is an external energy source that is continuously injected into the relevant system, and condition (b) allows the system to transform the absorbed energy into mechanical force during the interaction process, giving rise to non-reciprocal interactions that applies force to the center of mass of the two metal colloids. 
Notably, many of the known examples of non-reciprocal interactions in soft active systems are induced by such a mediating field: 
phoretic active matter
~\cite{Soto2014, Meredith2020} (where two colloids interact via the hydrodynamic interaction arising from the coupling to a chemically reactive medium), living matter~\cite{Tan2022, Uchida2010} (the self-propulsion or rotation of the agents affects the surrounding environments that affect the hydrodynamic interactions between the agents), and quorum sensing~\cite{Dinelli2023, Duan2023} (where cells interact by detecting and reacting to the released chemicals) all satisfy the above conditions (a) and (b).

In this paper, we propose to implement a solid-state analog of the above scenario to realize non-reciprocal interactions between electrons in a solid-state platform.
The electrons in solid-state systems often interact by exchanging bosonic excitations, such as phonons, spin waves, and density waves, etc. 
Our idea is to inject light that induces dissipation during this interaction process, 
satisfying the two conditions (a) and (b) mentioned above
and hence, gives rise to non-reciprocal interaction.

\begin{figure*}[ht]
\centering
\includegraphics[width=0.6\linewidth,keepaspectratio]{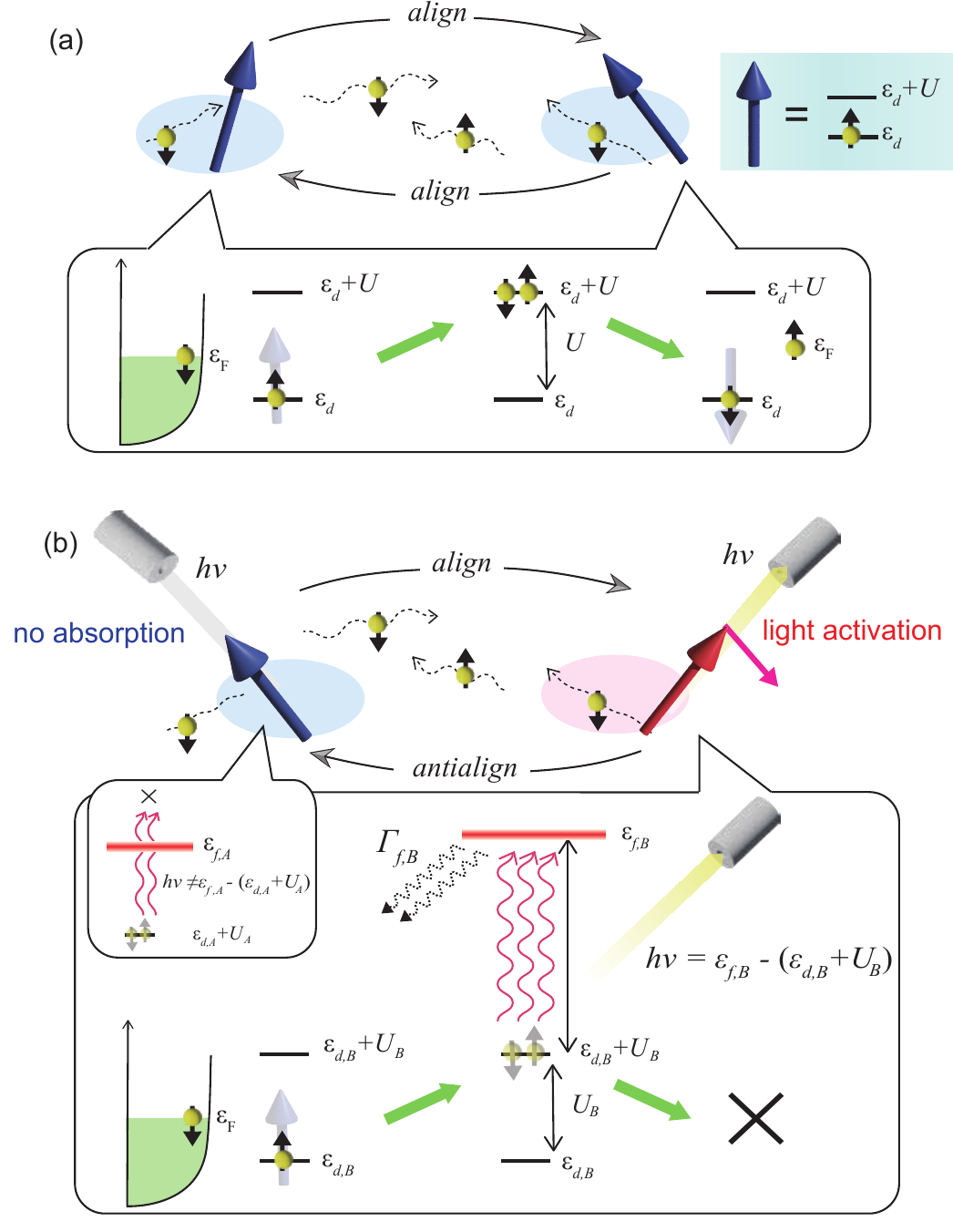}
\caption{
\textbf{Dissipation engineering RKKY interactions in magnetic metals.}
(a) Schematic description of magnetic metals.
It is composed of localized spins (thick blue arrows) and conduction electrons (yellow spheres with arrows attached).
The conduction electrons form a Fermi sea up to the Fermi energy $\varepsilon_{\rm F}$. 
The localized electrons are modeled as a two-level system with energy $\varepsilon_d$ with an on-site Coulomb interaction $U$.
The two types of electrons couple through a spin exchange coupling called the \textit{sd} coupling.
Here, the \textit{sd} spin exchange coupling arises through the second-order process where the conduction electron tunnels into the localized orbital to virtually excite to a double-occupied state and back, which involves spin-flip (the bottom panel).
Using the conduction electrons as a medium, a magnetic interaction between the localized spins arises (the so-called RKKY interaction).
[Note: there is another process where the localized electron first tunnels to the conduction band and back.]
As these perturbative processes lower the energy, the ground state configuration is the state that activates these processes the most.
(b) Our dissipation-engineering scheme is to inject light at a frequency $h\nu$ resonant with the double-occupied state and a higher-level state (say, an unoccupied higher-level f-orbital state at the energy $\varepsilon_{f, \rm B}$) that quickly dissipates with rate $\Gamma_{f, \rm B}$ in the figure.
Since the decay occurs only when the localized-conduction electron exchange process activates (which lowers the energy), 
this process selectively destroys the energetically favored states, giving rise to a light-induced torque (pink arrow in Fig.~\ref{fig: Light control of effective magnetic interactions}(b)) that applies opposite to Gilbert damping (green arrows in Fig.~\ref{fig: Light control of effective magnetic interactions}(a),(b)).
When this decay is turned on only to spin B but not A, the light-induced torque
is applied only to B spins and hence a non-reciprocal interaction emerges.
The lost electron is immediately resupplied from the surrounding environment.
}
\label{fig: second-order perturbation}
\end{figure*}

In this paper, for concreteness, we consider magnetic metals~\cite{Yosida1996} (Fig. \ref{fig: second-order perturbation}). 
These materials are composed of localized spins (responsible for magnetic properties) and conduction electrons that are free to move (responsible for metallic properties),
which couple through the spin-exchange coupling
called the \textit{sd} coupling~\cite{Kondo1964, Schrieffer1966} (Fig.~\ref{fig: second-order perturbation}(a)).
The \textit{sd} coupling arises 
through the exchange of electrons, where the conduction electrons tunnel to the localized orbital to virtually excite the localized electron to a double-occupied state (see the ``speech bubble'' in Fig.~\ref{fig: second-order perturbation}(a)).
Even in the case where the localized spins do not directly interact with each other, they may interact indirectly through this \textit{sd} coupling using the conduction electrons as the mediating field, giving rise to magnetic interactions known as the RKKY interactions \cite{Ruderman1954, Kasuya1956, Yosida1957}.

We propose to dissipation engineer the RKKY interaction by injecting light that has its energy $h\nu$  tuned to selectively turn on the tunneling from the double-occupied state to a higher-energy state that quickly dissipates (Fig. \ref{fig: second-order perturbation}(b)).
This introduces a finite lifetime to the virtual state, directly affecting the properties of the \textit{sd} coupling and hence the RKKY interaction.
The lost electron is immediately resupplied from the conduction electrons so that localized electrons are always singly occupied.

We show below 
from explicit microscopic calculations 
that this light-induced dissipation gives rise to a torque (pink arrows in Fig.~\ref{fig: Light control of effective magnetic interactions}(b)), which interestingly applies in the direction 
opposite to the energetically favored configuration.
This light-induced torque competes with the magnetic friction called the Gilbert damping~\cite{Landau1935, Gilbert2004} (green arrows) that relaxes the system to the ground state. 
When we choose the frequency of the light $h\nu$ in a way that it only resonates with a portion of the spins (red spins in Fig.~\ref{fig: Light control of effective magnetic interactions}(b)), this light-induced torque only applies to those activated spins, giving rise to non-reciprocal interactions.

Why does our dissipation scheme give rise to torque that applies in the opposite direction from the Gilbert damping?
This can be understood from the general picture we provide below.
Our light-injection scheme only dissipates the virtual state of the second-order process illustrated in Fig. \ref{fig: second-order perturbation}(b).
Hence, the decay channel is turned on only when this process is activated;
the configuration that activates the process more is the one that is likely to decay faster.
Note crucially that these second-order processes always lower the energy in equilibrium because, according to the second-order perturbation theory, the energy change due to this process is given by $\Delta E=\sum_m|v_{i,m}|^2/(E_i-E_m)<0$, where $E_{i(m)}$ is the energy of the initial (intermediate) state of the unperturbed system and $v_{i,m}$ is the matrix element between these states.
(Note that $E_i<E_m$.)
For example, the \textit{sd} exchange coupling is antiferromagnetic because the above process can only be activated when conduction and localized spins are antialigned due to the Pauli-blocking effect.
Similarly, the sign of RKKY interactions is determined by which configuration the perturbative processes activate the most frequently.
Therefore, the energetically favorable state is the state that experiences the strongest decay.
This results in dissipative interaction in the direction opposite from the friction towards the ground state arising due to the coupling with the surrounding environment, such as the substrate, bulk phonons, etc.
We remark that similar physics was discussed in Refs.~\cite{Nakagawa2020, Honda2023} in the context of cold atomic systems, where they also dissipation-engineered sign-reversal of (reciprocal) interactions, but 
see also Methods for their crucial differences arising from the absence of the surrounding environment.

Since the above scenario does not rely on features specific to magnetic systems, we expect our scheme to be equally relevant to a wide class of quantum materials~\cite{Mott1968, Imada1998, Kondo1963, Suh1959, Gorkov2016, Schooley1964}.


\subsection{Quantum master equation}

Our goal from here on is to develop a formalism that allows us to describe the effective spin dynamics of the open quantum system with dissipation induced by continuous light injection.
In this work, we consider magnetic metals modeled by the Anderson impurity model \cite{Anderson1961, Yosida1996} (Fig~\ref{fig: second-order perturbation}(a)),
described by the Hamiltonian $\hat H_{\rm A}=\hat H_0 + \hat H_{cd}$,
\begin{eqnarray}
\label{eq: Anderson 0}
    \hat H_0&=&\sum_{\bm k,\sigma=\uparrow,\downarrow}
    \varepsilon_{\bm k}
    \hat c^\dagger_{\bm k\sigma}\hat c_{\bm k\sigma}
    \\
    &+&
    \sum_a 
    \Big[
    \sum_{\sigma=\uparrow,\downarrow}
    \varepsilon_{d,a}
    \hat d^\dagger_{\sigma,a}
    \hat d_{\sigma,a}
    +U_a 
    \hat d^\dagger_{\uparrow,a}
    \hat d_{\uparrow,a}
    \hat d^\dagger_{\downarrow,a}
    \hat d_{\downarrow,a}
    \Big], 
    \nonumber
    \\
\label{eq: Anderson cd}
    \hat H_{cd}
    &=&
    \sum_{a,\sigma}
    \big[
    v_a
    \hat d_{\sigma,a}^\dagger
    \hat c_{\bm R_a\sigma} 
    +{\rm h.c.}
    \big]. 
\end{eqnarray}
Here, conduction electrons are modeled as free electrons,
where
$\hat c_{\bm k,\sigma}$
is the fermionic annihilation operator of the conduction electrons with spin $\sigma=\uparrow,\downarrow$ and momentum $\bm k$
and $\hat c_{\bm r,\sigma}=\sum_{\bm k}e^{i\bm k\cdot\bm r}\hat c_{\bm k,\sigma}$ is its Fourier transform.
$\varepsilon_{\bm k}
$ is the kinetic energy. 
The conduction electrons are assumed to be large enough to be always in thermal equilibrium, where its distribution is given by the Fermi distribution function $f(\varepsilon_{\bm k})=[e^{(\varepsilon_{\bm k}-
\varepsilon_{\rm F})/(k_{\rm B}T)}+1]^{-1}$ ($\varepsilon_{\rm F}$ is the Fermi energy and $k_{\rm B}$ is the Boltzmann constant) at low temperature $k_{\rm B}T\ll\varepsilon_{\rm F}$.
The localized electron at site $a$ is modeled as a spin-degenerate localized orbital of bare energy $\varepsilon_{d,a}$; if both spin states are occupied, the configuration is penalized by an on-site Coulomb energy $U_a$.
$v_a$ is the conduction-localized electron mixing (c-d mixing), and $\bm R_a$ is the position of the localized electron.
The energy level of localized electrons $\varepsilon_{d,a}$ sits below the Fermi energy $\varepsilon_{\rm F}$, but the double-occupied state $\varepsilon_{d,a}+U_a$ is above it.
As a result, the localized electrons are singly occupied in the steady state.
Below, we consider systems with strong enough Coulomb repulsion $U_a\gg v_a$ that justifies the perturbative treatment of the c-d mixing $v_a$ (See Supplemental Information (SI) Sec. III.C for a more precise condition).

As mentioned previously, we introduce a decay channel to localized electrons in the double-occupied state by injecting a laser that couples this state to a higher-energy state that quickly relaxes 
(see Fig.~\ref{fig: second-order perturbation}(b)).
This dissipative process can be safely regarded as a Markov process as long as the higher-energy state decays fast enough compared to its re-population rate, 
which is true in the range of interest (see Methods for details).
Such Markovian open quantum systems are generally described by 
the Gorini-Kossakowski-Sudarshan-Lindblad (GKSL) master equation \cite{Gorini1976, Lindblad1976} (also called the Lindblad master equation; see e.g. Ref. \cite{Gardiner2000} and SI Sec. I for a brief review),
given by
(where $\hat\rho$ is the reduced system density operator),
\begin{eqnarray}
    \label{eq: GKSL master equation dissipation engineering}
    \partial_t \hat\rho 
    &=&
    -i [\hat H_{\rm A},\hat \rho]
    +\sum_{a,\sigma} \kappa_a
    {\mathcal D}[\hat d_{\sigma,a}
    \hat P_{\uparrow\downarrow}^a]
    \hat\rho,
\end{eqnarray}
for our system.
The dissipator
${\mathcal D}[\hat L]\hat\rho
    =\hat L \hat\rho \hat L^\dagger 
    -\frac{1}{2}
    \{\hat L^\dagger\hat L,\hat\rho\}$
makes the time evolution non-unitary.
Here, $\hat P_{\uparrow\downarrow}^a$ is a projection operator onto the double-occupied state at site $a$, turning on a decay channel at the rate $\kappa_a$ 
only when site $a$ is double-occupied. 
We note that Eq.~\eqref{eq: GKSL master equation dissipation engineering} can be derived from a microscopic model that treats the time-dependent laser drive and a higher unoccupied level explicitly, as done in SI Sec. V.

We wish to derive the localized spin dynamics in the presence of light-induced dissipation $\kappa_a>0$.
In the equilibrium limit $\kappa_a\rightarrow 0$, a standard procedure to analyze the Anderson impurity model (Eq.~\eqref{eq: Anderson 0} and \eqref{eq: Anderson cd}) 
is to map the localized electrons in the fermionic picture to localized spins, which is performed 
by projecting out the virtual excited states that have fast oscillations~\cite{Schrieffer1966, Yosida1996}.
This incorporates the second-order process in terms of $v_a$ illustrated in 
Fig.~\ref{fig: second-order perturbation}(a).
Here, we perform the same procedure in spirit but employ it to the GKSL master equation~\eqref{eq: GKSL master equation dissipation engineering} \cite{Nakajima1958, Zwanzig1960, Li2014, Li2016, Hanai2021, Vanhoecke2024}, 
see 
Methods and 
SI Sec. I-III for details. 
It yields,
\begin{eqnarray}
    \label{eq: sd master equation}
    &&\partial_t\hat\rho
    =-i[\hat H_{\rm sd},
    \hat\rho 
    ]
    \\
    &&+\sum_a 
    \Big[
    \gamma_a 
    {\mathcal D}
    \big[
    \sum_\sigma
    \hat d_{\sigma,a}^\dagger
    \hat c_{\bm R_a,\sigma}
    \hat P_{\rm s}^a
    \big]  
    +\sum_\sigma
    \kappa_a {\mathcal {D}}[\hat d_{\sigma,a}\hat P_{\uparrow\downarrow}^a]\Big]
    \hat\rho.
    \nonumber
\end{eqnarray}
The first term on the right-hand side describes the coherent dynamics that have an identical form to those found in equilibrium
and the second and the third are the light-induced dissipative terms.
The \textit{sd} Hamiltonian~\cite{Kondo1964}
$\hat H_{\rm sd}
=-(1/2)\sum_a g_a 
\hat P_{\rm s}^a
[\hat {\bm \tau}(\bm R_a)
\cdot \hat {\bm S}_a ]
\hat P_{\rm s}^a$ 
describes the spin exchange coupling between the conduction and localized spins, where $\hat P_{\rm s}^a$ is the projection operator to singly-occupied localized electron states 
at site $a$.
Here, 
$(\hat{\bm S}_a)_i
=\sum_{\sigma,\sigma'} \hat d_{\sigma,a}^\dagger (\sigma_i)_{\sigma\sigma'}
\hat d_{\sigma',a}$ is the localized spins, $\bm \sigma=(\sigma_1,\sigma_2,\sigma_3)$ are the Pauli matrices, 
$\hat{\bm \tau}(\bm R_a)
=\sum_{\sigma,\sigma'}
(\hat c_{\bm R_a,\sigma'}^\dagger 
\bm \sigma_{\sigma',\sigma} 
\hat c_{\bm R_a,\sigma})
$ 
is the conduction spin at position $\bm R_a$,
and $g_a\simeq -|v_a|^2[(\varepsilon_{d,a}+U_a-\varepsilon_{\rm F})^{-1}+(\varepsilon_{\rm F}-\varepsilon_{d,a})^{-1}]<0$ (where $\kappa_a\ll U_a$ is assumed) is the \textit{sd} coupling strength that is antiferromagnetic.
(Note that $\varepsilon_{d,a}<\varepsilon_{\rm F}<\varepsilon_{d,a}+U_a$.)
As usual, we have assumed that only the excitations near the Fermi surface are responsible. 
We have also ignored the impurity potential of conduction electrons, as they play a minor role.

The second term is the emergent correlated dissipation with the rate $\gamma_a\simeq \kappa_a |v_a|^2 / (\varepsilon_{d,a}+U_a-\varepsilon_{\rm F})^2$ that arises from the interplay between the strong correlation effect and the light-induced decay. (See also Methods.) 
This term induces dissipative tunneling of electrons from the conduction band to the localized orbital when the electron $a$ is singly occupied.
The third term describes the decay of electrons from the double-occupied state at a much faster rate than the correlated tunneling ($\kappa_a\gg \gamma_a$), driving the system immediately back to the singly occupied state.


The emergent correlated dissipation (the second term in Eq.~\eqref{eq: sd master equation}) already captures the underlying mechanism of the dissipation-induced sign-reversal of interactions described in the previous section.
This can be seen from the localized spin dynamics that are derived
from the master equation~\eqref{eq: sd master equation}, 
\begin{eqnarray}
\label{eq: sd dynamics}
    \dot {{\bm S}}_a
    =
    g_a \avg{
    \hat{\bm S}_a
    \times\hat{\bm \tau}(\bm R_a) }
    -\gamma_a
    n {\bm S}_a
    +\gamma_a
    {\bm \tau}(\bm R_a),    
\end{eqnarray}
where 
$\langle \hat O \rangle ={\rm tr}[\hat\rho \hat O]$ represents the expectation value, $\bm S_a=\langle\hat{\bm S}_a\rangle, \bm\tau(\bm R_a)=\avg{\hat{\bm\tau}(\bm R_a)}$, and $n$ is the filling of the conduction electron.
The first and second terms describe the coherent dynamics arising from \textit{sd} interaction and the light-induced decay of the dipole moment of the localized spin, respectively. 
Here, the latter arises since the localized electrons that carry the spins are continuously lost and resupplied, which makes the (statistically averaged) spin dipole decay.
The third term is the emergent dissipative torque, which drives the localized spin toward alignment with the conduction spins.
This is the opposite of what is expected from energetics, where the \textit{sd} coupling is antiferromagnetic $g_a<0$.
We note that it was crucial for this light-induced dissipator in Eq.~\eqref{eq: sd master equation} to have the form ${\mathcal D}[\sum_\sigma \hat o_\sigma]$ instead of $\sum_\sigma{\mathcal D}[\hat o_\sigma]$ for the dissipative torque to appear, generating quantum entanglement between localized and conduction electrons that is important for non-reciprocal interactions to emerge; see SI Sec. III.C for details.

\subsection{Landau-Lifshitz-Gilbert equation with light-induced interactions}

The remaining task to obtain the effective interaction between the localized spins~\cite{Ruderman1954, Kasuya1956, Yosida1957} (i.e. the dissipation-engineered RKKY interaction) is to integrate out the conduction electron degrees of freedom that we regard as a (non-Markovian) bath.
We perform this by mapping the master equation \eqref{eq: sd master equation} to a Keldysh action \cite{Sieberer2016} that allows us to utilize field-theoretic approaches, taking into account non-adiabatic effects from the Fermi statistics of the conduction electrons. 
Once the Keldysh action is obtained, 
we integrate out the conduction electrons' degrees of freedom within the second-order perturbation in terms of \textit{sd} coupling $g_a$ and light-induced correlated decay rate $\gamma_a$ under the gradient approximation (i.e., Markov approximation plus a first-order non-Markovian correction).
We extract the localized spin dynamics in the saddle-point approximation from the obtained reduced Keldysh action consisting only of localized electrons' degrees of freedom.
We detail the procedure in 
Methods and 
SI Sec. IV.

The obtained semiclassical localized spin dynamics are
\begin{eqnarray}
    \label{eq: LLG equation}
    &&
    \dot{\bm S}_a
    =\sum_{b(\ne a)} 
    J_{ab}
    \bm S_a
    \times
    \bm S_b
    -
    \alpha_a
    \bm S_a
    \times
    \dot{\bm S}_a
    \nonumber\\
    &&\ \ \ \ \
    -\gamma_a n
    \bm S_a
    -\sum_{b
    (\ne a)
    }
    \Omega_{ab}
    \bm S_b,
\end{eqnarray}
which is one of the main results of this work.
We emphasize that all terms, including the Gilbert damping term, are obtained microscopically.
Here, for simplicity, we have assumed that we are at low temperature $(k_{\rm B} T/\varepsilon_{\rm F}\ll 1)$ and omitted the non-local non-Markovian (Gilbert-damping-like) terms~\cite{Kamra2017, Kamra2018, Reyes-Osorio2024} arising from the dissipative part of the spin wave propagator.

The first two terms on the right-hand side are the Landau-Lifshitz-Gilbert equation~\cite{Landau1935, Gilbert2004}, while the last two terms are terms generated through our controlled dissipation. 
The first term gives rise to the coherent precession motion around the effective magnetic field $\bm B_{\rm eff}=\sum_{b(\ne a)} J_{ab}(R_{ab})\bm S_b$,
where
$J_{ab}(=J_{ba})$ 
is the RKKY interaction strength given by $J_{ab}(R_{ab})=-9\pi 
[(g_a g_b)/\varepsilon_{\rm F}]
n^2
F(2k_{\rm F}R_{ab})
$ in the case of parabolic dispersion $\varepsilon_{\bm k}=\hbar^2\bm k^2/(2m)$ in three spatial dimensions, where $m$ is the conduction electron mass,
$k_{\rm F}=\sqrt{2m \varepsilon_{\rm F}}/\hbar$ is the Fermi momentum, 
$R_{ab}=|\bm R_a - \bm R_b|$ is the inter-spin distance, 
$F(x)=[-x\cos x+\sin x]/x^4$, 
and $\hbar$ is the Dirac constant.
The second term, obtained as the first-order correction to the Markov approximation~\cite{Reyes-Osorio2024}, is the Gilbert damping term describing the spins' magnetic friction~\cite{Landau1935, Gilbert2004}. 
This drives the system toward the ground state configuration when combined with the first term
(green arrows in Fig.~\ref{fig: Light control of effective magnetic interactions}).
Here, $\alpha_a$
is the Gilbert damping rate, which, in the parabolic dispersion case, reads $\alpha_a=(9\pi^2/2) n^2(g_a/\varepsilon_{\rm F})^2$.
The relaxation rate for such a process is $\gamma_{\rm Gil}^{ab}\sim \alpha_a |J_{ab}|$.

\begin{figure*}[t]
\centering
\includegraphics[width=1\linewidth,keepaspectratio]{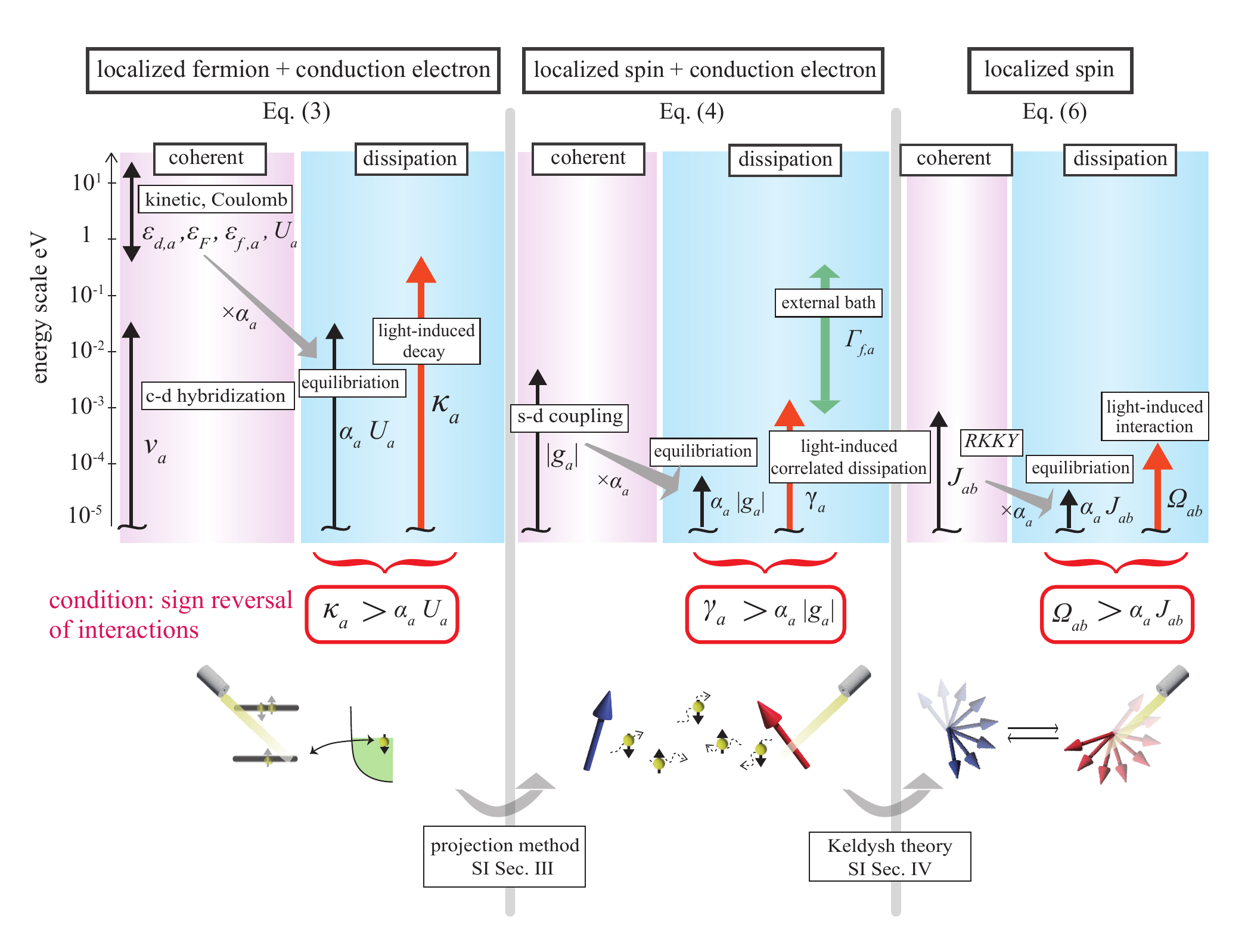}
\caption{
\textbf{Typical energy scales in different physical pictures.}
The energy scales in (left panel) localized electrons immersed in conduction electrons picture described by Eq.~\eqref{eq: GKSL master equation dissipation engineering}, (middle panel) localized spins immersed in conduction electrons picture described by Eq.~\eqref{eq: sd master equation}, and (right panel) in the interacting localized spin picture (where the conduction electrons are integrated out) described by Eq.~\eqref{eq: LLG equation}.
These pictures map from one to the other via the projection method (SI Sec. III) and Keldysh theory (SI Sec. IV).
Each panel lists the energy scales in the coherent (Hamiltonian) and dissipative dynamics.
The equilibration occurs at the timescale set by the energy scale in the Hamiltonian multiplied by the Gilbert damping rate $\alpha_a$.
This competes with the light-induced dissipation, and the sign reversal of interaction occurs when the latter exceeds the former. 
}
\label{fig: energy scale}
\end{figure*}

The light-induced dissipative interactions $\Omega_{ab}(R_{ab})=(\gamma_a/|g_a|) J_{ab}(R_{ab})\simeq (\kappa_a/U_a) J_{ab}(R_{ab})[\ne \Omega_{ba}]$ compete with this equilibration dynamics (pink arrows in Fig.~\ref{fig: Light control of effective magnetic interactions}(b)).
In addition to the unavoidable decay with rate $\gamma_a$ of the dipole moment described by the third term,
our light induces effective interactions that cannot be written as the derivative of the energy function.
(We briefly note that we have ignored the contribution from self-dissipative interaction $\Omega_{aa}$, which merely renormalizes the decay rate $\gamma_a$.)
This dissipative interaction drives the system towards a configuration opposite to the ground state configuration.
When $\kappa_a \gesim \alpha_a U_a$, this light-induced contribution ($\Omega_{ab}=\kappa_a/U_a\cdot J_{ab}$)
exceeds the Gilbert damping
($\gamma_{\rm Gil}^{ab}\sim\alpha_a J_{ab}$), 
causing the effective interaction to change its sign.
When this sign flip occurs to one of the spins but not the other, non-reciprocal interactions with effective opposite signs emerge, resulting in chase-and-runaway dynamics, the situation illustrated in Fig.~\ref{fig: Light control of effective magnetic interactions}(b).


In Methods, we estimate the pumping power $P$ required to achieve this regime as
\begin{eqnarray}
  \label{eq:laser power}
    P
    \gesim 
    \alpha_a \frac{2\pi U_a  \nu m_0 \epsilon_0 c}{ e^2 }
    \Gamma_{f,a}
\end{eqnarray}
using a Lorentz oscillator model,
where $e$ is the electron charge, $m_0$ is the electron mass, $\epsilon_0$ is the vaccuum dielectric constant, and $c$ is the speed of light.
Setting the typical values $U_a \sim 1{\rm eV}, \alpha_a \sim 10^{-2},\Gamma_{f,a}\sim 10{\rm meV},h\nu\sim 1{\rm eV}$, the required pump power is 
$P\gesim 10^8{\rm W/cm^2}$.
Not only is this pump power achievable, e.g. with Raman lasers with pulse duration of $O(10{\rm ns})$~\cite{Ferrara2020}
or even with a steady-state resource~\cite{Stenning2023}, but the heating effect should be minimal for magnetic metals. 
For instance, in Ref. \cite{Stanciu2007}, it was reported that the sample (GdFeCo) did not demagnetize until the pump power exceeded $P \sim 10^{10}{\rm W/cm^2}$ for the pulsed experiment, and in Ref. \cite{Stenning2023}, the ferromagnetic of the sample (NiFe) retained at least up to $P \sim 10^{8}{\rm W/cm^2}$ (3.5 mW power on a 580 nm diameter spot). 
For convenience for the readers, we have summarized the typical energy scales in Fig.~\ref{fig: energy scale}.

\begin{figure*}[t]
\centering
\includegraphics[width=0.7\linewidth,keepaspectratio]{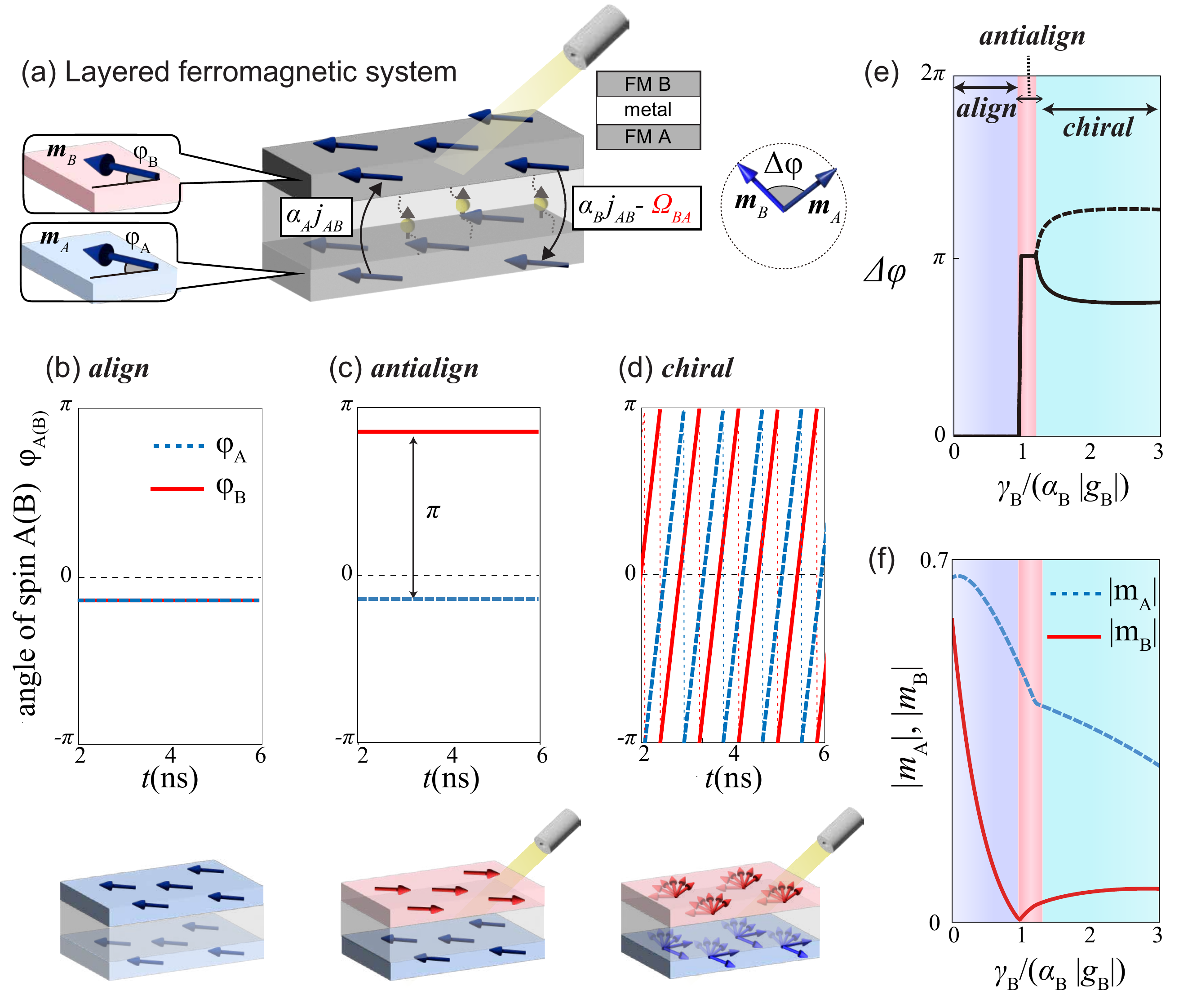}
\caption{
\textbf{Non-reciprocal phase transitions in light-activated layered ferromagnets.}
(a)
Layered ferromagnet composed of A and B layers separated by a non-magnetic metal exposed of light injection that activates the B layer. 
The interlayer RKKY interaction is mediated by the itinerant electrons in the non-magnetic metal layer, which is modified by light.
In particular, the magnetization in the A layer aligns with the rate $\alpha_{\rm A}j_{\rm AB}$ for ferromagnetic interlayer interactions $j_{\rm AB}>0$, while the B layer may align or antialign with layer A depending on the sign of the modified effective interaction $\alpha_{\rm B}j_{\rm AB}-\Omega_{\rm AB}$.
$\varphi_{\rm A(B)}$ is the orientation direction of the magnetization $\bm m_{\rm A(B)}$ of A(B)-layer ferromagnet.
(b)-(d) Different phases arising in this system.
(b) Aligned phase ($\Delta\varphi=\varphi_{\rm A}-\varphi_{\rm B}=0$) realized in the equilibrium limit $\gamma_{\rm B}=0$.
(c) Antialigned phase ($\Delta\varphi=\pi$) at 
$\gamma_{\rm B}/(\alpha_{\rm B}|g_{\rm B}|)=1.1$.
(d) Chiral phase
($\Delta\varphi\ne 0,\pi$) at 
$\gamma_{\rm B}/(\alpha_{\rm B}|g_{\rm B}|)=1.5$.
(e) The orientation angle difference $\Delta\varphi$
and (f) the magnitude of the magnetization $|\bm m_{\rm A,B}|$ as a function of $\gamma_{\rm B}$. 
We set $j_{\rm AB}=5{\rm meV},k_{\rm B}T=9{\rm meV}$, and the other parameters are the same as those used in Fig.~\ref{fig: Light control of effective magnetic interactions}(c). The simulations were run from random initial conditions and have checked that there is essentially no initial condition dependence on the final state.
}
\label{fig: many body}
\end{figure*}

\subsection{Non-reciprocal phase transitions}

So far, we have shown from microscopic calculations that one can generate non-reciprocal interactions with light.
An intriguing question is how such non-reciprocal interaction affects the collective properties of many-body systems. 
In this work, at the outset, we consider a simple setup illustrated in Fig. \ref{fig: many body}(a), where light is exposed to two layers of ferromagnets (that we label A and B) sandwich a non-magnetic metal. 
The light is tuned to activate only layer B spins.
In the absence of light, this is a type of MRAM device.
The itinerant electrons in the non-magnetic metal layer mediate the interlayer RKKY interaction~\cite{Yafet1987, Bruno1991, Stiles1999, Ostler2012}.
The light injection induces additional torque to the B layer applying oppositely from conventional RKKY interaction, giving rise to non-reciprocal interlayer interaction (Fig. \ref{fig: many body}(a)).

We wish to predict the magnetization dynamics of this system. 
Unfortunately, deriving the governing equation of the collective magnetization dynamics from a microscopic model is a highly non-trivial task that requires a beyond-saddle-point approximation we employed above (where a longitudinal relaxation is not incorporated)~\cite{Garanin1997}. 
Although deriving such a coarse-grained description from microscopics is an important challenge, here, we take a phenomenological approach below. 

We make the following observations: 
(a) In the absence of light, the magnetism $\bm m_{a={\rm A, B}}$ should converge to the known equilibrium value.
(b) Since the relaxation towards this state occurs through the Gilbert damping, their relaxation time for the $a$-layer is expected to be $\tau_{\rm Gil}^a = O([\alpha_a \sum_b j_{ab}]^{-1})$, where $j_{ab}(=j_{ba})$ is the interaction strength $J_{ab}$ multiplied by the number of spins a given spin couples to.
(c) When light is injected into the B layer, the light-induced torque $\Omega_{\rm BA}$ and a decay with the rate $\gamma_{\rm B}$ is introduced to B layer magnetization $\bm m_{\rm B}$.
This brings us to propose the following phenomenological meanfield description:
\begin{subequations}
\label{eq: non-reciprocal meanfield}
    \begin{align}
        &\dot{\bm m}_{\rm A}
        =\alpha_{\rm A}
        \bigg[
        -k_{\rm B} T \bm m_{\rm A}
        +\Big[
        1-\frac{(\bm h_{\rm eff}^{\rm A})^2}
        {3(k_{\rm B}T)^2}
        \Big]
        \bm h_{\rm eff}^{\rm A}
        \bigg]
        ,
        \\
        &\dot{\bm m}_{\rm B}
        =\alpha_{\rm B}
        \bigg[
        -k_{\rm B}T \bm m_{\rm B}
        +\Big[
        1-\frac{(\bm h_{\rm eff}^{\rm B})^2}
        {3(k_{\rm B}T)^2}
        \Big]
        \bm h_{\rm eff}^{\rm B}
        \bigg]
        \nonumber\\
        & \ \ \ \ \ \ \ 
        -\gamma_{\rm B} n
        \bm m_{\rm B}
        -\Omega_{\rm BA}
        \bm m_{\rm A},
    \end{align}
\end{subequations}
where $\bm h_{\rm eff}^a=\sum_{b={\rm A,B}} j_{ab}\bm m_b$ is the effective field applied to $\bm m_a$ and $\Omega_{\rm BA}=(\gamma_{\rm B}/|g_{\rm B}|)j_{\rm AB}$ is the light-induced torque.
For simplicity, we have assumed that 
the system is close enough to the disordered-ordered transition point that the Ginzburg-Landau expansion is justified
and the anisotropy is strong enough that $z$-component of the magnetization vanishes $m_a^z=0$.
In the absence of light $\gamma_{\rm B}=0$, the steady state $\dot {\bm m}_a=0$ reproduces the known result from the Weiss theory.
This is of the general form introduced in Ref.~\cite{Fruchart2021}.

Figure~\ref{fig: Light control of effective magnetic interactions}(c) shows the phase diagram obtained by simulating Eq.~\eqref{eq: non-reciprocal meanfield}.
In the absence of the light injection $\gamma_{\rm B}=0$, unsurprisingly, the magnetization orientation of the two layers aligns $\Delta\varphi=\varphi_{\rm A}-\varphi_{\rm B}=0$ (Fig.~\ref{fig: many body}(b)), where the orientation of the magnetism is defined by $\bm m_a = |\bm m_a|(\cos\varphi_a,\sin\varphi_a)$.
As one increases the laser power that increases $\gamma_{\rm B}$, the light-induced torque $\Omega_{\rm BA}$ weakens the ferromagnetic interaction, until it swaps the sign at
$\Omega_{\rm BA}\gesim \alpha_{\rm B}j_{\rm AB}$ or $\gamma_{\rm B}\gesim \alpha_{\rm B}|g_{\rm B}|$ (see Fig.~\ref{fig: many body}(e)).
This causes a transition from aligned $\Delta\varphi=0$ to antialigned configuration $\Delta\varphi=\pi$ (blue thin line in Fig.~\ref{fig: Light control of effective magnetic interactions}(c) and Fig.~\ref{fig: many body}(c)). 
Remarkably, the B layer completely demagnetizes at the transition point $|\bm m_{\rm B}|=0$, while the A layer is still ferromagnetic $\bm m_{\rm A}\ne 0$ (Fig.~\ref{fig: many body}(f)) even though the interlayer coupling is still present. 

As $\gamma_{\rm B}$ is further increased, 
the system exhibits a non-reciprocal phase transition \cite{Fruchart2021, Hanai2024} to a time-dependent chiral phase (see Fig.~\ref{fig: Light control of effective magnetic interactions}(c)) exhibiting a many-body chase-and-runaway motion (Fig.~\ref{fig: many body}(d)). 
The parity spontaneously breaks in this chiral phase, where the relative orientation angle converges to a state $\Delta\varphi=(\ne 0,\pi)$
that is not invariant under the parity operation $\Delta\varphi \rightarrow - \Delta\varphi$.
This cannot be understood from the Landau theory \cite{Fruchart2021}, where the critical point is characterized by the coalescence of the collective modes to the Nambu-Goldstone mode~\cite{Hanai2020, Zelle2023}.

One can estimate the necessary condition for the emergence of the chiral phase as (See SI Sec. VI for the derivation),
\begin{eqnarray}
    \frac{\alpha_{\rm B}}{\alpha_{\rm A}}
    \frac{g_{\rm B}^2}{j_{\rm AB}^2n^2}
    \lesssim 1
\end{eqnarray}
where we have assumed a small Gilbert damping rate of the B layer spins $\alpha_{\rm B}j_{\rm BB}\ll \alpha_{\rm A}j_{\rm AA}$ and $\alpha_{\rm B}\ll \gamma_{\rm B}/|g_{\rm B}|$. Plugging in the values used in Figs.~\ref{fig: Light control of effective magnetic interactions}(c) and \ref{fig: many body}(b)-(f), we find the required interlayer RKKY interaction strength to be $j_{\rm AB}\gesim 1 {\rm meV}$, in agreement with our numerics.
Since the electrons in the non-magnetic metal layer mediate the RKKY interaction, we estimate the interlayer RKKY interaction strength as
$j_{\rm AB}\sim g_{\rm A}g_{\rm B}\rho_{\rm spacer}$, 
where $\rho_{\rm spacer}$ is the density of states of the electrons in the non-magnetic spacer layer at the Fermi level.
The condition for the chiral phase is then derived as
\begin{eqnarray}
    \frac{\alpha_{\rm B}}{\alpha_{\rm A}}
    \frac{1}{g_{\rm A}^2 \rho_{\rm spacer}^2 n^2}
    \lesssim 1.
\end{eqnarray}
This implies that the larger the density of states of the spacer is, the more likely the chiral phase is achievable;
For instance, taking typical values $g_{\rm A}=0.1{\rm eV}, \alpha_{\rm A}=0.1, n=1$, and $\alpha_{\rm B}=0.001$, we expect the chiral phase may emerge by choosing the spacer material with $\rho_{\rm spacer}\gesim 1{\rm eV}^{-1}$.
(See SI Sec. VI for more details, including phase diagrams with different parameters.)

Interestingly, there are regions where this symmetry-broken phase expands when increasing the temperature, which is opposite from what is conventionally expected (see the dashed lines in Fig.~\ref{fig: Light control of effective magnetic interactions}(c) that show the phase boundary at a lower temperature).
This is a signature of order-by-disorder phenomena discussed in Ref.~\cite{Hanai2024}, where a direct analogy between the geometrically frustrated systems and non-reciprocal matter was drawn.

\section{Discussion}

In summary, we have proposed a scheme to dissipation-engineer non-reciprocal interactions with light.
We showed microscopically that the light injection to magnetic metals that introduces decay of a virtually excited state induces non-reciprocal interaction between localized spins. 
Applying this method to layered ferromagnets, we showed that a non-reciprocal phase transition to a time-dependent chiral phase emerges \cite{Fruchart2021, Hanai2024}.
The pump intensity required to achieve this is estimated to be within reach of the current experimental techniques.

The effect of non-reciprocal interactions on the collective properties of many-body systems is currently 
under heavy investigation in many different disciplines of science, 
ranging from 
active matter~\cite{Das2002,  Uchida2010, Bhatt2023,Soto2014, Meredith2020, Saha2019, Yifat2018, Parker2020, Zhang2021, Fruchart2021, Dinelli2023, Duan2023, Liu2024}, 
levitated particles~\cite{Ivlev2015, Rieser2022, Wu2025},
photonics~\cite{Wimmer2013}, robotics~\cite{Brandenbourger2019},
living matter~\cite{Tan2022, Basaran2024Nonreciprocal, Fujimori2019, Bhattacherjee2024},
open quantum systems~\cite{Ashida2021, Zhang2022, Metelmann2015, Hatano1996, Chiacchio2023, Nadolny2024}, 
ecology~\cite{Neubert1997, Kerr2002, Reichenbach2007, Rieger_eco1989, Allesina2012, Bunin2017}, and neuroscience~\cite{Wilson1972, Sompolinsky1986, Vreeswijk1996, Rieger1988}, to sociology~\cite{Hong2011}.
Particularly, recent works showed that non-reciprocal phase transitions exhibit unconventional critical phenomena associated with anomalously giant fluctuations 
\cite{Hanai2020}, fluctuation-induced first-order transition~\cite{Zelle2023}, and diverging entropy production~\cite{Suchanek2023a, Suchanek2023b, Suchanek2023c}, due to the unique feature that the criticality is driven by the coalescence of the modes to the Nambu-Goldstone mode.  
Furthermore, non-reciprocal interactions have been shown to give rise to emergent features such as odd elasticity~\cite{Scheibner2020, Fruchart2023} (anti-symmetric part of the static elastic modulus tensor), long-ranged order in two spatial dimensions \cite{Dadhichi2020, Loos2023} (in an apparent violation of the Hohenberg-Mermin-Wagner theorem), and phenomena analogous to those occurring in geometrically frustrated systems \cite{Hanai2024}.
Our dissipation-engineering scheme may allow exploring these non-reciprocal physics in solid-state platforms.


\begin{center}{\bf Methods}\end{center} 

\subsection{Outline of the derivation of Eq.~(4) and Eq.~(6)}

We provide here a brief outline of the derivation of Eq.~\eqref{eq: sd master equation} and Eq.~\eqref{eq: LLG equation} in the main text.
The full detail is provided in the SI Sec. III and IV, respectively.

\subsubsection{Derivation of Eq.~\eqref{eq: sd master equation}}

Our starting point to derive Eq.~\eqref{eq: sd master equation} is the quantum master equation~\eqref{eq: GKSL master equation dissipation engineering}, which, for convenience, we write it as 
\begin{eqnarray}
    \partial_t \hat\rho = {\mathcal L}\hat\rho,
\end{eqnarray}
where we have expressed the right-hand side of Eq.~\eqref{eq: GKSL master equation dissipation engineering} using a superoperator (an operator that acts on a matrix) called the Lindbladian ${\mathcal L}$.
We split the Lindbladian into two contributions ${\mathcal L}={\mathcal L}_0+{\mathcal L}_1$:
\begin{eqnarray}
    \label{eq: L1}
    {\mathcal L}_1\rho
    &=&-i [\hat H_{cd},\hat\rho]
\end{eqnarray}
is the contribution from the c-d mixing $\hat H_{cd}=\sum_{a,\sigma}
[v_a\hat d_{\sigma,a}^\dagger
\hat c_{\bm R_a\sigma} 
+{\rm h.c.}]
=\sum_a
\sum_{\bm k,\sigma} 
[v_a e^{i \bm k\cdot \bm R_a}
\hat d_{\sigma,a}^\dagger
\hat c_{\bm k\sigma} 
+{\rm h.c.}]$ 
that we treat as a perturbation. 
The rest ${\mathcal L}_0
={\mathcal L}_{c0}
+\sum_a {\mathcal L}_{d0, a}$ is the non-perturbative part, given by,
\begin{eqnarray}
    {\mathcal L}_{c0} \hat\rho 
    &=& -i\big[
    \sum_{\bm k,\sigma}
    \varepsilon_{\bm k}
    \hat c_{\bm k,\sigma}^\dagger 
    \hat c_{\bm k,\sigma},
    \hat\rho\big],
    \\
    {\mathcal L}_{d0,a} \hat\rho 
    &=& 
    -i \big[\big(
    \sum_\sigma
    \varepsilon_{d,a}
    \hat d^\dagger_{\sigma,a}
    \hat d_{\sigma,a}
    +U_a 
    \hat d^\dagger_{\uparrow,a}
    \hat d_{\uparrow,a}
    \hat d^\dagger_{\downarrow,a}
    \hat d_{\downarrow,a}
    \big),
    \hat \rho]
    \nonumber\\
    &+&
    \sum_{\sigma} 
    \kappa_a
    {\mathcal D}[\hat d_{\sigma,a}
    \hat P_{\uparrow\downarrow}^a]
    \hat\rho.
\end{eqnarray}

In the following, we take advantage of the property that our system has a separation of timescales
by dividing the double Hilbert space (where the density operator $\hat\rho$ lives in) into slow and fast degrees of freedom.
By perturbatively projecting out the latter~\cite{Schrieffer1966}, we obtain an effective low-energy description.
Specifically, we first divide the right (left) eigenstates $\hat r_n^{(0)}$ ($\hat l_n^{(0)}$) 
with the eigenvalue $\lambda_n^{(0)}$ 
of the non-perturbative Lindbladian, defined as ${\mathcal L}_0\hat r_n^{(0)}=\lambda_n^{(0)}\hat r_n^{(0)}$ 
($\hat l_n^{(0)\dagger} {\mathcal L}_0 = \hat l_n^{(0)\dagger} \lambda_n^{(0)}$),
to slow ($n\in \mathfrak{s}$) and fast ($n\in \mathfrak{f}$) degrees of freedom ($|\lambda_{n\in\mathfrak{s}}^{(0)}|\ll||\lambda_{n\in\mathfrak{f}}^{(0)}|$).
The perturbative Lindbladian ${\mathcal L}_1$ couples the slow and fast modes.
Then, as derived in SI Sec. I.A.2, we perturbatively project out the fast degrees of freedom to yield the effective low-energy Lindbladian,
\begin{eqnarray}
\label{eq: effective Lindbladian matrix element}
    &&
    ({\mathcal L}_{\rm eff})_{n_l,n_r}
    \equiv
    (\hat l_{n_l}^{(0)},{\mathcal L_{\rm eff}\hat r_{n_r}^{(0)}})
    ={\rm tr}[
    \hat l_{n_l}^{(0)\dagger}{\mathcal L_{\rm eff}\hat r_{n_r}^{(0)}}
    ]
    \nonumber\\
    &&= 
    {\rm tr}[
    \hat l_{n_l}^{(0)\dagger}{\mathcal L_0\hat r_{n_r}^{(0)}}
    ]
    +
    {\rm tr}[
    \hat l_{n_l}^{(0)\dagger}{\mathcal L_1\hat r_{n_r}^{(0)}}]
    \nonumber\\
    &&-
    \sum_{m\in \mathfrak{f}}
    \frac{
    {\rm tr}[\hat l_{n_l}^{(0)\dagger} 
    {\mathcal L}_1
    \hat r_m^{(0)}]
    {\rm tr}[\hat l_m^{(0)\dagger}
    {\mathcal L}_1
    \hat r_{n_r}^{(0)}]
    }
    {\lambda_m^{(0)}}
    +O(({\mathcal L}_1)^3).
\end{eqnarray}
Here, $(\hat A,\hat B)={\rm tr}[\hat A^\dagger \hat B]$ is the Hilbert-Schmidt inner product and $\hat r_{n_r}^{(0)}$ ($\hat l_{n_l}^{(0)}$) is the right (left) eigenstates that form the basis of the slow degrees of freedom ($n_r,n_l\in \mathfrak{s}$).
The first, second, and third terms on the rightmost side are the zeroth, first, and second-order contribution in terms of ${\mathcal L}_1$, respectively.
In the third term, the sum is taken over the fast degrees of freedom.
Note how the third term has a similar form to the familiar second-order Rayleigh-Schr\"odinger perturbation theory, which is given by the matrix element ${\rm tr}[\hat l_{n_l}^{(0)\dagger} 
{\mathcal L}_1\hat r_m^{(0)}]
{\rm tr}[\hat l_m^{(0)\dagger}
{\mathcal L}_1
\hat r_{n_r}^{(0)}]$
divided by the eigenvalue of the intermediate state $\lambda_m^{(0)}$.
Equation~\eqref{eq: effective Lindbladian matrix element} is consistent with the so-called Lindblad perturbation theory~\cite{Li2014, Li2016, Hanai2021}.

In our problem, first note that the localized and conduction electrons are decoupled in the non-perturbative Lindbladian ${\mathcal{L}}_0={\mathcal L}_{c0}
+\sum_a {\mathcal L}_{d0, a}$
and therefore
the right eigenstate is expressed as a direct product $\hat r_{n_r}^{(0)}
=(\prod_a \hat r_{a,n_r}^{d(0)})
\otimes
\hat r_{n_r}^{c(0)}$ of the right eigenstates of  ${\mathcal L}_{d0,a}$ and ${\mathcal L}_{c0}$ described by $\hat r_{a,n_r}^{d(0)}$ and $\hat r_{n_r}^{c(0)}$,  respectively.
For the conduction electrons,
we will always be considering low-temperature states that have their conduction electrons in their ground state that forms a Fermi sea $\hat r_{n_r}^{c(0)}=\ket{F}\bra{F}$, where $\ket{F}=\prod_{\varepsilon_{\bm k}<\varepsilon_{\rm F}}
\prod_{\sigma=\uparrow,\downarrow}
\hat c^\dagger_{\bm k,\sigma}\ket{0}$. 
For the localized electrons, we regard the eigenstates with singly occupied state
as slow degrees of freedom, i.e.,
$\{ 
\ket{\uparrow}_a\bra{\uparrow}_a,
\ket{\uparrow}_a\bra{\downarrow}_a,
\ket{\downarrow}_a\bra{\uparrow}_a,
\ket{\downarrow}_a\bra{\downarrow}_a,
\}$ where 
$\ket{\sigma}_a=\hat d_{\sigma,a}^\dagger\ket{\varnothing}_a$ is a singly occupied state and $\ket{\varnothing}_a$ is a vacant state.
We also regard eigenstates that are diagonal in the Fock basis as slow modes for the localized electrons (which includes states like $\ket{\uparrow\downarrow}_a\bra{\uparrow\downarrow}_a$, where $\ket{\uparrow\downarrow}_a$ is a double-occupied state) as they do not involve fast coherent dynamics.
The rest, such as $\ket{\uparrow\downarrow}_a\bra{\uparrow}_a$ and $\ket{\varnothing}_a\bra{\downarrow}_a$, 
are fast degrees of freedom.

Among these slow degrees of freedom, we are mainly interested in the states where the localized electron is singly occupied,
i.e., $\hat r_{n_r}^{(0)}=
\prod_a
\ket{\sigma_a}_a\bra{\sigma_a'}_a \otimes \ket{F}\bra{F}$ ($\sigma_a,\sigma_a'=\uparrow,\downarrow$).
In this case, note that the c-d mixing ${\mathcal L}_1$ transfers the state into a state where
(a) the localized electron is double-occupied and a hole is excited in the conduction band [the process illustrated in Fig.~\ref{fig: second-order perturbation}]
or
(b) the localized electron is vacant and a particle is excited in the conduction band.
Since these processes excite the system to a fast mode, the first-order contribution (the second term in the rightmost side of Eq.~\eqref{eq: effective Lindbladian matrix element}) is absent and the leading term is second-order.
The second-order contribution (the third term in Eq.~\eqref{eq: effective Lindbladian matrix element}) arises from the processes where the intermediate state involves 
states with eigenvalues $\lambda_{{\rm (a)}\pm}^{(0)}=
\pm i (\varepsilon_{\bm k}-\varepsilon_{d,a} - U_a) - \kappa_a/2$ from the process (a)
and 
$\lambda_{{\rm (b)}\pm}^{(0)}=
\pm i (\varepsilon_{d,a}-\varepsilon_{\bm k})$ from the process (b).
The real part of the process (a) ${\rm Re}\lambda_{{\rm (a)}\pm}=-\kappa_a/2$ reflects the light-induced decay that turns on in the double-occupied state.
Assuming further that only excitation near the Fermi surface contributes $\varepsilon_{\bm k}\approx\varepsilon_{\rm F}$,
this yields, as detailed in SI Sec. III,
\begin{eqnarray}
    \label{eq: Leff singly occupied}
    {\mathcal L}_{\rm eff}^{\rm sd}
    (\hat P_{\rm s}^a
     \hat \rho
     \hat P_{\rm s}^a)
    &=& -i[\hat H_{\rm sd},
    \hat\rho 
    ]
    \nonumber\\
    &+&\sum_a 
    \gamma_a 
    {\mathcal D}
    \big[
    \sum_\sigma
    \hat d_{\sigma,a}^\dagger
    \hat c_{\bm R_a,\sigma}
    \hat P_{\rm s}^a
    \big]  
    \hat\rho.
\end{eqnarray}
Here, the \textit{sd} coupling 
\begin{eqnarray}
    g_a = -
    |v_a|^2
    \bigg[
        \frac{\varepsilon_{d,a}+U_a-\varepsilon_{\rm F}}
        {(\varepsilon_{d,a}+U_a-\varepsilon_{\rm F})^2+\frac{\kappa_a^2}{4}}
        +
        \frac{1}{\varepsilon_{\rm F}-\varepsilon_{d,a}}
    \bigg]
\end{eqnarray}
in the \textit{sd} Hamiltonian $\hat H_{\rm sd}
=-(1/2)\sum_a g_a 
\hat P_{\rm s}^a
[\hat {\bm \tau}(\bm R_a)
\cdot \hat {\bm S}_a ]
\hat P_{\rm s}^a$
and the correlated dissipation
\begin{eqnarray}
    \gamma_a = \frac{|v_a|^2 \kappa_a}{(\varepsilon_{d,a}+U_a -\varepsilon_{\rm F})^2 +\frac{\kappa_a^2}{4}},
\end{eqnarray}
are given by the imaginary and real part, 
respectively, of $|v_a|^2[\lambda_{{\rm (a)}+}^{-1}+\lambda_{{\rm (b)}+}^{-1}]$ that arise from the two processes (a) and (b).
The expression valid at regimes $\kappa_a\ll \varepsilon_{\rm F},\varepsilon_{d, a}, U_a$ is reported in the main text.

The correlated dissipation (the second term in Eq.~\eqref{eq: Leff singly occupied}) adds an electron to the localized orbital such that the state transfers to a double-occupied state.
This is quickly returned to a singly-occupied state via the light-induced decay with rate $\kappa_a$, which can readily be seen from the effective Lindbladian applied to $\hat r_{n_r}^{(0)}
=\ket{\uparrow\downarrow}_a\bra{\uparrow\downarrow}_a\otimes\ket{F}\bra{F}$ as,
\begin{eqnarray}
    \label{eq: Leff double occupied}
    {\mathcal L}_{\rm eff}^{\rm sd}
    (
    \hat P_{\uparrow\downarrow}^a
    \hat\rho
    \hat P_{\uparrow\downarrow}^a
    )
    =\sum_{a,\sigma} 
    \kappa_a
    {\mathcal D}[
    \hat d_{\sigma,a}
    \hat P_{\uparrow\downarrow}^a
    ]\hat\rho,
\end{eqnarray}
where we have ignored the contribution to the coherent dynamics since we are not interested in the details of the double-occupied state.
When applied to a vacant state $\hat r_{n_r}^{(0)}=\prod_a \ket{\varnothing}_a\bra{\varnothing}_a\otimes\ket{F}\bra{F}$, we find ${\mathcal L}_{\rm eff}^{\rm sd}(\hat P_\varnothing^a\hat\rho\hat P_\varnothing^a)=0$ ($\hat P_\varnothing^a$ is a projection operator to a vacant state), where again, we have ignored the contribution to the coherent dynamics.
Summing up these results 
gives the desired Eq.~\eqref{eq: sd master equation} in the main text.

\subsubsection{Derivation of Eq.~\eqref{eq: LLG equation}}

We next integrate out the conduction electron degrees of freedom to derive the RKKY interactions between the localized spins modified by light.
As emphasized in the main text, it is crucial to consider the non-adiabatic (non-Markovian) effect arising from the Fermi distribution function of the conduction electrons.
A useful approach to take such effect into account is to analyze a generating function called the Keldysh partition function, defined as~\cite{Kamenev2023, Sieberer2016, Sieberer2023} (See SI Sec. I for a brief review.),
\begin{eqnarray}
    Z&\equiv&{\rm tr}[\hat\rho(t_f)]={\rm tr}\big[e^{{\mathcal L}_{\rm eff}^{\rm sd}
    (t_f-t_0)}
    \hat\rho(t_0)\big],
\end{eqnarray}
for the master equation~\eqref{eq: sd master equation}. 
We expand the time evolution operator $e^{{\mathcal L}_{\rm eff}(t_f-t_0)}$ in terms of fermionic coherent states into a product of infinitesimally short time interval, similarly to the path integral formalism in quantum mechanics.
Unlike in quantum mechanics (that deals with wave functions $\ket\psi$) that involves one Grassmann field $\psi(t)$ per degree of freedom, however, as we are dealing with the dynamics of the density matrix $\hat\rho$ that lives in the double Hilbert space, each degree of freedom is assigned with two fields $\psi_+(t)$ and $\psi_-(t)$ that loosely describes the time evolution of the ket and bra space, respectively.
For our system (Eq.~\eqref{eq: sd master equation}), the Keldysh partition function is given by,
\begin{eqnarray}
    Z
    &=&\int {\mathcal D}
    (d_+,\bar d_+,d_-,\bar d_-)
    {\mathcal D}(c_+, \bar c_+, c_-, \bar c_-)
    e^{i S}
\end{eqnarray}
where
$S[d_+,\bar d_+,d_-,\bar d_-,c_+, \bar c_+, c_-, \bar c_-]=S_d^0[d,\bar d]+S_c^0[c,\bar c]+S_{\rm sd}^{\rm coh}[c,\bar c,d,\bar d]+S_{\rm sd}^{\rm dis}[c,\bar c,d,\bar d]$ is the so-called Keldysh action, given by
\begin{eqnarray}
    &&S_d^{0}[d,\bar d]
    =
    \int dt 
    \sum_{s=\pm}
    \sum_{a,\sigma}
    s
    \bar d_{\sigma,a}^s(t)
    i\partial_t 
    d_{\sigma,a}^s(t)
    \\
    &&S_c^0[c,\bar c]
    =
    \int dt
    \sum_{s=\pm}\sum_{\bm k,\sigma}
    s
    \nonumber\\
    && \ \ \ \ \ \ \ \ \ \
    \times
    \Big[
    \bar c_{\bm k,\sigma}^s (t)
    i\partial_t 
    c_{\bm k,\sigma}^s(t)
    -\varepsilon_{\bm k}
    \bar c_{\bm k,\sigma}^s (t)
    c_{\bm k,\sigma}^s (t)
    \Big], \\
    &&S_{\rm sd}^{\rm coh}[c,\bar c,d,\bar d]
    = - \int dt
    \sum_{s=\pm} s 
    \sum_a 
    \sum_{\bm k,\bm q}
    (-g_a) e^{i\bm q \cdot\bm R_a}
    \nonumber\\
    && \ \ \ \ \ \ \ \ \ \ \ \ \
    \times
    \sum_{\sigma,\sigma'}
    \bar d_{\sigma,a}^s(t)
    \bar c_{\bm k+\bm q,\sigma'}^s
    (t)
    c_{\bm k,\sigma}^s(t)
    d_{\sigma',a}^s(t),
    \\
    &&S_{\rm sd}^{\rm dis}
    [c,\bar c,d,\bar d]
    =
    -i
    \int dt 
    \sum_a  
    \sum_{\bm k,\bm q}
    \gamma_a
    e^{i\bm q\cdot\bm R_a}
    \nonumber\\
    &&\times 
    \sum_{\sigma,\sigma'}
    \Big[
    \bar c_{\bm k+\bm q,\sigma'}^-(t)
    d_{\sigma',a}^-
    (t)
    \bar d_{\sigma,a}^+(t)
    c_{\bm k,\sigma}^+(t)
    \nonumber\\
    &&
    -\frac{1}{2}
    \bar c_{\bm k+\bm q,\sigma'}^+(t_{+\delta})
    d_{\sigma',a}^+(t_{+\delta})
    \bar d_{\sigma,a}^+(t_{-\delta})
    c_{\bm k,\sigma}^+(t_{-\delta})
    \nonumber\\
    &&
    -\frac{1}{2}
    \bar c_{\bm k+\bm q,\sigma'}^-(t_{-\delta})
    d_{\sigma',a}^-(t_{-\delta})
    \bar d_{\sigma,a}^-(t_{+\delta})
    c_{\bm k,\sigma}^-(t_{+\delta})
    \Big],
\end{eqnarray}
Here, $c_{\bm k,\sigma}^\pm$ and $d_{\sigma,a}^\pm$ are the Grassmann variables of conduction and localized electrons, respectively, and $t_{\pm \delta}=t\pm 0^+$.

Since the Keldysh action $S$ is quadratic in terms of $(c,\bar c)$, 
one can analytically integrate out the conduction electron degrees of freedom to obtain the effective action $S_{\rm eff}[d,\bar d]$ defined as 
$Z
\equiv \int{\mathcal D}(d,\bar d)e^{iS_{\rm eff}[d,\bar d]}$.
As detailed in SI Sec IV, the effective action
within the second-order perturbation in terms of $g_a$ and $\gamma_a$ (with several additional assumptions detailed in SI Sec. IV)
reads 
$S_{\rm eff}[d,\bar d]=S_d^0[d,\bar d] + S_{\gamma}[d,\bar d]
+ S_M[d,\bar d]$, where 
\begin{eqnarray}            
    &&S_\gamma[d,\bar d] 
    =i \int dt
    \sum_{a,\sigma}
    \gamma_{a} n
    \big[  
    d_{\sigma,a}^-
    (t)
    \bar d_{\sigma,a}^+ 
    (t)
    \nonumber\\
    &&-\frac{1}{2} 
    d_{\sigma,a}^+
    (t_{+\delta})
    \bar d_{\sigma,a}^+    
    (t_{-\delta})   
    -
    \frac{1}{2}
    d_{\sigma,a}^-
    (t_{-\delta})   
    \bar d_{\sigma,a}^-        
    (t_{+\delta})   
    \big],
\end{eqnarray}
is the first-order contribution  and $S_M[d,\bar d] =S^{\rm coh}_{\rm RKKY}[d,\bar d]
+S_{\rm Gilbert}[d,\bar d]
+S^{\rm neq}_{\rm RKKY}[d,\bar d]$ is the second-order contribution, with
\begin{eqnarray}
    S^{\rm coh}_{\rm RKKY}[d,\bar d]
    &=&
    \int dt \sum_{a,b}
    \frac{J_{a,b}(\bm R_{a,b})}{2}
    \sum_{j=0}^3
    \sum_{s=\pm}s
    \hat m_{a,j}^{s,s}
    \hat m_{b,j}^{s,s},
    \nonumber\\
    \\
    S_{\rm Gilbert}[d,\bar d]
    &=&
    - 
    \sum_a
    \frac{\alpha_a}{4}
    \int dt    
    \sum_{j=0}^3
    \hat m_{a,j}^{+,+}(t)
    \partial_t     
    \hat m_{a,j}^{-,-}(t),
    \nonumber\\
    \\
    S^{\rm neq}_{\rm RKKY}[d,\bar d]
    &=&
    i
    \int dt     
    \sum_{a,b}
    \frac{\Omega_{a,b}(\bm R_{a,b})}{2}
    \nonumber\\
    &\times &
    \sum_{j=0}^3
    \Big[
    \hat m_{a,j}^{+,+} 
    \hat m_{b,j}^{+,+}
    +    \hat m_{a,j}^{-,-} \hat m_{b,j}^{-,-}
    \nonumber\\
    && \ \ \ \ \ \ \  
    -    \hat m_{a,j}^{+,-} \hat m_{b,j}^{+,+}
    -
    \hat m_{a,j}^{-,-} \hat m_{b,j}^{+,-}
   \Big].
\end{eqnarray}
Here, 
$\hat m_{a,j}^{l_1,l_2}[d,\bar d]
=\sum_{\mu,\nu=\uparrow,\downarrow}
\bar d_{\mu,a}^{l_1}
\hat \sigma_j^{\mu\nu}
d_{\nu, a}^{l_2}~(l_1,l_2=+,-)$ is a localized spin written in terms of Grassmann variables, 
and
\begin{eqnarray}
    J_{a,b}(\bm R_{a,b})
    &=&
    -\frac{|g_a| |g_b|}{2} 
    \sum_{\bm k,\bm q}\cos(\bm q\cdot \bm R_{a,b})
    \frac{f_+ -f_-}{\varepsilon_+ -\varepsilon_-},
    \\
    \Omega_{a,b}(\bm R_{a,b})
    &=&
    -\frac{\gamma_a |g_b|}{2} 
    \sum_{\bm k,\bm q}\cos(\bm q\cdot \bm R_{a,b})
    \frac{f_+ -f_-}{\varepsilon_+ -\varepsilon_-},
    \\
    \alpha_a
    &=&    
    -4\pi g_a^2
    \sum_{\bm k,\bm q}
    \frac{f_+ -f_-}{\varepsilon_+-\varepsilon_-}
    \delta(\varepsilon_+-\varepsilon_-),
\end{eqnarray} 
with $\varepsilon_\pm=\varepsilon_{\bm k\pm\bm q/2}$ and $f_\pm=f(\varepsilon_{\bm k\pm\bm q/2})$.
$J_{a,b}(\bm R_{a,b})$ is identical to the well-known form of the RKKY interaction strength~\cite{Yosida1996}.
In calculating $S_M$, we have employed a gradient approximation, i.e. a Markovian approximation ($S^{\rm coh}_{\rm RKKY}$ and $S^{\rm neq}_{\rm RKKY}$) plus the first-order correction to it ($S_{\rm Gilbert}$).

The physical meaning of each term becomes clear by deriving the equation of motion of the spins. 
To do this, we introduce a set of auxiliary fields $m$ and Lagrange multipliers $\lambda$ as
\begin{eqnarray}
\label{eq: effective action Lagrange multiplier}
    Z&=&    \int {\mathcal D}[m]
    e^{i S_M[m]}
    \nonumber\\
    &\times&
    \int {\mathcal D}[\lambda]
    \int {\mathcal D}[d,\bar d] 
    e^{i S_d^0[d,\bar d] + i S_\gamma[d,\bar d] }
    e^{iS_\lambda[\lambda,m,\hat m[d,\bar d]]}   
    \nonumber\\
    &\equiv&    
    \int {\mathcal D}[m]
    \int {\mathcal D}[\lambda]
    e^{i S_M[m]}
    e^{iS_B^\lambda[\lambda,m]}
\end{eqnarray}
with
\begin{eqnarray}
    &&S_\lambda[\lambda,m,\hat m[d,\bar d]]
    =
    \int dt    
    \sum_a
    \sum_{l_1,l_2=q,c}
    \sum_{j=0}^3
    \lambda_{a,j}^{l_1,l_2}(t)
    \nonumber\\
    &&
    \ \ \ \ \ \ \ \ \ \ \ \ \ \ \ \ \ \ 
    \times
    \big[
    m_{a,j}^{l_1,l_2}(t)
    -\hat m_{a,j}^{l_1,l_2}[d(t),\bar d(t)]
    \big].
\end{eqnarray}
As detailed in SI Sec. IV.C, taking the saddle-point approximation of Eq.~\eqref{eq: effective action Lagrange multiplier} as $\frac{\delta{S_M^{\rm eff}[\lambda,m]}}{\delta{\lambda}_{a,j}^{l_1,l_2}}=\frac{\delta{S_M^{\rm eff}[\lambda,m]}}{\delta m_{a,j}^{l_1,l_2}}=0$ 
(where $S_M^{\rm eff}[\lambda,m]=S_M[m]+S_B^\lambda[\lambda,m]$)
gives 
\begin{eqnarray}
    &&\partial_t
    \bm S_a
    =
    -\gamma_a n
    \bm S_a
    -\sum_{b
    (\ne a)
    }
    \Omega_{a,b}(\bm R_{a,b})
    \bm S_b(t)
    \nonumber\\
    &&-
    \Big[
    \sum_b 
    J_{a,b}(\bm R_{a,b})
    \bm S_{b}(t)
    -
    \alpha_a
    \bm {\dot S}_a(t)
    \Big]
    \times 
    \bm S_{a}(t),
\end{eqnarray}
where $S_{a,j}
=m_{a,j}^{++}(t)
=m_{a,j}^{--}(t)
=\langle
\sum_{\sigma,\sigma'=\uparrow,\downarrow}
\hat d_{\sigma,a}^\dagger(t) \sigma_j^{\sigma,\sigma'} \hat d_{a,\sigma'}(t)\rangle$
and $m_{a,j}^{+-}(t)
=m_{a,j}^{-+}(t)=0$,
which is the desired Eq.~\eqref{eq: LLG equation} in the main text.
The first, second, third, and fourth terms on the right-hand side arise from $S_\gamma, S_{\rm RKKY}^{\rm neq}, S_{\rm RKKY}^{\rm coh}$, and $S_{\rm Gilbert}$, respectively.

\subsection{Estimation of the required power}

Below, we estimate the required laser power $P$ to realize the sign-inversion of the interactions, which occurs when the decay rate of the double-occupied state $\kappa_a$ exceeds $\alpha_a U_a$ (see main text and Fig.~\ref{fig: energy scale}).
Our scheme considers the situation where the double-occupied (at energy $\varepsilon_{d, a}+U_a$) and higher-level states (at energy $\varepsilon_{f, a}$) are coupled through the injected laser. 
We assume that the higher-level state is localized and dissipates with the rate $\Gamma_{f ,a}$, so one can model it with a Lorentz oscillator model~\cite{Dressel2002}.

\begin{figure*}[t]
\centering
\includegraphics[width=0.8\linewidth,keepaspectratio]{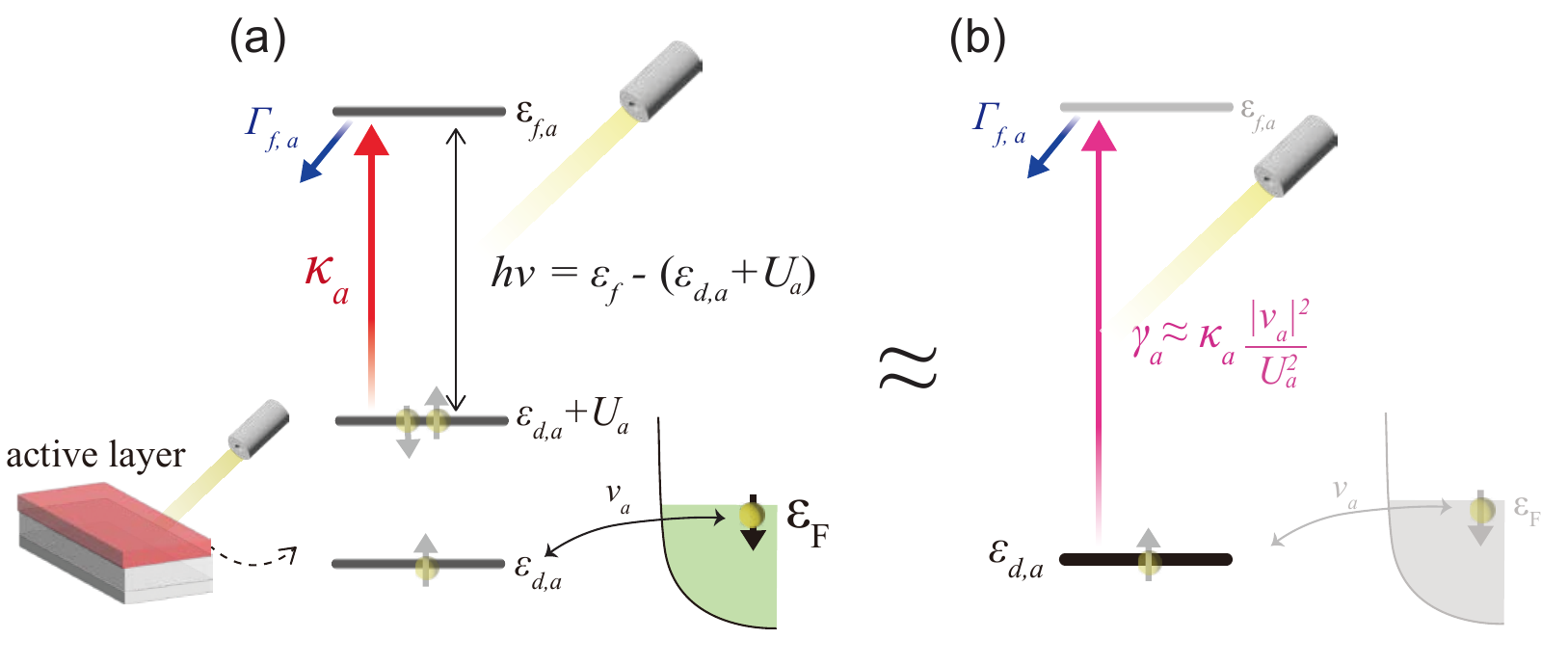}
\caption{
\textbf{Light-injection induced dissipation and their energy scales.}
(a) The double-occupied (at the energy $\varepsilon_{d, a}+U_a$) and the higher energy states (at the energy $\varepsilon_{f, a}$) are coupled by the injection of a resonant laser $h\nu=\varepsilon_{f, a} - (\varepsilon_{d,a} + U_a)$.
The higher-energy state dissipates with the rate $\Gamma_{f, a}$.
The localized electrons are typically in a single-occupied state but may virtually excite once in a while to a double-occupied state via the c-d mixing $v_a$.
Note that no electrons decay in the absence of c-d mixing $v_a=0$ because they are always in the single-occupied state.
(b) Localized spin picture obtained after projecting out the double-occupied states (see Eq.~\eqref{eq: sd dynamics}).
In this picture, one finds that the effective transfer rate from the localized electron to the higher-energy state is given by $\gamma_a \approx \kappa_a |v_a|^2/U_a^2$.
We require $\gamma_a \ll \Gamma_{f, a}$ to justify the Markov approximation.
}
\label{fig: dissipation engineering scheme methods}
\end{figure*}

When a laser with the pump power $P$ is injected into the material, the dissipation causes the energy loss of the laser intensity
if the system is in a double-occupied state. 
The lost energy density per unit time and volume $W$ is given by
\begin{eqnarray}
W  = \frac{1}{2} \epsilon_0 \omega \chi''(\omega) |E|^2 = \frac{ \omega \chi'' (\omega) }{c} P, 
\end{eqnarray}
where $\omega=2\pi\nu$ is the laser frequency.
Here, we have expressed the pump power $P=\frac{1}{2} c \epsilon_0 |E|^2$ in terms of speed of light $c$, vacuum dielectric constant $\epsilon_0$, and electric field $E$.
The absorption susceptibility
$\chi'' (\omega)$ is computed according to the Lorentz oscillator model as
\begin{eqnarray}
\chi'' (\omega)= \frac{n e^2}{\epsilon_0 m_0}  \frac{\omega \Gamma_{f,a}}{(\omega^2_0 -\omega^2)^2+ \omega^2\Gamma^2_{f,a}} 
\simeq \frac{n e^2}{\epsilon_0 m_0}  \frac{1}{\omega_0 \Gamma_{f,a} }.
\end{eqnarray}
Here, 
$\omega_0$ is the resonant frequency (which, in our case, corresponds to $\hbar \omega_0 = \varepsilon_{f, a} -(\varepsilon_{d, a} + U_a)$),
$n$ is the number of 
electrons per unit volume, and $m_0$ is the electron mass.
In the second equality, we have set the laser frequency to be on resonance $h\nu = \hbar\omega = \hbar\omega_0$.

The decay rate of the double-occupied state $\kappa_a$ per electron is estimated as
\begin{eqnarray}
    \kappa_a = \frac{W}{n\cdot \omega_0}.
\end{eqnarray}
This needs to be larger than $\alpha_a U_a$ to achieve the regime for showing laser-induced switching of interactions.
This condition is given by,
\begin{eqnarray}
    \kappa_a =\frac{ \chi'' (\omega_0) P }{n \cdot c } \gesim \alpha_a U_a. \end{eqnarray}
This yields the condition,
\begin{eqnarray}
P \gesim \frac{ \alpha_a U_a  n \cdot c }{ \chi'' (\omega_0) } 
= \alpha_a  \frac{U_a \omega_0 m_0 c \epsilon_0 } {e^2}\Gamma_{f,a} ,
\end{eqnarray}
shown in Eq.~\eqref{eq:laser power}.

\subsection{Justification of Markov approximation}

For the Markov approximation to be valid, the relaxation rate of the bath must be much faster than the timescale of the system dynamics. 
We argue here that this is likely to be justified in the range of interest at realistic parameters for magnetic metals.

In our setup, the double-occupied state couples to the higher-level state with the decay rate $\Gamma_{f,a}$ via a laser injection tuned to be resonant with the two states.
The higher-level state can be regarded as our ``external bath'' in the context of open quantum systems. 
As explained above, this process gives rise to the decay rate $\kappa_a$ once the site $a$ is double occupied (see Fig.~\ref{fig: dissipation engineering scheme methods}(a)).
Note crucially that, as illustrated in Fig.~\ref{fig: dissipation engineering scheme methods}(b), this is different from \textit{the rate at which the relevant system pumps an electron to this high-energy state} because the relevant system transfers into a double-occupied state only once in a while when the conduction electron tunnels to the localized orbital.
For example, when the c-d mixing is absent $v_a=0$, there would be no electron transfer from the relevant system to the higher-energy state.
The relevant transfer rate of a localized electron from the relevant system to the higher-energy state
is estimated to be $\gamma_a \simeq \kappa_a |v_a|^2/U_a^2$, which we have derived in Eq.~\eqref{eq: sd master equation}.

For the Markov approximation to be valid, the dissipation rate of this higher energy state (i.e. the ``external bath'') $\Gamma_{f,a}$ must be much faster than the supply rate to this state $\gamma_a \ll \Gamma_{f, a}$. 
This is because when the higher-level state is occupied,
the Pauli blocking effect would suppress the decay.
A slow relaxation of the occupancy of the state would lead to a non-Markovian effect.

It should be relatively easy to satisfy this Markov condition at the regime of interest.
We are interested in the regime where we see the sign-reversal of the RKKY interactions, which happens when the dissipation rate exceeds $\kappa_a\gesim \alpha_a U_a$ or 
$\gamma_a\gesim \alpha_a |g_a|$
(see the discussion above Eq.~\eqref{eq:laser power}).
In the case $\alpha_a = 10^{-2}$ and $|g_a|=10{\rm meV}$,
this sets the condition, 
$\gamma_a \gesim 0.1 {\rm meV}$.
This required dissipation rate is less than the typical value of linewidth $\Gamma_{f, a}$, satisfying the justification condition for the Markov approximation, $\gamma_a\ll \Gamma_{f, a}$.

Putting the conditions together, 
\begin{eqnarray}
    \alpha_a U_a
    \frac{|v_a|^2}{U_a^2}
    \lesssim 
    \frac{P e^2}{\epsilon_0 m_0 c}  
    \frac{1}{\omega_0 \Gamma_{f,a} }
    \frac{|v_a|^2}{U_a^2}
    =
    \gamma_a 
\ll \Gamma_{f, a}.
\end{eqnarray}
This shows that the required pump power $P$ to achieve sign inversion becomes less by choosing the higher-energy state that has a longer lifetime (i.e. smaller $\Gamma_{f, a}$) as shown in Eq.~\eqref{eq:laser power} but small $\Gamma_{f, a}$ makes it more difficult to satisfy the Markovian condition 
$\gamma_a 
\ll \Gamma_{f, a}$.
We remark that a smaller c-d mixing $|v_a|/U_a$ helps satisfy the Markovian condition $\gamma_a\ll \Gamma_{f, a}$ without modifying the condition for the sign inversion, which physically makes sense because small $v_a$ makes the double-occupied state rarer and hence rarer for the electrons to escape from the relevant system.

\subsection{Comparison to cold atom experiments}
It is interesting to compare our proposal relevant to solid-state systems to the recent cold atom experiment of dissipative Fermi Hubbard model \cite{Nakagawa2020, Honda2023}, where they demonstrated a dynamic sign reversal of interactions. 
In their experiment, similar to our situation, 
they introduced a controlled decay channel that is activated only when the sites are double occupied, causing both atoms to decay whenever they are on the same site. 
They demonstrated that this engineered dissipation decreases the anti-ferromagnetic correlation (present in the ground state) and increases the ferromagnetic correlation, again analogous to our sign reversal of effective interactions.
We briefly note, however, that the possibility of implementing non-reciprocal interactions was not explored in their work.

In contrast to our solid-state case, where the coupling to the environment is unavoidable,
their cold atomic systems are almost perfectly isolated from the environment other than the decay to the vacuum they purposely introduced.
This fundamental difference critically impacts the resulting dynamics.
First, in our proposal, the surrounding environment (i.e. the conduction band) immediately compensates for the lost electrons such that localized orbitals are always singly occupied, while their atomic system only has a loss channel. 
As a result, their dissipation stops activating when the atoms stop colliding, resulting in a strong initial state dependence on the final configuration.
This is in stark contrast to ours, where no initial-state dependence is present and even exhibits persistent time-dependent to a collective chase and runaway phase due to non-reciprocal interaction.
Second, the Gilbert damping (present in our system) is absent in the cold atomic systems, as there is no environment where the atoms can dissipate their spin angular momentum.
The competition between the friction (that drives the system toward equilibrium) and the light-induced dissipative interaction is hence a unique feature of our proposal for solid-state systems.

\subsection{Data availability}
The data generated in this study have been deposited in the Figshare database under the accession code 10.6084/m9.figshare.26334430.

\subsection{Code availability}
The code that generated the results in the main text and Supplementary Information is available.

\bibliography{main}

\begin{center}{\bf Acknowledgement}\end{center} 

We thank Aashish Clerk, 
Romain Daviet, 
Sebastian Diehl, 
Takayuki Kurihara, 
James Harden,
Peter B. Littlewood,
Sriram Ramaswamy, 
Shuntaro Tani,
Kento Uchida,
and 
Carl Philipp Zelle
for discussions. 
RH was supported by a Grant in Aid for
Transformative Research Areas
(No. 25H01364),
for Scientific Research (B) (General)
(No. 25K00935),
and for Research Activity Start-up from JSPS in Japan (No. 23K19034) and the National Research Foundation (NRF) funded by the Ministry of Science of Korea (Grant No. RS-2023-00249900).
DO was supported by a Grant-in-Aid
for Scientific Research (C)
(No. 25K07184)
and for Early-Career Scientists from JSPS in Japan (No. 21K13882).
RT was supported by a Grant-in-Aid for
Transformative Research Areas
(No. 25H01248), 
for Early-Career Scientists (No. 22K14003) and for Research Activity Start-up (No. 20K22328) from JSPS in Japan.

\subsection{Author contributions}
RH conceived and designed this study, performed analytical calculations, and wrote the draft.
RH and RT estimated the required power needed for a non-reciprocal phase transition to arise. 
RH, DO, and RT contributed to designing an experimentally feasible setup, discussed the results, and commented on the manuscript.

\subsection{Competing interests}

The authors declare no competing interests.

\clearpage
\newpage

\setcounter{section}{0}

\onecolumngrid
\renewenvironment{widetext}{}{}

\begin{center}{\bf SUPPLEMENTARY INFORMATION}\end{center} 
\setcounter{equation}{0}
\setcounter{figure}{0}
\renewcommand{\thefigure}{S\arabic{figure}}
\renewcommand{\theequation}{S\arabic{equation}}

This Supplemental Information (SI) provides full details of our formalism.

To make our manuscript self-contained, 
in Sec.~\ref{SIsec: review of open quantum system}, we briefly review the formalism of open quantum systems, focusing on those used in this work. 
We provide theoretical methods to treat Markovian (Sec.~\ref{SIsubsec: Markov}) and non-Markovian (Sec.~\ref{SIsubsec: Keldysh}) baths.

In Sec.~\ref{SIsec: model}, we introduce our model.

In Sec.~\ref{SIsec: sd}, we derive 
Eq.~(4) in the main text, which describes the effective localized spin dynamics immersed in conduction electrons with light-induced decay. In the derivation, we use the results from the projection method introduced in Sec.~\ref{SIsubsec: projection}.

In Sec.~\ref{SIsec: LLG}, we derive the Landau-Lifshitz-Gilbert equation with light-induced interactions (Eq.~(6) in the main text), which is our main result of this work. We use the Keldysh formalism introduced in Sec.~\ref{SIsubsec: Keldysh} in the derivation.

In Sec.~\ref{SIsec: microscopic derivation of GKSL equation}, we give a microscopic derivation of the starting point of our analysis, the GKSL equation (Eq. (3) in the main text), from a model that explicitly considers the laser drive and the higher-energy level that induces the controlled dissipation.

In Sec. \ref{SIsec: chiral regime}, we give details of the physics for the case where we applied our method to layered ferromagnets. We provide the conditions for the emergence of the chiral phase and additional data that provide the time dependence of the magnetization in the full time range. 

\section{
Mini-review:
Theoretical description of open quantum systems}
\label{SIsec: review of open quantum system}

This section briefly reviews the theoretical description of open quantum systems. 
Open quantum systems are systems where the relevant system is coupled to baths that model the effect of the surrounding environment.
When the relevant system is (weakly) attached to a single thermal bath, the system reaches a thermal equilibrium with the bath.
In contrast, when the relevant system is attached to two (or more) baths, an energy current may constantly flow from one bath to the system that dissipates to the other baths, driving the state into a nonequilibrium steady state.
The latter is the situation we will be interested in in this work.

The full microscopic information of the total system composed of the relevant system plus the baths, described by the total Hamiltonian $\hat H_{\rm tot}=\hat H_s + \hat H_b + \hat H_{sb}$, is encoded in the density operator of the full system $\hat\rho_{\rm tot}$.
Here, $\hat H_{s(b)}$ is the relevant system (bath) Hamiltonian that is composed of operators acting on the relevant system (bath), and $\hat H_{sb}$ is the system-bath coupling.
The bath Hamiltonian $\hat H_b$ may contain multiple baths, in which case drives the system into a nonequilibrium steady state.
The dynamics are
governed by Liouville's equation,
\begin{eqnarray}
\label{SIeq: Liouville's equation}
    \partial_t\hat\rho_{\rm tot} = -i [\hat H_{\rm tot},\hat\rho_{\rm tot}].
\end{eqnarray}

We will be interested in the properties of 
the reduced-density matrix 
$\hat\rho={\rm Tr}_{\rm B}\hat\rho_{\rm tot}$, 
where the baths' degrees of freedom are traced out.
When the relevant system is weakly attached to a \textit{single} thermal bath, the system generically reaches an equilibrium state, which is fully characterized by eigenenergy $E_n$ and eigenstates $\ket{\phi_n}$ of the system Hamiltonian $\hat H_s$ ($\hat H_s\ket{\phi_n}=E_n\ket{\phi_n}$), thanks to the equilibrium statistical mechanics that tells us that the reduced density operator $\hat\rho$ reaches the canonical ensemble (when attached to a thermal bath at the temperature $T$), 
\begin{eqnarray}
    \hat\rho_{\rm eq}=\frac{1}{Z}e^{-\hat H_s/(k_{\rm B}T)}
    =\frac{1}{Z}
    \sum_n e^{-E_n/(k_{\rm B}T)}
    \ket{\phi_n}\bra{\phi_n}. 
\end{eqnarray}
Here, $Z={\rm tr}[e^{-\hat H_s/(k_{\rm B}T)}]$ is the partition function, where ${\rm tr}[\cdots]$ is the trace over the relevant system degrees of freedom, and $k_{\rm B}$ is the Boltzmann constant.
Therefore, 
once we compute the eigenenergy and the eigenstates, the density operator $\hat\rho$ is fully determined
and it is not necessary to compute Liouville's equation~\eqref{SIeq: Liouville's equation} directly.

However, for systems attached to \textit{multiple} baths, the system generically converges to non-equilibrium systems, where such a general relation is unknown. 
Therefore, one is required to directly analyze the dynamics of the density operator $\hat\rho={\rm Tr}_{\rm B}\hat\rho_{\rm tot}$. This is indeed what we will do in the following.

Roughly speaking, there are two types of baths, i.e. Markovian and non-Markovian baths.
Markovian baths are baths where they relax fast enough such that the evolution of the relevant system is determined solely by the state of at the instantaneous time; memory effects are absent.
Non-Markovian baths are the baths where such memory effects cannot be ignored. 
As we explain in Secs. \ref{SIsec: sd} and \ref{SIsec: LLG}, our dissipation-engineering scheme requires treating both types of baths and therefore, we give a short review of their theoretical description of both cases.

This section is organized as follows.  
In Sec. \ref{SIsubsec: Markov}, we review the theoretical treatment of the quantum master equation relevant to systems attached to Markovian baths.
After introducing the Gorini-Kossakowski-Sudarshan-Lindblad (GKSL) master equation~\cite{Gorini1976, Lindblad1976}, which is the most general description of Markov dynamics of the reduced density operator $\hat\rho$,
we describe the general properties of the GKSL master equation in Sec.~\ref{SIsubsec: Markov}. 
We then introduce a projection method we use to derivate Eq.~(4) in the main text.
In Sec. \ref{SIsubsec: Keldysh}, we introduce the so-called Keldysh formalism, which allows us to treat non-Markovian open systems on a field-theoretical basis.
This will be used to derive Eq.~(6) in the main text.

\subsection{Markov environment: the GKSL master equation}
\label{SIsubsec: Markov}

In this section, we will be concerned with situations where the system is weakly attached to a bath that relaxes fast enough such that the relevant system dynamics can be regarded as Markovian. 
One strategy to treat this system is to start from Liouville's equation~\eqref{SIeq: Liouville's equation} and trace out the bath degrees of freedom.   
One can microscopically derive the governing equation for reduced density matrix $\hat\rho(t)$ dynamics,  e.g., by employing a second-order Born-Markov approximation in terms of system-bath coupling strength~(see e.g., \cite{Breuer2007}).
Here, we take a more phenomenological approach, where we take advantage of the property that
in Markovian dynamics, by definition,
the density matrix $\hat\rho(t+\Delta t)$ is determined solely by transforming 
the density matrix $\hat\rho(t)$ [i.e., $\hat\rho(t+\Delta t)$ does not depend on the density matrix before the time $t$] and that
the density operator must fulfill certain mathematical properties \cite{Breuer2007, Gardiner2000}.

Let us denote such a transformation as $\hat\rho(t+\Delta t)={\mathcal V}\hat\rho(t)$, where  ${\mathcal V}$ operates on the density matrix $\hat\rho$. [Operators that act on operators are called \textit{superoperators}.]
There are several properties that this operation must satisfy, arising from demanding that the reduced operator $\hat\rho(t+\Delta t)$ retains all the properties density operators must satisfy. 
The first property is the trace-preservation, i.e., 
${\rm tr}[\hat\rho(t+\Delta t)]
={\rm tr}[{\mathcal V}\hat\rho(t)]
={\rm tr}[\hat\rho(t)](=1)$,
which is required  
because the eigenvalues $p_n$ of the density operator $\hat\rho=\sum_n p_n\ket{n}\bra{n}$ describes the probability of being in the state $\ket{n}$, which must sum to unity $\sum_n p_n = 1$.

The second property is a property called complete positivity. 
Positivity is the property that all the eigenvalues $p_n$ are positive, which is again required because $p_n$ is the probability and must be positive.
The operation ${\mathcal V}$ is required to satisfy a stronger condition called complete positivity, where not only the reduced density operator $\hat\rho$ but the total system $\hat\rho_{\rm tot}$ stays positive after the operation $[{\mathcal V}\otimes{\mathcal I}]\hat\rho_{\rm tot}$, where ${\mathcal V}$ operates on the relevant system and
${\mathcal I}$, the identity superoperator, 
acts on the bath degrees of freedom.
See e.g., \cite{Breuer2007} for details.
The map that satisfies both these properties is called the CPTP map.

Under this condition, it has been shown~\cite{Gorini1976, Lindblad1976} that this operation must have the form (see e.g. Ref.~\cite{Breuer2007} for the proof),
\begin{eqnarray}
    \hat\rho(t+\Delta t)
    ={\mathcal V}\hat\rho(t)
    =\hat\rho(t)
    -i [\hat H, \hat\rho(t)]\Delta t
    +\sum_{\ell}
    \Big[
    \hat L_{\ell}\hat\rho(t)
    \hat L_\ell^\dagger 
    -\frac{1}{2}\{
    \hat L_\ell^\dagger 
    \hat L_\ell,
    \hat\rho(t)\}
    \Big]
    \Delta t
    +
    O((\Delta t)^2)
\end{eqnarray}
or in the $\Delta t\rightarrow 0$ limit,
\begin{eqnarray}
    \label{SIeq: GKSL master equation}
    \partial_t \hat\rho 
    &=& {\mathcal L}\hat\rho
    =-i[\hat H,\hat\rho]
    +\sum_\ell 
    \kappa_\ell
    {\mathcal D}[\hat L_\ell]\hat\rho.
\end{eqnarray}
This is called the GKSL or Lindblad master equation~\cite{Gorini1976, Lindblad1976}. 
Here, the superoperator ${\mathcal L}$ is called the Lindbladian.
$\hat H$ is the Hamiltonian of the system which determines the coherent dynamics. 
The dissipator, 
${\mathcal D}[\hat L]\hat\rho
    =\hat L \hat\rho \hat L^\dagger 
    -\frac{1}{2}
    \{\hat L^\dagger\hat L,\hat\rho\}$,
is the term that makes the time evolution non-unitary.
The so-called jump operator $\hat L_\ell$ consists of operators acting on the system, which
describes how the environment interacts with the system with the dissipation rate $\kappa_\ell$ 
in channel $\ell$.
For instance,
suppose the system consists of bosonic/fermionic particles and
one takes $\hat L_\ell= \hat a$ (where $\hat a$ is an annihilation operator of a particle in the system) and $\kappa_\ell=\kappa$. 
This corresponds to the case where the environment induces one-body loss from the system with the loss rate $\kappa$. 
When one takes $\hat L=\hat a^\dagger$, on the other hand, it describes the situation where the particles are pumped into the system.
Taking $\hat L=\hat a^2$ means two-body loss, etc.


\subsubsection{
General properties of the GKSL master equation}
\label{SIsubsec: vectorization}

Here we review some general properties of the GKSL master equation \eqref{SIeq: GKSL master equation} relevant to our study.
To formulate them, it is useful to express the GKSL master equation in terms of the vectorized form to make explicit connections to the more familiar Schr\"odinger equation.
The vectorization (Choi-Jamiolkowski isomorphism) is chosen to be performed by the replacement \cite{Gardiner2000}, 
\begin{eqnarray}
    \ket{i}\bra{j}
    \rightarrow\ket{j}\otimes\ket{i}.
\end{eqnarray}
For example, an operator $\hat A=\sum_{i,j}a_{i,j}\ket{i}\bra{j}$ is vectorized as
\begin{eqnarray}
    {\rm vec}(\hat\rho)
    =\sum_{i,j}a_{i,j}\ket{j}\otimes\ket{i}.
\end{eqnarray}
Their dual 
$\hat A^\dagger = \sum_{i,j}a_{i,j}^*\ket{j}\bra{i}$
is vectorized as
\begin{eqnarray}
    [{\rm vec}(\hat A)]^\dagger
    =\sum_{i,j}a_{i,j}^*\bra{j}\otimes\bra{i}.
\end{eqnarray}

This choice of vectorization is useful, thanks to the following properties.
First, the Hilbert-Schmidt inner product $(\hat A, \hat B)=
    {\rm{tr}}[{\hat A^\dagger\hat B}]$ can be expressed as
\begin{eqnarray}
    \label{SIeq: Hilbert-Schmidt inner product}
    (\hat A, \hat B)
    &=&
    {\rm{tr}}[{\hat A^\dagger\hat B}]
    =\sum_{i,j,i',j'}
    a_{i,j}^* b_{i',j'}{\rm tr}[(\ket{i}\bra{j})^\dagger
    (\ket{i'}\bra{j'})
    ]
    =\sum_{i,j}
    a_{i,j}^* b_{i,j}
    \nonumber\\
    &=&
    \sum_{i,j,i',j'}
    (a_{j,i}^*\bra{i}\otimes\bra{j})
    (b_{i',j'}
    \ket{j'}\otimes\ket{i'})
    =[{\rm vec}(\hat A)]^\dagger {\rm vec}(\hat B)
\end{eqnarray}
which is just the inner product of the two vectors.
Another useful property is the relation 
\begin{eqnarray}
    {\rm vec}
    (\hat A\hat B\hat C)
    =(\hat C^{\mathrm T}\otimes\hat A){\rm vec}({\hat B}),
\end{eqnarray}
which can be shown similarly by decomposing the matrix in terms of the chosen eigenbasis.
This relation implies (where $\hat I=\sum_i \ket{i}\otimes\bra{i}$ is the identity operator)
\begin{eqnarray}
    {\rm vec}(\hat A\hat\rho\hat B)&=&
    (\hat B^{\mathrm T}\otimes\hat A)
    {\rm vec}(\hat\rho)
    \\
    {\rm vec}(\hat A\hat\rho)&=&
    {\rm vec}(\hat A\hat\rho \hat I)
    =(\hat I
    \otimes\hat A){\rm vec}
    (\hat\rho)
    \\
    {\rm vec}(\hat\rho\hat A)&=&
    {\rm vec}(\hat I\hat\rho\hat A)
    =(\hat A^{\mathrm T}
    \otimes \hat I)
    {\rm vec}(\hat\rho),
\end{eqnarray}
allowing us to express the GKSL master equation \eqref{SIeq: GKSL master equation} in the vectorized form,
\begin{eqnarray}
\label{SIeq: vectorized Lindblad equation}
    \partial_t \rm{vec}(\hat\rho) =\breve{\mathcal L}\rm{vec}(\hat\rho),
\end{eqnarray}
where
\begin{eqnarray}            \breve{\mathcal{L}}
    =-i(\hat I\otimes \hat H
    -\hat H^{\mathrm T}\otimes \hat I)
    +\sum_\ell \kappa_\ell
    \Big[
        \hat L_\ell^*\otimes \hat L_\ell
        -\frac{1}{2}\hat I\otimes 
        \hat L_\ell^\dagger \hat L_\ell
        -\frac{1}{2}(\hat L_\ell^\dagger\hat L_\ell)^{\mathrm T}
        \otimes \hat I
    \Big].
\end{eqnarray}
Equation~\eqref{SIeq: vectorized Lindblad equation} formally has a similar form to the Schr\"odinger equation of quantum mechanics. 
However, unlike in the Schr\"odinger equation, the operator $\check{\mathcal L}$ that governs the dynamics is non-Hermitian.
In those cases, it is useful to characterize the $\check{\mathcal{L}}$ with \textit{both} the right and left eigenvectors ($\hat r_n$ and $\hat l_n$, respectively),
\begin{eqnarray}   
    \breve{\mathcal L}{\rm{vec}}(\hat r_n)
    =\lambda_n {\rm{vec}}(\hat r_n),
    \qquad
    \breve{\mathcal L}^{\dagger}{\rm{vec}}(\hat l_n)
    =    \lambda_n^* {\rm{vec}}(\hat l_n),
\end{eqnarray}
where the eigenvalues $\lambda_n$ are sorted as $\lambda_{N-1}\le\lambda_{N-2}\le...\le\lambda_2\le\lambda_1\le \lambda_0$.
Note that in Hermitian cases, $\hat r_n = \hat l_n$ but this does not generally hold otherwise.
In the operator form, these read
\begin{eqnarray}
    {\mathcal L}\hat r_n = \lambda_n \hat r_n,
    \qquad
    \hat l_n^\dagger {\mathcal L} = \hat l_n^\dagger \lambda_n,
\end{eqnarray}
which form a biorthogonal and complete basis, 
\begin{eqnarray}
    {\rm vec}(\hat l_n)^\dagger 
    {\rm vec} (\hat r_m) = \delta_{nm},
    \qquad
    \sum_n {\rm vec} (\hat r_n){\rm vec}(\hat l_n)^\dagger 
    =\hat I.
    \label{SIeq: completeness orthogonality}
\end{eqnarray}
Here, we used the property that an inner product is properly defined (Eq.~\eqref{SIeq: Hilbert-Schmidt inner product}). 
One can then decompose the density operator into eigenmodes of $\breve{\mathcal L}$:
\begin{eqnarray}
    {\rm vec}(\hat\rho(t))
    =e^{\breve{\mathcal L}t}
    {\rm vec}(\hat\rho(0))
    =\sum_n e^{\lambda_n t}
    {\rm vec} (\hat r_n)
    {\rm vec} (\hat l_n)^\dagger \hat \rho(0),
\end{eqnarray}
which is equivalent to
\begin{eqnarray}
    \hat\rho(t)
    =e^{{\mathcal L}t}\hat\rho(0)
    =\sum_n e^{\lambda_n t}
    \hat r_n {\rm tr}[\hat l_n^\dagger \hat\rho(0)].
\end{eqnarray}
As one sees from here, $-{\rm Re}\lambda_n$ characterizes the decay rate of the mode $n$ and ${\rm Im}\lambda_n$ characterizes its oscillation frequency.

We remark that the GKSL master equation \eqref{SIeq: GKSL master equation} is guaranteed to have at least one eigenstate with a vanishing eigenvalue $\lambda_0=0$, corresponding to the steady state $\partial_t\hat\rho_{\rm ss}=0$.
This can be shown from the trace-preserving property of the GKSL master equation that the dynamics assure ${\rm tr}\hat\rho(t)=1$ at arbitrary times,
\begin{eqnarray}
    \partial_t {\rm tr}\hat\rho
    ={\rm tr}[{\mathcal L}\hat \rho] = 0.
\end{eqnarray}
In the vectorized GKSL master equation, this reads
\begin{eqnarray}
    0={\rm vec}(\hat I)^\dagger
    \breve{\mathcal L}
    {\rm vec}
    (\hat \rho).
\end{eqnarray}
Here, we have used the relation ${\rm tr}(\hat A)={\rm vec}(\hat I)^\dagger {\rm vec}(\hat A)$ where ${\rm vec}(\hat I)=\sum_i\ket{i}\otimes\ket{i}$.
For this to be true for arbitrary $\hat\rho$, 
\begin{eqnarray}
    0={\rm vec}(I)^\dagger
    \breve{\mathcal L}
\end{eqnarray}
must be satisfied.
This assures the presence of a left eigenstate given by
\begin{eqnarray}
\label{SIeq: left zero eigenvector}
    \hat l_0 = \hat I,
\end{eqnarray}
with zero eigenvalue $\lambda_0=0$.
The right eigenstate is the state satisfying
\begin{eqnarray}
    {\mathcal L}(\hat r_0)=0,
\end{eqnarray}
which is nothing but a steady state $\hat r_0=\hat \rho_{ss}$.
Using this property, we find
\begin{eqnarray}
    \hat\rho(t)
    =
    \hat\rho_{ss}
    +\sum_{n\ne 0}
    e^{\lambda_n t}
    \hat r_n {\rm tr}[\hat l_n^\dagger \hat\rho(0)].
\end{eqnarray}
Here, the coefficient in front of $\hat\rho_{ss}$ is unity ${\rm tr}[\hat l_0^\dagger\hat\rho(0)]=1$, which can be shown from orthogonality relation~\eqref{SIeq: completeness orthogonality} that implies ${\rm tr}[\hat l_m \hat r_n]=\delta_{n,m}$
and $\hat l_0=\hat I$.
One also finds that the $n\ne 0$ right eigenstates are traceless, from ${\rm tr}[\hat r_n]={\rm tr}[\hat l_0^\dagger \hat r_n]=0~(n\ne 0)$.

\subsubsection{Projection method}
\label{SIsubsec: projection}

There are many situations in condensed matter physics and quantum optics where there is a separation of timescales in the problem. 
In such cases, it is often useful to project out the higher-energy degrees of freedom to obtain an effective low-energy description~\cite{Schrieffer1966, Nakajima1958, Zwanzig1960}. 
These approaches have been proven useful in describing, e.g., electrons in a potential perturbed by impurities can be described by an effective $k\cdot p$ theory \cite{Luttinger1955}; the Fermi-Hubbard model at the half-filling can be mapped to an anti-ferromagnetically interacting spin model~\cite{Manousakis1991}; the Anderson impurity model can be mapped to an sd exchange coupling model \cite{Schrieffer1966} that has successfully described the Kondo effect \cite{Kondo1964}.

Here, we extend this concept to an open quantum system, which will be used in Sec.~\ref{SIsec: sd}. 
We develop here a projection method that is consistent with the so-called Lindblad perturbation theory \cite{Li2014, Li2016, Hanai2021}, which is similar to the generalized Schrieffer-Wolff transformation technique \cite{Cirac1992, Reiter2012, Kessler2012} in spirit.
We consider a Lindbladian that can be split into the non-perturbative and perturbative part ${\mathcal L}={\mathcal L}_0+\epsilon{\mathcal L}_1$ (where $\epsilon=1$ is a book-keeping constant).
The non-perturbative Lindbladian master equation,
\begin{eqnarray}   
    \partial_t {\rm{vec}}(\hat\rho^{(0)})
    =\breve{\mathcal L}_0\rm{vec}(\hat\rho^{(0)})
\end{eqnarray}
is assumed to have a separation of timescales.
We label a set of eigenvectors $\hat r_n^{(0)} [n\in \mathfrak{s(f)}]$ with small [in magnitude] eigenvalues $|\lambda_{n\in \mathfrak{s}}^{(0)}| (< |\lambda_{n\in \mathfrak{f}}^{(0)}|)$ as slow (fast) modes,
where 
$\hat r_n^{(0)}(\hat l_n^{(0)})$ is the right (left) eigenstate of the non-perturbative Lindbladian defined as ${\mathcal L}_0\hat r_n^{(0)}=\lambda_n^{(0)}\hat r_n^{(0)}$ 
($\hat l_n^{(0)\dagger} {\mathcal L}_0 = \hat l_n^{(0)\dagger} \lambda_n^{(0)}$) with the eigenvalue $\lambda_n^{(0)}$.
We split the density operator into slow and fast part $\hat\rho^{(0)}=\hat\rho_s^{(0)}+\hat\rho_f^{(0)}$. 
Here, $\hat\rho_s^{(0)}=
{\mathcal P}\hat\rho^{(0)}
=\sum_{n\in \mathfrak{s}} 
c_n\hat r_n^{(0)}$ 
is the density operator spanned projected to the slow variable space
(where ${\mathcal P}\hat\rho=\sum_{n\in \mathfrak{s}}
\hat r_n^{(0)}
{\rm tr}[\hat l_n^{(0)\dagger}\hat\rho]$ is a projection operator onto the slow space)
and 
$\hat\rho_f^{(0)}={\mathcal Q}\hat\rho^{(0)}
=({\mathcal I} - {\mathcal P})\hat\rho^{(0)}
=\sum_{n\in \mathfrak{f}}
c_n\hat r_n^{(0)}$ is the fast variable space
(where 
${\mathcal Q}\hat\rho=\sum_{n\in \mathfrak{f}}
\hat r_n^{(0)}
{\rm tr}[\hat l_n^{(0)\dagger}\hat\rho]$ is a projection operator onto the fast variable space
and ${\mathcal I}$ is an identity superoperator).
The non-perturbative GKSL master equation $\partial_t \hat\rho_s^{(0)} + \partial_t \hat\rho_f^{(0)} 
= {\mathcal L}_0\hat\rho_s^{(0)}+ {\mathcal L}_0\hat\rho_f^{(0)}$ reads
\begin{eqnarray}
    \partial_t
    \begin{pmatrix}
        \breve{\mathcal P}
        {\rm vec}(\hat\rho^{(0)})\\
        \breve{\mathcal Q}
        {\rm vec}(\hat\rho^{(0)})
    \end{pmatrix}
    = 
    \begin{pmatrix}
        \breve{\mathcal P}
        \breve{\mathcal L}_0& 0 \\
        0 & 
        \breve{\mathcal Q}\breve{\mathcal L}_0
    \end{pmatrix}
    \begin{pmatrix}
        \breve{\mathcal P}
        {\rm vec}(\hat\rho^{(0)})\\
        \breve{\mathcal Q}
        {\rm vec}(\hat\rho^{(0)})
    \end{pmatrix}
\end{eqnarray}
in the vectorized representation, where we used the property that these two spaces do not talk to each other via ${\mathcal L}_0$.
Here, we have introduced a projection operator in the vectorized representation, ${\rm vec}({\mathcal P}\hat\rho)\equiv \breve{\mathcal P}{\rm vec}(\hat\rho)$ and similarly for $\breve{\mathcal Q}$.

Let us now add back the perturbation ${\mathcal L}_1$. In such a case, we find 
\begin{eqnarray}
    \partial_t
    \begin{pmatrix}
        \breve{\mathcal P}
        {\rm vec}(\hat\rho)\\
        \breve{\mathcal Q}
        {\rm vec}(\hat\rho)
    \end{pmatrix}
    &\approx&
    \begin{pmatrix}
        \partial_t
       \breve{\mathcal P}        {\rm vec}
        (\hat\rho)\\
        0
    \end{pmatrix}
    = 
    \begin{pmatrix}
        \breve{\mathcal L}_{ss} & \breve{\mathcal L}_{sf} \\
        \breve{\mathcal L}_{fs} & \breve{\mathcal L}_{ff}
    \end{pmatrix}
    \begin{pmatrix}
        \breve{\mathcal P}
        {\rm vec}(\hat\rho)\\
        \breve{\mathcal Q}
        {\rm vec}(\hat\rho)
    \end{pmatrix}
    \nonumber\\
    &=&    
    \begin{pmatrix}
        \breve{\mathcal P}
        \breve{\mathcal L}
        & \epsilon
        \breve{\mathcal P}
        \breve{\mathcal L}_1
        \\
        \epsilon
        \breve{\mathcal Q}
        \breve{\mathcal L}_1
        & 
        \breve{\mathcal Q}
        \breve{\mathcal L}
    \end{pmatrix}
    \begin{pmatrix}
        \breve{\mathcal P}
        {\rm vec}(\hat\rho)\\
        \breve{\mathcal Q}
        {\rm vec}(\hat\rho)
    \end{pmatrix}
\end{eqnarray}
where in the second equality, we have done an adiabatic approximation for the fast modes (i.e., the fast modes relax fast enough such that the fast variables adiabatically follow the slow variables).
This leads to
\begin{eqnarray}
    \breve{\mathcal Q}
    {\rm vec}(\hat\rho) \approx - (\breve{\mathcal L}_{ff})^{-1}
    \breve{\mathcal L}_{fs}
    \breve{\mathcal P}    
    {\rm vec}(\hat\rho),
\end{eqnarray}
therefore,
\begin{eqnarray}
    \partial_t 
    \breve{\mathcal P}
    {\rm vec}(\hat\rho)
    =\breve{\mathcal L}_{\rm eff}
    \breve{\mathcal P}
    {\rm vec}(\hat\rho)
    =[\breve{\mathcal L}_{ss}
    -\breve{\mathcal L}_{sf}
    (\breve{\mathcal L}_{ff})^{-1}
    \breve{\mathcal L}_{fs}
    ]
    \breve{\mathcal P}
    {\rm vec}(\hat\rho)
\end{eqnarray}
or
\begin{eqnarray}
    \breve{\mathcal L}_{\rm eff}
    &=&\breve{\mathcal P}
    \breve{\mathcal L}
    \breve{\mathcal P}
    -\epsilon^2 
    \breve{\mathcal P}
    \breve{\mathcal L}_1
    \breve{\mathcal Q}
    (\breve{\mathcal L})^{-1}
    \breve{\mathcal Q}
    \breve{\mathcal L}_1
    \breve{\mathcal P}
    \nonumber\\
    &=&
    \breve{\mathcal P}
    \bigg[
    \breve{\mathcal L}_0
    +\epsilon
    \breve{\mathcal L}_1
    -\epsilon^2
    \sum_{m\in \mathfrak{f}}
    \frac{
    \breve{\mathcal L}_1
    {\rm vec}(\hat r_m^{(0)})
    {\rm vec}(\hat l_m^{(0)})^\dagger
    \breve{\mathcal L}_1
    }
    {\lambda_m^{(0)}}
    \bigg]
    \breve{\mathcal P}
    +O(\epsilon^3)
\end{eqnarray}

The matrix element of the effective Lindbladian $({\mathcal L}_{\rm eff})_{n_l\in \mathfrak{s},n_r\in \mathfrak{s}}
\equiv(\hat l_{n_l\in \mathfrak{s}}^{(0)},{\mathcal L_{\rm eff}\hat r_{n_r\in \mathfrak{s}}^{(0)}})$, which contains all the information about the system in the slow dynamics of interest,
reads 
\begin{eqnarray}
    \label{SIeq: effective Lindbladian matrix element}
    ({\mathcal L}_{\rm eff})_{n_l,n_r}
    &\equiv&(\hat l_{n_l}^{(0)},{\mathcal L_{\rm eff}\hat r_{n_r}^{(0)}})
    ={\rm vec}(\hat l_{n_l}^{(0)})^\dagger
    \breve{\mathcal L}_{\rm eff}
    {\rm vec}(\hat r_{n_r}^{(0)})
    \nonumber\\
    &=&
    ({\mathcal L}_0)_{n_l,n_r}
    +\epsilon
    ({\mathcal L}_1)_{n_l,n_r}
    -\epsilon^2
    \sum_{m\in \mathfrak{f}}
    \frac{
    {\rm tr}[\hat l_{n_l}^{(0)\dagger} 
    {\mathcal L}_1
    \hat r_m^{(0)}]
    {\rm tr}[\hat l_m^{(0)\dagger}
    {\mathcal L}_1
    \hat r_{n_r}^{(0)}]
    }
    {\lambda_m^{(0)}}
    +O(\epsilon^3).
\end{eqnarray}
Equation \eqref{SIeq: effective Lindbladian matrix element} is the central relation that will be used below to derive
the effective spin-exchange dynamics between the localized and conduction electrons (Eq. (4) in the main text). 

\subsection{Non-Markov environment: Keldysh theory}
\label{SIsubsec: Keldysh}

The previous section discussed open quantum systems whose dynamics are assumed to be Markovian.
However, there are situations where such assumptions are inappropriate, where the baths do not relax fast enough compared to the relevant system and the dynamics of the baths play a role as a result.
A prominent example is a many-body system attached to a thermal bath.
In this case, the relevant system generically approaches an equilibrium state, where general properties such as the fluctuation-dissipation theorem~\cite{Kubo1966} should be satisfied.
However, these relations cannot be satisfied when we neglect the non-Markovianity arising from Fermi/Bose distribution functions \cite{Sieberer2015}. 
As we will be interested in situations where we continuously tune from equilibrium to non-equilibrium, a formalism that includes such effects is desired.


This subsection briefly reviews the Keldysh approach that can incorporate such effects. (For further details, we refer to e.g.,  Refs.~\cite{Kamenev2023, Sieberer2016, Sieberer2023}.) 
In this approach, we describe the dynamics of the density matrix $\hat\rho$ in terms of a path integral, which enables us to use field-theoretical techniques such as perturbative diagrammatic calculations, renormalization group methods, and makes the connection to the semi-classical limit explicit. 
This will prove useful for obtaining the equation of motion of observables and will be used in the analysis performed in Sec.~\ref{SIsec: LLG}.

\subsubsection{Keldysh formalism for closed systems}

Let us start by formulating the Keldysh theory for closed systems described by the Hamiltonian $\hat H$.
The Liouville equation reads, 
\begin{equation}
    i
    \partial_t
    \hat\rho= [\hat H, \hat\rho],
\end{equation}
which has a formal solution,
\begin{eqnarray}
    \hat\rho(t)
    =
    \hat U(t,t_0)
    \hat\rho(t_0)
    \hat U^{-1}(t,t_0),
\end{eqnarray}
with the time evolution operator given by,
\begin{eqnarray}
    \hat U(t,t_0)=e^{-i\hat H(t-t_0)}.
\end{eqnarray}
We set the initial state to be a thermal state at temperature $T$,
\begin{eqnarray}
    \hat\rho(t=t_0)
    =\frac{1}{Z_0}\sum_n e^{-E_n/(k_{\rm B}T)}
    \ket{n}\bra{n}.
\end{eqnarray}
Here, $E_n$ is an eigenenergy, $\ket{n}$ is the eigenstate, and $Z_0=\sum_n e^{-E_n/(k_{\rm B}T)}$.
We define the Keldysh partition function as,
\begin{eqnarray}
\label{SIeq: Keldysh partition function free fermion}
    &&
    Z \equiv {\rm tr}\hat\rho(t_f)
    ={\rm tr}
    [\hat U(t_f,t_0)
    \hat\rho(t_0)
    \hat U^{-1}(t_f,t_0)].
\end{eqnarray}
Although this quantity itself is trivially computed as unity $Z=1$, the functional integral form of this quantity serves as a generating function that gives us useful macroscopic information we will be interested in.

To illustrate how this formalism works, 
we will focus on a simple, single-mode fermion system $\hat H=\varepsilon_0\hat \psi^\dagger \hat \psi$, where $\hat \psi$ is a fermionic annihilation operator.
Similarly to the path integral formalism in quantum mechanics, we expand the time evolution operator into a product of infinitesimally short time interval $\delta_t = (t_f-t_0)/N$ (with $N\rightarrow\infty$) and replace the annihilation operators $\hat\psi_j$ ($j$ labels discretized time $t_j =j \delta_t$) to a Grassmann variable in the fermionic case $\psi_i$ with coherent states
\begin{eqnarray}
    \hat \psi_j \ket{\psi_j}=\psi_j \ket{\psi_j},
    \qquad
    \bra{\psi_j}\hat\psi_j^\dagger
    =\bra{\psi_j} \psi_j^*.
\end{eqnarray}
by inserting the completeness relation
\begin{eqnarray}
    \hat 1 
    =\int d\psi_j d\psi_j^*
    e^{-\psi_j^*\psi_j}
    \ket{\psi_j}\bra{\psi_j}.
\end{eqnarray}
We obtain,
\begin{eqnarray}
    Z&=&
    Z_0 ^{-1}
    \int     
    \prod_n d\bar\psi_{+,n} d\psi_{+,n} 
    d\bar\psi_{-,n} d\psi_{-,n}
    \nonumber\\
    &\times & 
    e^{\sum_j 
    [\bar\psi_{j,+}
    (\psi_{j-1,+}-\psi_{j,+}-i\delta_t \varepsilon_0 \psi_{j-1,+})
    +\bar\psi_{j,-}
    (\psi_{j-1,-}
    -\psi_{j,-}
    +i\delta_t \varepsilon_0  \psi_{j-1,-})]
    }
    e^{-\bar\psi_{1,+}\psi_{1,+}
    -\bar\psi_{1,-}\psi_{1,-}
    +\zeta \bar\psi_{1,+}\psi_{N,-}
    +\bar\psi_{1,-}\psi_{N,+}}
    \nonumber\\
    &\equiv&
    \int {\mathcal D}[\psi_+,\bar\psi_+,\psi_-,\bar\psi_-]
    e^{i S[\psi_+,\bar\psi_+,\psi_-,\bar\psi_-]},
\end{eqnarray}
where we have used
\begin{eqnarray}
    \langle \psi_{+,n+1}|\psi_{+,n}\rangle
    =e^{\bar\psi_{+,n+1}\psi_{+,n}},
    \qquad
    \langle -\psi_{-,n}|-\psi_{-,n+1}\rangle
    =e^{\bar\psi_{-,n}\psi_{-,n+1}}
\end{eqnarray}
and $Z_0^{-1}=    \bra{\psi_+(t_0)}\hat\rho(t_0)
    \ket{-\psi_-(t_0)}$.
Here, the action has the form
$S=\bar\psi G^{-1}\psi$, where
we have introduced a set of coherent states for both the time evolution of ket ($+$) and bra ($-$), collectively denoted as
\begin{eqnarray}
    \psi = 
    \begin{pmatrix}
        \psi_{1,+} &
        \cdots &
        \psi_{N,+} &
        \psi_{1,-} & 
        \cdots  &
        \psi_{N,-}
    \end{pmatrix}
    ^{\rm T},
    \qquad
    \bar\psi
    =
    \begin{pmatrix}
        \psi_{1,+} &
        \cdots &
        \psi_{N,+} &
        \psi_{1,-} & 
        \cdots  &
        \psi_{N,-}
    \end{pmatrix}
\end{eqnarray}
with $\zeta = - e^{-\varepsilon_0/(k_{\rm B}T)}$
and the kernel is given by,
\begin{eqnarray}
    G^{-1}=
    \begin{pmatrix}
        (G^{-1})_{++} & (G^{-1})_{+-}  \\
        (G^{-1})_{-+} & (G^{-1})_{--} 
    \end{pmatrix}  
\end{eqnarray}
with
\begin{eqnarray}
    (G^{-1})_{++} 
    &=& i
    \begin{pmatrix}
        1  & 0 & 0 & \cdots & 0\\
        -a_+  & 1 & 0 & \cdots & 0 \\
        0 & -a_+  & 1 & \cdots & 0 \\
        0 & 0 & \ddots & \ddots & 0 \\
        0 & 0 & 0 & -a_+ & 1
    \end{pmatrix},
    \qquad
    (G^{-1})_{--}     
    = i
    \begin{pmatrix}
        1  & 0 & 0 & \cdots & 0\\
        -a_-  & 1 & 0 & \cdots & 0 \\
        0 & -a_-  & 1 & \cdots & 0 \\
        0 & 0 & \ddots & \ddots & 0 \\
        0 & 0 & 0 & -a_- & 1
    \end{pmatrix},
    \\
    (G^{-1})_{+-} 
    &=& i
    \begin{pmatrix}
        0  & 0 & 0 & \cdots & -\zeta \\
        0  & 0 & 0 & \cdots & 0 \\
        0 & 0  & 0 & \cdots & 0 \\
        0 & 0 & \ddots & \ddots & 0 \\
        0 & 0 & 0 & 0 & 0
    \end{pmatrix},
    \qquad
    (G^{-1})_{-+} 
    =   i
    \begin{pmatrix}
        0  & 0 & 0 & \cdots & 1 \\
        0  & 0 & 0 & \cdots & 0 \\
        0 & 0  & 0 & \cdots & 0 \\
        0 & 0 & \ddots & \ddots & 0 \\
        0 & 0 & 0 & 0 & 0
    \end{pmatrix},
\end{eqnarray}
where $a_\pm = 1 \mp i\varepsilon_0 \delta_t$.
The action in the continuous-time limit  $N\rightarrow\infty$ reads,
\begin{eqnarray}
\label{SIeq: action free fermion +-}
    S &=& \int dt dt'
    \begin{pmatrix}
        \bar\psi_+(t) &
        \bar\psi_-(t)
    \end{pmatrix}
    \begin{pmatrix}
        G_{++}(t,t') & 
        G_{+-}(t,t') \\
        G_{-+}(t,t') & 
        G_{--}(t,t')
    \end{pmatrix}^{-1}
    \begin{pmatrix}
        \psi_+(t') \\
        \psi_-(t')
    \end{pmatrix}
    \nonumber\\
    &=& \int dt 
    \begin{pmatrix}
        \bar\psi_+(t) &
        \bar\psi_-(t)
    \end{pmatrix}
    \begin{pmatrix}
        i\partial_t - \varepsilon_0 & 
        0 \\
        0 & 
        -(i\partial_t - \varepsilon_0)
    \end{pmatrix}
    \begin{pmatrix}
        \psi_+(t) \\
        \psi_-(t)
    \end{pmatrix}.
\end{eqnarray}

Taking the inverse of $G^{-1}$, one obtains in the continuous-time limit $N\rightarrow\infty$ (see e.g. Ref.~\cite{Kamenev2023} for derivation),
\begin{subequations}    
    \label{SIeq: G0 +-rep}
    \begin{align}
        G_{+-}(t, t')
        &= i 
        f(\varepsilon_0)
        e^{-i\varepsilon_0(t-t')}, \\
        G_{-+}(t, t')
        &= i  
        (1+    f(\varepsilon_0)) 
        e^{-i\varepsilon_0(t-t')}, \\
        G_{++}(t, t')
        &= \theta(-t) G_{+-}(t, t') 
        +\theta(t) G_{-+}(t, t') ,\\
        G_{--}(t, t')
        &= \theta(t) G_{+-}(t, t') 
        +\theta(-t) G_{-+}(t, t').
    \end{align}
\end{subequations}
where $f(\omega)=1/(e^{\omega/(k_{\rm B}T)}+ 1)$ is the Fermi distribution function and $\theta(t)$ is a step function.
Their Fourier transformation is given by,
\begin{subequations}    
    \label{SIeq: G0 +-rep Fourier}
    \begin{align}
        G_{+-}(\omega)
        &=
        2\pi i \delta(\omega-\varepsilon_0) f(\varepsilon_0), 
        \\
        G_{-+}(\omega)
        &= -2\pi i
        \delta(\omega-\varepsilon_0)
        (1-f(\varepsilon_0)),
        \\
        G_{++}(\omega)
        &= \frac{1-f(\varepsilon_0)}{\omega - \varepsilon_0+i\delta}
        +\frac{f(\varepsilon_0)}{\omega - \varepsilon_0-i\delta},
        \\
        G_{--}(\omega)
        &= -\frac{f(\varepsilon_0)}{\omega - \varepsilon_0+i\delta}
        -\frac{1-f(\varepsilon_0)}{\omega - \varepsilon_0-i\delta}.
    \end{align}
\end{subequations} 
These are related to the various two-point functions as  \cite{Kamenev2023, Sieberer2016, Sieberer2023}, 
\begin{eqnarray}
    G_{+-}(t, t')
    &=&-i\avg{\psi_+(t)\bar\psi_-(t')}
    =-i\int {\mathcal D}
    (\psi_+, \bar\psi_+, \psi_-, \bar\psi_-)
    \psi_+(t)\bar\psi_-(t')
    e^{iS}
    =
    i
    \avg{
    \hat\psi^\dagger(t')
    \hat\psi        (t)
    }
    \\
    G_{-+}(t, t')
    &=&-i\avg{\psi_-(t)\bar\psi_+(t')}
    =-i\int {\mathcal D}
    (\psi_+, \bar\psi_+, \psi_-, \bar\psi_-)
    \psi_-(t)\bar\psi_+(t')
    e^{iS}
    =
    -i
    \avg{
    \hat\psi        (t)
    \hat\psi^\dagger(t')
    }
    \\
    G_{++}(t, t')
    &=&-i\avg{\psi_+(t)\bar\psi_+(t')}
    =-i\int {\mathcal D}
    (\psi_+, \bar\psi_+, \psi_-, \bar\psi_-)
    \psi_+(t)\bar\psi_+(t')
    e^{iS}
    =
    -i
    \avg{{\mathcal T}
    \hat\psi        (t)
    \hat\psi^\dagger(t')
    }
    \\
    G_{--}(t, t')
    &=&-i\avg{\psi_-(t)\bar\psi_-(t')}
    =-i\int {\mathcal D}
    (\psi_+, \bar\psi_+, \psi_-, \bar\psi_-)
    \psi_-(t)\bar\psi_-(t')
    e^{iS}
    =
    -i
    \avg{\bar{\mathcal T}
    \hat\psi        (t)
    \hat\psi^\dagger(t')
    }
\end{eqnarray}
where ${\mathcal T}$ and $\bar{\mathcal T}$ are the time-ordered and anti-time-ordered operators, respectively.

It is often useful to change the variables via a ``Wick rotation'',
\begin{eqnarray}
    \label{SIeq: quantum-classical psi}
    \psi_c = \frac{1}{\sqrt{2}}
           (\psi_+ + \psi_-), 
    \qquad
    \psi_q = \frac{1}{\sqrt{2}}
           (\psi_+ - \psi_-), 
\end{eqnarray}
to transform the action into,
\begin{eqnarray}
\label{SIeq: action free fermion RAK}
    S &=& \int dt dt'
    \begin{pmatrix}
        \bar\psi_q(t) &
        \bar\psi_c(t)
    \end{pmatrix}
    \begin{pmatrix}
        G_{qq}(t,t') & 
        G_{qc}(t,t') \\
        G_{cq}(t,t') & 
        G_{cc}(t,t')
    \end{pmatrix}^{-1}
    \begin{pmatrix}
        \psi_q(t') \\
        \psi_c(t')
    \end{pmatrix}
\end{eqnarray}
where the kernel in this representation is related to the correlation functions and response functions as,
\begin{subequations}
    \begin{align}
        G^{\rm K}(t,t')
        \equiv
        G_{cc}(t, t')
        &=-i\avg{\psi_c(t)\bar\psi_c(t')}
        =-i\int {\mathcal D}
        (\psi_q, \bar\psi_q, \psi_c, \bar\psi_c)
        \psi_c(t)\bar\psi_c(t')
        e^{iS}
        =
        -i
        \avg{
        [
        \hat\psi        (t),
        \hat\psi^\dagger(t')
        ]
        }
        \\
        G^{\rm R}(t,t')
        \equiv
        G_{cq}(t, t')
        &=-i\avg{\psi_c(t)\bar\psi_q(t')}
        =-i\int {\mathcal D}
        (\psi_q, \bar\psi_q, \psi_c, \bar\psi_c)
        \psi_c(t)\bar\psi_q(t')
        e^{iS}
        =
        -i\theta(t-t')
        \avg{
        \{
        \hat\psi        (t),
        \hat\psi^\dagger(t')
        \}
        }
        \\
        G^{\rm A}(t,t')
        \equiv
        G_{qc}(t, t')
        &=-i\avg{\psi_q(t)\bar\psi_c(t')}
        =-i\int {\mathcal D}
        (\psi_q, \bar\psi_q, \psi_c, \bar\psi_c)
        \psi_q(t)\bar\psi_c(t')
        e^{iS}
        =
        i\theta(t'-t)
        \avg{
        \{
        \hat\psi        (t),
        \hat\psi^\dagger(t')
        \}
        }
        \\
        0
        =
        G_{qq}(t, t')
        &=-i\avg{\psi_q(t)\bar\psi_q(t')}
        =-i\int {\mathcal D}
        (\psi_q, \bar\psi_q, \psi_c, \bar\psi_c)
        \psi_q(t)\bar\psi_q(t')
        e^{iS}.        
    \end{align}
\end{subequations}
Here, 
$G^{\rm R}, G^{\rm A}$, and $G^{\rm K}$ are called the retarded, advanced, and Keldysh Green's functions, respectively.
Their Fourier transform is given by,
\begin{subequations}
    \label{SIeq: Greens function, single fermion, RAK}
    \begin{align}
        G^{\rm R}(\omega)
        &=\frac{1}{\omega - \varepsilon_0 + i\delta}, 
        \\
        G^{\rm A}(\omega)
        &=\frac{1}{\omega - \varepsilon_0 - i\delta}, 
        \\
        G^{\rm K}(\omega) 
        &=-\pi i (1-2 f(\omega))
        \delta(\omega - \varepsilon_0).         
        \end{align}   
\end{subequations}
Physically, $G^{\rm R}(\omega)[=G^{\rm A*}(\omega)]$ describes the response to an external field, while $G^{\rm K}(\omega)$ describes the correlation function.
One can check that the fluctuation-dissipation relation for fermions~\cite{Kubo1966}
\begin{eqnarray}
    \label{SIeq: FDT}
    G^{\rm K}(\omega) 
    = (1-2f(\omega))
    [G^{\rm R}(\omega)
    -G^{\rm A}(\omega)]
\end{eqnarray}
is satisfied, as it should in general closed equilibrium systems.

It is worth pointing out that the action Eq.~\eqref{SIeq: action free fermion +-} (or Eq.~\eqref{SIeq: action free fermion RAK}) vanishes when one puts $\psi_+=\psi_-$ (or equivalently, $\psi_q=0$).
This so-called causality structure originates from the trace-preserving property of the governing dynamics \cite{Kamenev2023, Sieberer2016, Sieberer2023}.
This can be seen by noting that, replacing $\psi_+$ to $\psi_-$ implies moving the time evolution operator $\hat U(t_f,t_0)$ that time-evolves the ket in the Keldysh partition function~\eqref{SIeq: Keldysh partition function free fermion} to the bra-dynamics,
using the cyclic property of the trace. 

It is straightforward to generalize this result to many-body systems. 
An example relevant to our study is a free electron gas described by the Hamiltonian $\hat H=\sum_{\bm k}\varepsilon_{\bm k}\hat c^\dagger_{\bm k}\hat c_{\bm k}$.
The action is given by,
\begin{eqnarray}
    S=\int dt dt'\sum_{\bm k} 
    \begin{pmatrix}
        \bar\psi_{\bm k}^+(t) &
        \bar\psi_{\bm k}^-(t)
    \end{pmatrix}
    \begin{pmatrix}
        G_{\bm k}^{++}(t,t') & 
        G_{\bm k}^{+-}(t,t') \\
        G_{\bm k}^{-+}(t,t') & 
        G_{\bm k}^{--}(t,t')
    \end{pmatrix}^{-1}
    \begin{pmatrix}
        \psi_{\bm k}^+(t') \\
        \psi_{\bm k}^-(t')
    \end{pmatrix}
\end{eqnarray}
where
\begin{eqnarray}
    \begin{pmatrix}
        G_0^{++}(\bm k,\omega) 
        &
        G_0^{+-}(\bm k,\omega) 
        \\        
        G_0^{-+}(\bm k,\omega)
        &
        G_0^{--}(\bm k,\omega) 
    \end{pmatrix}
    =\begin{pmatrix}
        \big[
        \frac{1-f_{\bm k}}
        {\omega-\varepsilon_{\bm k}+i\delta}
        +\frac{f_{\bm k}}
        {\omega-\varepsilon_{\bm k}-i\delta}   
        \big] &
        2\pi i \delta(\omega-\varepsilon_{\bm k})
        f_{\bm k}
        \\
        -2\pi i \delta(\omega-\varepsilon_{\bm k})
        (1-f_{\bm k})
        &
        \big[
        -\frac{f_{\bm k}}{\omega-\varepsilon_{\bm k}+i\delta}
        -
        \frac{1-f_{\bm k}}{\omega-\varepsilon_{\bm k}-i\delta}
        \big] 
    \end{pmatrix}.
\end{eqnarray}

\subsubsection{Open quantum system attached to a thermal bath}

Keldysh formalism is useful for formulating open quantum systems.
To illustrate this, we will show an example of a single-mode fermion attached to a thermal bath modeled by a collection of free fermions.
The Hamiltonian $\hat H=\hat H_s + \hat H_b + \hat H_{sb}$ is given by,
\begin{eqnarray}
    \hat H_s = \varepsilon_0 
    \hat \psi^\dagger \psi,
    \qquad
    \hat H_b = \sum_\mu \omega_\mu 
    \hat\phi_\mu^\dagger \phi_\mu,
    \qquad
    \hat H_{sb} 
    = \sum_\mu 
    [v_\mu \hat\phi_\mu^\dagger \hat\psi+ {\rm h.c.}],
\end{eqnarray}
where $\hat\phi_\mu$ and $\omega_\mu$ are the fermionic annihilation operator and the energy of bath fermions at mode $\mu$, respectively. 
The Keldysh partition function can be obtained for this system as,
\begin{eqnarray}
    Z = \int {\mathcal D}[\psi_+,\bar\psi_+,\psi_-,\bar\psi_-,
    \phi_+,\bar\phi_+,\phi_-,\bar\phi_-]
    e^{i S_{\rm tot}[\psi_+,\bar\psi_+,\psi_-,\bar\psi_-,
    \phi_+,\bar\phi_+,\phi_-,\bar\phi_-]},
\end{eqnarray}
where the total action is composed of three parts $S_{\rm tot}=S_0+S_{b}+S_{sb}$, which are given by,
\begin{eqnarray}
    S_0 &=& 
    \int dt dt'
        \begin{pmatrix}
        \bar\psi_+(t) &
        \bar\psi_-(t)
    \end{pmatrix}
    \begin{pmatrix}
        G_{++}(t,t') & 
        G_{+-}(t,t') \\
        G_{-+}(t,t') & 
        G_{--}(t,t')
    \end{pmatrix}^{-1}
    \begin{pmatrix}
        \psi_+(t') \\
        \psi_-(t')
    \end{pmatrix},
    \\
    S_b &=& 
        \sum_\mu
        \int dt dt'
        \begin{pmatrix}
        \bar\phi_+^\mu(t) &
        \bar\phi_-^\mu(t)
    \end{pmatrix}
    \begin{pmatrix}
        B_{++}^\mu(t,t') & 
        B_{+-}^\mu(t,t') \\
        B_{-+}^\mu(t,t') & 
        B_{--}^\mu(t,t')
    \end{pmatrix}^{-1}
    \begin{pmatrix}
        \phi_+^\mu(t') \\
        \phi_-^\mu(t')
    \end{pmatrix},
    \\
    S_{sb}
    &=&
    \sum_\mu
    \int dt
    [
    (v_\mu\bar\phi_+(t)\psi_+(t) + {\rm c.c.})
    -
    (v_\mu\bar\phi_-(t)\psi_-(t) + {\rm c.c.})
    ].
\end{eqnarray}
Here, $\phi_\pm, \bar\phi_\pm$ are the Grassmann variables of the baths,
$G_{s,s'}(s,s'=+,-)$ is given by Eq.~\eqref{SIeq: G0 +-rep}, and
\begin{eqnarray}
    &&\begin{pmatrix}
        B_{++}^\mu(t,t') & 
        B_{+-}^\mu(t,t') \\
        B_{-+}^\mu(t,t') & 
        B_{--}^\mu(t,t')
    \end{pmatrix}
    \nonumber\\
    &&=
    \begin{pmatrix}
        i \theta(-t) 
        f(\omega_\mu)
        e^{-i\varepsilon_0(t-t')} 
        +i \theta(t) 
        (1+    f(\omega_\mu)) 
        e^{-i\omega_\mu(t-t')}
        &
        i 
        f(\omega_\mu)
        e^{-i\omega_\mu(t-t')}
        \\
         i  
        (1+    f(\omega_\mu)) 
        e^{-i\omega_\mu(t-t')} &
        i \theta(t) f(\omega_\mu)
        e^{-i\omega_\mu(t-t')} 
        + i \theta(-t) 
        (1+    f(\omega_\mu)) 
        e^{-i\omega_\mu(t-t')}
    \end{pmatrix}
    \nonumber\\
\end{eqnarray}
are the various two-point correlation functions of the baths.

Since the total action $S_{\rm tot}$ is quadratic in terms of the bath Grassmann variables $\phi_\pm$, one can analytically integrate out the bath degrees of freedom as,
\begin{eqnarray}
    Z = 
    \int {\mathcal D}[\psi_+,\bar\psi_+,\psi_-,\bar\psi_-] e^{i S_{\rm eff}[\psi_+,\bar\psi_+,\psi_-,\bar\psi_-]},
\end{eqnarray}
where 
\begin{eqnarray}
    e^{i S_{\rm eff}[\psi_+,\bar\psi_+,\psi_-,\bar\psi_-]}
    &=&e^{i S_0[\psi_+,\bar\psi_+,\psi_-,\bar\psi_-]} 
    \int {\mathcal D}[\phi_+,\bar\phi_+,\phi_-,\bar\phi_-] e^{i S_b[\phi_+,\bar\phi_+,\phi_-,\bar\phi_-] + S_{sb}[\psi_+,\bar\psi_+,\psi_-,\bar\psi_-,\phi_+,\bar\phi_+,\phi_-,\bar\phi_-]}
    \nonumber\\
    &=&e^{i S_0[\psi_+,\bar\psi_+,\psi_-,\bar\psi_-]}
    e^{i\Delta S[\psi_+,\bar\psi_+,\psi_-,\bar\psi_-]}.
\end{eqnarray}
The contribution arising from the attachment of the bath $\Delta S$ is computed as,
\begin{eqnarray}
    \Delta S[\psi_+,\bar\psi_+,\psi_-,\bar\psi_-]
    &=&
    \sum_\mu |v_\mu|^2
    \int dt dt'
        \begin{pmatrix}
        \bar\psi_+(t) &
        \bar\psi_-(t)
    \end{pmatrix}
    \begin{pmatrix}
        B_{++}^\mu(t, t') & 
        B_{+-}^\mu(t, t') \\
        B_{-+}^\mu(t, t') & 
        B_{--}^\mu(t, t')
    \end{pmatrix}
    \begin{pmatrix}
        \psi_+(t') \\
        \psi_-(t')
    \end{pmatrix},
    \nonumber\\
    &=&\sum_\mu |v_\mu|^2
    \int \frac{d\omega}{2\pi}
    \begin{pmatrix}
        \bar\psi_q(\omega) &
        \bar\psi_c(\omega)
    \end{pmatrix}
    \begin{pmatrix}
        0  & 
        \frac{1}{\omega-\omega_\mu-i\delta} \\
        \frac{1}{\omega-\omega_\mu+i\delta} & 
        -\pi i(1-2f(\omega))\delta(\omega-\omega_\mu)
    \end{pmatrix}
    \begin{pmatrix}
        \psi_q(\omega) \\
        \psi_c(\omega)
    \end{pmatrix},
    \nonumber\\
    &\simeq&
    \int \frac{d\omega}{2\pi}
    \begin{pmatrix}
        \bar\psi_q(\omega) &
        \bar\psi_c(\omega)
    \end{pmatrix}
    \begin{pmatrix}
        0 & 
        i\frac{\gamma(\omega)}{2} \\
        -i\frac{\gamma(\omega)}{2} & 
        i\frac{\gamma(\omega)}{2}(1-2 f(\omega))
    \end{pmatrix}
    \begin{pmatrix}
        \psi_q(\omega) \\
        \psi_c(\omega)
    \end{pmatrix}
\end{eqnarray}
Here, in the second equality, we have moved to the quantum-classical representation Eq.~\eqref{SIeq: quantum-classical psi}.
In the third, we have ignored the contribution from the real part of the propagators (the ``Lamb shift'') and have introduced a dissipation rate  $\gamma(\omega) = \pi \sum_\mu |v_\mu|^2\delta(\omega-\omega_\mu)$,
where we have taken the spectrum of the bath fermions to be a continuum.

The effective action $S_{\rm eff}=S_0+\Delta S$ is therefore summarized as,
\begin{eqnarray}
    S &=& \int dt dt'
    \begin{pmatrix}
        \bar\psi_q(\omega) &
        \bar\psi_c(\omega)
    \end{pmatrix}
    \begin{pmatrix}
        0 & 
        G^{\rm A}(\omega) \\
        G^{\rm R}(\omega) & 
        G^{\rm K}(\omega)
    \end{pmatrix}^{-1}
    \begin{pmatrix}
        \psi_q(\omega) \\
        \psi_c(\omega)
    \end{pmatrix},
\end{eqnarray}
where the correlation functions and response functions are given by,
\begin{eqnarray}
    G^{\rm R}(\omega)
    &=&\frac{1}{\omega - \varepsilon_0 + i\gamma(\omega)}, 
    \\
    G^{\rm A}(\omega)
    &=&\frac{1}{\omega - \varepsilon_0 - i\gamma(\omega)}, 
    \\
    G^{\rm K}(\omega) 
    &=&-i \pi (1-2 f(\omega))
    \frac{1}{\pi}
    \frac{\gamma(\omega)}{
    (\omega - \varepsilon_0)^2+\gamma^2(\omega)
    }.     
\end{eqnarray}
Compared with the closed system counterpart (Eq.~\eqref{SIeq: Greens function, single fermion, RAK}), 
the delta-functions of the spectrum in Eq.~\eqref{SIeq: Greens function, single fermion, RAK} is replaced by a Lorentzian. This shows that the coupling to the bath gives rise to dissipation in the relevant system.

Remarkably, the fluctuation-dissipation theorem (Eq.~\eqref{SIeq: FDT}) is still satisfied with this attachment to the thermal bath.
This illustrates how the Keldysh theory rightly describes how an attachment to a thermal bath thermalizes the relevant system.
(In fact, this can be understood as a result of a certain symmetry of the Keldysh action that is respected in the thermal equilibrium, see Ref.~\cite{Sieberer2015}.)
We remark that this could not have been captured in the GKSL formalism discussed in Sec.~\ref{SIsubsec: Markov}~\cite{Sieberer2015}, where the Markovian approximation ignores the dynamical effects arising from the Fermi distribution function.


\subsubsection{Keldysh technique applied to the GKSL master equation}
\label{SIsubsubsec: Keldysh-GKSL}

So far, we have considered the case where the relevant system is attached to a single bath, reproducing the results known from the equilibrium statistical mechanics.
The real advantage of using the Keldysh approach is that it can treat nonequilibrium systems arising from attaching to multiple baths.

In our study, we will be concerned with a system (localized electrons) attached to two baths.
One of the baths is Markovian (the higher-energy state) 
and the other is non-Markovian (conduction electrons).
As we have reviewed in Sec.~\ref{SIsubsec: Markov}, 
once we integrated out the Markovian bath,
the reduced density operator $\hat\rho$ is governed by 
the GKSL equation~\eqref{SIeq: GKSL master equation}, reproduced below for convenience:
\begin{eqnarray}
    \label{SIeq: GKSL master equation 2}
    \partial_t \hat\rho 
    &=& {\mathcal L}\hat\rho
    =-i[\hat H,\hat\rho]
    +\sum_\ell 
    \kappa_\ell
    {\mathcal D}[\hat L_\ell]\hat\rho,
\end{eqnarray}
with ${\mathcal D}[\hat L]\hat\rho
    =\hat L \hat\rho \hat L^\dagger 
    -\frac{1}{2}
    \{\hat L^\dagger\hat L,\hat\rho\}$.
Therefore, we will have in mind the situation where the Hamiltonian $\hat H$ is composed of the relevant system and the non-Markovian bath (but the formalism below is general).

Following Refs.~\cite{Sieberer2016, Sieberer2023}, 
we formulate a Keldysh path integral formalism for the GKSL equation to treat this type of open system. 
As done in the previous subsections, 
we define the Keldysh action 
\begin{eqnarray}
    Z&=&{\rm tr}[\hat\rho(t_f)]={\rm tr}\big[e^{{\mathcal L}(t_f-t_0)}\hat\rho(t_0)\big]
    =
    {\rm tr}
    \Big[\lim_{N\rightarrow\infty}
    \prod_{n=0}^N
    \hat\rho_n
    \Big]
\end{eqnarray}
where we have discretized the GKSL equation~\eqref{SIeq: GKSL master equation 2} as
\begin{eqnarray}
    \hat \rho_{n+1} = (\hat 1 + \delta_t{\mathcal L})[\hat\rho_n]+O(\delta_t^2).
\end{eqnarray}
with $\delta_t = (t_f - t_0)/N$ and defined $\hat\rho_0\equiv\hat\rho(t_0)$.

Let us restrict ourselves, for simplicity, to a single-mode fermion system.
As before, we expand the Keldysh partition function into coherent states. 
This can be computed once $\bra{\psi_{+,n+1}}
    \hat\rho_{n+1}
    \ket{-\psi_{-,n+1}}
    =\bra{\psi_{+,n+1}}
    (\hat 1+\delta_t {\mathcal L})[\hat\rho_{n}]
    \ket{-\psi_{-,n+1}}$ is calculated. 
Using
\begin{eqnarray}
    &&
    \bra{\psi_{+,n+1}}
    {\mathcal L}\hat\rho_n
    \ket{
    -
    \psi_{-,n+1}}
    \nonumber\\
    &&
    =\int 
    d\bar\psi_{+,n} d\psi_{+,n} 
    d\bar\psi_{-,n} d\psi_{-,n}
    e^{-\bar\psi_{+,n}\psi_{+,n}
       -\bar\psi_{-,n}\psi_{-,n}}
    \bra{\psi_{+,n}}
    \hat\rho_n
    \ket{
    -
    \psi_{-,n+1}}
    \bra{\psi_{+,n+1}}
    {\mathcal L}
    [\ket{\psi_{+,n}}
    \bra{-\psi_{-,n}}]
    \ket{-\psi_{-,n+1}}
    \nonumber\\
    &&=\int 
    d\bar\psi_{+,n} d\psi_{+,n} 
    d\bar\psi_{-,n} d\psi_{-,n}
    e^{-\bar\psi_{+,n}\psi_{+,n}
       -\bar\psi_{-,n}\psi_{-,n}}
    \bra{\psi_{+,n}}
    \hat\rho_n
    \ket{-\psi_{-,n}}
    \nonumber\\
    &\times &
    \Bigg[
    (-i) 
    \Big[
    \bra{\psi_{+,n+1}}
    \hat H 
    \ket{\psi_{+,n}}
    \langle -\psi_{-,n}|
    -\psi_{-,n+1}\rangle
    -    
    \langle \psi_{+,n+1}|\psi_{+,n}\rangle
    \bra{-\psi_{-,n}}
    \hat H 
    \ket{-\psi_{-,n+1}}
    \Big]
    \nonumber\\
    &&
    +\sum_\ell \kappa_\ell
    \Big[
    \bra{\psi_{+,n+1}}
    \hat L_\ell
    \ket{\psi_{+,n}}
    \bra{-\psi_{-,n}}
    \hat L_\ell^\dagger
    \ket{-\psi_{-,n+1}}
    \nonumber\\
    &&
    \ \ \ \ \ \ \ \ \ \ 
    -\frac{1}{2}
    \Big[
    \bra{\psi_{+,n+1}}
    \hat L_{\ell}^\dagger
    \hat L_{\ell}
    \ket{\psi_{+,n}}
    \langle -\psi_{-,n}|-\psi_{-,n+1}\rangle
    +   
    \langle \psi_{+,n+1}|\psi_{+,n}\rangle
    \bra{-\psi_{-,n}}
    \hat L_{\ell}^\dagger
    \hat L_{\ell}
    \ket{-\psi_{-,n+1}}
    \Big]
    \Bigg]
    \nonumber\\
    &&=\int 
    d\bar\psi_{+,n} d\psi_{+,n} 
    d\bar\psi_{-,n} d\psi_{-,n}
    e^{-(\bar\psi_{+,n}-
        \bar\psi_{+,n+1})
        \psi_{+,n}
       -\bar\psi_{-,n}
        (\psi_{-,n}-\psi_{-,n+1})}
    \bra{\psi_{+,n}}
    \hat\rho_n
    \ket{-\psi_{-,n}}
    \nonumber\\
    &\times &
    \Bigg[
    (-i) 
    \Big[
    H_+(\psi_{+,n},\bar\psi_{+,n})
    -    
    H_-(\psi_{-,n},\bar\psi_{-,n})
    \Big]
    +\sum_\ell \kappa_\ell
    \Big[
    L_{+,n,\ell}
    (-\bar L_{-,n,\ell})
    -\frac{1}{2}
    \big(
    \bar L_{+,n,\ell}   
    L_{+,n,\ell}  
    +
    \bar L_{-,n,\ell}   
    L_{-,n,\ell}  
    \big)
    \Big]
    \Bigg]
\end{eqnarray}
where
\begin{eqnarray}
    &&H_+(\psi_{+,n},\bar\psi_{+,n})
    = \bra{\psi_{+,n}}
    \hat H
    \ket{\psi_{+,n}},   
    \quad
    H_-(\psi_{-,n},\bar\psi_{-,n})
    = \bra{-\psi_{-,n}}
    \hat H
    \ket{-\psi_{-,n}},  
    \\
    &&
    L_{+,n,\ell}  
    = \bra{\psi_{+,n}}
    \hat L_{\ell}
    \ket{\psi_{+,n}},
    \qquad
    L_{-,n,\ell}  
    = \bra{-\psi_{-,n}}
    \hat L_{\ell}
    \ket{-\psi_{-,n}},
\end{eqnarray}
This yields, 
\begin{eqnarray}
    \bra{\psi_{+,n+1}}
    \hat\rho_{n+1}
    \ket{-\psi_{-,n+1}}
    &=&\int 
    d\bar\psi_{+,n} d\psi_{+,n} 
    d\bar\psi_{-,n} d\psi_{-,n}
    e^{-(\bar\psi_{+,n}-
        \bar\psi_{+,n+1})
        \psi_{+,n}
       -\bar\psi_{-,n}
        (\psi_{-,n}-\psi_{-,n+1})}
    e^{\delta_t{\mathcal L}
    [\psi_+,\bar\psi_+,\psi_-,\bar\psi_-]}
    \nonumber\\
\end{eqnarray}
with
\begin{eqnarray}
    {\mathcal L}[\psi_+,\bar\psi_+,\psi_-,\bar\psi_-]
    &=&(-i) 
    \Big[
    H_+(\psi_{+,n},\bar\psi_{+,n})
    -    
    H_-(\psi_{-,n},\bar\psi_{-,n})
    \Big]
    \nonumber\\
    &+&\sum_\ell \kappa_\ell
    \Big[
    L_{+,n,\ell}
    (-\bar L_{-,n,\ell})
    -\frac{1}{2}
    \big(
    \bar L_{+,n,\ell}   
    L_{+,n,\ell}  
    +
    \bar L_{-,n,\ell}   
    L_{-,n,\ell}  
    \big)
    \Big]
\end{eqnarray}
Here, we have assumed 
that the Hamiltonian $\hat H$, the dissipators $\hat L_\ell$, and $\hat L_\ell^\dagger \hat L_{\ell}$ are normal ordered.
We note that there are subtleties in this normal ordering procedure, where the causality structure may be broken~\cite{Sieberer2014}. 
This can be mitigated by introducing an infinitesimal time difference that enforces the ordering of the operators, see e.g. Appendix A in Ref.~\cite{Sieberer2023} for details.
Taking the continuous limit,
the Keldysh partition function
can be computed as
\begin{eqnarray}
    Z=\int {\mathcal D}[\psi_+,\bar\psi_+,\psi_-,\bar\psi_-]
    e^{i S[\psi_+,\bar\psi_+,\psi_-,\bar\psi_-]},
\end{eqnarray}
with the Keldysh action $S$ given by,
\begin{eqnarray}
\label{SIeq: action GKSL}
    S[\psi_+,\bar\psi_+,\psi_-,\bar\psi_-]
    &=&\int dt 
    \Big[
    \bar \psi_+(t)
    (i\partial_t - H_+(t))
    \psi_+(t)
    -
    \bar \psi_-(t)
    (i\partial_t - H_-(t))
    \psi_+(t)
    \nonumber\\
    &+&
    \sum_\ell 
    \kappa_\ell
    \big[
    L_{+,\ell}     (t)
    \bar L_{-,\ell}(t)
    -\frac{1}{2}
    (\bar L_{+,\ell}(t_{+\delta}) L_{+,\ell}(t_{-\delta})
    +\bar L_{-,\ell}(t_{-\delta}) L_{-,\ell}(t_{+\delta}) )
    \big]
    \Big]
\end{eqnarray}
Here, we have introduced  $t_{\pm\delta}=t\pm 0^+$ to ensure the causality structure of the action, i.e., $S[\psi_+=\psi_-,\bar\psi_+=\bar\psi_-]=0$. 
Equation~\eqref{SIeq: action GKSL} that gives the relation between the GKSL equation~\eqref{SIeq: GKSL master equation 2} and the Keldysh action $S$ is the central relation used in Sec.~\ref{SIsec: LLG}.

\section{The model}
\label{SIsec: model}


Equipped with the theoretical tools introduced in Sec.~\ref{SIsec: review of open quantum system}, we formulate below our dissipation-engineering scheme to control local spin-spin interactions in magnetic metals. 
This section summarizes our model.

As illustrated schematically in Fig. 2(a) in the main text, this system is composed of localized electrons immersed in a Fermi sea of conduction electrons.
The localized electrons are responsible for magnetic properties, while the conduction electrons are responsible for metallic properties.
To manipulate the effective spin-spin interactions between the localized sites (Fig.~2(b) in the main text), we inject a laser light that gives rise to dissipation to the double-occupied state at site $a$ at the rate $\kappa_a$.
By tuning the frequency of the laser $h\nu$ to be resonant with a higher-level state (say, upper-level $f$-orbital state) that has a linewidth $\Gamma_{f, a}$, (Fig.~2(c) in the main text),
we turn on tunneling from the double-occupied state to the higher-level state ($a$ labels the localized electron sites).

This process can be regarded as a Markov process, as long as the dissipation rate of the higher-level state $\Gamma_{f,a}$ (which is the timescale of the memory of the bath) is larger than its supply rate.
A detailed discussion on this point is provided in Methods. 
As explained in Sec.~\ref{SIsubsec: Markov}, the time evolution of reduced system density operator $\hat\rho$ of such Markov system is generally governed by the GKSL master equation. 
In our system, it reads
\begin{eqnarray}
    \label{SIeq: GKSL master equation dissipation engineering}
    \partial_t\hat\rho
    &=&
    {\mathcal L}\hat\rho
    \nonumber\\
    &=&
    -i [\hat H,\hat \rho]
    +\sum_{a,\sigma} \kappa_a
    {\mathcal D}[\hat d_{\sigma,a}
    \hat P_{\uparrow\downarrow}^a]
    \hat\rho
    +\sum_{a,\sigma} \delta_a 
    {\mathcal D}[\hat d_{\sigma,a}^\dagger
    \hat P_{\varnothing}^a]
    \hat\rho.
\end{eqnarray}
The first term describes the coherent dynamics of the magnetic metals and the second term is the light-induced dissipation.
We have added the third term phenomenologically to ensure the steady state of localized electrons is singly occupied at each site.
(We have omitted the third term in the main text, as they only play a role in ensuring the state is in the singly occupied state.)

The magnetic metal is modeled by the Anderson impurity model $\hat H=\hat H_0+ \hat H_{cd}$ consisting of two parts. 
The unperturbed part $\hat H_0
=\hat H_{c0}+\hat H_{d0}$ describes the conduction electrons and the localized electrons, given respectively by,
\begin{eqnarray}
    \label{SIeq: Hc0}
    \hat H_{c0}
    &=&
    \sum_{\bm k,\sigma=\uparrow,\downarrow}
    \varepsilon_{\bm k}
    \hat c^\dagger_{\bm k\sigma}\hat c_{\bm k\sigma},
    \\
    \label{SIeq: Hd0}
    \hat H_{d0}
    &=&
    \sum_a 
    \Big[
    \sum_{\sigma=\uparrow,\downarrow}
    \varepsilon_{d,a}
    \hat d^\dagger_{\sigma,a}
    \hat d_{\sigma,a}
    +U_a 
    \hat d^\dagger_{\uparrow,a}
    \hat d_{\uparrow,a}
    \hat d^\dagger_{\downarrow,a}
    \hat d_{\downarrow,a}
    \Big].
\end{eqnarray}
Here, the conduction electrons are modeled as free electrons, where $\hat c_{\bm k,\sigma}$ is a fermionic annihilation operator of conduction electrons with momentum $\bm k$ and spin $\sigma=\uparrow,\downarrow$ and $\varepsilon_{\bm k}$ is the kinetic energy.
We assume that the conduction electrons' system size is large enough to ensure they are always in thermal equilibrium.
Its distribution obeys the Fermi distribution function $f(\varepsilon_{\bm k})=[e^{(\varepsilon_{\bm k}-\varepsilon_{\rm F})/(k_{\rm B}T)}+1]^{-1}$ that is characterized by temperature $T$ and the Fermi energy $\varepsilon_{\rm F}$.
We will focus on low-temperature regime $k_{\rm B}T\ll \varepsilon_{\rm F}$.
The localized electrons are modeled as a two-level system with energy $\varepsilon_{d, a}$ and an on-site Coulomb repulsion $U_a>0$ at site $a$.
Here, $\hat d_{\sigma,a}$ is a fermionic annihilation operator of localized electrons at site $a$.
These two types of electrons couple through the second part of the Hamiltonian $\hat H$, namely the c-d mixing $\hat H_{cd}$, where the electrons in the conduction band tunnel at rate $v_a$ to the localized orbital and vice versa, described by
\begin{eqnarray}
    \label{SIeq: Hcd}
    \hat H_{cd} 
    &=&\sum_{a,\sigma}
    \big[
    v_a
    \hat d_{\sigma,a}^\dagger
    \hat c_{\bm R_a\sigma} 
    +{\rm h.c.}
    \big]
    =\sum_a
    \sum_{\bm k,\sigma} 
    \big[   
    v_a    
    e^{i \bm k\cdot \bm R_a}
    \hat d_{\sigma,a}^\dagger
    \hat c_{\bm k\sigma} 
    +{\rm h.c.}
    \big], 
\end{eqnarray}
Here, $\hat c_{\bm r,\sigma}=\sum_{\bm k}e^{i\bm k\cdot\bm r}\hat c_{\bm k,\sigma}$ is the Fourier transform of $\hat c_{\bm k,\sigma}$ and $\bm R_a$ is the position of the local electron at site $a$. 
In what follows, we restrict our interest to systems where the Coulomb repulsion is strong enough ($U_a\gg v_a$) to treat $\hat H_{cd}$ as a perturbation. 

The second term in the GKSL master equation~\eqref{SIeq: GKSL master equation dissipation engineering} 
describes the light-induced loss of an electron in the double-occupied state.
The localized electron at site $a$ can take four states: it can be vacant $\ket{\varnothing}_a$, singly occupied with $\uparrow (\downarrow)$-spin $\ket{\uparrow(\downarrow)}_a
=\hat d_{a,\uparrow(\downarrow)}^\dagger\ket{\varnothing}$, or double-occupied occupied $\ket{\uparrow\downarrow}_a
=\hat d_{a,\uparrow}^\dagger
\hat d_{a,\downarrow}^\dagger
\ket{\varnothing}_a$.
$\hat P_{\uparrow\downarrow}^a$ at site $a$ projects the state to the double-occupied state $\ket{\uparrow\downarrow}_a$,
enforcing the loss to activate only when site $a$ is double-occupied.
It will become clear in the following that this selective decay of localized electrons effectively modifies the interactions between the localized spins.

The third term
($\delta_a\rightarrow 0^+$) is phenomenologically added to drive the system towards a singly occupied state ($\ket{\uparrow}_a$ or $\ket{\downarrow}_a$), where $\hat P_\varnothing^a$ is the projection operator to the vacant state $\ket{\varnothing}$ at site $a$.

We note that the GKSL equation~\eqref{SIeq: GKSL master equation dissipation engineering} (which serves as our starting point) can be derived microscopically from a Hamiltonian directly modeling the laser injection as a time-dependent drive that couples the double-occupied state and a higher order state. This is detailed in Sec. \ref{SIsec: microscopic derivation of GKSL equation}.


\section{Effective localized spin dynamics immersed in conduction electrons with light-induced correlated dissipation}
\label{SIsec: sd}

To proceed, we resort to the projection method introduced in Sec.~\ref{SIsubsec: projection} to perturbatively derive the effective low-energy dynamics given by Eq.~(4) in the main text.
We split the Lindbladian ${\mathcal L}={\mathcal L}_0+{\mathcal L}_1$ in Eq.~\eqref{SIeq: GKSL master equation dissipation engineering} to non-perturbative (${\mathcal L}_0$) and perturbative part (${\mathcal L}_1$) as follows:
\begin{eqnarray}
    \label{SIeq: L0}
    {\mathcal L}_0\hat\rho
    &=&-i [\hat H_{c0}+\hat H_{d0},\hat \rho]
    +\sum_{a,\sigma} 
    \kappa_a
    {\mathcal D}[\hat d_{\sigma,a}
    \hat P_{\uparrow\downarrow}^a]
    \hat\rho
    +\sum_{a,\sigma} 
    \delta_a 
    {\mathcal D}[\hat d_{\sigma,a}^\dagger
    \hat P_{\varnothing}^a]
    \hat\rho,\\
    \label{SIeq: L1}
    {\mathcal L}_1\rho
    &=&-i [\hat H_{cd},\hat\rho].
\end{eqnarray}
The perturbative part ${\mathcal L}_1$, i.e. the c-d mixing, gives a contribution at most $O(v_a / U_a)$, which is assumed to be small.
In this section, 
we will use the relation Eq.~\eqref{SIeq: effective Lindbladian matrix element} to derive the low-energy effective Lindbladian of our system (Eq.~\eqref{SIeq: GKSL master equation dissipation engineering}) within the second-order perturbation in ${\mathcal L}_1$, which we reproduce below for convenience:
\begin{eqnarray}
\label{SIeq: effective Lindbladian matrix element 2}
    ({\mathcal L}_{\rm eff})_{n_l,n_r}
    &\equiv&
    (\hat l_{n_l}^{(0)},{\mathcal L_{\rm eff}\hat r_{n_r}^{(0)}})
    ={\rm tr}[
    \hat l_{n_l}^{(0)\dagger}{\mathcal L_{\rm eff}\hat r_{n_r}^{(0)}}
    ]
    \nonumber\\
    &= &
    {\rm tr}[
    \hat l_{n_l}^{(0)\dagger}{\mathcal L_0\hat r_{n_r}^{(0)}}
    ]
    +
    {\rm tr}[
    \hat l_{n_l}^{(0)\dagger}{\mathcal L_1\hat r_{n_r}^{(0)}}]
    -
    \sum_{m\in \mathfrak{f}}
    \frac{
    {\rm tr}[\hat l_{n_l}^{(0)\dagger} 
    {\mathcal L}_1
    \hat r_m^{(0)}]
    {\rm tr}[\hat l_m^{(0)\dagger}
    {\mathcal L}_1
    \hat r_{n_r}^{(0)}]
    }
    {\lambda_m^{(0)}}
    +O(({\mathcal L}_1)^3).
\end{eqnarray}
Here, $\hat r_{n_r}^{(0)}$ and $\hat l_{n_l}^{(0)}$ are the right and left eigenstates, respectively, that form a basis of the slow variable space we define below. 
$(\hat A,\hat B)={\rm tr}[\hat A^\dagger \hat B]$ is the Hilbert-Schmidt inner product (see Eq.~\eqref{SIeq: Hilbert-Schmidt inner product}).
In the third term, the sum is taken over the fast degrees of freedom, also defined below.

\subsection{Characterization of non-perturbative Lindbladian ${\mathcal L}_0$}
\label{SIsec: L0}

As is clear from the expression of Eq.~\eqref{SIeq: effective Lindbladian matrix element 2}, the first step 
to derive the effective low-energy Lindbladian 
is to characterize the (left) right eigenstates $\hat r_n^{(0)}(\hat l_n^{(0)})$ and eigenvalues of 
the unperturbed Lindbladian ${\mathcal L}_0$ (Eq.~\eqref{SIeq: L0}).
This can be done exactly since the non-perturbative Lindbladian 
${\mathcal L}_0={\mathcal L}_{c0}
+\sum_a {\mathcal L}_{d0, a}$ is composed of free conduction electrons and localized electrons that are completely decoupled, where
\begin{eqnarray}
    {\mathcal L}_{c0} \hat\rho 
    &=& -i\big[
    \sum_{\bm k,\sigma}
    \varepsilon_{\bm k}
    \hat c_{\bm k,\sigma}^\dagger 
    \hat c_{\bm k,\sigma},
    \hat\rho\big],
    \\
    {\mathcal L}_{d0,a} \hat\rho 
    &=& 
    -i \big[\big(
    \sum_\sigma
    \varepsilon_{d,a}
    \hat d^\dagger_{\sigma,a}
    \hat d_{\sigma,a}
    +U_a 
    \hat d^\dagger_{\uparrow,a}
    \hat d_{\uparrow,a}
    \hat d^\dagger_{\downarrow,a}
    \hat d_{\downarrow,a}
    \big),
    \hat \rho]
    +\sum_{\sigma} 
    \kappa_a
    {\mathcal D}[\hat d_{\sigma,a}
    \hat P_{\uparrow\downarrow}^a]
    \hat\rho
    +\sum_{\sigma} 
    \delta_a 
    {\mathcal D}[\hat d_{\sigma,a}^\dagger
    \hat P_{\varnothing}^a]
    \hat\rho.
\end{eqnarray}
The eigenvectors can be expressed as a direct product, 
i.e. $\hat r_n^{(0)}
=
\hat r_n^{c(0)}\otimes\prod_a\hat r_{a,n}^{d(0)}$ and 
$\hat l_n^{(0)}
=\hat l_n^{c(0)}\otimes
\prod_a\hat l_{a,n}^{d(0)}$.
Here, $\hat r_n^{c(0)}$ and $\hat l_n^{c(0)}$ are the right and left eigenstates, respectively, of the (non-perturbative) conduction electron Lindbladian ${\mathcal L}_{c0}$ that satisfies, 
\begin{eqnarray}
    {\mathcal L}_{c0} 
    \hat r_n^{c(0)}
    =\lambda_n^{c(0)}
    \hat r_n^{c(0)},
    \qquad
    {\mathcal L}_{c0}^{\dagger} 
    \hat l_n^{c(0)}
    =\lambda_n^{c(0)*}
    \hat l_n^{c(0)},
\end{eqnarray}
with an eigenvalue $\lambda_n^{c(0)}$.
Similarly, $\hat r_{a,n}^{d(0)}$ and $\hat l_{a,n}^{d(0)}$ are the right and left eigenstates, respectively, with an eigenvalue $\lambda_{a,n}^{d(0)}$ for the localized electron part ${\mathcal L}_{d0, a}$ 
\begin{eqnarray}
    {\mathcal L}_{d0,a} 
    \hat r_{a,n}^{d(0)}
    =\lambda_{a,n}^{d(0)}
    \hat r_{a,n}^{d(0)},
    \qquad
    {\mathcal L}_{d0,a}^{\dagger} 
    \hat l_{a,n}^{d(0)}
    =\lambda_{a,n}^{d(0)*}
    \hat l_{a,n}^{d(0)}.
\end{eqnarray}
The total eigenvalue $\lambda_n^{(0)}$ is given by their sum $\lambda_n^{(0)}=\lambda_n^{c(0)}+\sum_a \lambda_{a,n}^{d(0)}$.

\subsubsection{Characterization of ${\mathcal L_{d0,a}}$}
\label{SIsubsubsec: Ld0}

Let us first characterize ${\mathcal L_{d0,a}}$. 
It is helpful to point out that the superoperator 
${\mathcal N_{a,\sigma}^d}\hat \rho = 
[\hat d_{\sigma,a}^\dagger \hat d_{\sigma,a},\hat \rho]$ commutes with ${\mathcal L_{d0,a}}$,
implying that the two superoperators share a set of eigenvectors.
The eigenvector of the superoperator ${\mathcal N_{a,\sigma}^d}$ is the outer product of Fock states with corresponding eigenvalue $m_\uparrow(m_\downarrow)=0,\pm 1$.
Here, the eigenvalue $m_\uparrow(m_\downarrow)=0,\pm 1$ is given by the number of $\sigma=\uparrow(\downarrow)$-spin electrons in ket minus those in bra.
Therefore, the right and left eigenvectors of 
${\mathcal L_{d0,a}}$ can be cast into a block-diagonal form
${\mathcal L_{d0,a}}=\otimes_{m_\uparrow,m_\downarrow}
{\mathcal L_{d0,a,(m_\uparrow,m_\downarrow)}}$, where $m_\sigma$ is the eigenvalue of ${\mathcal N}_{a,\sigma}^d$.
The simultaneous right and left eigenstates of ${\mathcal L_{d0,a}}$ and ${\mathcal N}_{a,\sigma}^d$ are given respectively by
\begin{eqnarray}
\label{SIeq: Ld0 eigenvectors}
    {\mathcal L}_{d0,a,(m_\uparrow,m_\downarrow)}
    \hat r_{a,\ell,(m_\uparrow,m_\downarrow)}^{d(0)}
    =    
    \lambda_{a,\ell,(m_\uparrow,m_\downarrow)}^{d(0)}
    \hat r_{a,\ell,(m_\uparrow,m_\downarrow)}^{d(0)}, 
    \qquad
    {\mathcal L}_{d0,a,(m_\uparrow,m_\downarrow)}^\dagger
    \hat l_{a,\ell,(m_\uparrow,m_\downarrow)}^{d(0)}
    =    
    \lambda_{a,\ell,(m_\uparrow,m_\downarrow)}^{d(0)*}
    \hat l_{a,\ell,(m_\uparrow,m_\downarrow)}^{d(0)}.
\end{eqnarray}
For example, $(m_\uparrow,m_\downarrow)=(0,0)$ sector has the right and left eigenmode of the form
\begin{eqnarray}
    \label{SIeq: Ld0 mup=0 mdn = 0 right}
    \hat r_{a,\ell,m_\uparrow=0,m_\downarrow=0}^{d(0)}
    &=&
    r_{a,\ell,(m_\uparrow=0,m_\downarrow=0),\varnothing}
    ^{(0)}
    \ket{\varnothing}\bra{\varnothing}
    +    r_{a,\ell,(m_\uparrow=0,m_\downarrow=0),\uparrow}
    ^{(0)}
    \ket{\uparrow}\bra{\uparrow}
    \nonumber\\
    &+&    r_{a,\ell,(m_\uparrow=0,m_\downarrow=0),\downarrow}
    ^{(0)}
    \ket{\downarrow}
    \bra{\downarrow}
    + r_{a,\ell,(m_\uparrow=0,m_\downarrow=0),\uparrow\downarrow}
    ^{(0)}
    \ket{\uparrow\downarrow}
    \bra{\uparrow\downarrow},
    \\
    \label{SIeq: Ld0 mup=0 mdn = 0 left}
    \hat l_{a,\ell,m_\uparrow=0,m_\downarrow=0}^{d(0)}
    &=&
    l_{a,\ell,(m_\uparrow=0,m_\downarrow=0),\varnothing}
    ^{(0)}
    \ket{\varnothing}\bra{\varnothing}
    +    l_{a,\ell,(m_\uparrow=0,m_\downarrow=0),\uparrow}
    ^{(0)}
    \ket{\uparrow}\bra{\uparrow}
    \nonumber\\
    &+&    l_{a,\ell,(m_\uparrow=0,m_\downarrow=0),\downarrow}
    ^{(0)}
    \ket{\downarrow}
    \bra{\downarrow}
    + l_{a,\ell,(m_\uparrow=0,m_\downarrow=0),\uparrow\downarrow}
    ^{(0)}
    \ket{\uparrow\downarrow}
    \bra{\uparrow\downarrow}.
\end{eqnarray}
Similarly, the right and left eigenvectors have the form
\begin{eqnarray}
    \label{SIeq: Ld0 mup=1 mdn = 0 right}
    \hat r_{a,\ell,(m_\uparrow=1,m_\downarrow=0)}^{d(0)}
    &=&
    r_{a,\ell,(m_\uparrow=1,m_\downarrow=0),\varnothing}
    ^{(0)}
    \ket{\uparrow}\bra{\varnothing}
    +
    r_{a,\ell,(m_\uparrow=1,m_\downarrow=0),\downarrow}
    ^{(0)}
    \ket{\uparrow\downarrow}
    \bra{\downarrow},
    \\
    \label{SIeq: Ld0 mup=1 mdn = 0 left}
    \hat l_{a,\ell,(m_\uparrow=1,m_\downarrow=0)}^{d(0)}
    &=&
    l_{a,\ell,(m_\uparrow=1,m_\downarrow=0),\varnothing}
    ^{(0)}
    \ket{\uparrow}\bra{\varnothing}
    +   l_{a,\ell,(m_\uparrow=1,m_\downarrow=0),\downarrow}
    ^{(0)}
    \ket{\uparrow\downarrow}
    \bra{\downarrow},
\end{eqnarray}
for $(m_\uparrow,m_\downarrow)=(1,0)$,
\begin{eqnarray}
    \label{SIeq: Ld0 mup=-1 mdn = 0 right}
    \hat r_{a,\ell,(m_\uparrow=-1,m_\downarrow=0)}^{d(0)}
    &=&
    r_{a,\ell,(m_\uparrow=-1,m_\downarrow=0),\uparrow}
    ^{(0)}
    \ket{\varnothing}\bra{\uparrow}
    +
    r_{a,\ell,(m_\uparrow=-1,m_\downarrow=0),\uparrow\downarrow}
    ^{(0)}
    \ket{\downarrow}
    \bra{\uparrow\downarrow},
    \\
    \label{SIeq: Ld0 mup=-1 mdn = 0 left}
    \hat l_{a,\ell,(m_\uparrow=-1,m_\downarrow=0)}^{d(0)}
    &=&
    l_{a,\ell,(m_\uparrow=-1,m_\downarrow=0),\uparrow}
    ^{(0)}
    \ket{\varnothing}\bra{\uparrow}
    +   l_{a,\ell,(m_\uparrow=-1,m_\downarrow=0),\uparrow\downarrow}
    ^{(0)}
    \ket{\downarrow}
    \bra{\uparrow\downarrow},
\end{eqnarray}
for $(m_\uparrow,m_\downarrow)=(-1,0)$, and so on.

Let us start with the analysis of the eigenvectors of $(m_\uparrow,m_\downarrow)=(0,0)$-sector, which is often called the $T_1$-modes that physically correspond to the population dynamics of the states.
In this sector, one finds by plugging the form Eqs.~\eqref{SIeq: Ld0 mup=0 mdn = 0 right} and \eqref{SIeq: Ld0 mup=0 mdn = 0 left} into Eqs.~\eqref{SIeq: Ld0 eigenvectors},
\begin{eqnarray}
    \bm M_a 
    \bm r_{a,\ell,(m_\uparrow=0,m_\downarrow=0)}^{(0)}
    &=&\lambda_{a,\ell,(m_\uparrow=0,m_\downarrow=0)}^{(0)}
    \bm r_{a,\ell,(m_\uparrow=0,m_\downarrow=0)}^{(0)},
    \qquad
    \bm r_{a,\ell,(m_\uparrow=0,m_\downarrow=0)}^{(0)}=    \begin{pmatrix}
        r_{a,\ell,(m_\uparrow=0,m_\downarrow=0),\varnothing}^{(0)} \\
        r_{a,\ell,(m_\uparrow=0,m_\downarrow=0),\uparrow}^{(0)}
        \\
        r_{a,\ell,(m_\uparrow=0,m_\downarrow=0),\downarrow}^{(0)}
        \\
        r_{a,\ell,(m_\uparrow=0,m_\downarrow=0),\uparrow\downarrow}^{(0)}
    \end{pmatrix}, 
    \\
    (\bm l_{a,\ell,(m_\uparrow=0,m_\downarrow=0)}^{(0)})^{\mathsf T}
    \bm M_a 
    &=&\lambda_{a,\ell,(m_\uparrow=0,m_\downarrow=0)}^{(0)}
    (\bm l_{a,\ell,(m_\uparrow=0,m_\downarrow=0)}^{(0)})^{\mathsf T},
    \qquad
    \bm l_{a,\ell,(m_\uparrow=0,m_\downarrow=0)}^{(0)}=    \begin{pmatrix}
        l_{a,\ell,(m_\uparrow=0,m_\downarrow=0),\varnothing}^{(0)} \\
        l_{a,\ell,(m_\uparrow=0,m_\downarrow=0),\uparrow}^{(0)}
        \\
        l_{a,\ell,(m_\uparrow=0,m_\downarrow=0),\downarrow}^{(0)}
        \\
        l_{a,\ell,(m_\uparrow=0,m_\downarrow=0),\uparrow\downarrow}^{(0)}
    \end{pmatrix}, 
\end{eqnarray}
with 
\begin{eqnarray}
    \bm M_a = 
    \begin{pmatrix}
        -2\delta_a & 0 & 0 & 0 \\
        \delta_a & 0 & 0 & \kappa_a
        \\
        \delta_a & 0 & 0 & \kappa_a
        \\
        0 & 0 & 0 & -2\kappa_a
    \end{pmatrix}.
\end{eqnarray}
Note that the Hamiltonian $\hat H$ gives no contribution to this Fock-space diagonal subspace, where the dynamical matrix $\bm M_a$ contains no contribution from $\hat H$. 
This is due to the property that $\hat H$ is diagonal in the Fock space.

The first four rows of Table \ref{SItable: eigenstates localized electrons} report the eigenvalues and the right and left eigenvectors in the $(m_\uparrow,m_\downarrow)=(0,0)$ sector, computed by diagonalizing $\bm M_a$. 
Note that the first two eigenvectors are degenerate with zero eigenvalues. 
This is because both spin configurations of singly occupied state ($\ket{\uparrow}$ and $\ket{\downarrow}$) do not experience any dissipation in ${\mathcal L}_{d0, a}$.
We also note that this zero-eigenvalue degenerate space of the left eigenvectors contains the identity operator $\hat I$, as is expected from the general result (Eq. \eqref{SIeq: left zero eigenvector}).

We now move on to the analysis of other sectors $(m_\uparrow, m_\downarrow)\ne (0,0)$ (often called the $T_2$-modes).
Noting that the jump term $\hat d_{\sigma,a}\hat P_{\uparrow\downarrow}\hat r_{a,\ell,(m_\uparrow,m_\downarrow)}\hat P_{\uparrow\downarrow}\hat d_{\sigma,a}^\dagger = 0~((m_\uparrow, m_\downarrow)\ne (0,0))$ 
does not contribute in this sector because of the presence of the projection operator $\hat P_{\uparrow\downarrow}$,
the eigenvalues and eigenstates are given as a simple beamsplitter operator, see Table~\ref{SItable: eigenstates localized electrons}.
This can be readily checked. 
For example, for the $(m_\uparrow, m_\downarrow)= (1,0)$ sector, ${\mathcal L}_{d0}\hat r_{a,\ell,(m_\uparrow=1,m_\downarrow=0)}$ can be compute as,
\begin{eqnarray}
    &&
    {\mathcal L}_{d0}
    (r_{a,\ell,(m_\uparrow=1,m_\downarrow=0),\varnothing}
    ^{(0)}
    \ket{\uparrow}\bra{\varnothing}
    +
    r_{a,\ell,(m_\uparrow=1,m_\downarrow=0),\downarrow}
    ^{(0)}
    \ket{\uparrow\downarrow}
    \bra{\downarrow})
    \nonumber\\
    &&=
    \Big(-i\varepsilon_{d,a}+\frac{\delta_a}{2}\Big)
    r_{a,\ell,(m_\uparrow=1,m_\downarrow=0),\varnothing}
    ^{(0)}
    \ket{\uparrow}\bra{\varnothing}
    +\Big(-i(\varepsilon_{d,a}+U_a)-\frac{\kappa_a}{2}\Big)
    r_{a,\ell,(m_\uparrow=1,m_\downarrow=0),\downarrow}
    ^{(0)}
    \ket{\uparrow\downarrow}
    \bra{\downarrow}
\end{eqnarray}
which is diagonal, giving the eigenvalues and eigenstates listed in the seventh and ninth row of Table~\ref{SItable: eigenstates localized electrons}.

As mentioned earlier, while the $T_1$-modes (the $(m_\uparrow,m_\downarrow )=(0,0)$ sector) only involve dissipation, the $T_2$-modes (the $(m_\uparrow,m_\downarrow )\ne(0,0)$ sector) involves coherence between the two different states. 
The oscillation frequencies of the $T_2$-modes are the difference between the energies of two states, which are of order $O(\varepsilon_{d, a}, U_a)$.
We regard the first six (the last six) eigenvalues on Table~\ref{SItable: eigenstates localized electrons} as the slow (fast) modes.
We will perturbatively project out the fast modes to obtain the effective dynamics including the effect of the c-d mixing ${\mathcal L}_1$.

\begin{table}[t]
    \centering
    \caption{Eigenvalues and eigenstates of ${\mathcal L}_{d0,a}$}
    \label{SItable: eigenstates localized electrons}
    \begin{tabular}{c|c|c|c}
        \hline
         &
        Eigenvalue $\lambda_{a,n}^{d(0)}$ & Right eigenstate $\hat r_{a,n}^{d(0)}$
        & Left eigenstate $\hat l_{a,n}^{d(0)}$
        \\
        \hline
        \multirow{6}{*}{slow} & $0$      & 
        $\frac{1}{2}[\ket{\uparrow}\bra{\uparrow}
        +\ket{\downarrow}\bra{\downarrow}]$
        & 
        $\hat I$  
        \\
        & $0$ & 
        $\frac{1}{2}[\ket{\uparrow}\bra{\uparrow}
        -\ket{\downarrow}\bra{\downarrow}]$ 
        & 
        $\frac{1}{2}[\ket{\uparrow}\bra{\uparrow}
        -\ket{\downarrow}\bra{\downarrow}]$
        \\
        & $-2\delta_a$ &
        $\ket{\uparrow}\bra{\uparrow}
        +\ket{\downarrow}\bra{\downarrow}
        -2\ket{\varnothing}
        \bra{\varnothing}$ & 
        $-\frac{1}{2}\ket{\varnothing}
        \bra{\varnothing}
        $ 
        \\             
        & $
        -2\kappa_a$ 
        & 
        $\ket{\uparrow\downarrow}
        \bra{\uparrow\downarrow}
        -\frac{1}{2}
        \ket{\uparrow}
        \bra{\uparrow}
        -\frac{1}{2}
        \ket{\downarrow}
        \bra{\downarrow}$ 
        &         
        $\ket{\uparrow\downarrow}
        \bra{\uparrow\downarrow}$
        \\
        & $0$ &
        $\ket{\uparrow}\bra{\downarrow}$ & 
        $\ket{\uparrow}\bra{\downarrow}$ \\ 
        & $0$ &
        $\ket{\downarrow}\bra{\uparrow}$ & 
        $\ket{\downarrow}\bra{\uparrow}$\\      
        \hline
        \multirow{6}{*}{fast} &
        $ -i \varepsilon_{d,a}
        -\frac{\delta_a}{2}$ 
        & $
        \ket{\sigma}\bra{\varnothing}$ & 
        $
        \ket{\sigma}\bra{\varnothing}$
        \\ 
        & $i\varepsilon_{d,a}
        -\frac{\delta_a}{2}$ & 
        $
        \ket{\varnothing}\bra{\sigma}$ 
        &
        $
        \ket{\varnothing}\bra{\sigma}$
        \\
        & $ -i (\varepsilon_{d,a}+U_a) 
        - \frac{\kappa_a}{2}$ 
        & $
        \ket{\uparrow\downarrow}\bra{\sigma}$ & 
        $
        \ket{\uparrow\downarrow}\bra{\sigma}$
        \\
        & $ i (\varepsilon_{d,a}+U_a) 
        - \frac{\kappa_a}{2}$ 
        & $
        \ket{\sigma}\bra{\uparrow\downarrow}$ & 
        $
        \ket{\sigma}\bra{\uparrow\downarrow}$
        \\
        & $ -i (2\varepsilon_{d,a}+U_a) 
        - \frac{\kappa_a}{2}$ 
        & $
        \ket{\uparrow\downarrow}\bra{\varnothing}$ & 
        $
        \ket{\uparrow\downarrow}\bra{\varnothing}$
        \\
        & $ i (2\varepsilon_{d,a}+U_a) 
        - \frac{\kappa_a}{2}$ 
        & $
        \ket{\varnothing}\bra{\uparrow\downarrow}$ & 
        $
        \ket{\varnothing}\bra{\uparrow\downarrow}$
        \\
        \hline
    \end{tabular}
\end{table}

\subsubsection{Characterization of ${\mathcal L_{c0}}$}
\label{SIsubsubsec: Lc0}

Let us next analyze ${\mathcal L}_{c0}$. 
As noted earlier, we assume that the system size of the conduction electrons is large enough such that they are always in the equilibrium state.
Assuming further that the system is at low temperature $k_{\rm B}T\ll\varepsilon_{\rm F}$, the steady state (${\mathcal L}_{c0}\hat\rho_{c, ss}^{(0)}=0$) is approximately given by a Fermi sphere $\hat\rho_{c, ss}^{(0)}
    =\ket{F}\bra{F}$,  where
\begin{eqnarray}
    \ket{F}=\prod_{\varepsilon_{\bm k}<\varepsilon_{\rm F}}
    \prod_{\sigma=\uparrow,\downarrow}
    \hat c^\dagger_{\bm k,\sigma}\ket{0},
\end{eqnarray}
and $\ket{0}$ is a vacuum state.
Since ${\mathcal L}_{c0}$ is composed of free electrons, one can obtain all the eigenvalues and eigenstates by adding or subtracting electrons from this state to both ket and bra space.
For example, by creating or annihilating an electron above or below the Fermi sphere, given respectively by,
\begin{eqnarray}
    \ket{\bm k\sigma} 
    \equiv 
    \hat c_{\bm k,\sigma}^\dagger \ket{F}
    ~(\varepsilon_{\bm k}>\varepsilon_{\rm F}),
    \qquad
    \ket{\overline{\bm k\sigma}} 
    \equiv \hat c_{\bm k,\sigma} \ket{F}~(\varepsilon_{\bm k}<\varepsilon_{\rm F})
\end{eqnarray}
then
\begin{subequations}
    \label{SIeq: Lc0 right eigenstates charged}
    \begin{align}
        {\mathcal L}_{c0}
        (\ket{\bm k,\sigma}\bra{F}) 
        &= -i\varepsilon_{\bm k} 
        (\ket{\bm k\sigma}\bra{F}),
        \\
        {\mathcal L}_{c0}
        (\ket{F}\bra{\bm k,\sigma}) 
        &= +i\varepsilon_{\bm k} 
        (\ket{F}\bra{\bm k,\sigma}),
        \\
        {\mathcal L}_{c0}
        (\ket{\overline{\bm k\sigma}}\bra{F}) 
        &= +i\varepsilon_{\bm k}
        (\ket{\overline{\bm k\sigma}}\bra{F}),
        \\
        {\mathcal L}_{c0}
        (\ket{F}\bra{\overline{\bm k\sigma}}) 
        &= -i\varepsilon_{\bm k}
        (\ket{F}\bra{\overline{\bm k\sigma}}), 
    \end{align}
\end{subequations}
gives a set of eigenvalues and right eigenstates of ${\mathcal L}_{c0}$.
The left eigenstate is identical to the right eigenstate, which is immediate from the property ${\mathcal L}_{c0}^\dagger ={\mathcal L}_{c0}^*$.
Similarly, by further adding one more particle or a hole to these states, one obtains further sets of eigenvalues and right eigenstates of ${\mathcal L}_{c0}$,
\begin{subequations}
    \label{SIeq: Lc0 right eigenstates neutral}
    \begin{align}        
        {\mathcal L}_{c0}
        (\ket{\bm k\sigma,\overline{\bm k'\sigma'}}\bra{F}) 
        &= -i(\varepsilon_{\bm k} -\varepsilon_{\bm k'})
        (\ket{\bm k\sigma,\overline{\bm k'\sigma'}}\bra{F}),
        \\
        {\mathcal L}_{c0}
        (\ket{F}\bra{\bm k,\sigma,\overline{\bm k'\sigma'}}) 
        &= +i(\varepsilon_{\bm k} -\varepsilon_{\bm k'})
        (\ket{F}\bra{\bm k,\sigma,\overline{\bm k'\sigma'}}),
        \\
        {\mathcal L}_{c0}
        (\ket{\bm k\sigma}\bra{\bm k'\sigma'}) 
        &= -i(\varepsilon_{\bm k}-\varepsilon_{\bm k'})
        (\ket{\bm k\sigma}\bra{\bm k'\sigma'}),
        \\
        {\mathcal L}_{c0}
        (\ket{\overline{\bm k\sigma}}\bra{\overline{\bm k'\sigma'}}) 
        &= +i(\varepsilon_{\bm k}-\varepsilon_{\bm k'})
        (\ket{\overline{\bm k\sigma}}\bra{\overline{\bm k'\sigma'}}),
        \end{align}
\end{subequations}
etc., where
\begin{eqnarray}
    \ket{\bm k\sigma,\overline{\bm k'\sigma'}}
    \equiv
    \hat c_{\bm k\sigma}^\dagger
    \hat c_{\bm k'\sigma'}
    \ket{F}
    ~(\varepsilon_{\bm k}>\varepsilon_{\rm F},\varepsilon_{\bm k'}<\varepsilon_{\rm F}).
\end{eqnarray}
One can continue the same procedure to obtain all sets of eigenvalues and eigenstates.

Note crucially that, at the low-temperature regime we are interested in, the energy of the created/removed electrons from the Fermi sphere are generically near the Fermi surface, i.e., $\varepsilon_{\bm k},\varepsilon_{\bm k'}\sim \varepsilon_{\rm F}$.
Therefore, the eigenvalues of the eigenstates with different total charges of excitation in the ket and bra space (such as those in Eq.~\eqref{SIeq: Lc0 right eigenstates charged}) are at least $O(\varepsilon_{\rm F})$, while those with the same charge in ket and bra space (such as those in Eq.~\eqref{SIeq: Lc0 right eigenstates neutral}) are generically small.
We will regard the latter (former) as the slow (fast) modes in the next subsection.

\subsection{Derivation of Eq.~(4) in the main text}
Having derived the right and left eigenstates of the non-perturbative Lindbladian ${\mathcal L}_0$, 
we are now in the position to derive the effective low-energy master equation (Eq.~(4) in the main text) 
that incorporates the effect of c-d mixing ${\mathcal L}_1$ within the second-order perturbation.
In what follows, we will explicitly compute the matrix element of the low-energy effective Lindbladian $({\mathcal L}_{\rm eff,sd})_{n_l,n_r}
=(\hat l_{n_l},{\mathcal L}_{\rm eff,sd} \hat r_{n_r}^{(0)})
={\rm tr}[\hat l_{n_l}^{(0)\dagger}{\mathcal L}_{\rm eff,sd} \hat r_{n_r}^{(0)}]$ (see Eq.~\eqref{SIeq: effective Lindbladian matrix element 2}).

We will mainly be interested in the singly-occupied case, where we take the right eigenstate $\hat r_{n_r}^{(0)}$
as a singly occupied state for the localized electrons and a Fermi sphere for the conduction electron,
given by ($\sigma_+, \sigma_- = \uparrow,\downarrow$)
\begin{eqnarray}
    \hat r_{n_r}^{(0)}
    =\hat r_{(\sigma_+;F), (\sigma_-;F)}^{(0)}
    =\ket{\sigma_+}\ket{F}
    \bra{\sigma_-}\bra{F}.
\end{eqnarray}
This state has a zero eigenvalue in the unperturbed Lindbladian, i.e., ${\mathcal L}_0\hat r^{(0)}_{(\sigma_+;F), (\sigma_-;F)}=0$. 
For simplicity, we have temporarily dropped the localized electron label $a$ and will focus on a single impurity problem (as the extension to the multiple impurity case is straightforward).
We will discuss the matrix elements that involves $T_1$ modes with vacant $\ket{\varnothing}\bra{\varnothing}$ and double occupied state $\ket{\uparrow\downarrow}\bra{\uparrow\downarrow}$ later.
The left eigenstate $\hat l_{n_l}^{(0)}$ is taken from the slow variable space defined in the previous section, see Table~\ref{SItable: eigenstates localized electrons} and the final paragraph of Sec.~\ref{SIsubsubsec: Lc0}.
One can readily check that the zeroth order contribution in Eq.~\eqref{SIeq: effective Lindbladian matrix element 2} vanishes, i.e.,
$({\mathcal L}_{\rm eff,sd})_{n_l,[(\sigma_+;F),(\sigma_-;F)]}^{(0)}={\rm tr}[\hat l_{n_l}^{(0)\dagger}{\mathcal L}_0 
\hat r_{(\sigma_+;F),(\sigma_-;F)}^{(0)}]=0$.

For later convenience, we find it useful to introduce the c-d mixing operators,
\begin{eqnarray}
    \hat V_{d\Leftarrow c}^{a\bm k\sigma} 
    =v_a e^{i \bm k\cdot \bm R_a}
    \hat d_{\sigma,a}^\dagger
    \hat c_{\bm k\sigma},
    \qquad
    \hat V_{c\Leftarrow d}^{a\bm k\sigma} 
    =v_a^* e^{-i \bm k\cdot \bm R_a}
    \hat c_{\bm k\sigma}^\dagger
    \hat d_{\sigma,a}
    (=(\hat V_{d\Leftarrow c}^{a\bm k\sigma})^\dagger ),
\end{eqnarray}
that simplifies the notation of perturbative Lindbladian to
\begin{eqnarray}
    {\mathcal L}^{(1)}\rho
    &=&-i [\hat H_{cd},\hat\rho]
    =
    -i\sum_a
    \sum_{\bm k,\sigma} 
    \big[
    \hat V_{d\Leftarrow c}^{a\bm k\sigma} 
    +\hat V_{c\Leftarrow d}^{a\bm k\sigma}
    ,
    \hat\rho
    \big].
\end{eqnarray}

The state $\hat r^{(0)}_{\sigma_+,\sigma_-; F}$ is perturbed with ${\mathcal L}^{(1)}$ as, 
\begin{eqnarray}
    \label{SIeq: perturbed with L1}
    {\mathcal L}^{(1)} 
    \hat r_{(\sigma_+;F), (\sigma_-;F)}^{(0)}
    &=&-i
    \sum_{\bm k,\sigma}
    \big[
    (
    \hat V_{d\Leftarrow c}^{a\bm k\sigma}
    +
    \hat V_{c\Leftarrow d}^{a\bm k\sigma}
    )
    (\ket{\sigma_+}\ket{F}
    \bra{\sigma_-}\bra{F})
    -(\ket{\sigma_+}\ket{F}
    \bra{\sigma_-}\bra{F})
    (
    \hat V_{d\Leftarrow c}^{a\bm k\sigma\dagger}
    +
    \hat V_{c\Leftarrow d}^{a\bm k\sigma\dagger}
    )
    \big]
    \nonumber\\
    &=&
    -i 
    \sum_{\bm k,\sigma}
    \Big[
    \big[v_a e^{i\bm k\cdot\bm R_a}
    \delta_{\sigma_+,-\sigma}
    \delta_{\bm k\in F}(\ket{\uparrow\downarrow}
    \ket{\overline{\bm k\sigma}}
    \bra{\sigma_-}\bra{F})
    +
    v_a^* e^{-i\bm k\cdot\bm R_a}
    \delta_{\sigma_+,\sigma}
    \delta_{\bm k\in \overline{F}}
    (\ket{\varnothing}
    \ket{\bm k\sigma}
    \bra{\sigma_-}\bra{F})
    \big]
    \nonumber\\
    && \ \ 
    -\big[
    (\ket{\sigma_+}\ket{F}
    \bra{\uparrow\downarrow}
    \bra{\overline{\bm k\sigma}})
    v_a^* e^{-i\bm k\cdot\bm R_a}
    \delta_{\sigma_-,-\sigma}
    \delta_{\bm k\in F}
    +
    (\ket{\sigma_+}\ket{F}
    \bra{\varnothing}\bra{\bm k\sigma})
    v_a e^{i\bm k\cdot\bm R_a}
    \delta_{\sigma_-,\sigma}
    \delta_{\bm k\in \overline{F}}
    \big]
    \Big]
\end{eqnarray}
where we have introduced
$\delta_{\bm k\in F}=1(0)$ when $\bm k\in F$ is an (un)occupied state, for simplicity of notation.
We have also introduced the notation $\uparrow\equiv -\downarrow$ and $\downarrow\equiv -\uparrow$, again for simplicity of notation.
Notice how the c-d mixing ${\mathcal L}_1$ transfers the state in the slow variable space to the fast variable state.
The localized electron state is transferred from a singly occupied state in the $(m_\uparrow,m_\downarrow)=(0,0)$, $(m_\uparrow,m_\downarrow)=(1,-1)$ or $(m_\uparrow,m_\downarrow)=(-1,1)$ sector
to a different sector, i.e., 
$(m_\uparrow,m_\downarrow)=(\pm 1,0)$, $(m_\uparrow,m_\downarrow)=(0,\pm1)$ sector.
This property immediately tells us that the first-order contribution in Eq.~\eqref{SIeq: effective Lindbladian matrix element 2} vanishes, i.e., $({\mathcal L}_{\rm eff,sd})_{n_l,[(\sigma_+;F),(\sigma_-;F)]}^{(1)}={\rm tr}[\hat l_{n_l}^{(0)\dagger}{\mathcal L}_1 
\hat r_{(\sigma_+;F),(\sigma_-;F)}^{(0)}]=0$.

We, therefore, concentrate below on the second-order contribution in Eq.~\eqref{SIeq: effective Lindbladian matrix element 2},
\begin{eqnarray}
    &&({\mathcal L}_{\rm eff,sd})_{n_l,[(\sigma_+;F),(\sigma_-;F)]}^{(2)}
    =-\sum_{m\in \mathfrak{f}}[\lambda_m^{(0)}]^{-1}
    (\hat l_{n_l}^{(0)},{\mathcal L}^{(1)}\hat r_m^{(0)})
    (\hat l_m^{(0)},{\mathcal L}^{(1)}
    \hat r_{(\sigma_+;F), (\sigma_-;F)}^{(0)}),
\end{eqnarray}
where the sum is taken over the fast modes defined in Secs.~\ref{SIsubsubsec: Ld0} and \ref{SIsubsubsec: Lc0}.
A straightforward calculation using the results derived in Sec.~\ref{SIsubsubsec: Ld0} and \ref{SIsubsubsec: Lc0} yields,
\begin{eqnarray}
\label{SIeq: Leff(2) 4 terms}
    &&({\mathcal L}_{\rm eff,sd})_{n_l,[(\sigma_+;F),(\sigma_-;F)]}^{(2)}
    =\sum_{\bm k,\bm k',\sigma,\sigma'}
    \nonumber\\
    &&
    \times\Bigg[
    \frac{(\hat l_{n_l}^{(0)},
    \hat V^{a\bm k'\sigma'}_{c\Leftarrow d}
    \hat r_{(\uparrow\downarrow;(\overline{\bm k,-\sigma_+})),(\sigma_-;F)}^{(0)}
    -\hat r_{(\uparrow\downarrow;(\overline{\bm k,-\sigma_+})),(\sigma_-;F)}^{(0)}
    \hat V^{a\bm k'\sigma'\dagger}_{d\Leftarrow c}
    )
    (\hat l_{(\uparrow\downarrow;(\overline{\bm k,-\sigma_+})),(\sigma_-;F)}^{(0)},
    \hat V^{a\bm k\sigma}_{d\Leftarrow c}
    \hat r_{(\sigma_+;F),(\sigma_-;F)}^{(0)})
    }
    {-i(\varepsilon_{d,a}+U_a-\varepsilon_{\bm k})-\frac{\kappa_a}{2}}
    \nonumber\\
    &&-\frac{(\hat l_{n_l}^{(0)},
    \hat V^{a\bm k'\sigma'}_{d\Leftarrow c}
    \hat r_{(\sigma_+;F),(\uparrow\downarrow;(\overline{\bm k,-\sigma_-}))}^{(0)}
    -\hat r_{(\sigma_+;F),(\uparrow\downarrow;(\overline{\bm k,-\sigma_-}))}^{(0)}
    \hat V^{a\bm k'\sigma'\dagger}_{c\Leftarrow d}
    )
    (\hat l_{(\sigma_+;F),
    (\uparrow\downarrow;(\overline{\bm k,-\sigma_-}))}^{(0)},
    \hat r_{(\sigma_+;F),(\sigma_-;F)}^{(0)}
    \hat V^{a\bm k\sigma\dagger}_{d\Leftarrow c}
    )}
    {i(\varepsilon_{d,a}+U_a-\varepsilon_{\bm k})-\frac{\kappa_a}{2}}
    \nonumber\\
    &&+\frac{(\hat l_{n_l}^{(0)},
    \hat V^{a\bm k'\sigma'}_{d\Leftarrow c}
    \hat r_{(\varnothing;(\bm k,\sigma_+)),(\sigma_-;F)}^{(0)}
    -
    \hat r_{(\varnothing;(\bm k,\sigma_+)),(\sigma_-;F)}^{(0)}
    \hat V^{a\bm k'\sigma'\dagger}_{c\Leftarrow d}
    )
    (\hat l_{(\varnothing;(\bm k,\sigma_+)),(\sigma_-;F)}^{(0)},
    \hat V^{a\bm k\sigma}_{c\Leftarrow d}
    \hat r_{(\sigma_+;F),(\sigma_-;F)}^{(0)})}
    {-i(\varepsilon_{\bm k}-\varepsilon_{d,a})
    -\frac{\delta_a}{2}}
    \nonumber\\
    &&-\frac{(\hat l_{n_l}^{(0)},
    \hat V^{a\bm k'\sigma'}_{c\Leftarrow d}
    \hat r_{(\sigma_+;F),(\varnothing;(\bm k,\sigma_-))}^{(0)}
    -
    \hat r_{(\sigma_+;F),(\varnothing;(\bm k,\sigma_-))}^{(0)}
    \hat V^{a\bm k'\sigma'\dagger}_{d\Leftarrow c})
    ( \hat l_{(\sigma_+;F),(\varnothing;(\bm k,\sigma_-))}^{(0)},
    \hat r_{(\sigma_+;F),(\sigma_-;F)}^{(0)}
    \hat V^{a\bm k\sigma\dagger}_{c\Leftarrow d}
    )}
    {i(\varepsilon_{\bm k}-\varepsilon_{d,a})
    -\frac{\delta_a}{2}}
    \Bigg].
\end{eqnarray}
Here, we have introduced the notation for the intermediate states,
\begin{subequations}
\label{SIeq: intermediate states}
    \begin{align}
        \hat r^{(0)}_{(\uparrow\downarrow;(\overline{\bm k,-\sigma_+})),(\sigma_-; F)}
        =\hat l^{(0)}_{(\uparrow\downarrow;(\overline{\bm k,-\sigma_+})),(\sigma_-; F)}
        &=\ket{\uparrow\downarrow}\ket{\overline{\bm k,-\sigma_+}}
        \bra{\sigma_-}\bra{F},    
        \\
        \hat r^{(0)}_{(\sigma_+; F),(\uparrow\downarrow;(\overline{\bm k,-\sigma_-}))}
        =\hat l^{(0)}_{(\sigma_+; F),(\uparrow\downarrow;(\overline{\bm k,-\sigma_-}))}
        &=\ket{\sigma_+}\ket{F}
        \bra{\uparrow\downarrow}\bra{\overline{\bm k,-\sigma_-}},    
        \\
        \hat r^{(0)}_{(\varnothing;(\bm k,\sigma_+)),(\sigma_-; F)}
        = \hat l^{(0)}_{(\varnothing;(\bm k,\sigma_+)),(\sigma_-; F)}
        &=\ket{\varnothing}
        \ket{\bm k,\sigma_+}
        \bra{\sigma_-}\bra{F},    
        \\
        \hat r^{(0)}_{(\sigma_+; F),(\varnothing;(\bm k,-\sigma_-))}
        = \hat l^{(0)}_{(\sigma_+; F),(\varnothing;(\bm k,-\sigma_-)}
        &=\ket{\sigma_+}\ket{F}
        \bra{\varnothing}\bra{\bm k,-\sigma_-}.
    \end{align}
\end{subequations}

Let us focus first on the first two terms that involve a double-occupied state as their intermediate state, which we denote $({\mathcal L}_{\rm eff,sd})^{(2{\rm A})}_{n_l,[(\sigma_+;F),(\sigma_-;F)]}$:
\begin{eqnarray}
    &&({\mathcal L}_{\rm eff,sd})^{(2{\rm A})}_{n_l,[(\sigma_+;F),(\sigma_-;F)]}
    =\sum_{\bm k,\bm k',\sigma,\sigma'}
    \nonumber\\
    &&
    \times\Bigg[
    \frac{(\hat l_{n_l}^{(0)},
    \hat V^{a\bm k'\sigma'}_{c\Leftarrow d}
    \hat r_{(\uparrow\downarrow;(\overline{\bm k,-\sigma_+})),(\sigma_-;F)}^{(0)}
    -\hat r_{(\uparrow\downarrow;(\overline{\bm k,-\sigma_+})),(\sigma_-;F)}^{(0)}
    \hat V^{a\bm k'\sigma'\dagger}_{d\Leftarrow c}
    )
    (\hat l_{(\uparrow\downarrow;(\overline{\bm k,-\sigma_+})),(\sigma_-;F)}^{(0)},
    \hat V^{a\bm k\sigma}_{d\Leftarrow c}
    \hat r_{\sigma_-;F}^{(0)})
    }
    {-i(\varepsilon_{d,a}+U_a-\varepsilon_{\bm k})-\frac{\kappa_a}{2}}
    \nonumber\\
    &&-\frac{(\hat l_{n_l}^{(0)},
    \hat V^{a\bm k'\sigma'}_{d\Leftarrow c}
    \hat r_{(\sigma_+;F),(\uparrow\downarrow;(\overline{\bm k,-\sigma_-}))}^{(0)}
    -\hat r_{(\sigma_+;F),(\uparrow\downarrow;(\overline{\bm k,-\sigma_-}))}^{(0)}
    \hat V^{a\bm k'\sigma'\dagger}_{c\Leftarrow d}
    )
    (\hat l_{(\sigma_+;F),
    (\uparrow\downarrow;(\overline{\bm k,-\sigma_-}))}^{(0)},
    \hat r_{(\sigma_+;F),(\sigma_-;F)}^{(0)}
    \hat V^{a\bm k\sigma\dagger}_{d\Leftarrow c}
    )}
    {i(\varepsilon_{d,a}+U_a-\varepsilon_{\bm k})-\frac{\kappa_a}{2}}
    \Bigg].
\end{eqnarray}
One can proceed with the calculation as, 
\begin{eqnarray}
    &&({\mathcal L}_{\rm eff,sd})^{(2{\rm A})}_{n_l,[(\sigma_+;F),(\sigma_-;F)]}
    =\sum_{\bm k,\bm k',\sigma,\sigma'}
    \nonumber\\
    &&\times\Bigg[
    \frac{1}{-i(\varepsilon_{d,a}+U_a -\varepsilon_{\bm k})-\frac{\kappa_a}{2}}
    {\rm tr}\big[\hat l_{n_l}^{(0)\dagger}
    \big\{
    \hat V^{a\bm k'\sigma'}_{c\Leftarrow d}
    \ket{\uparrow\downarrow}
    \ket{\overline{\bm k,-\sigma_+};F}
    \bra{\sigma_-}\bra{F}
    - \ket{\uparrow\downarrow}
    \ket{\overline{\bm k,-\sigma_+};F}
    \bra{\sigma_-}\bra{F}
    \hat V^{a\bm k'\sigma'\dagger}_{d\Leftarrow c}
    \big\}
    \big]
    \nonumber\\
    &&\ \ \ \ \ \ \ \ \ \ \ \ \ \ \ \ \ \ \ \ \ \ \ \ \ \ \ \ \ \ \ \ \ \ \ 
    \times
    {\rm tr}\big[
    \ket{\sigma_-}\ket{F}
    \bra{\uparrow\downarrow}
    \bra{\overline{\bm k,-\sigma_+};F}
    \hat V^{a\bm k\sigma}_{d\Leftarrow c}
    \hat r_{(\sigma_+;F), (\sigma_-;F)}^{(0)}
    \big]
    \nonumber\\
    &&-\frac{1}{i(\varepsilon_{d,a}+U_a -\varepsilon_{\bm k})-\frac{\kappa_a}{2}}
    {\rm tr}\big[\hat l_{n_l}^{(0)\dagger}
    \big\{
    \hat V^{a\bm k'\sigma'}_{d\Leftarrow c}
    \ket{\sigma_+}\ket{F}
    \bra{\uparrow\downarrow}\bra{\overline{\bm k,-\sigma_-};F}
    - \ket{\sigma_+}\ket{F}
    \bra{\uparrow\downarrow}\bra{\overline{\bm k,-\sigma_-};F}
    \hat V^{a\bm k'\sigma'\dagger}_{c\Leftarrow d}
    \big\}
    \big]
    \nonumber\\
    &&\ \ \ \ \ \ \ \ \ \ \ \ \ \ \ \ \ \ \ \ \ \ \ \ \ \ \ \ \ \ \ \ \ \ \ 
    \times
    {\rm tr}\big[
    \ket{\uparrow\downarrow}
    \ket{\overline{\bm k,-\sigma_-};F}
    \bra{\sigma_+}\bra{F}
    \hat r_{(\sigma_+;F), (\sigma_-;F)}^{(0)}
    \hat V^{a\bm k\sigma\dagger}_{d\Leftarrow c}
    \big]
    \Bigg]
    \nonumber\\
    &&=-\sum_{\bm k,\bm k',\sigma,\sigma'}
    \nonumber\\
    &&\times\Bigg[
    \frac{
    \bra{\sigma_-}\bra{F}
    \big[
    \hat l_{n_l}^{(0)\dagger}
    \hat V^{a\bm k'\sigma'}_{c\Leftarrow d}
    - 
    \hat V^{a\bm k'\sigma'\dagger}_{d\Leftarrow c}
    \hat l_{n_l}^{(0)\dagger}
    \big]
    \ket{\uparrow\downarrow}
    \ket{\overline{\bm k,-\sigma_+};F}
    \bra{\uparrow\downarrow}
    \bra{\overline{\bm k,-\sigma_+};F}
    \hat V^{a\bm k\sigma}_{d\Leftarrow c}
    \hat r_{(\sigma_+;F), (\sigma_-;F)}^{(0)}
    \ket{\sigma_-}\ket{F}   
    }
    {i(\varepsilon_{d,a}+U_a -\varepsilon_{\bm k})
    +\frac{\kappa_a}{2}}
    \nonumber\\
    &&-\frac{
    \bra{\uparrow\downarrow}\bra{\overline{\bm k,-\sigma_-};F}
    \big[
    \hat l_{n_l}^{(0)\dagger}
    \hat V^{a\bm k'\sigma'}_{d\Leftarrow c}
    - 
    \hat V^{a\bm k'\sigma'\dagger}_{c\Leftarrow d}
    \hat l_{n_l}^{(0)\dagger}
    \big]
    \ket{\sigma_+}\ket{F}
    \bra{\sigma_+}\bra{F}
    \hat r_{(\sigma_+;F), (\sigma_-;F)}^{(0)}
    \hat V^{a\bm k\sigma\dagger}_{d\Leftarrow c}
    \ket{\uparrow\downarrow}
    \ket{\overline{\bm k,-\sigma_-};F}
    }
    {-i(\varepsilon_{d,a}+U_a -\varepsilon_{\bm k})+\frac{\kappa_a}{2}}
    \Bigg]
    \nonumber\\
    &&=-\sum_{\bm k,\bm k',\sigma,\sigma'}
    \Bigg[
    \frac{
    {\rm tr}
    \Big[
    \big[
    \hat l_{n_l}^{(0)\dagger}
    \hat V^{a\bm k'\sigma'}_{c\Leftarrow d}
    - 
    \hat V^{a\bm k'\sigma'\dagger}_{d\Leftarrow c}
    \hat l_{n_l}^{(0)\dagger}
    \big]
    \hat V^{a\bm k\sigma}_{d\Leftarrow c}
    \hat r_{(\sigma_+;F), (\sigma_-;F)}^{(0)}
    \Big]
    }
    {i(\varepsilon_{d,a}+U_a -\varepsilon_{\bm k})
    +\frac{\kappa_a}{2}}
    -\frac{
    {\rm tr}
    \Big[
    \big[
    \hat l_{n_l}^{(0)\dagger}
    \hat V^{a\bm k'\sigma'}_{d\Leftarrow c}
    - 
    \hat V^{a\bm k'\sigma'\dagger}_{c\Leftarrow d}
    \hat l_{n_l}^{(0)\dagger}
    \big]
    \hat r_{(\sigma_+;F), (\sigma_-;F)}^{(0)}
    \hat V^{a\bm k\sigma\dagger}_{c\Leftarrow d}
    \big]
    }
    {-i(\varepsilon_{d,a}+U_a -\varepsilon_{\bm k})+\frac{\kappa_a}{2}}
    \Bigg]
    \nonumber\\
\end{eqnarray}
where we have substituted the definition of the Hilbert-Schmidt inner product (Eq.~\eqref{SIeq: Hilbert-Schmidt inner product}) and the right and left eigenstates of the intermediate states (Eq.~\eqref{SIeq: intermediate states}) in the first equality, 
took the traces in the second, 
and rewrote the expression in terms of the trace over the integrand in the third.
The advantage of writing $({\mathcal L}_{\rm eff,sd})^{(2{\rm A})}_{n_l,[(\sigma_+;F),(\sigma_-;F)]}$ in this form becomes clear by separating the coefficients to the real and imaginary part as,
\begin{eqnarray}
    &&({\mathcal L}_{\rm eff,sd})^{(2{\rm A})}_{n_l,[(\sigma_+;F),(\sigma_-;F)]}
    =-\sum_{\bm k,\bm k',\sigma,\sigma'}
    \nonumber\\
    &&\times
    \Bigg[
    \frac{-i(\varepsilon_{d,a}+U_a -\varepsilon_{\bm k})
    }
    {(\varepsilon_{d,a}+U_a -\varepsilon_{\bm k})^2+\frac{\kappa_a^2}{4}}
    {\rm tr}
    \bigg[
    \hat l_{n_l}^{(0)\dagger}
    \Big(
    \hat V^{a\bm k'\sigma'}_{c\Leftarrow d}
    \hat V^{a,\bm k,\sigma}_{d\Leftarrow c}
    \hat r_{(\sigma_+;F), (\sigma_-;F)}^{(0)}
    - 
    \hat r_{(\sigma_+;F), (\sigma_-;F)}^{(0)}
    \hat V^{a,\bm k,\sigma\dagger}_{d\Leftarrow c}
    \hat V^{a\bm k'\sigma'\dagger}_{c\Leftarrow d}
    \Big)
    \bigg]
    \nonumber\\
    && 
    \ \ \  +
    \frac{\kappa_a}{(\varepsilon_{d,a}+U_a -\varepsilon_{\bm k})^2
    +\frac{\kappa_a^2}{4}}
    \nonumber\\
    && \ \ \ 
    \times
    {\rm tr}
    \bigg[
    \hat l_{n_l}^{(0)\dagger}
    \Big(
    \frac{1}{2}
    \hat V^{a\bm k'\sigma'}_{c\Leftarrow d}
    \hat V^{a,\bm k,\sigma}_{d\Leftarrow c}
    \hat r_{(\sigma_+;F), (\sigma_-;F)}^{(0)}
    +       
    \frac{1}{2}
    \hat r_{(\sigma_+;F), (\sigma_-;F)}^{(0)}
    \hat V^{a,\bm k,\sigma\dagger}_{d\Leftarrow c}
    \hat V^{a\bm k'\sigma'\dagger}_{c\Leftarrow d}
    - 
    \hat V^{a,\bm k,\sigma}_{d\Leftarrow c}
    \hat r_{(\sigma_+;F), (\sigma_-;F)}^{(0)}
    \hat V^{a\bm k'\sigma'\dagger}_{d\Leftarrow c}
    \Big)
    \bigg]
    \Bigg]
    \nonumber\\
    &&\equiv(\hat l_{n_l}^{(0)}, {\mathcal L}_{\rm eff,sd}^{(2{\rm A})}
    \hat r_{(\sigma_+;F),(\sigma_-;F)}^{(0)})
\end{eqnarray}
where we have used the cyclic property of the trace ${\rm tr}[\hat A\hat B]={\rm tr}[\hat B\hat A]$.
By further employing 
an approximation that the excitation of the conduction electrons occurs near the Fermi surface
$\varepsilon_{\bm k}\simeq \varepsilon_{\rm F}$,
the effective Lindbladian ${\mathcal L}_{\rm eff,sd}^{(2{\rm A})}$ is simplified 
to a GKSL form (Eq.~\eqref{SIeq: GKSL master equation}), 
\begin{eqnarray}
    \label{SIeq: Leff(2A)}
    {\mathcal L}_{\rm eff,sd}^{(2{\rm A})}
    \hat \rho
    &=&-i[\hat H_{\rm eff}^{({\rm A})},
    \hat\rho
    ]
    + 
    \frac{\kappa_a}{(\varepsilon_{d,a}+U_a -\varepsilon_{\rm F})^2
    +\frac{\kappa_a^2}{4}}
    {\mathcal D}
    \Big[
    \sum_{\bm k,\sigma}
    \hat V_{d\Leftarrow c}^{a\bm k\sigma}
    \hat P_{\rm s}
    \Big]
    \hat\rho
    \nonumber\\
    &=&-i[\hat H_{\rm eff}^{({\rm A})},
    \hat\rho
    ]
    + 
    \gamma_a
    {\mathcal D}
    \Big[
    \sum_{\bm k,\sigma}
    \hat d_{\sigma,a}^\dagger
    \hat c_{\bm k,\sigma}
    \hat P_{\rm s}
    e^{i \bm k \cdot\bm R_a}
    \Big]
    \hat\rho
    \nonumber\\
    &=&-i[\hat H_{\rm eff}^{({\rm A})},
    \hat\rho
    ]
    + 
    \gamma_a
    {\mathcal D}
    \Big[
    \sum_{\sigma}
    \hat d_{\sigma,a}^\dagger
    \hat c_{\bm R_a,\sigma}
    \hat P_{\rm s}
    \Big]
    \hat\rho.
\end{eqnarray}
Here, $\hat P_{\rm s}^a$ is a projection operator to a singly-occupied state,
\begin{eqnarray}
    \gamma_a = \frac{|v_a|^2 \kappa_a}{(\varepsilon_{d,a}+U_a -\varepsilon_{\rm F})^2 +\frac{\kappa_a^2}{4}},
\end{eqnarray}
is the light-induced dissipation rate, and the Hamiltonian part $\hat H_{\rm eff}^{\rm (A)}$ is given by,
\begin{eqnarray}
    \hat H_{\rm eff}^{\rm (A)}
    &=&\sum_{\bm k,\bm k',\sigma,\sigma'}
    \frac{-(\varepsilon_{d,a}+U_a -\varepsilon_{\bm k})}
    {(\varepsilon_{d,a}+U_a -\varepsilon_{\bm k})^2
    +\frac{\kappa_a^2}{4}}
    \hat P_{\rm s}^a
    \hat V^{a\bm k'\sigma'}_{c\Leftarrow d}
    \hat V^{a,\bm k,\sigma}_{d\Leftarrow c}
    \hat P_{\rm s}^a
    \nonumber\\
    &\approx&
    \frac{-(\varepsilon_{d,a}+U_a -\varepsilon_{\rm F})}
    {(\varepsilon_{d,a}+U_a -\varepsilon_{\rm F})^2
    +\frac{\kappa_a^2}{4}}
    |v_a|^2
    \sum_{\bm k,\bm k',\sigma,\sigma'}
    e^{i(\bm k-\bm k')\cdot\bm R_a}
    \hat P_{\rm s}^a
    \hat c^\dagger_{\bm k'\sigma'}\hat d_{\sigma',a} 
    \hat d^\dagger_{\sigma,a}
    \hat c_{\bm k,\sigma}
    \hat P_{\rm s}^a.
\end{eqnarray}

The remaining two terms of $({\mathcal L}_{\rm eff,sd})_{n_l,[(\sigma_+;F),(\sigma_-;F)]}^{(2)}$ in Eq.~\eqref{SIeq: Leff(2) 4 terms}, given by,
\begin{eqnarray}
    &&({\mathcal L}_{\rm eff,sd})_{n_l,[(\sigma_+;F),(\sigma_-;F)]}^{(2{\rm B})}
    =\sum_{\bm k,\bm k',\sigma,\sigma'}
    \nonumber\\
    &&\times
    \Bigg[\frac{(\hat l_{n_l}^{(0)},
    \hat V^{a\bm k'\sigma'}_{d\Leftarrow c}
    \hat r_{(\varnothing;(\bm k,\sigma_+)),(\sigma_-;F)}^{(0)}
    -
    \hat r_{(\varnothing;(\bm k,\sigma_+)),(\sigma_-;F)}^{(0)}
    \hat V^{a\bm k'\sigma'\dagger}_{c\Leftarrow d}
    )
    (\hat l_{(\varnothing;(\bm k,\sigma_+)),(\sigma_-;F)}^{(0)},
    \hat V^{a\bm k\sigma}_{c\Leftarrow d}
    \hat r_{(\sigma_+;F),(\sigma_-;F)}^{(0)})}
    {-i(\varepsilon_{\bm k}-\varepsilon_{d,a})
    -\frac{\delta_a}{2}}
    \nonumber\\
    &&-\frac{(\hat l_{n_l}^{(0)},
    \hat V^{a\bm k'\sigma'}_{c\Leftarrow d}
    \hat r_{(\sigma_+;F),(\varnothing;(\bm k,\sigma_-))}^{(0)}
    -
    \hat r_{(\sigma_+;F),(\varnothing;(\bm k,\sigma_-))}^{(0)}
    \hat V^{a\bm k'\sigma'\dagger}_{d\Leftarrow c})
    ( \hat l_{(\sigma_+;F),(\varnothing;(\bm k,\sigma_-))}^{(0)},
    \hat r_{(\sigma_+;F),(\sigma_-;F)}^{(0)}
    \hat V^{a\bm k\sigma\dagger}_{c\Leftarrow d}
    )}
    {i(\varepsilon_{\bm k}-\varepsilon_{d,a})
    -\frac{\delta_a}{2}}
    \Bigg],
\end{eqnarray}
can be similarly computed. We report the result as follows:
\begin{eqnarray}
    &&({\mathcal L}_{\rm eff,sd})_{n_l,[(\sigma_+;F),(\sigma_-;F)]}
    ^{(2{\rm B})}
    =\Big(
    \hat l_{n_l}^{(0)},{\mathcal L}_{\rm eff,sd}^{\rm (2B)}
    \hat r^{(0)}_{(\sigma_+;F),(\sigma_-;F)}
    \Big )
    = ({\mathcal L}_{\rm eff,sd})_{n_l,[(\sigma_+;F),(\sigma_-;F)]}^{\rm (2B)}
\end{eqnarray}
with
\begin{eqnarray}
\label{SIeq: Leff(2B)}
    {\mathcal L}_{\rm eff,sd}^{\rm (2B)}
    \hat\rho
    &=&-i[\hat H_{\rm eff}^{\rm (B)},
    \hat\rho
    ]
    \nonumber\\
    &+&\sum_{\bm k,\bm k'}
    \sum_{\sigma,\sigma'}
    \pi|v_a|^2\delta(\varepsilon_{\bm k}-\varepsilon_{d,a})
    e^{i(\bm k-\bm k')\cdot\bm R_a}
    \bigg[
    \hat c_{\bm k,\sigma}^\dagger
    \hat d_{\sigma,a}
    \hat P_{\rm s}^a
    \hat\rho
    \hat P_{\rm s}^a
    \hat d_{\sigma',a}^\dagger
    \hat c_{\bm k',\sigma'}
    -\frac{1}{2}
    \{    
    \hat P_{\rm s}^a
    \hat d_{\sigma',a}^\dagger
    \hat c_{\bm k,\sigma'}
    \hat c_{\bm k',\sigma}^\dagger
    \hat d_{\sigma,a}
    \hat P_{\rm s}^a
    ,\hat\rho
    \}
    \bigg]
\end{eqnarray}
and
\begin{eqnarray}
    \hat H_{\rm eff}^{\rm (B)}
    &=&
    -\sum_{\bm k,\bm k',\sigma,\sigma'}
    \frac{1}
    {\varepsilon_{\bm k}-\varepsilon_{d,a}}
    \hat P_{\rm s}^a
    \hat V^{a\bm k'\sigma'}_{d\Leftarrow c}
    \hat V^{a,\bm k,\sigma}_{c\Leftarrow d}
    \hat P_{\rm s}^a
    \approx 
    -   \frac{|v_a|^2}
    {\varepsilon_{\rm F}-\varepsilon_{d,a}}
    \sum_{\bm k,\bm k',\sigma,\sigma'}
    e^{-i(\bm k-\bm k')\cdot \bm R_a}
    \hat P_{\rm s}^a
    \hat d^\dagger_{\sigma'a}\hat c_{\bm k'\sigma'}
    \hat c^\dagger_{\bm k\sigma}
    \hat d_{\sigma,a}
    \hat P_{\rm s}^a.
\end{eqnarray}
The second term in Eq.~\eqref{SIeq: Leff(2B)} vanishes for the following reasons. The term contains a delta function, which enforces the created conduction electron to have energy at $\varepsilon_{\bm k}=\varepsilon_{d,a}$. 
This lies below the Fermi surface $\varepsilon_{\rm F}$, which is forbidden due to the Pauli blocking effect.

Adding the two contributions up and reintroducing the sum over the localized electron sites $a$, we obtain $({\mathcal L}_{\rm eff,sd})_{n_l,[(\sigma_+;F),(\sigma_-;F)]}
    ^{(2)}$, with
\begin{eqnarray}
\label{SIeq: Leff(2) single}
    {\mathcal L}_{\rm eff,sd}^{(2)}
    \hat\rho
    &=&-i[\hat H_{\rm eff},
    \hat\rho
    ]
    + 
    \sum_a
    \gamma_a
    {\mathcal D}
    \Big[
    \sum_{\sigma}
    \hat d_{\sigma,a}^\dagger
    \hat c_{\bm R_a,\sigma}
    \hat P_{\rm s}^a
    \Big]
    \hat\rho.
\end{eqnarray}
Here, the total Hamiltonian $\hat H_{\rm eff}=\hat H_{\rm eff}^{\rm (A)}+\hat H_{\rm eff}^{\rm (B)}$.
The Hamiltonian can be rewritten into a sum of two contributions $\hat H_{\rm eff}=\hat H_{\rm sd}+\hat H_{\rm imp}$.
The \textit{sd} Hamiltonian describes the exchange coupling between the conduction electrons and localized spins,
\begin{eqnarray}
    \hat H_{\rm sd}
    &=&\sum_a
    |v_a|^2
    \bigg[
        \frac{\varepsilon_{d,a}+U_a-\varepsilon_{\rm F}}
        {(\varepsilon_{d,a}+U_a-\varepsilon_{\rm F})^2+\frac{\kappa_a^2}{4}}
        +
        \frac{1}{\varepsilon_{\rm F}-\varepsilon_{d,a}}
    \bigg]
    \sum_{\bm k,\bm k'}
    e^{i(\bm k-\bm k')\cdot\bm R_a}
    \sum_{\sigma,\sigma'}
    \hat P_{\rm s}^a
    \hat c^\dagger_{\bm k'\sigma'}
    \hat c_{\bm k\sigma}
    d^\dagger_{\sigma,a}
    \hat d_{\sigma',a}
    \hat P_{\rm s}^a
    \nonumber\\
    &= &
    -
    \sum_a
    \frac{g_a}{2}
    \sum_{\sigma,\sigma'}
    \sum_{i=0}^3
    \hat P_{\rm s}^a
    (\hat c^\dagger_{\bm R_a,\sigma'}
    \sigma_i^{\sigma,\sigma'}
    \hat c_{\bm R_a\sigma})
    (d^\dagger_{\sigma,a}
    \sigma_i^{\sigma,\sigma'}
    \hat d_{\sigma',a})    
    \hat P_{\rm s}^a
    \nonumber\\
    &= &
    -
    \sum_a
    \frac{g_a}{2}
    \sum_{\sigma,\sigma'}
    \hat P_{\rm s}^a
    [\bm \tau(\bm R_a) \cdot \bm S_a]
    \hat P_{\rm s}^a    
    +{\rm const.},
\end{eqnarray}
where
\begin{eqnarray}
    g_a = -
    |v_a|^2
    \bigg[
        \frac{\varepsilon_{d,a}+U_a-\varepsilon_{\rm F}}
        {(\varepsilon_{d,a}+U_a-\varepsilon_{\rm F})^2+\frac{\kappa_a^2}{4}}
        +
        \frac{1}{\varepsilon_{\rm F}-\varepsilon_{d,a}}
    \bigg]
    (<0)
\end{eqnarray}
is an antiferromagnetic \textit{sd} coupling, 
\begin{eqnarray}
    \bm S_a=
    \begin{pmatrix}
        \hat d_{\uparrow,a}^\dagger &
        \hat d_{\downarrow,a}^\dagger
    \end{pmatrix}
    \bm \sigma
    \begin{pmatrix}
        \hat d_{\uparrow,a} \\
        \hat d_{\downarrow,a}
    \end{pmatrix}
    =(S_a^1,S_a^2,S_a^3)
\end{eqnarray}
is the localized spins (with $\bm \sigma=(\sigma_1,\sigma_2,\sigma_3)$ being the Pauli matrices), and 
\begin{eqnarray}
    \bm \tau(\bm R_a)=
    \begin{pmatrix}
        \hat c_{\bm R_a,\uparrow}^\dagger &
        \hat c_{\bm R_a,\downarrow}^\dagger
    \end{pmatrix}
    \bm \sigma
    \begin{pmatrix}
        \hat c_{\bm R_a,\uparrow}  \\
        \hat c_{\bm R_a,\downarrow}
    \end{pmatrix}
    =(\tau_1(\bm R_a),\tau_2(\bm R_a),\tau_3(\bm R_a))
\end{eqnarray}
is the conduction spin at position $\bm R_a$.
The impurity Hamiltonian $\hat H_{\rm imp}$ is given by
\begin{eqnarray}
    \hat H_{\rm imp}
    &=&-\sum_a
    \sum_{\bm k,\bm k'}
    |v_a|^2
    \frac{\varepsilon_{d,a}+U_a-\varepsilon_{\rm F}}
    {(\varepsilon_{d,a}+U_a-\varepsilon_{\rm F})^2+\frac{\kappa_a^2}{4}}
    \sum_\sigma
    e^{i(\bm k-\bm k')\cdot\bm R_a}
    \hat P_{\rm s}^a
    \hat c^\dagger_{\bm k'\sigma}
    \hat c_{\bm k\sigma}
    \hat P_{\rm s}^a
    -
    \sum_a\sum_{\bm k}
    |v_a|^2
    \frac{1}{\varepsilon_{\rm F}-\varepsilon_{d,a}}
    \sum_{\sigma}
    \hat P_{\rm s}^a
    \hat d^\dagger_{\sigma,a}
    \hat d_{\sigma,a}
    \hat P_{\rm s}^a
    \nonumber\\
    &=&
    \sum_a
    \sum_{\sigma}
    \bigg[
    g^{c,a}_{\rm imp}
    \hat P_{\rm s}^a
    \hat c^\dagger_{\bm R_a,\sigma}
    \hat c_{\bm R_a,\sigma}
    \hat P_{\rm s}^a
    +    g_{\rm imp}^{d,a}
    \hat P_{\rm s}^a
    \hat d^\dagger_{\sigma,a}
    \hat d_{\sigma,a}
    \hat P_{\rm s}^a
    \bigg].
\end{eqnarray}
with
\begin{eqnarray}
    g^{c,a}_{\rm imp}&=&
    -|v_a|^2
    \frac{\varepsilon_{d,a}+U_a-\varepsilon_{\rm F}}
    {(\varepsilon_{d,a}+U_a-\varepsilon_{\rm F})^2+\frac{\kappa_a^2}{4}},
    \\
    g_{\rm imp}^{d,a}
    &=&
    -|v_a|^2
    \frac{1}{\varepsilon_{\rm F}-\varepsilon_{d,a}}.
\end{eqnarray}
In the equilibrium limit $\kappa_a\rightarrow 0$, these reproduce the conventional \textit{sd} Hamiltonian.


A similar calculation can be performed for the contribution for the vacant $\ket{\varnothing}\bra{\varnothing}$ and double-occupied $\ket{\uparrow\downarrow}\bra{\uparrow\downarrow}$ states.
The effective Lindbladian reads, respectively,
\begin{eqnarray}
\label{SIeq: Leff2 vacant}
    &&
    {\mathcal L}_{\rm eff,sd}^{(2)}(
    \hat P_\varnothing^a
    \hat\rho
    \hat P_\varnothing^a
    )
    = \sum_{\bm k,\bm k'}
    \sum_{\sigma,\sigma'}
    \pi|v_a|^2
    \delta(\varepsilon_{\bm k}-\varepsilon_{d,a})
    e^{i(\bm k-\bm k')\cdot\bm R_a}
    \bigg[
    \hat d_{\sigma,a}^\dagger
    \hat c_{\bm k,\sigma}
    \hat P_\varnothing^a
    \hat\rho
    \hat P_\varnothing^a
    \hat c_{\bm k',\sigma'}^\dagger
    \hat d_{\sigma',a}
    -\frac{1}{2}
    \{    
    \hat P_\varnothing^a
    \hat d_{\sigma',a}^\dagger
    \hat c_{\bm k,\sigma'}
    \hat c_{\bm k',\sigma}^\dagger
    \hat d_{\sigma,a}
    \hat P_\varnothing^a
    ,\hat\rho
    \}
    \bigg]
    \nonumber\\
    &&=
    \sum_{\sigma,\sigma'}
    \gamma_{{\rm dis},a,\varnothing}
    \sum_{\bm q}
    e^{i\bm q\cdot\bm R_a}
    \bigg[
    \hat d_{\sigma,a}^\dagger
    \hat c_{\bm k_{d,a},\sigma}
    \hat P_\varnothing^a
    \hat\rho
    \hat P_\varnothing^a
    \hat c_{\bm k_{d,a}-\bm q,\sigma'}^\dagger
    \hat d_{\sigma',a}
    -\frac{1}{2}
    \{    
    \hat P_\varnothing^a
    \hat d_{\sigma',a}^\dagger
    \hat c_{\bm k_{d,a},\sigma'}
    \hat c_{\bm k_{d,a}-\bm q,\sigma}^\dagger
    \hat d_{\sigma,a}
    \hat P_\varnothing^a
    ,\hat\rho
    \}
    \bigg],
    \\
\label{SIeq: Leff2 double}
    &&{\mathcal L}_{\rm eff,sd}^{(2)}
    (\hat P_{\uparrow\downarrow}^a
    \hat\rho
    \hat P_{\uparrow\downarrow}^a)
    =\kappa_a
    \sum_{\sigma}
    {\mathcal D}
    [
    \hat d_{\sigma,a}
    \hat P_{\uparrow\downarrow}^a
    ]
    \hat\rho
    \nonumber\\
    &&+
    \sum_{\bm k,\bm k'}
    \sum_{\sigma,\sigma'}
    \pi|v_a|^2\delta(\varepsilon_{\bm k}-\varepsilon_{d,a}-U_a)
    e^{i(\bm k-\bm k')\cdot\bm R_a}
    \bigg[
    \hat c_{\bm k,\sigma}^\dagger
    \hat d_{\sigma,a}
    \hat P_{\uparrow\downarrow}^a
    \hat\rho
    \hat P_{\uparrow\downarrow}^a
    \hat d_{\sigma',a}^\dagger
    \hat c_{\bm k',\sigma'}
    -\frac{1}{2}
    \{    
    \hat P_{\uparrow\downarrow}^a
    \hat d_{\sigma',a}^\dagger
    \hat c_{\bm k,\sigma'}
    \hat c_{\bm k',\sigma}^\dagger
    \hat d_{\sigma,a}
    \hat P_{\uparrow\downarrow}^a
    ,\hat\rho
    \}
    \bigg]
    \nonumber\\
    &&=
    \kappa_a
    \sum_{\sigma}
    {\mathcal D}
    [
    \hat d_{\sigma,a}
    \hat P_{\uparrow\downarrow}^a
    ]
    \hat\rho
    \nonumber\\
    &&+
    \sum_{\sigma,\sigma'}
    \gamma_{{\rm dis},a,\uparrow\downarrow}
    \sum_{\bm q}
    e^{i\bm q\cdot\bm R_a}
    \bigg[
    \hat c_{\bm k_{d,a}^U,\sigma}^\dagger
    \hat d_{\sigma,a}
    \hat P_{\uparrow\downarrow}^a
    \hat\rho
    \hat P_{\uparrow\downarrow}^a
    \hat d_{\sigma',a}^\dagger
    \hat c_{\bm k_{d,a}^U-\bm q,\sigma'}
    -\frac{1}{2}
    \{    
    \hat P_{\uparrow\downarrow}^a
    \hat d_{\sigma',a}^\dagger
    \hat c_{\bm k_{d,a}^U,\sigma'}
    \hat c_{\bm k_{d,a}^U-\bm q,\sigma}^\dagger
    \hat d_{\sigma,a}
    \hat P_{\uparrow\downarrow}^a
    ,\hat\rho
    \}
    \bigg].
\end{eqnarray}
Here, as we are only interested in the dynamics of singly occupied states, we omitted the contribution to the coherent dynamics in the vacant or double-occupied state.
\begin{eqnarray}
    \gamma_{{\rm dis},a,\varnothing}
    =\pi|v_a|^2
    \sum_{\bm k}\delta(\varepsilon_{\bm k}-\varepsilon_{d,a}),
    \qquad
    \gamma_{{\rm dis},a,\uparrow\downarrow}
    =\pi|v_a|^2
    \sum_{\bm k}\delta(\varepsilon_{\bm k}-\varepsilon_{d,a}-U_a).
\end{eqnarray}
are the dissipation rate of the vacant and double-occupied states, and
$\bm k_d$ and $\bm k_{d,a}^U$ are momenta that satisfies, respectively,
\begin{eqnarray}
    \varepsilon_{\bm k_{d,a}}=\varepsilon_{d,a},
    \qquad
    \varepsilon_{\bm k_{d,a}^U}=\varepsilon_{d,a}+U_a.
\end{eqnarray}
In contrast to the singly occupied state counterparts, (the second term of Eq.~\eqref{SIeq: Leff(2B)}), the Pauli blocking effect does not play a role in these terms and therefore they do not vanish. 

Summing up all the contributions (Eqs.~\eqref{SIeq: Leff(2) single}, \eqref{SIeq: Leff2 vacant}, \eqref{SIeq: Leff2 double}), we finally obtain,
\begin{eqnarray}
    \label{SIeq: sd master equation full}
    &&\partial_t\hat\rho
    =-i[\hat H_{\rm sd}
    +\hat H_{\rm imp},
    \hat\rho 
    ]
    +\sum_a 
    \Big[
    \gamma_a 
    {\mathcal D}
    \big[
    \sum_\sigma
    \hat d_{\sigma,a}^\dagger
    \hat c_{\bm R_a,\sigma}
    \hat P_{\rm s}^a
    \big]  
    +\sum_\sigma
    \kappa_a {\mathcal {D}}[\hat d_{\sigma,a}\hat P_{\uparrow\downarrow}^a]
    \Big]
    \hat\rho
    \nonumber\\
    && +\sum_{\sigma,\sigma'}
    \gamma_{{\rm dis},a,\varnothing}
    \sum_{\bm q}
    e^{i\bm q\cdot\bm R_a}
    \bigg[
    \hat d_{\sigma,a}^\dagger
    \hat c_{\bm k_{d,a},\sigma}
    \hat P_{\rm s}^a
    \hat\rho
    \hat P_{\rm s}^a
    \hat c_{\bm k_{d,a}-\bm q,\sigma'}^\dagger
    \hat d_{\sigma',a}
    -\frac{1}{2}
    \{    
    \hat P_{\rm s}^a
    \hat d_{\sigma',a}^\dagger
    \hat c_{\bm k_{d,a},\sigma'}
    \hat c_{\bm k_{d,a}-\bm q,\sigma}^\dagger
    \hat d_{\sigma,a}
    \hat P_{\rm s}^a
    ,\hat\rho
    \}
    \bigg]
    \nonumber\\
    &&+
    \sum_{\sigma,\sigma'}
    \gamma_{{\rm dis},a,\uparrow\downarrow}
    \sum_{\bm q}
    e^{i\bm q\cdot\bm R_a}
    \bigg[
    \hat c_{\bm k_{d,a}^U,\sigma}^\dagger
    \hat d_{\sigma,a}
    \hat P_{\uparrow\downarrow}^a
    \hat\rho
    \hat P_{\uparrow\downarrow}^a
    \hat d_{\sigma',a}^\dagger
    \hat c_{\bm k_{d,a}^U-\bm q,\sigma'}
    -\frac{1}{2}
    \{    
    \hat P_{\uparrow\downarrow}^a
    \hat d_{\sigma',a}^\dagger
    \hat c_{\bm k_{d,a}^U,\sigma'}
    \hat c_{\bm k_{d,a}^U-\bm q,\sigma}^\dagger
    \hat d_{\sigma,a}
    \hat P_{\uparrow\downarrow}^a
    ,\hat\rho
    \}
    \bigg].
\end{eqnarray}
In the main text in Eq.~(4), we have reported the description without the final two terms, as their role is merely to retain the state in a singly occupied state and they do not play any role in the singly occupied sector we are interested in (but see the remarks below).
We have also dropped the impurity Hamiltonian $\hat H_{\rm imp}$. 
This completes the derivation of Eq.~(4) in the main text.


\subsection{Remarks on Eq.~(4) in the main text and Eq.~\eqref{SIeq: sd master equation full}}
Two remarks are in order.

The first remark is on the condition for justifying the second-order perturbation we employed to derive Eq.~(4) in the main text.
The second-order perturbation is justified when its contribution, roughly given by $\sum_{\bm k}|v|^2/(\varepsilon_{\bm k}-\varepsilon_d+U+i\delta)\sim -|v|^2/U - i\gamma_{\rm dis}$ in the typical case $\varepsilon_d\sim\varepsilon_{\rm F}\sim U$ 
(where we have omitted the site index $a$ and $\gamma_{\rm dis}\sim \gamma_{{\rm dis},\varnothing},\gamma_{{\rm dis},\uparrow\downarrow}$),
is dominant over the higher-order terms.
The fourth-order contribution (note that the third-order contribution is absent for a similar reason to why the first-order contribution is absent) is roughly estimated as 
$|v|^4(\sum_{\bm k}1/(\varepsilon_{\bm k}-\varepsilon_d+U+i\delta)))^{3}$
which has the magnitude of the real part 
$O(|v|^4/U^3, \gamma_{\rm dis}^3/|v|^2)$ and 
the imaginary part $O((|v|^2/U^2)\gamma_{\rm dis}, \gamma_{\rm dis}^2/U)$.
By comparing with the second-order contribution, one finds that the second-order perturbation is justified when $|v|/U, \gamma_{\rm dis}/U, \gamma_{\rm dis}/|v|\ll 1$.
Noting that $\gamma_{\rm dis}\sim |v|^2/W$ (where $W$ is the bandwidth), the final condition reads $|v|/W\ll 1$.

This situation for the dissipation rate $\gamma_{\rm dis}$ (where it must be small compared to $U$), is in stark contrast to the situation for $\kappa$, where the second-order perturbation is better for larger $\kappa$. This is because the timescale of the intermediate state becomes faster for larger $\kappa$.

Secondly, in the light-induced dissipator in Eq.~\eqref{SIeq: sd master equation full}, 
the sum over the spin configuration $\sigma=\uparrow,\downarrow$ is taken \textit{within} the jump operator,
${\mathcal D}[\sum_\sigma \hat d_{\sigma,a}^\dagger
\hat c_{\bm R_a,\sigma}]\hat\rho$, instead of having a form $\sum_\sigma {\mathcal D}[\hat d_{\sigma,a}^\dagger
\hat c_{\bm R_a,\sigma}]\hat\rho$.
This form is quite crucial to induce tunneling between different spin configurations in a \textit{correlated} manner, giving rise to dissipative coupling between different spin configurations between the conduction and localized spins (the last term in Eq.~(5) in the main text).
This can be seen explicitly by operating the dissipator to a singly occupied state. 
For example, when the dissipator of this form is applied to the down-spin state, i.e.,  $\ket{\downarrow}\bra{\downarrow}$,
(where we omit the site index $a$ for simplicity)
\begin{eqnarray}
    &&{\mathcal D}[
    (\hat d_{\uparrow}^\dagger
    \hat c_{\uparrow}
    +\hat d_{\downarrow}^\dagger
    \hat c_{\downarrow})
    ] 
    (\ket{\downarrow}\ket{F}
    \bra{\downarrow}\bra{F})
    =
    (\hat d_{\uparrow}^\dagger\hat c_{\uparrow}
    +\hat d_{\downarrow}^\dagger
    \hat c_{\downarrow})
    (\ket{\downarrow}\ket{F}
    \bra{\downarrow}\bra{F})
    (\hat c_{\uparrow}^\dagger
    \hat d_{\uparrow} 
    +\hat c_{\downarrow}^\dagger
    \hat d_{\downarrow})
    \nonumber\\
    &&
    -\frac{1}{2}
    (\hat c_{\uparrow}^\dagger
    \hat d_{\uparrow}
    +\hat c_{\downarrow}^\dagger
    \hat d_{\downarrow})
    (\hat d_{\uparrow}^\dagger
    \hat c_{\uparrow}
    +\hat d_\downarrow^\dagger
    \hat c_{\downarrow})
    (\ket{\downarrow}\ket{F}
    \bra{\downarrow}\bra{F})
    -\frac{1}{2}
    (\ket{\downarrow}\ket{F}
    \bra{\downarrow}\bra{F})
    (\hat c_{\uparrow}^\dagger
    \hat d_{\uparrow}
    +\hat c_{\downarrow}^\dagger
    \hat d_{\downarrow})
    (\hat d_{\uparrow}^\dagger
    \hat c_{\uparrow}
    +\hat d_\downarrow^\dagger
    \hat c_{\downarrow})
    \nonumber\\
    &&=
    \ket{\uparrow\downarrow}
    \ket{\overline{c,\uparrow}}
    \bra{\uparrow\downarrow}
    \bra{\overline{c,\uparrow}}
    -\hat c_{\uparrow}^\dagger
    \hat c_{\uparrow}
    (\ket{\downarrow}\ket{F}
    \bra{\downarrow}\bra{F})
    +\frac{1}{2}
    \hat c_{\downarrow}^\dagger
    \hat c_{\uparrow}
    (\ket{\uparrow}\ket{F}
    \bra{\downarrow}\bra{F})
    )
    +\frac{1}{2}
    (\ket{\downarrow}\ket{F}
    \bra{\uparrow}\bra{F})
    \hat c_{\uparrow}^\dagger
    \hat c_{\downarrow}
    \nonumber\\
    &&=
    \ket{\uparrow\downarrow}
    \ket{\overline{c,\uparrow}}
    \bra{\uparrow\downarrow}
    \bra{\overline{c,\uparrow}}
    -n
    (\ket{\downarrow}\ket{F}
    \bra{\downarrow}\bra{F})
    +\frac{1}{2}
    \hat c_{\downarrow}^\dagger
    \hat c_{\uparrow}
    (\ket{\uparrow}\ket{F}
    \bra{\downarrow}\bra{F})
    )
    +\frac{1}{2}
    (\ket{\downarrow}\ket{F}
    \bra{\uparrow}\bra{F})
    \hat c_{\uparrow}^\dagger
    \hat c_{\downarrow}
\end{eqnarray}
where $n=(1/2)\bra{F}\sum_\sigma \hat c_\sigma^\dagger \hat c_\sigma\ket{F}$ and $\ket{\overline{c,\uparrow}}=\hat c_\uparrow\ket{F}$.
As is clear from the last two terms, this dissipation involves a spin-flip 
\begin{eqnarray}
    \ket{\downarrow}\bra{\downarrow}
    \rightarrow 
    \ket{\uparrow}\bra{\downarrow},
    \ket{\downarrow}\bra{\uparrow}.
\end{eqnarray}
This is in stark different to the case where the sum over the spins $\sigma$ is outside the jump operator, 
$\sum_\sigma{\mathcal D}[ \hat d_{\sigma,a}^\dagger
\hat c_{\bm R_a,\sigma}]\hat\rho$, in which the $\uparrow$ and $\downarrow$ spins would decay independently. Starting again with the down spin configuration, one finds 
\begin{eqnarray}
    &&
    \sum_\sigma
    {\mathcal D}[
    \hat d_{\sigma}^\dagger
    \hat c_{\sigma}]
    (\ket{\downarrow}\ket{F}
    \bra{\downarrow}\bra{F})
    ={\mathcal D}[
    \hat d_{\uparrow}^\dagger
    \hat c_{\uparrow}]
    (\ket{\downarrow}\ket{F}
    \bra{\downarrow}\bra{F})
\end{eqnarray}
where it is clear that no spin flips are involved.

\section{Landau-Lifshitz-Gilbert equation with light-induced interactions}
\label{SIsec: LLG}

In the previous section, we have derived an effective low-energy description of the localized electron coupled to conduction electrons (Eq.~(4) in the main text or Eq.~\eqref{SIeq: sd master equation full}).
In this section, we derive the equation of motion of the localized spins [Eq.~(6) in the main text] interacting through the RKKY interaction~\cite{Ruderman1954, Kasuya1956, Yosida1957}) modified by the light injection, by integrating out the conduction electrons, treating them as a non-Markovian bath.

To incorporate the non-Markovian effect arising from the Fermi distribution function of conduction electrons, we employ the Keldysh theory introduced in Sec.~\ref{SIsubsec: Keldysh}.
Using Eq.~\eqref{SIeq: action GKSL},
the Keldysh partition function of Eq.~(4) in the main text is given by,
\begin{eqnarray}
    Z=\int{\mathcal D}(d,\bar d)
    {\mathcal D} (c,\bar c)
    e^{iS[c,\bar c,d,\bar d]}
\end{eqnarray}
where the effective action is the sum of three parts:
\begin{eqnarray}
    S[c,\bar c,d,\bar d]=S_d^0[d,\bar d]+S_c^0[c,\bar c]+S_{\rm sd}[c,\bar c,d,\bar d]    
\end{eqnarray}
Here, $d,\bar d$ and $c,\bar c$ are Grassmann variables and
\begin{eqnarray}
    S_d^{0}[d,\bar d]
    &=&
    \int dt 
    \sum_{s=\pm}
    \sum_{a,\sigma}
    s
    \bar d_{\sigma,a}^s(t)
    i\partial_t 
    d_{\sigma,a}^s(t)
    \\
    S_c^0[c,\bar c]
    &=&
    \int dt
    \sum_{s=\pm}\sum_{\bm k,\sigma}
    s\Big[
    \bar c_{\bm k,\sigma}^s (t)
    i\partial_t 
    c_{\bm k,\sigma}^s(t)
    -\varepsilon_{\bm k}
    \bar c_{\bm k,\sigma}^s (t)
    c_{\bm k,\sigma}^s (t)
    \Big], \\
    S_{\rm sd}^{\rm coh}[c,\bar c,d,\bar d]
    &=& - \int dt
    \sum_{s=\pm} s 
    \sum_a 
    \sum_{\bm k,\bm q}
    (-g_a) e^{i\bm q \cdot\bm R_a}
    \sum_{\sigma,\sigma'}
    \bar d_{\sigma,a}^s(t)
    \bar c_{\bm k+\bm q,\sigma'}^s
    (t)
    c_{\bm k,\sigma}^s(t)
    d_{\sigma',a}^s(t)
    \nonumber\\
    &=& \int dt
    \sum_{s=\pm} s 
    \sum_a 
    \sum_{\bm k,\bm q}
    g_a e^{i\bm q \cdot\bm R_a}
    \sum_{\sigma,\sigma'}
    \bar c_{\bm k+\bm q,\sigma'}^s
    (t)
    c_{\bm k,\sigma}^s(t)
    [
    \delta_{\sigma\sigma'}
    -d_{\sigma',a}^s(t_{s\delta})
    \bar d_{\sigma,a}^s(t_{-s\delta})
    ]
    \\
    S_{\rm sd}^{\rm dis}
    [c,\bar c,d,\bar d]
    &=&
    -i
    \int dt 
    \sum_a  
    \sum_{\bm k,\bm q}
    \gamma_a
    e^{i\bm q\cdot\bm R_a}
    \sum_{\sigma,\sigma'}
    \Big[
    \bar c_{\bm k+\bm q,\sigma'}^-(t)
    d_{\sigma',a}^-
    (t)
    \bar d_{\sigma,a}^+(t)
    c_{\bm k,\sigma}^+(t)
    \nonumber\\
    &&
    -\frac{1}{2}
    \bar c_{\bm k+\bm q,\sigma'}^+(t_{+\delta})
    d_{\sigma',a}^+(t_{+\delta})
    \bar d_{\sigma,a}^+(t_{-\delta})
    c_{\bm k,\sigma}^+(t_{-\delta})
    -\frac{1}{2}
    \bar c_{\bm k+\bm q,\sigma'}^-(t_{-\delta})
    d_{\sigma',a}^-(t_{-\delta})
    \bar d_{\sigma,a}^-(t_{+\delta})
    c_{\bm k,\sigma}^-(t_{+\delta})
    \Big]
\end{eqnarray}
with $S_{\rm sd} = S_{\rm sd}^{\rm coh}+S_{\rm sd}^{\rm dis}$. 
$s=\pm$ labels the forward and backward contours.
We have added or subtracted an infinitesimal $t_{\pm \delta}=t\pm 0^+$ to avoid subtleties in normal ordering, see, e.g. Refs.~\cite{Sieberer2016, Sieberer2014, Sieberer2023} for details. 
For the simplicity of notation, we have omitted the projection operator to the singly occupied state and omitted the contribution from the double-occupied state.
The actions involving conduction electrons can be rewritten in a compact form,
\begin{eqnarray}
    &&S_c[d,\bar d,c,\bar c]
    \equiv S_c^0[c,\bar c]+S_{\rm sd}[d,\bar d,c,\bar c]
    =\int dt 
    \sum_{\bm q,\bm k}
    \bar\Psi_{\bm k+\bm q}^{\mathrm{T}}(t)
    \hat{\mathbb{M}}_{\bm k+\bm q,\bm k}[d(t),\bar d(t)]
    \Psi_{\bm k}(t)
\end{eqnarray}
by introducing a spinor-Keldysh representation,
\begin{eqnarray}
    &&\hat{\mathbb{M}}_{\bm k+\bm q,\bm k}(t)
    =(\hat{\mathbb{G}}_0^{-1}(t))_{\bm k}
    \delta_{\bm q,\bm 0}
    -{\bm\Sigma}_{\bm k+\bm q,\bm k}[d(t),\bar d(t)],
    \qquad
    \Psi_{\bm k}(t)=
    \begin{pmatrix}
        c_{\bm k,\uparrow}^+(t) \\
        c_{\bm k,\downarrow}^+(t) \\
        c_{\bm k,\uparrow}^-(t) \\
        c_{\bm k,\downarrow}^-(t) 
    \end{pmatrix}.
\end{eqnarray}
Here,
$\hat{\mathbb{G}}_0^{-1}(t)_{\bm k}$ is a free Green function in the Keldysh formalism, where its Fourier transform is given by (see Eq.~\eqref{SIeq: G0 +-rep Fourier}) \cite{Kamenev2023},
\begin{eqnarray}
    \mathbb{G}_0(\bm k,\omega)
    =\begin{pmatrix}
        G_0^{++}(\bm k,\omega)\bm 1_{2\times 2} 
        &
        G_0^{+-}(\bm k,\omega)\bm 1_{2\times 2} 
        \\        
        G_0^{-+}(\bm k,\omega)\bm 1_{2\times 2} 
        &
        G_0^{--}(\bm k,\omega)\bm 1_{2\times 2} 
    \end{pmatrix}
    =\begin{pmatrix}
        \big[
        \frac{1-f_{\bm k}}
        {\omega-\varepsilon_{\bm k}+i\delta}
        +\frac{f_{\bm k}}
        {\omega-\varepsilon_{\bm k}-i\delta}   
        \big]\bm 1_{2\times 2} &
        2\pi i \delta(\omega-\varepsilon_{\bm k})
        f_{\bm k}\bm 1_{2\times 2}
        \\
        -2\pi i \delta(\omega-\varepsilon_{\bm k})
        (1-f_{\bm k})\bm 1_{2\times 2}
        &
        \big[
        -\frac{f_{\bm k}}{\omega-\varepsilon_{\bm k}+i\delta}
        -
        \frac{1-f_{\bm k}}{\omega-\varepsilon_{\bm k}-i\delta}
        \big]\bm 1_{2\times 2} 
    \end{pmatrix},
    \nonumber\\
\end{eqnarray}
where $f_{\bm k}=[e^{(\varepsilon_{\bm k}-\varepsilon_{\rm F})/(k_{\rm B}T)}+1]^{-1}$ is the Fermi distribution function and 
$\bm 1_{2\times 2}$ is a 2-by-2 identity operator.
The self-energy is given by,
\begin{eqnarray}
    \bm\Sigma_{\bm k+\bm q,\bm k}
    [d(t),\bar d(t)]
    =\begin{pmatrix}
        \bm\Sigma_{\bm k+\bm q,\bm k}^{++}    
        [d(t),\bar d(t)]&
        \bm\Sigma_{\bm k+\bm q,\bm k}^{+-} 
        [d(t),\bar d(t)]\\
        \bm\Sigma_{\bm k+\bm q,\bm k}^{-+}
        [d(t),\bar d(t)] &
        \bm\Sigma_{\bm k+\bm q,\bm k}^{--}
        [d(t),\bar d(t)]
    \end{pmatrix}
\end{eqnarray}
with 
\begin{eqnarray}
\label{SIeq: Sigma++ momentum}
    (\bm\Sigma_{\bm k+\bm q,\bm k}^{++}[d(t),\bar d(t)])_{\sigma',\sigma}
    &=&
    -\sum_a
    \bigg[
    g_a
    e^{i\bm q\cdot\bm R_a}
    \delta_{\sigma\sigma'}
    -    
    \tilde g_a
    e^{i\bm q\cdot\bm R_a}
    d_{\sigma',a}^+
    (t_{+\delta}) 
    \bar d_{\sigma,a}^+
    (t_{-\delta}) 
    \bigg],
    \\
\label{SIeq: Sigma-- momentum}
    (\bm\Sigma_{\bm k+\bm q,\bm k}^{--}[d(t),\bar d(t)])_{\sigma',\sigma}
    &=&
    \sum_a
    \bigg[
    g_a
    e^{i\bm q\cdot\bm R_a}
    \delta_{\sigma\sigma'}
    - 
    \tilde g_a^*
    e^{i\bm q\cdot\bm R_a}
    d_{\sigma',a}^-
    (t_{-\delta}) 
    \bar d_{\sigma,a}^-(t_{+\delta})
    \bigg],
    \\
\label{SIeq: Sigma-+ momentum}
    (\bm\Sigma_{\bm k+\bm q,\bm k}^{-+}[d(t),\bar d(t)])_{\sigma',\sigma}
    &=&
    i    \sum_a
    \gamma_a
    e^{i\bm q\cdot\bm R_a}
    d_{\sigma',a}^-
    (t) 
    \bar d_{\sigma,a}^+
    (t),
    \\
\label{SIeq: Sigma+- momentum}
    (\bm\Sigma_{\bm k+\bm q,\bm k}^{+-}[d(t),\bar d(t)])_{\sigma',\sigma}
    &=& 0
\end{eqnarray}
where
\begin{eqnarray}
    \tilde g_a
    &=&g_a
    -i\frac{\gamma_a}{2}.
\end{eqnarray}
Below, for simplicity, we omit the first term in Eqs.~\eqref{SIeq: Sigma++ momentum} and \eqref{SIeq: Sigma-- momentum}
(which should only quantitatively shift the magnitude of dephasing).
\begin{eqnarray}
    &&\bm\Sigma_{\bm k+\bm q,\bm k}^{++}[d(t),\bar d(t)]
    =
    \sum_a
    e^{i\bm q\cdot\bm R_a}
    \tilde g_a
    \mathbbm{D}_a^{++}(t),
    \\
    &&
    \bm\Sigma_{\bm k+\bm q,\bm k}^{--}[d(t),\bar d(t)]
    =-
    \sum_a
    e^{i\bm q\cdot\bm R_a}
    \tilde g_a^*
    \mathbbm{D}_a^{--}(t),
    \\
    &&\bm\Sigma_{\bm k+\bm q,\bm k}^{-+}[d(t),\bar d(t)]
    =i\sum_a
    e^{i\bm q\cdot\bm R_a}
    \gamma_a
    \mathbbm{D}_a^{-+}(t),
    \\
    &&\bm\Sigma_{\bm k+\bm q,\bm k}^{+-}[d(t),\bar d(t)]
    =0,
\end{eqnarray}
where
\begin{eqnarray}
    \mathbbm{D}_a^{++}(t)
    &=&\begin{pmatrix}
        d_{\uparrow,a}^+
        (t_{+\delta})
        \bar d_{\uparrow,a}^+ 
        (t_{-\delta})
        & 
        d_{\uparrow,a}^+
        (t_{+\delta})
        \bar d_{\downarrow,a}^+ 
        (t_{-\delta})
        \\
        d_{\downarrow,a}^+
        (t_{+\delta})
        \bar d_{\uparrow,a}^+
        (t_{-\delta})
        & 
        d_{\downarrow,a}^+
        (t_{+\delta})
        \bar d_{\downarrow,a}^+
        (t_{-\delta})
    \end{pmatrix},
    \\
    \mathbbm{D}_a^{--}(t)
    &=&\begin{pmatrix}
        d_{\uparrow,a}^-
        (t_{-\delta})
        \bar d_{\uparrow,a}^-
        (t_{+\delta})
        & 
        d_{\uparrow,a}^-
        (t_{-\delta})
        \bar d_{\downarrow,a}^-
        (t_{+\delta})
        \\
        d_{\downarrow,a}^-
        (t_{-\delta})
        \bar d_{\uparrow,a}^-
        (t_{+\delta})
        & 
        d_{\downarrow,a}^-
        (t_{-\delta})
        \bar d_{\downarrow,a}^-
        (t_{+\delta})
    \end{pmatrix},
    \\
    \mathbbm{D}_a^{-+}(t)
    &=&\begin{pmatrix}
        d_{\uparrow,a}^-
        (t)
        \bar d_{\uparrow,a}^+
        (t)
        & 
        d_{\uparrow,a}^-
        (t)
        \bar d_{\downarrow,a}^+
        (t)
        \\
        d_{\downarrow,a}^-
        (t)
        \bar d_{\uparrow,a}^+
        (t)
        & 
        d_{\downarrow,a}^-
        (t)
        \bar d_{\downarrow,a}^+
        (t)
    \end{pmatrix}.
\end{eqnarray}
For later use, we Fourier transform these self-energies as,
\begin{eqnarray}
    \bm\Sigma^{++}(\bm r_1,\bm r_2)[d(t),\bar d(t)]
    &=&
    \sum_a
    \delta((\bm r_1+\bm r_2)/2-\bm R_a)    
    \tilde g_a
    \mathbbm{D}_a^{++}(t)    
    \\
    \bm\Sigma^{--}(\bm r_1,\bm r_2)[d(t),\bar d(t)]
    &=&
    \sum_a
    \delta((\bm r_1+\bm r_2)/2-\bm R_a)    
    (-\tilde g_a)
    \mathbbm{D}_a^{--}(t)
    ,
    \\
    \bm\Sigma^{-+}(\bm r_1,\bm r_2)[d(t),\bar d(t)]
    &=&
    i\sum_a
    \gamma_a
    \delta((\bm r_1+\bm r_2)/2-\bm R_a)    
    \mathbbm{D}_a^{-+}(t),
    \\
    \bm\Sigma^{+-}(\bm r_1,\bm r_2)[d(t),\bar d(t)]
    &=&0,
\end{eqnarray}
where
\begin{eqnarray}
    \bm\Sigma(\bm r_1,\bm r_2,t)
    =\sum_{\bm q}\sum_{\bm k}
    e^{i\bm q\cdot(\bm r_1+\bm r_2)/2}
    e^{i\bm k\cdot(\bm r_1-\bm r_2)}
    \bm\Sigma_{\bm k+\bm q,\bm k}(t).
\end{eqnarray}

We now perform the Grassmann integral over the conduction electrons, regarding them as the (non-Markovian) environment for localized electrons, to obtain the effective dynamics of the localized spin dynamics:
\begin{eqnarray}
    Z=\int{\mathcal D}(d,\bar d)e^{iS_d[d,\bar d]}
    \int {\mathcal D}(c,\bar c) e^{iS_c[d,\bar d,c,\bar c]}
    \equiv \int{\mathcal D}(d,\bar d)e^{iS_{\rm eff}[d,\bar d]}
\end{eqnarray}
where the effective action $S_{\rm eff}[d,\bar d]=S_d^0[d,\bar d]+\Delta S_{\rm eff}[d,\bar d]$ reads
\begin{eqnarray}
    e^{i\Delta S_{\rm eff}[d,\bar d]}
    &=&{\rm det}\Big[
    (-i)
    (\mathbbm{G}_0^{-1}-\bm\Sigma[d,\bar d])    
    \Big]
    \nonumber\\
    &\simeq&
    {\rm exp}\Big[
    -{\rm Tr}
    \big[
    \bm\Sigma[d,\bar d]
    \mathbbm{G}_0
    \big]
    -\frac{1}{2}
    {\rm Tr}
    \big[
    \bm\Sigma[d,\bar d]
    \mathbbm{G}_0
    \bm\Sigma[d,\bar d]
    \mathbbm{G}_0
    \big]\Big]
    +{\rm const.}
    \nonumber\\
    &\equiv&{\rm exp}\Big[
    i\Delta S_{\rm eff}^{(1)}[d,\bar d]
    + i\Delta S_{\rm eff}^{(2)}[d,\bar d]
    \Big]
    +{\rm const.}
\end{eqnarray}
Here, the first-order contribution is 
$\Delta S_{\rm eff}^{(1)}[d,\bar d]
    =i{\rm Tr}
    \big[
    \bm\Sigma
    \mathbbm{G}_0
    \big]$
and the second-order contribution is given by,
$\Delta S_{\rm eff}^{(2)}[d,\bar d]
    =
    \frac{i}{2}
    {\rm Tr}
    \big[
    \bm\Sigma[d,\bar d]
    \mathbbm{G}_0
    \bm\Sigma[d,\bar d]
    \mathbbm{G}_0
    \big]$,
where ${\rm Tr}[\bm A]=\int d\bm r_1\int d\bm r_2\int dt_1 \int dt_2 {\rm tr}_\sigma{\rm tr}_s[A(\bm r_1,t_1;\bm r_2, t_2)]\delta(\bm r_1-\bm r_2)\delta(t_1-t_2)$
and 
${\rm tr}_{\sigma(s)}[\cdots]$ describes the trace in the spin (Keldysh) space.

\subsection{First-order correction: Decay of dipole moment}

The first-order correction reads
\begin{eqnarray}
    \Delta S_{\rm eff}^{(1)}[d,\bar d]
    &=&i{\rm Tr}
    \big[
    \bm\Sigma
    \mathbbm{G}_0
    \big]
    =i{\rm tr}_{\sigma,s}
    \int dt_1 \int d\bm r_1 
    \int dt_2 \int d\bm r_2
    \big[\bm\Sigma(\bm r_1,\bm r_2,t_1)\delta(t_1-t_2)
    \mathbbm{G}_0 (\bm r_2-\bm r_1,t_2-t_1+\delta)
    \big]
    \nonumber\\
    &=&
    i\int dt'
    \sum_{\bm k}
    {\rm tr}_{\sigma,s}
    \big[\bm\Sigma_{\bm k,\bm k}(t')
    \mathbbm{G}_0 (-\bm k,t=\delta) 
    \big]
    =    i\int dt
    \sum_{\bm k}
    \int \frac{d\omega}{2\pi} 
    e^{i\omega s\delta}
    {\rm tr}_{\sigma,s}
    \big[    
    \bm\Sigma_{\bm k,\bm k}(t)
    \mathbbm{G}_0 (\bm k,\omega) 
    \big]
    \nonumber\\    
    &=&    i\int dt
    \sum_{\bm k}
    \int \frac{d\omega}{2\pi} 
    e^{i\omega s\delta}
    {\rm tr}_{\sigma}
    \big[    
    \bm\Sigma_{\bm k,\bm k}^{++}(t)
    \mathbbm{G}_0^{++} (\bm k,\omega) 
    +\bm\Sigma_{\bm k,\bm k}^{--}(t)
    \mathbbm{G}_0^{--} (\bm k,\omega) 
    +\bm\Sigma_{\bm k,\bm k}^{-+}(t)
    \mathbbm{G}_0^{+-} (\bm k,\omega) 
    \big]
    \nonumber\\    
    &=& \int dt
    \sum_a 
    \sum_\sigma
    \bigg[    
    -g_a
    \big[
    d_{\sigma,a}^+   
    (t_{+\delta})
    \bar d_{\sigma,a}^+
    (t_{-\delta})
    -
    d_{\sigma,a}^-    
    (t_{-\delta})
    \bar d_{\sigma,a}^- 
    (t_{+\delta})
    \big]
    \nonumber\\
    && \ \ \ \ \ \ \ \ \ \ \ \ \ 
    -i
    \gamma_a n
    \Big[  
    d_{\sigma,a}^-
    (t)
    \bar d_{\sigma,a}^+ 
    (t)
    -\frac{1}{2} 
    d_{\sigma,a}^+
    (t_{+\delta})
    \bar d_{\sigma,a}^+    
    (t_{-\delta})   
    -
    \frac{1}{2}
    d_{\sigma,a}^-
    (t_{-\delta})   
    \bar d_{\sigma,a}^-        
    (t_{+\delta})   
    \Big]
    \bigg]
\end{eqnarray}
where we have used
\begin{eqnarray}
    \int_{-\infty}^\infty d\omega
    e^{i\omega\delta}
    \mathbbm{G}_0^{++} (\bm k,\omega)
    = 2 \pi i f_{\bm k}, \\
    \int_{-\infty}^\infty d\omega
    e^{-i\omega\delta}
    \mathbbm{G}_0^{--} (\bm k,\omega)
    = 2 \pi i f_{\bm k}, \\
    \int_{-\infty}^\infty d\omega
    \mathbbm{G}_0^{+-} (\bm k,\omega)
    = 2 \pi i f_{\bm k},
\end{eqnarray}
and $\sum_{\bm k}f_{\bm k}=n$, where $n$ is the filling.
Comparing this action with 
Eq.~\eqref{SIeq: action GKSL}, we find that this corresponds to the equation of motion $\partial_t\hat\rho={\mathcal L}^{(1)}_{\rm eff}\hat\rho$ governed by a superoperator,  
\begin{eqnarray}
    {\mathcal L}^{(1)}_{\rm eff}
    \hat\rho
    =-i\sum_a
    \sum_\sigma
    [\hat P_{\rm s}
    g_a
    \hat d_{\sigma,a} \hat d_{\sigma,a}^\dagger
    \hat P_{\rm s},
    \hat\rho]
    +\sum_a
    \sum_\sigma
    \gamma_a n
    {\mathcal D}[\hat d_{\sigma,a}^\dagger
    \hat P_{\rm s}]
    \hat\rho,
\end{eqnarray}
which are the Hartree energy shift (which we ignore below) and an onsite pumping of localized electrons.
The latter gives rise to a decay of magnetic dipole moment, which can be seen from the equation of motion for the averaged magnetic dipole $\avg{\hat {\bm S}_a}={\rm tr}[\hat\rho
\hat {\bm {S}}_a]$,
\begin{eqnarray}
    \avg{\dot S_a^i}
    &=&{\rm tr}[(\partial_t\hat\rho) 
    \hat S_a^i]
    \nonumber\\
    &=&
    \gamma_a
    {\rm tr}
    \big[
    -\frac{1}{2}
    \sum_\nu 
    \hat d_{\nu,a}\hat d_{\nu,a}^\dagger 
    \hat\rho
    \hat S_a^i
    -\frac{1}{2}
    \hat\rho
    \sum_\nu 
    \hat d_{\nu,a}\hat d_{\nu,a}^\dagger 
    \hat S_a^i
    +
    \sum_\nu 
    \hat d_{\nu,a}^\dagger 
    \hat\rho
    \hat d_{\nu,a}
    \hat S_a^i
    \big]
    \nonumber\\
    &=&
    \gamma_a
    \Big[
    -\frac{1}{2}
    \avg{
    \hat S_a^i
    \sum_\nu 
    \hat d_{\nu,a}\hat d_{\nu,a}^\dagger
    }
    -\frac{1}{2}
    \avg{
    \sum_\nu 
    \hat d_{\nu,a}\hat d_{\nu,a}^\dagger 
    \hat S_a^i
    }
    +
    \avg{
    \sum_\nu 
    \hat d_{\nu,a}
    \hat S_a^i
    \hat d_{\nu,a}^\dagger 
    }
    \Big]    
    \nonumber\\
    &=&
    \gamma_a
    \Big[
    -\frac{1}{2}
    \avg{
    \hat S_a^i
    }
    -\frac{1}{2}
    \avg{
    \hat S_a^i
    }
    +
    \avg{
    \sum_\nu 
    \sum_{\mu'\nu'}
    \sigma_i^{\mu'\nu'}
    (\delta_{\mu'\nu}
    -\hat d_{\mu',a}^\dagger
    \hat d_{\nu,a})
    (\delta_{\nu'\nu}-
    \hat d_{\nu,a}^\dagger 
    \hat d_{\nu',a}
    )
    }
    \Big]    
    \nonumber\\
    &=&
    \gamma_a
    \Big[
    -\avg{\hat S_a^i}
    +
    \sum_{\mu'\nu'}
    \avg{
    -(\hat d_{\mu',a}^\dagger
    \sigma_i^{\mu'\nu'}
    \hat d_{\nu',a})
    -
   ( \hat d_{\mu',a}^\dagger 
    \sigma_i^{\mu'\nu'}
    \hat d_{\nu',a})
    +2(\hat d_{\mu',a}^\dagger
    \sigma_i^{\mu'\nu'}
    \hat d_{\nu',a})
    }
    \Big]    
    \nonumber\\
    &=&
    -\gamma_a \avg{S_a^i}.
\end{eqnarray}

\subsection{Second-order correction: RKKY interactions and Gilbert damping}

We now move on to the analysis of the second-order perturbation contribution $S_{\rm eff}^{(2)}$, our central interest, 
which gives rise to the light-modified RKKY interaction and the Gilbert damping:
\begin{eqnarray}
    &&\Delta S_{\rm eff}^{(2)}[d,\bar d]
    =
    \frac{i}{2}
    {\rm Tr}
    \big[
    \bm\Sigma[d,\bar d]
    \mathbbm{G}_0
    \bm\Sigma[d,\bar d]
    \mathbbm{G}_0
    \big]
    \nonumber\\
    &&=    \frac{i}{2}
    \int dt  dt'
    \int d\bm r_1 d\bm r_2 d\bm r_3 d\bm r_4
    {\rm tr}_{\sigma,s}
    \bigg[
    \bm\Sigma(\bm r_1,\bm r_2;t)
    \mathbbm{G}_0(\bm r_2-\bm r_3,t-t')
    \bm\Sigma(\bm r_3,\bm r_4,t')
    \mathbbm{G}_0(\bm r_4-\bm r_1;t'-t)
    \bigg]
    \nonumber\\
    &&=    \frac{i}{2}
    \int dt  dt'
    \sum_{\bm q,\bm k}
    {\rm tr}_{\sigma,s}
    \bigg[
    \bm\Sigma_{\bm k+\bm q,\bm k}(t)
    \mathbbm{G}_0(\bm k+\bm q/2,t-t')
    \bm\Sigma_{\bm k-\bm q,\bm k}(t')
    \mathbbm{G}_0(\bm k-\bm q/2;t'-t)
    \bigg].
\end{eqnarray}
Recall that the self-energy $\bm \Sigma[d(t),\bar d(t)]$ (the Green's function ${\mathbbm G}_0$) is composed of (conduction) localized electrons. 
The timescale of the dynamics of $\bm \Sigma$ is therefore much slower than those of ${\mathbbm G}_0$.
Taking advantage of this property, we perform a gradient expansion as 
\begin{eqnarray}
    &&\Delta S_{\rm eff}^{(2)}[d,\bar d]
    = \frac{i}{2}
    \int dt_g  d\tau
    \sum_{\bm q,\bm k}
    {\rm tr}_{\sigma,s}
    \bigg[
    \bm\Sigma_{\bm k+\bm q,\bm k}(t_g+\tau/2)
    \mathbbm{G}_0(\bm k+\bm q/2,\tau)
    \bm\Sigma_{\bm k-\bm q,\bm k}(t_g-\tau/2)
    \mathbbm{G}_0(\bm k-\bm q/2;-\tau)
    \bigg]
    \nonumber\\
    &&\approx
    S_{\rm eff}^{\rm RKKY}[d,\bar d] 
    + S_{\rm Gilbert}[d,\bar d].
\end{eqnarray}
Here, ($\delta\rightarrow 0^+$)
\begin{eqnarray}
    &&S_{\rm eff}^{\rm RKKY}[d,\bar d] 
    = \frac{i}{2}
    \int dt_g  
    d\tau
    \sum_{\bm q,\bm k}
    {\rm tr}_{\sigma,s}
    \bigg[
    \bm\Sigma_{\bm k+\bm q,\bm k}(t_g+\delta)
    \mathbbm{G}_0(\bm k+\bm q/2,\tau)
    \bm\Sigma_{\bm k-\bm q,\bm k}(t_g-\delta)
    \mathbbm{G}_0(\bm k-\bm q/2;-\tau)
    \bigg]
\end{eqnarray}
is the Markovian limit contribution that, as we will show below, gives rise to the RKKY interaction and its correction from the light-induced dissipation.
The first-order non-Markovian correction is given by,
\begin{eqnarray}
    S_{\rm Gilbert}[d,\bar d]
    = \frac{i}{2}
    \int dt_g  d\tau
    \frac{\tau}{2}
    \sum_{\bm q,\bm k}
    \Bigg[
    {\rm tr}_{\sigma,s}
    \bigg[
    [\partial_{t_g}
    \bm\Sigma_{\bm k+\bm q,\bm k}(t_g+\delta)
    ]
    \mathbbm{G}_0(\bm k+\bm q/2,\tau)
    \bm\Sigma_{\bm k-\bm q,\bm k}(t_g-\delta)
    \mathbbm{G}_0(\bm k-\bm q/2;-\tau)
    \bigg]
    \nonumber\\
    - {\rm tr}_{\sigma,s}
    \bigg[
    \bm\Sigma_{\bm k+\bm q,\bm k}(t_g+\delta)
    \mathbbm{G}_0(\bm k+\bm q/2,\tau)
    [\partial_{t_g}
    \bm\Sigma_{\bm k-\bm q,\bm k}(t_g-\delta)
    ]
    \mathbbm{G}_0(\bm k-\bm q/2;-\tau)
    \bigg]
    \Bigg]
    \nonumber\\
\end{eqnarray}
which will be shown below to give rise to the Gilbert damping.

\subsubsection{Markovian contribution: RKKY interaction}

First, consider the Markovian contribution. 
Taking the trace in the Keldysh space and using the property that $\bm \Sigma^{+-}=0$,
\begin{eqnarray}
\label{SIeq: Seff RKKY 1}
    S_{\rm eff}^{\rm RKKY}[d,\bar d]
    &&=    \frac{i}{2}
    \int dt_g
    \int
    d\tau
    \sum_{\bm q,\bm k}
    {\rm tr}_{\sigma}
    \bigg[
    \bm\Sigma_{\bm k+\bm q,\bm k}^{++}(t_g+\delta)
    \bm\Sigma_{\bm k-\bm q,\bm k}^{++}(t_g-\delta)
    G_0^{++}(\bm k+\bm q/2,\tau)
    G_0^{++}(\bm k-\bm q/2;-\tau)
    \nonumber\\
    && \ \ \ \ \ \ \ \ \ \ \ \ \ \ \ \ \ \ \ \ \ \ \ \ \ 
    +
    \bm\Sigma_{\bm k+\bm q,\bm k}^{++}(t_g+\delta)
    \bm\Sigma_{\bm k-\bm q,\bm k}^{-+}(t_g-\delta)
    G_0^{+-}(\bm k+\bm q/2,\tau)
    G_0^{++}(\bm k-\bm q/2;-\tau)
    \nonumber\\    
    && \ \ \ \ \ \ \ \ \ \ \ \ \ \ \ \ \ \ \ \ \ \ \ \ \ 
    +
    \bm\Sigma_{\bm k+\bm q,\bm k}^{++}(t_g+\delta)
    \bm\Sigma_{\bm k-\bm q,\bm k}^{--}(t_g-\delta)
    G_0^{+-}(\bm k+\bm q/2,\tau)
    G_0^{-+}(\bm k-\bm q/2;-\tau)
    \nonumber\\
    && \ \ \ \ \ \ \ \ \ \ \ \ \ \ \ \ \ \ \ \ \ \ \ \ \ 
    +
    \bm\Sigma_{\bm k+\bm q,\bm k}^{-+}(t_g+\delta)
    \bm\Sigma_{\bm k-\bm q,\bm k}^{++}(t_g-\delta)
    G_0^{++}(\bm k+\bm q/2,\tau)
    G_0^{+-}(\bm k-\bm q/2;-\tau)
    \nonumber\\    
    && \ \ \ \ \ \ \ \ \ \ \ \ \ \ \ \ \ \ \ \ \ \ \ \ \ 
    +
    \bm\Sigma_{\bm k+\bm q,\bm k}^{-+}(t_g+\delta)
    \bm\Sigma_{\bm k-\bm q,\bm k}^{-+}(t_g-\delta)
    G_0^{+-}(\bm k+\bm q/2,\tau)
    G_0^{+-}(\bm k-\bm q/2;-\tau)
    \nonumber\\    
    && \ \ \ \ \ \ \ \ \ \ \ \ \ \ \ \ \ \ \ \ \ \ \ \ \ 
    +
    \bm\Sigma_{\bm k+\bm q,\bm k}^{-+}(t_g+\delta)
    \bm\Sigma_{\bm k-\bm q,\bm k}^{--}(t_g-\delta)
    G_0^{+-}(\bm k+\bm q/2,\tau)
    G_0^{--}(\bm k-\bm q/2;-\tau)
    \nonumber\\    
    && \ \ \ \ \ \ \ \ \ \ \ \ \ \ \ \ \ \ \ \ \ \ \ \ \ 
    +
    \bm\Sigma_{\bm k+\bm q,\bm k}^{--}(t_g+\delta)
    \bm\Sigma_{\bm k-\bm q,\bm k}^{++}(t_g-\delta)
    G_0^{-+}(\bm k+\bm q/2,\tau)
    G_0^{+-}(\bm k-\bm q/2;-\tau)
    \nonumber\\
    && \ \ \ \ \ \ \ \ \ \ \ \ \ \ \ \ \ \ \ \ \ \ \ \ \ 
    +
    \bm\Sigma_{\bm k+\bm q,\bm k}^{--}(t_g+\delta)
    \bm\Sigma_{\bm k-\bm q,\bm k}^{-+}(t_g-\delta)
    G_0^{--}(\bm k+\bm q/2,\tau)
    G_0^{+-}(\bm k-\bm q/2,-\tau)
    \nonumber\\
    && \ \ \ \ \ \ \ \ \ \ \ \ \ \ \ \ \ \ \ \ \ \ \ \ \ 
    +
    \bm\Sigma_{\bm k+\bm q,\bm k}^{--}(t_g+\delta)
    \bm\Sigma_{\bm k-\bm q,\bm k}^{--}(t_g-\delta)
    G_0^{--}(\bm k+\bm q/2,\tau)
    G_0^{--}(\bm k-\bm q/2;-\tau)
    \bigg].
\end{eqnarray}
The conduction electron Green's functions can be written as \cite{Kamenev2023},
\begin{subequations}
\label{SIeq: Green's function lesser greater}
    \begin{align}    
    &G_0^{++}(\bm k,t-t')
    =\theta(t-t')
    G_0^>(\bm k,t-t')
    +
    \theta(t' - t)
    G_0^<(\bm k,t-t'),
    \\
    &G_0^{--}(\bm k,t-t')
    =\theta(t'-t)
    G_0^>(\bm k,t-t')
    +
    \theta(t - t')
    G_0^<(\bm k,t-t'),
    \\
    &G_0^{+-}(\bm k,t-t')
    =
    G_0^<(\bm k,t-t'),
    \\
    &G_0^{-+}(\bm k,t-t')
    =
    G_0^>(\bm k,t-t'),
    \end{align}
\end{subequations}
where the lesser ($s=<$) and greater ($s=>$) Green's function are collectively expressed as (in Fourier space),
\begin{eqnarray}
    G_0^s(\bm k,\omega)
    =F^s_{\bm k}
    [G_0^{\rm R}(\bm k,\omega)
    -G_0^{\rm A}(\bm k,\omega)]
\end{eqnarray}
with
\begin{eqnarray}
    &&G_0^{\rm R}(\bm k,\omega)=\frac{1}{\omega-\varepsilon_{\bm k}+i\delta},
    \qquad
    G_0^{\rm A}(\bm k,\omega)=\frac{1}{\omega-\varepsilon_{\bm k}-i\delta},
    \\
    &&F^<(\bm k)= - f_{\bm k},
    \qquad
    F^>(\bm k)= 1 - f_{\bm k}.
\end{eqnarray} 
Substituting this expression into Eq.~\eqref{SIeq: Seff RKKY 1}, 
we encounter integrals of the form
($s_1,...,s4=+,-$ and $s_a,s_b = <,>$),
\begin{eqnarray}
    \int_{-\infty}^\infty d\tau
    \theta(\tau)
    G_0^{s_a}(\bm k+\bm q/2,\tau)
    G_0^{s_b}(\bm k-\bm q/2;-\tau)
    &=&i
    \frac{F_{+}^{s_a} F_{-}^{s_b}}
    {\varepsilon_+ -\varepsilon_-  -i\delta},
    \\
    \int_{-\infty}^\infty d\tau
    \theta(-\tau)
    G_0^{s_a}(\bm k+\bm q/2,\tau)
    G_0^{s_b}(\bm k-\bm q/2;-\tau)
    &=&
    -i
    \frac{F_{+}^{s_a} F_{-}^{s_b}}{\varepsilon_+ - \varepsilon_- + i\delta},
\end{eqnarray}
where
\begin{eqnarray}
    \varepsilon_\pm \equiv \varepsilon_{\bm k\pm\bm q/2},
    \qquad
    F_\pm^s \equiv F^s_{\bm k\pm\bm q/2}.
\end{eqnarray}

Using these relations, one obtains,
\begin{eqnarray}
    &&S^{\rm RKKY}_{\rm eff}[d,\bar d]
    \approx    
    -\frac{1}{2}
    \int dt
    \nonumber\\
    &&
    \times 
    \sum_{\bm q,\bm k}
    {\rm tr}_{\sigma}
    \bigg[
    \frac{F_{+}^{>} F_{-}^{<}}
    {\varepsilon_+ - \varepsilon_- - i\delta} 
    \bm\Sigma_{\bm k+\bm q,\bm k}^{++}(t + \delta )
    \bm\Sigma_{\bm k-\bm q,\bm k}^{++}(t - \delta)  
    -    \frac{F_{+}^{<} F_{-}^{>}}
    {\varepsilon_+ - \varepsilon_- + i\delta} 
    \bm\Sigma_{\bm k+\bm q,\bm k}^{++}(t - \delta)
    \bm\Sigma_{\bm k-\bm q,\bm k}^{++}(t + \delta)  
    \nonumber\\
    &&+
    \frac{F_{+}^{<} F_{-}^{<}}
    {\varepsilon_+ - \varepsilon_- - i\delta} 
    \bm\Sigma_{\bm k+\bm q,\bm k}^{++}(t + \delta )
    \bm\Sigma_{\bm k-\bm q,\bm k}^{-+}(t - \delta)  
    -    \frac{F_{+}^{<} F_{-}^{>}}
    {\varepsilon_+ - \varepsilon_- + i\delta} 
    \bm\Sigma_{\bm k+\bm q,\bm k}^{++}(t - \delta)
    \bm\Sigma_{\bm k-\bm q,\bm k}^{-+}(t + \delta)  
    \nonumber\\    
    &&
    +
    \frac{F_{+}^{<} F_{-}^{>}}
    {\varepsilon_+ - \varepsilon_- - i\delta} 
    \bm\Sigma_{\bm k+\bm q,\bm k}^{++}
    (t + \delta)
    \bm\Sigma_{\bm k-\bm q,\bm k}^{--}
    (t - \delta)
    -
    \frac{F_{+}^{<} F_{-}^{>}}
    {\varepsilon_+ - \varepsilon_- + i\delta} 
    \bm\Sigma_{\bm k+\bm q,\bm k}^{++}(t -\delta)
    \bm\Sigma_{\bm k-\bm q,\bm k}^{--}(t +\delta)
    \nonumber\\
    && +
    \frac{F_{+}^{>} F_{-}^{<}}
    {\varepsilon_+ - \varepsilon_- - i\delta} 
    \bm\Sigma_{\bm k+\bm q,\bm k}^{-+}(t + \delta)
    \bm\Sigma_{\bm k-\bm q,\bm k}^{++}
    (t - \delta)
    -    \frac{F_{+}^{<} F_{-}^{<}}
    {\varepsilon_+ - \varepsilon_- + i\delta} 
    \bm\Sigma_{\bm k+\bm q,\bm k}^{-+}(t -\delta)
    \bm\Sigma_{\bm k-\bm q,\bm k}^{++}(t +\delta)
    \nonumber\\    
    && 
    +\frac{F_{+}^{<} F_{-}^{>}}
    {\varepsilon_+ - \varepsilon_- - i\delta} 
    \bm\Sigma_{\bm k+\bm q,\bm k}^{-+}(t + \delta)
    \bm\Sigma_{\bm k-\bm q,\bm k}^{--}(t - \delta)
    -    \frac{F_{+}^{<} F_{-}^{<}}
    {\varepsilon_+ - \varepsilon_- + i\delta} 
    \bm\Sigma_{\bm k+\bm q,\bm k}^{-+}(t -\delta)
    \bm\Sigma_{\bm k-\bm q,\bm k}^{--}(t +\delta)
    \nonumber\\
    && +
    \frac{F_{+}^{<} F_{-}^{<}}
    {\varepsilon_+ - \varepsilon_- - i\delta} 
    \bm\Sigma_{\bm k+\bm q,\bm k}^{--}(t + \delta)
    \bm\Sigma_{\bm k-\bm q,\bm k}^{-+}(t - \delta)
    -\frac{F_{+}^{>} F_{-}^{<}}
    {\varepsilon_+ - \varepsilon_- + i\delta} 
    \bm\Sigma_{\bm k+\bm q,\bm k}^{--}(t -\delta)
    \bm\Sigma_{\bm k-\bm q,\bm k}^{-+}(t  +\delta)
    \nonumber\\
    && +
    \frac{F_{+}^{<} F_{-}^{>}}
    {\varepsilon_+ - \varepsilon_- - i\delta} 
    \bm\Sigma_{\bm k+\bm q,\bm k}^{--}(t + \delta)
    \bm\Sigma_{\bm k-\bm q,\bm k}^{--}(t - \delta)
    -    \frac{F_{+}^{>} F_{-}^{<}}
    {\varepsilon_+ - \varepsilon_- + i\delta} 
    \bm\Sigma_{\bm k+\bm q,\bm k}^{--}(t-\delta)
    \bm\Sigma_{\bm k-\bm q,\bm k}^{--}(t +\delta)
    \nonumber\\    
    && +
    \frac{F_{+}^{<} F_{-}^{<}}
    {\varepsilon_+ - \varepsilon_- - i\delta}
    \bm\Sigma_{\bm k+\bm q,\bm k}^{-+}(t + \delta)
    \bm\Sigma_{\bm k-\bm q,\bm k}^{-+}(t - \delta)
    -    \frac{F_{+}^{<} F_{-}^{<}}
    {\varepsilon_+ - \varepsilon_- + i\delta}     
    \bm\Sigma_{\bm k+\bm q,\bm k}^{-+}(t -\delta)
    \bm\Sigma_{\bm k-\bm q,\bm k}^{-+}(t +\delta)
    \nonumber\\    
    && +
    \frac{F_{+}^{>} F_{-}^{<}}
    {\varepsilon_+ - \varepsilon_- - i\delta}
    \bm\Sigma_{\bm k+\bm q,\bm k}^{--}(t + \delta)
    \bm\Sigma_{\bm k-\bm q,\bm k}^{++}(t - \delta)
    -    \frac{F_{+}^{>} F_{-}^{<}}
    {\varepsilon_+ - \varepsilon_- + i\delta}
    \bm\Sigma_{\bm k+\bm q,\bm k}^{--}(t -\delta)
    \bm\Sigma_{\bm k-\bm q,\bm k}^{++}(t +\delta)
    \bigg]
\end{eqnarray}
We proceed by plugging in the explicit form of $\Sigma$, 
\begin{eqnarray}
    &&\sum_{\bm k,\bm q}
    \frac{F_{+}^{s_a} F_{-}^{s_b}}
    {\varepsilon_+ - \varepsilon_- \mp i\delta}
    \bm\Sigma_{\bm k+\bm q,\bm k}^{s_1 s_2}(t \pm \delta)
    \bm\Sigma_{\bm k-\bm q,\bm k}^{s_3 s_4}(t \mp \delta)
    \nonumber\\
    &&=
    \sum_{a,b}
    \sum_{\bm k,\bm q}
    \frac{F_{+}^{s_a} F_{-}^{s_b}}{\varepsilon_+ - \varepsilon_- \mp i\delta}
    e^{i\bm q\cdot(\bm R_a-\bm R_b)}
    \bm \Sigma_{a}^{s_1,s_2}(t\pm \delta)
    \bm \Sigma_{b}^{s_3,s_4}(t\mp \delta)
    \nonumber\\
    &&=
    \sum_{a,b}
    \sum_{\bm k,\bm q}
    \frac{F_{+}^{s_a} F_{-}^{s_b}}{\varepsilon_+ - \varepsilon_- \mp i\delta}
    \cos(\bm q\cdot \bm R_{a,b})
    \bm \Sigma_{a}^{s_1,s_2}(t\pm \delta)
    \bm \Sigma_{b}^{s_3,s_4}(t\mp \delta)
    \label{SIeq: sum k q Sigma Sigma}
\end{eqnarray}
where $\bm R_{a,b}=\bm R_a-\bm R_b$ and
\begin{eqnarray}
    \bm \Sigma_{\bm k+\bm q,\bm k}^{s_1,s_2}(t)
    = \sum_a 
    e^{i\bm q\cdot \bm R_{a}}
    \bm \Sigma_{a}^{s_1,s_2}(t)
\end{eqnarray}
with
\begin{eqnarray}
    \bm \Sigma_{a}^{++}(t)
    &=& 
    \Big(
    g_a - i\frac{\gamma_a}{2}
    \Big)
    \mathbbm{D}_{a}^{++}(t),
    \\
    \bm \Sigma_{a}^{--}(t)
    &=& 
    - \Big(
    g_a + i\frac{\gamma_a}{2}
    \Big)
    \mathbbm{D}_{a}^{--}(t),
    \\
    \bm \Sigma_{a}^{-+}(t)
    &=& i\gamma_a
    \mathbbm{D}_{a}^{-+}(t),    
    \\
    \bm \Sigma_{a}^{+-}(t)
    &=& 0.
\end{eqnarray}
In the final equality of Eq.~\eqref{SIeq: sum k q Sigma Sigma}, we have used the property that 
$\varepsilon_\pm \equiv\varepsilon_{\bm k\pm\bm q/2}, F_\pm^s \equiv F^s_{\bm k\pm\bm q/2}$, and $\bm \Sigma_{a}^{s,s'}$ are symmetric under the transformation
$(\bm k, \bm q)\rightarrow(-\bm k,-\bm q)$.

After a lengthy but straightforward calculation, we arrive at,
\begin{eqnarray}
    &&
    S_{\rm eff}^{\rm RKKY}[d,\bar d]
    =
    \int dt
    \sum_{a,b}
    \bigg[
    J_{a,b}(\bm R_{a,b})
    \Big(
    {\rm tr}_{\sigma}
    \big[
    \mathbbm{D}_{a}^{++}(t+\delta)
    \mathbbm{D}_{b}^{++}(t-\delta)
    \big]
    -
    {\rm tr}_{\sigma}
    \big[
    \mathbbm{D}_{a}^{--}(t-\delta)
    \mathbbm{D}_{b}^{--}(t+\delta)
    \big]
    \Big)
    \nonumber\\
    &&
    +i
    \Lambda_{a,b}(\bm R_{a,b})
    \Big(
    {\rm tr}_{\sigma}
    \big[
    \mathbbm{D}_{a}^{++}(t+\delta)
    \mathbbm{D}_{b}^{++}(t-\delta)
    \big]
    +
    {\rm tr}_{\sigma}
    \big[
    \mathbbm{D}_{a}^{--}(t-\delta)
    \mathbbm{D}_{b}^{--}(t+\delta)
    \big]
    \Big)
    \nonumber\\
    &&
    +[
    H_{a,b}(\bm R_{a,b})
    -i\Gamma_{a,b}(\bm R_{a,b})
    ]
    {\rm tr}_\sigma
    \big[
    \mathbbm{D}_{a}^{--}(t)
    \mathbbm{D}_{b}^{++}(t)
    \big]
    \nonumber\\    &&
    +
    P_{a,b}(\bm R_{a,b})
    \Big(
    {\rm tr}_\sigma
    \big[
    \mathbbm{D}_{a}^{-+}(t+\delta)
    \mathbbm{D}_{b}^{++}(t-\delta)
    \big]
    -
    {\rm tr}_\sigma 
    \big[
    \mathbbm{D}_{b}^{--}(t-\delta)
    \mathbbm{D}_{a}^{-+}(t+\delta)
    \big]
    \Big)
    \nonumber\\
    &&
    -i
    \Omega_{a,b}(\bm R_{a,b})
    \Big(
    {\rm tr}_\sigma
    \big[
    \mathbbm{D}_{a}^{-+}(t+\delta)
    \mathbbm{D}_{b}^{++}(t-\delta)
    \big]
    +
    {\rm tr}_\sigma 
    \big[
    \mathbbm{D}_{b}^{--}(t-\delta)
    \mathbbm{D}_{a}^{-+}(t+\delta)
    \big]
    \Big)
    \nonumber\\   
    &&
    -
    W_{a,b}(\bm R_{a,b})
    \Big(
    {\rm tr}_\sigma
    \big[
    \mathbbm{D}_{a}^{++}(t+\delta)
    \mathbbm{D}_{b}^{-+}(t-\delta)
    \big]
    -
    {\rm tr}_\sigma 
    \big[
    \mathbbm{D}_{b}^{-+}(t-\delta)
    \mathbbm{D}_{a}^{--}(t+\delta)
    \big]
    \Big)
    \nonumber\\
    &&
    -i
    \Phi_{a,b}(\bm R_{a,b})
    \Big(
    {\rm tr}_\sigma
    \big[
    \mathbbm{D}_{a}^{++}(t+\delta)
    \mathbbm{D}_{b}^{-+}(t-\delta)
    \big]    
    +    
    {\rm tr}_\sigma 
    \big[
    \mathbbm{D}_{b}^{-+}(t-\delta)
    \mathbbm{D}_{a}^{--}(t+\delta)
    \big]
    -
    2{\rm tr}_\sigma 
    \big[
    \mathbbm{D}_{a}^{-+}(t+\delta)
    \mathbbm{D}_{b}^{-+}(t-\delta)
    \big]
    \Big)
    \bigg]
    \nonumber\\
\end{eqnarray}
where
\begin{eqnarray}
    J_{a,b}(\bm R_{a,b})
    &=&f_{\rm RKKY}(\bm R_{a,b})
    \Big(
    g_a g_b - \frac{\gamma_a \gamma_b}{4}
    \Big)
    -    \frac{1}{2}
    m_{\rm dis}(\bm R_{a,b})    
    [(-g_a) \gamma_b +(-g_b) \gamma_a],
    \\
    \Lambda_{a,b}(\bm R_{a,b})
    &=&
    m_{\rm dis}(\bm R_{a,b})    
    \Big(
    g_a g_b - \frac{\gamma_a \gamma_b}{4}
    \Big)
    + \frac{1}{2}
    f_{\rm RKKY}(\bm R_{a,b})
    [(-g_a) \gamma_b +(-g_b)\gamma_a], 
    \\
    H_{a,b}(\bm R_{a,b})-i\Gamma_{a,b}(\bm R_{a,b})
    &=&    m_{\rm dis}(\bm R_{a,b})
    \Big[
    [(-g_a) \gamma_b - (-g_b) \gamma_a]
    -2 i
    \Big(
    g_a g_b + \frac{\gamma_a\gamma_b}{4}
    \Big)
    \Big],
    \\
    P_{a,b}(\bm R_{a,b})
    &=&    f_{\rm RKKY}(\bm R_{a,b})
    \frac{\gamma_a
    \gamma_b}{2}
    +m_{\rm dis}(\bm R_{a,b})
    \gamma_a
    (-g_b) , 
    \\
    \Omega_{a,b}(\bm R_{a,b})
    &=& 
    f_{\rm RKKY}(\bm R_{a,b})
    \gamma_a
    (-g_b) 
    - m_{\rm dis}(\bm R_{a,b})
    \frac{\gamma_a
    \gamma_b}{2},
    \\
    W_{a,b}(\bm R_{a,b})
    &=&
    h_{\rm dis}(\bm R_{a,b})
    (-g_a) \gamma_b,
    \\
    \Phi_{a,b}(\bm R_{a,b})
    &=& h_{\rm dis}(\bm R_{a,b})
    \frac{\gamma_a \gamma_b}{2}. 
\end{eqnarray}
and
\begin{eqnarray}
    f_{\rm RKKY}(\bm R_{a,b})
    &\equiv &
    -
    \frac{1}{2}
    \sum_{\bm k,\bm q}\cos(\bm q\cdot\bm R_{a,b})
    \frac{F_{+}^{>} F_{-}^{<} - F_{+}^{<} F_{-}^{>}}
    {\varepsilon_+ - \varepsilon_-}
    =-
    \frac{1}{2}
    \sum_{\bm q}
    \cos(\bm q\cdot\bm R_{a,b})
    \frac{(1-f_+)(-f_-) - (-f_+) (1-f_-))}
    {\varepsilon_+ - \varepsilon_-}   
    \nonumber\\
    &=&
    -
    \frac{1}{2}
    \sum_{\bm k,\bm q}
    \cos(\bm q\cdot\bm R_{a,b})
    \frac{f_+ - f_- }
    {\varepsilon_+ - \varepsilon_-} ,
    \\
    m_{\rm dis}(\bm R_{a,b})
    &\equiv &
    -
    \frac{1}{2}
    \sum_{\bm k,\bm q}
    \cos(\bm q\cdot\bm R_{a,b})
    \pi\delta(\varepsilon_+ - \varepsilon_-)
    (F_+^> F_-^< + F_+^< F_-^>)
    \nonumber\\
    &=&
    \frac{1}{2}
    \sum_{\bm k,\bm q}
    \cos(\bm q\cdot\bm R_{a,b})
    \pi\delta(\varepsilon_+ - \varepsilon_-)
    ((1-f_+) f_- + f_+ (1-f_-))  ,
    \\
    h_{\rm dis}(\bm R_{a,b})
    &\equiv &
    \frac{1}{2}
    \sum_{\bm k,\bm q}
    \cos(\bm q\cdot\bm R_{a,b})
    2\pi\delta(\varepsilon_+ - \varepsilon_-)
    F_+^< F_-^< 
    =
    \frac{1}{2}
    \sum_{\bm k,\bm q}
    \cos(\bm q\cdot\bm R_{a,b})
    2\pi\delta(\varepsilon_+ - \varepsilon_-)
    f_+ f_-
\end{eqnarray}
Here, $J_{a,b}(\bm R_{a,b})$ is identical to the well-known form of the RKKY interaction in the equilibrium limit $\gamma_a=0$. 
We note $m_{\rm dis}(\bm R_{a,b})$ 
vanishes at $T=0$.

One can check that this action has the causality structure $S_{\rm eff}^{\rm RKKY}[d,\bar d]|_{d_+= d_-,\bar d_+=\bar d_-}=0$~\cite{Kamenev2023}, implying the trace conserving property $\partial_t{\rm tr}[\hat\rho] =0$ \cite{Sieberer2016}.

Let us first show that the conventional RKKY interaction \cite{Ruderman1954, Kasuya1956, Yosida1957} recovers in the equilibrium limit $\gamma_a=0$.
In this limit, the effective action reads,
\begin{eqnarray}
    &&
    S_{\rm eq}^{\rm RKKY}[d,\bar d]
    =\int dt 
    \sum_{a,b}
    \bigg[
    J_{a,b}(\bm R_{a,b})
    \Big(
    {\rm tr}_{\sigma}
    \big[
    \mathbbm{D}_{a}^{++}(t+\delta)
    \mathbbm{D}_{b}^{++}(t-\delta)
    \big]
    -
    {\rm tr}_{\sigma}
    \big[
    \mathbbm{D}_{a}^{--}(t-\delta)
    \mathbbm{D}_{b}^{--}(t+\delta)
    \big]
    \Big)
    \nonumber\\
    &&
    \ \ \ \ \ \ \ \ \ \ \ \ \ \ \ \ \ \ \ \ \ 
    + i
    g_a g_b m_{\rm dis}(\bm R_{a,b})
    \Big(
    {\rm tr}_{\sigma}
    \big[
    \mathbbm{D}_{a}^{++}(t+\delta)
    \mathbbm{D}_{b}^{++}(t-\delta)
    \big]
    +
    {\rm tr}_{\sigma}
    \big[
    \mathbbm{D}_{a}^{--}(t-\delta)
    \mathbbm{D}_{b}^{--}(t+\delta)
    \big]
    \Big)
    \nonumber\\
    &&
    \ \ \ \ \ \ \ \ \ \ \ \ \ \ \ \ \ \ \ \ \ 
    -2i
    g_a g_b
    m_{\rm dis}(\bm R_{a,b})
    {\rm tr}_\sigma
    \big[
    \mathbbm{D}_{a}^{--}(t)
    \mathbbm{D}_{b}^{++}(t)
    \big]
    \bigg].
\end{eqnarray}
Using the relation,
\begin{eqnarray}
    {\rm tr}_{\sigma}[
    \mathbbm{D}^{s_1,s_2}_a(t)
    \mathbbm{D}^{s_3,s_4}_b(t)
    ]
    &=&{\rm tr}_\sigma
    \bigg[
    \begin{pmatrix}
        d_{\uparrow,a}^{s_1}
        \bar d_{\uparrow,a}^{s_2} & 
        d_{\uparrow,a}^{s_1}
        \bar d_{\downarrow,a}^{s_2} \\
        d_{\downarrow,a}^{s_1}
        \bar d_{\uparrow,a}^{s_2} & 
        d_{\downarrow,a}^{s_1}
        \bar d_{\downarrow,a}^{s_2} 
    \end{pmatrix}
    \begin{pmatrix}
        d_{\uparrow,b}^{s_3}
        \bar d_{\uparrow,b}^{s_4} & 
        d_{\uparrow,b}^{s_3}
        \bar d_{\downarrow,b}^{s_4} \\
        d_{\downarrow,b}^{s_3}
        \bar d_{\uparrow,b}^{s_4} & 
        d_{\downarrow,b}^{s_3}
        \bar d_{\downarrow,b}^{s_4} 
    \end{pmatrix}
    \bigg]
    \nonumber\\
    &=&d_{\uparrow,a}^{s_1}
        \bar d_{\uparrow,a}^{s_2}
        d_{\uparrow,b}^{s_3}
        \bar d_{\uparrow,b}^{s_4}      
        +d_{\downarrow,a}^{s_1}
        \bar d_{\downarrow,a}^{s_2} 
        d_{\downarrow,b}^{s_3}
        \bar d_{\downarrow,b}^{s_4}
        +d_{\uparrow,a}^{s_1}
        \bar d_{\downarrow,a}^{s_2}  
        d_{\downarrow,b}^{s_3}
        \bar d_{\uparrow,b}^{s_4}
        +        d_{\downarrow,a}^{s_1}
        \bar d_{\uparrow,a}^{s_2}
        d_{\uparrow,b}^{s_3}
        \bar d_{\downarrow,b}^{s_4}    
    \nonumber\\
    &=&
    \frac{1}{2}
        \sum_{i=0}^3
        \sum_{\mu ,\nu =\uparrow,\downarrow}
        \sum_{\mu',\nu'=\uparrow,\downarrow}
        (d_{\nu,a}^{s_1}
        \sigma_i^{\mu\nu} \bar d_{\mu,a}^{s_2})
        (d_{\nu',b}^{s_3}
        \sigma_i^{\mu'\nu'}
        \bar d_{\mu',b}^{s_4} ),  
\end{eqnarray}
we get
\begin{eqnarray}
    S^{\rm RKKY}_{\rm eq}[d,\bar d]
    &=&
    \int dt \sum_{a,b}
    \sum_{j=0}^3
    \bigg[
    \frac{J_{a,b}(\bm R_{a,b})}{2}
    \sum_{s=\pm}s
    \hat m_{a,j}^{s,s}
    \hat m_{b,j}^{s,s}
    \nonumber\\
    &&+\frac{i}{2}
    g_a g_b m_{\rm dis}(\bm R_{a,b})
    \Big[
    \hat m_{a,j}^{+,+}
    \hat m_{b,j}^{+,+}
    +
    \hat m_{a,j}^{-,-}
    \hat m_{b,j}^{-,-}
    -2 \hat m_{a,j}^{-,-}
    \hat m_{b,j}^{+,+}
    \Big]
    \bigg]
\end{eqnarray}
where ($l_1,l_2=+,-$ )
\begin{eqnarray}
    \hat m_{a,j}^{l_1,l_2}[d,\bar d]
    =\sum_{\mu,\nu=\uparrow,\downarrow}
    \bar d_{\mu,a}^{l_1}
    \hat \sigma_j^{\mu\nu}
    d_{\nu,a}^{l_2}.
\end{eqnarray}
We find by comparing with Eq.~\eqref{SIeq: action GKSL} that the corresponding Liouville equation is,
\begin{eqnarray}
    \partial_t \hat\rho 
    =-i [\hat H_{\rm RKKY}, \hat\rho]
    +\sum_{j=1}^3 
    {\mathcal D}
    \Big[\sum_{a}g_a \sqrt{m_{\rm dis}(\bm R_{a,b})}
    \hat S_{a,j}\Big]
    \hat\rho
\end{eqnarray}
with
$\hat H_{\rm RKKY}= \sum_{a,b}(J_{a,b}/2)\bm {\hat S}_a\cdot \bm {\hat S}_b$
and $(\hat{\bm S}_a)_j
=\sum_{\sigma,\sigma'} \hat d_{\sigma,a}^\dagger (\sigma_j)_{\sigma\sigma'}
\hat d_{\sigma',a}$ is the localized spin operator.
This is identical to those found in Ref.~\cite{Rikitake2005},
where the dissipator describes the decoherence effect on the spins that become non-zero at finite temperature $T>0$.


Having checked that our formalism reproduces the known results in the equilibrium limit,
we now add back the light-induced terms.
For simplicity, we restrict ourselves to $T=0$ and the case $\gamma_a\ll |g_a|$, corresponding to the case where $\kappa_a\ll U_a$.
Noting that $m_{\rm dis}(\bm R_{a,b})$ vanishes at the zero temperature limit, the effective Keldysh action simplifies to
\begin{eqnarray}
    &&
    S_{\rm eff}^{\rm RKKY}[d,\bar d]
    =
    \int dt
    \sum_{a,b}
    \bigg[
    J_{a,b}(\bm R_{a,b})
    \Big(
    {\rm tr}_{\sigma}
    \big[
    \mathbbm{D}_{a}^{++}(t+\delta)
    \mathbbm{D}_{b}^{++}(t-\delta)
    \big]
    -
    {\rm tr}_{\sigma}
    \big[
    \mathbbm{D}_{a}^{--}(t-\delta)
    \mathbbm{D}_{b}^{--}(t+\delta)
    \big]
    \Big)
    \nonumber\\
    &&
    +i
    \frac{\Omega_{a,b}(\bm R_{a,b})+\Omega_{ba}(\bm R_{a,b})}{2}
    \Big(
    {\rm tr}_{\sigma}
    \big[
    \mathbbm{D}_{a}^{++}(t+\delta)
    \mathbbm{D}_{b}^{++}(t-\delta)
    \big]
    +
    {\rm tr}_{\sigma}
    \big[
    \mathbbm{D}_{a}^{--}(t-\delta)
    \mathbbm{D}_{b}^{--}(t+\delta)
    \big]
    \Big)
    \nonumber\\
    &&
    -i
    \Omega_{a,b}(\bm R_{a,b})
    \Big(
    {\rm tr}_\sigma
    \big[
    \mathbbm{D}_{a}^{-+}(t+\delta)
    \mathbbm{D}_{b}^{++}(t-\delta)
    \big]
    +
    {\rm tr}_\sigma 
    \big[
    \mathbbm{D}_{b}^{--}(t-\delta)
    \mathbbm{D}_{a}^{-+}(t+\delta)
    \big]
    \Big)
    \nonumber\\   
    &&
    -
    W_{a,b}(\bm R_{a,b})
    \Big(
    {\rm tr}_\sigma
    \big[
    \mathbbm{D}_{a}^{++}(t+\delta)
    \mathbbm{D}_{b}^{-+}(t-\delta)
    \big]
    -
    {\rm tr}_\sigma 
    \big[
    \mathbbm{D}_{b}^{-+}(t-\delta)
    \mathbbm{D}_{a}^{--}(t+\delta)
    \big]
    \Big)
    \bigg]  
    \nonumber\\
    &=&
    \int dt \sum_{a,b}
    \frac{J_{a,b}(\bm R_{a,b})}{2}
    \sum_{j=0}^3
    \sum_{s=\pm}s
    \hat m_{a,j}^{s,s}
    \hat m_{b,j}^{s,s}
    \nonumber\\
    &+&
    i
    \int dt     
    \sum_{a,b}
    \frac{\Omega_{a,b}(\bm R_{a,b})}{2}
    \sum_{j=0}^3
    \Big[
    \hat m_{a,j}^{+,+} m_{b,j}^{+,+}
    +    \hat m_{a,j}^{-,-} m_{b,j}^{-,-}
    -    \hat m_{a,j}^{+,-} m_{b,j}^{+,+}
    -
    \hat m_{a,j}^{-,-} m_{b,j}^{+,-}
   \Big]
\end{eqnarray}
where (note that $g_a<0$)
\begin{eqnarray}
    J_{a,b}(\bm R_{a,b})
    &=&
    g_a g_b  
    f_{\rm RKKY}(\bm R_{a,b}),
    \\
    \Omega_{a,b}(\bm R_{a,b})
    &=&
    \gamma_a |g_b|
    f_{\rm RKKY}(\bm R_{a,b}).
\end{eqnarray}
In the final equality, we have ignored the contribution arising from the term proportional to $W_{a,b}$ since they only contribute to the coherent motion, 
which is subdominant $o(J_{a,b})$ in the regime of interest $\kappa_a\ll U_a$ (or $\gamma_a \ll |g_a|)$.


\subsubsection{Non-Markovian correction: Gilbert damping}

We next analyze the non-Markovian correction $S^{\rm Gilbert}_{\rm eff}[d,\bar d]$.
Taking the trace over the Keldysh space,
\begin{eqnarray}
    &&S_{\rm Gilbert}[d,\bar d]
    = \frac{i}{2}
    \int dt_g  d\tau
    \frac{\tau}{2}
    \sum_{\bm q,\bm k}
    \Bigg[
    {\rm tr}_{\sigma,s}
    \bigg[
    [\partial_{t_g}
    \bm\Sigma_{\bm k+\bm q,\bm k}^{++}
    (t_g+\delta)
    ]
    \mathbbm{G}_0^{++}
    (\bm k+\bm q/2,\tau)
    \bm\Sigma_{\bm k-\bm q,\bm k}^{++}
    (t_g-\delta)
    \mathbbm{G}_0^{++}
    (\bm k-\bm q/2;-\tau)
    \bigg]
    \nonumber\\
    &&
    \ \ \ \ \ \ \ \ \ \ \ \ \ \ \ \ \ \ \ \ \ \ \ \ \ \ \ \ \ \ \ \ \ \ \ \ \ \ \ \ \ \
    - {\rm tr}_{\sigma,s}
    \bigg[
    \bm\Sigma_{\bm k+\bm q,\bm k}^{++}
    (t_g+\delta)
    \mathbbm{G}_0^{++}
    (\bm k+\bm q/2,\tau)
    [\partial_{t_g}
    \bm\Sigma_{\bm k-\bm q,\bm k}^{++}
    (t_g-\delta)
    ]
    \mathbbm{G}_0^{++}
    (\bm k-\bm q/2;-\tau)
    \bigg]
    \nonumber\\
    &&
    \ \ \ \ \ \ \ \ \ \ \ \ \ \ \ \ \ \ \ \ \ \ \ \ \ \ \ \ \ \ \ \ \ \ \ \ \ \ \ \ \ \
    +
    {\rm tr}_{\sigma,s}
    \bigg[
    [\partial_{t_g}
    \bm\Sigma_{\bm k+\bm q,\bm k}^{++}
    (t_g+\delta)
    ]
    \mathbbm{G}_0^{+-}
    (\bm k+\bm q/2,\tau)
    \bm\Sigma_{\bm k-\bm q,\bm k}^{--}
    (t_g-\delta)
    \mathbbm{G}_0^{-+}
    (\bm k-\bm q/2;-\tau)
    \bigg]
    \nonumber\\
    &&
    \ \ \ \ \ \ \ \ \ \ \ \ \ \ \ \ \ \ \ \ \ \ \ \ \ \ \ \ \ \ \ \ \ \ \ \ \ \ \ \ \ \
    - {\rm tr}_{\sigma,s}
    \bigg[
    \bm\Sigma_{\bm k+\bm q,\bm k}^{++}
    (t_g+\delta)
    \mathbbm{G}_0^{+-}
    (\bm k+\bm q/2,\tau)
    [\partial_{t_g}
    \bm\Sigma_{\bm k-\bm q,\bm k}^{--}
    (t_g-\delta)
    ]
    \mathbbm{G}_0^{-+}
    (\bm k-\bm q/2;-\tau)
    \bigg]   
    \nonumber\\
    &&
    \ \ \ \ \ \ \ \ \ \ \ \ \ \ \ \ \ \ \ \ \ \ \ \ \ \ \ \ \ \ \ \ \ \ \ \ \ \ \ \ \ \
    +
    {\rm tr}_{\sigma,s}
    \bigg[
    [\partial_{t_g}
    \bm\Sigma_{\bm k+\bm q,\bm k}^{--}
    (t_g+\delta)
    ]
    \mathbbm{G}_0^{-+}
    (\bm k+\bm q/2,\tau)
    \bm\Sigma_{\bm k-\bm q,\bm k}^{++}
    (t_g-\delta)
    \mathbbm{G}_0^{+-}
    (\bm k-\bm q/2;-\tau)
    \bigg]
    \nonumber\\
    &&
    \ \ \ \ \ \ \ \ \ \ \ \ \ \ \ \ \ \ \ \ \ \ \ \ \ \ \ \ \ \ \ \ \ \ \ \ \ \ \ \ \ \
    - {\rm tr}_{\sigma,s}
    \bigg[
    \bm\Sigma_{\bm k+\bm q,\bm k}^{--}
    (t_g+\delta)
    \mathbbm{G}_0^{-+}
    (\bm k+\bm q/2,\tau)
    [\partial_{t_g}
    \bm\Sigma_{\bm k-\bm q,\bm k}^{++}
    (t_g-\delta)
    ]
    \mathbbm{G}_0^{+-}
    (\bm k-\bm q/2;-\tau)
    \bigg]    
    \nonumber\\
    &&
    \ \ \ \ \ \ \ \ \ \ \ \ \ \ \ \ \ \ \ \ \ \ \ \ \ \ \ \ \ \ \ \ \ \ \ \ \ \ \ \ \ \
    +
    {\rm tr}_{\sigma,s}
    \bigg[
    [\partial_{t_g}
    \bm\Sigma_{\bm k+\bm q,\bm k}^{--}
    (t_g+\delta)
    ]
    \mathbbm{G}_0^{--}
    (\bm k+\bm q/2,\tau)
    \bm\Sigma_{\bm k-\bm q,\bm k}^{--}
    (t_g-\delta)
    \mathbbm{G}_0^{--}
    (\bm k-\bm q/2;-\tau)
    \bigg]
    \nonumber\\
    &&
    \ \ \ \ \ \ \ \ \ \ \ \ \ \ \ \ \ \ \ \ \ \ \ \ \ \ \ \ \ \ \ \ \ \ \ \ \ \ \ \ \ \
    - {\rm tr}_{\sigma,s}
    \bigg[
    \bm\Sigma_{\bm k+\bm q,\bm k}^{--}
    (t_g+\delta)
    \mathbbm{G}_0^{--}
    (\bm k+\bm q/2,\tau)
    [\partial_{t_g}
    \bm\Sigma_{\bm k-\bm q,\bm k}^{--}
    (t_g-\delta)
    ]
    \mathbbm{G}_0^{--}
    (\bm k-\bm q/2;-\tau)
    \bigg]    
    \Bigg]
    \nonumber\\
\end{eqnarray}
Here, we have omitted the contribution from the light-induced dissipation by setting $\Sigma^{-+}=0$, assuming that they are small compared to those already present in their absence.
Similarly to the case of Markovian contribution $S^{\rm RKKY}_{\rm eff}[d,\bar d]$, 
since the conduction electron Green's functions can be split into a sum over lesser and greater Green's function (Eq.~\eqref{SIeq: Green's function lesser greater}),
we encounter forms like
\begin{eqnarray}
    \int_{-\infty}^\infty d\tau
    \tau
    \theta(\tau)
    G_0^{s_a}(\bm k+\bm q/2,\tau)
    G_0^{s_b}(\bm k-\bm q/2;-\tau)
    &=&
    \frac{1}{2}
    \frac{   F_+^{s_a} F_-^{s_b}
    }
    {(\varepsilon_+ -\varepsilon_- -i\delta)^2},
    \\
    \int_{-\infty}^\infty d\tau
    \tau
    \theta(-\tau)
    G_0^{s_a}(\bm k+\bm q/2,\tau)
    G_0^{s_b}(\bm k-\bm q/2;-\tau) 
    &=& -\frac{1}{2}
    \frac{    F_+^{s_a} F_-^{s_b}
    }{(\varepsilon_+ -\varepsilon_- +i\delta)^2} .
\end{eqnarray}
Using these relations, we obtain,
\begin{eqnarray}
    &&S^{\rm Gilbert}_{\rm eff}[d,\bar d]
    =
    \frac{i}{2}
    \int dt    
    \nonumber\\
    &&\times
    \sum_{\bm k, \bm q}
    {\rm tr}_\sigma
    \bigg[
    \frac{   F_+^{>} F_-^{<}}
    {(\varepsilon_+ -\varepsilon_- -i\delta)^2}
    [
    \partial_{t}
    \bm\Sigma_{\bm k+\bm q,\bm k}^{++}(t + \delta )
    ]
    \bm\Sigma_{\bm k-\bm q,\bm k}^{++}(t - \delta)
    +
    \frac{F_+^{<} F_-^{>}    }
    {(\varepsilon_+-\varepsilon_- +i\delta)^2}
    \bm\Sigma_{\bm k+\bm q,\bm k}^{++}(t - \delta)
    [\partial_{t}    
    \bm\Sigma_{\bm k-\bm q,\bm k}^{++}(t + \delta)
    ]
    \nonumber\\
    &&
    +
    \frac{   F_+^{<} F_-^{>}}
    {(\varepsilon_+ -\varepsilon_- -i\delta)^2}
    [
    \partial_{t}
    \bm\Sigma_{\bm k+\bm q,\bm k}^{--}(t + \delta )
    ]
    \bm\Sigma_{\bm k-\bm q,\bm k}^{--}(t - \delta)
    +   
    \frac{F_+^{>} F_-^{<}    }
    {(\varepsilon_+-\varepsilon_- +i\delta)^2}
    \bm\Sigma_{\bm k+\bm q,\bm k}^{--}(t - \delta)
    [\partial_{t}    
    \bm\Sigma_{\bm k-\bm q,\bm k}^{--}(t + \delta)
    ]
    \nonumber\\
    &&+
    \frac{   F_+^{<} F_-^{>}}
    {(\varepsilon_+ -\varepsilon_- -i\delta)^2}
    [
    \partial_{t}
    \bm\Sigma_{\bm k+\bm q,\bm k}^{++}(t + \delta )
    ]
    \bm\Sigma_{\bm k-\bm q,\bm k}^{--}(t - \delta)
    +
    \frac{F_+^{<} F_-^{>}    }
    {(\varepsilon_+-\varepsilon_- +i\delta)^2}
    \bm\Sigma_{\bm k+\bm q,\bm k}^{++}(t - \delta)
    [\partial_{t}    
    \bm\Sigma_{\bm k-\bm q,\bm k}^{--}(t + \delta)
    ]    
    \nonumber\\
    &&+
    \frac{   F_+^{>} F_-^{<}}
    {(\varepsilon_+ -\varepsilon_- -i\delta)^2}
    [
    \partial_{t}
    \bm\Sigma_{\bm k+\bm q,\bm k}^{--}(t + \delta )
    ]
    \bm\Sigma_{\bm k-\bm q,\bm k}^{++}(t - \delta)
    + 
    \frac{F_+^{>} F_-^{<}    }
    {(\varepsilon_+-\varepsilon_- +i\delta)^2}
    \bm\Sigma_{\bm k+\bm q,\bm k}^{--}(t - \delta)
    [\partial_{t}    
    \bm\Sigma_{\bm k-\bm q,\bm k}^{++}(t + \delta)
    ]
    \bigg].
\end{eqnarray}
Let us further ignore the non-local contribution \cite{Kamra2017, Kamra2018, Reyes-Osorio2024} for simplicity, by taking into account contribution only from $\bm q=0$ in the self-energy, i.e., $\Sigma_{\bm k,\bm k\pm\bm q}\approx
\Sigma_{\bm k,\bm k}$.
In this case, using $
    \int dt {\rm tr}_\sigma
    \big[
    \bm\Sigma_{\bm k,\bm k}^{++}(t + \delta)
    [\partial_{t}    
    \bm\Sigma_{\bm k,\bm k}^{++}(t - \delta)
    ]
    \big]
    =\int dt {\rm tr}_\sigma \big[\bm\Sigma_{\bm k,\bm k}^{--}(t - \delta)
    [\partial_{t}    
    \bm\Sigma_{\bm k,\bm k}^{--}(t + \delta)
    ]
    \big]
    =0$, the effective action simplifies to 
\begin{eqnarray}
    &&
    S^{\rm Gilbert}_{\rm eff}[d,\bar d]
    =
    \sum_{\bm k,\bm q}
    [F_+^{<} F_-^{>}-F_+^{>} F_-^{<}]
    {\rm Im}\bigg[
    \frac{1}
    {(\varepsilon_+ -\varepsilon_- -i\delta)^2}
    \bigg]
    \int dt    
    \Big[
    {\rm tr}_\sigma
    \big[
    \bm\Sigma_{\bm k,\bm k}^{++}(t)
    [\partial_{t}
    \bm\Sigma_{\bm k,\bm k}^{--}(t)
    ]
    \big]
    \Big]
    \nonumber\\
    &&=
    -
    \sum_a
    \frac{\alpha_a}{2}
    \int dt    
    \Big[
    {\rm tr}_\sigma
    \big[
    {\mathbbm D}_a^{++}(t)
    [\partial_{t}
    {\mathbbm D}_a^{--}(t)
    ]
    \big]
    \Big]
    \nonumber\\
    &&=
    - 
    \sum_a
    \frac{\alpha_a}{4}
    \int dt    
    \sum_{j=0}^3
    (\bar d_{\mu,a}^+\sigma_j^{\mu\nu}d_{\nu,a}^+)
    [\partial_{t}
    (\bar d_{\mu',a}^-\sigma_j^{\mu'\nu'}d_{\nu',a}^-)
    ]
\end{eqnarray}
where
\begin{eqnarray}
    \alpha_a
    &=&
    2
    g_a^2
    \sum_{\bm k,\bm q}
    [F_+^{<} F_-^{>}-F_+^{>} F_-^{<}]
    {\rm Im}\bigg[
    \frac{1}
    {(\varepsilon_+ -\varepsilon_- -i\delta)^2}
    \bigg]
    =    -
    4
    \pi g_a^2
    \sum_{\bm k,\bm q}
    \frac{f_+ -f_-}{\varepsilon_+-\varepsilon_-}
    \delta(\varepsilon_+-\varepsilon_-)
\end{eqnarray}
For parabolic dispersion $\varepsilon_{\bm k}=\bm k^2/(2m)$ at three spatial dimensions, one obtains the form,
\begin{eqnarray}
    &&\alpha_a = 
    -
    4
    \pi g_a^2
    \frac{2\pi (V/N)^2}{(4\pi^2)^2} 
    \int_0^\infty dk k^2 \int_0^\infty dq q^2 
    \int_{-\pi}^\pi d\theta \cos\theta 
    \frac{f_{\bm k+\bm q} -f_{\bm k}}{\varepsilon_{\bm k+\bm q}-\varepsilon_{\bm k}}
    \delta(\varepsilon_{\bm k+\bm q}-\varepsilon_{\bm k})
    \nonumber\\
    &&
    = 
    \frac{9\pi^2}{2}
    \frac{n_e^2}{(N/V)^2}
    \frac{g_a^2}
    {\varepsilon_{\rm F}^2}
    =
    \frac{9\pi^2}{2} 
    n^2
    \frac{g_a^2}
    {\varepsilon_{\rm F}^2}
\end{eqnarray}
where $n_e=k_{\rm F}^3/(3\pi^2)$ is the electron density 
(where $k_{\rm F}$ is the Fermi momentum), $N$ is the number of sites, $V$ is the volume, and 
$n=n_e/(N/V)$ is the filling of the conduction electrons.



\subsection{Semiclassical approximation: Derivation of Eq. (6) in the main text}

Summarizing all terms derived above, the Keldysh action $Z=\int {\mathcal D}[d,\bar d]e^{iS_{\rm eff}[d,\bar d]}$ reads, 
\begin{eqnarray}
    S_{\rm eff}[d,\bar d]=S_d^0[d,\bar d] 
    + S_\gamma[d,\bar d] + S_M[d,\bar d]
\end{eqnarray}
with 
\begin{eqnarray}
    S_M[d,\bar d]
    =S^{\rm coh}_{\rm RKKY}[d,\bar d]
    +S_{\rm Gilbert}[d,\bar d]
    +S^{\rm neq}_{\rm RKKY}[d,\bar d].
\end{eqnarray}
Here,
\begin{eqnarray}
    S_d^0[d,\bar d]+ S_\gamma[d,\bar d] 
    &=&\int dt
    \sum_{a,\sigma}
    \bigg[
    \Big[
    \sum_{s=\pm}s 
    \bar d_{\sigma,a}^s
    i\partial_t 
    d_{\sigma,a}^s
    \Big]
    + i\gamma_{a} n
    \big[  
    d_{\sigma,a}^-
    (t)
    \bar d_{\sigma,a}^+ 
    (t)
    -\frac{1}{2} 
    d_{\sigma,a}^+
    (t_{+\delta})
    \bar d_{\sigma,a}^+    
    (t_{-\delta})   
    -
    \frac{1}{2}
    d_{\sigma,a}^-
    (t_{-\delta})   
    \bar d_{\sigma,a}^-        
    (t_{+\delta})   
    \big]
    \bigg]
    \nonumber\\
    &=&
    \int dt \sum_{a,\sigma}
    \Big[
    \big[
    \bar d_{\sigma,a}^q
    i\partial_t 
    d_{\sigma,a}^c
    +\bar d_{\sigma,a}^c
    i \partial_t d_{\sigma,a}^q 
     \big]
    -i\gamma_{a} n
    \big[  
    \bar d_{\sigma,a}^q
    d_{\sigma,a}^q
    +\frac{1}{2} 
    \bar d_{\sigma,a}^q    d_{\sigma,a}^c
    -
    \frac{1}{2}
    \bar d_{\sigma,a}^c  
    d_{\sigma,a}^q
    \big]
    \Big]
    \\
    S^{\rm coh}_{\rm RKKY}[d,\bar d]
    &=&
    \int dt \sum_{a,b}
    \frac{J_{a,b}(\bm R_{a,b})}{2}
    \sum_{j=0}^3
    \sum_{s=\pm}s
    \hat m_{a,j}^{s,s}
    \hat m_{b,j}^{s,s}
    \nonumber\\
    &=&
    \int dt \sum_{a,b}
    \frac{J_{a,b}(\bm R_{a,b})}{4}    
    \sum_{j=0}^3
    \Big[
    (\hat m_{a,j}^{cq}
    +    \hat m_{a,j}^{qc}
    )
    (\hat m_{b,j}^{cc}
    +    \hat m_{b,j}^{qq}
    )
    \Big],
    \\
    S_{\rm Gilbert}[d,\bar d]
    &=&
    - 
    \sum_a
    \frac{\alpha_a}{4}
    \int dt    
    \sum_{j=0}^3
    \hat m_{a,j}^{+,+}(t)
    \partial_t     
    \hat m_{a,j}^{-,-}(t)
    \nonumber\\
    &=&
    -\sum_a
    \frac{\alpha_a}{4}
    \int dt 
    \sum_{j=0}^3
    \big[
    \hat m_{a,j}^{c,q}(t)
    +
    \hat m_{a,j}^{q,c}(t)
    \big]
    \partial_t
    \big[
    \hat m_{a,j}^{c,c}(t)
    +
    \hat m_{a,j}^{q,q}(t)
    \big]
    \\
    S^{\rm neq}_{\rm RKKY}[d,\bar d]
    &=&
    i
    \int dt     
    \sum_{a,b}
    \frac{\Omega_{a,b}(\bm R_{a,b})}{2}
    \sum_{j=0}^3
    \Big[
    \hat m_{a,j}^{+,+} \hat m_{b,j}^{+,+}
    +    \hat m_{a,j}^{-,-} \hat m_{b,j}^{-,-}
    -    \hat m_{a,j}^{+,-} \hat m_{b,j}^{+,+}
    -
    \hat m_{a,j}^{-,-} \hat m_{b,j}^{+,-}
   \Big]
   \nonumber\\
    &=&
    i
    \int dt \sum_{a,b}
    \frac{\Omega_{a,b}(\bm R_{a,b})}{4}
    \sum_{j=0}^3
    \Big[
    (\hat m_{a,j}^{q,c} 
    -\hat m_{a,j}^{c,q}
    +2
    \hat m_{a,j}^{q,q})
    (    \hat m_{b,j}^{c,c}
    +   
    \hat m_{b,j}^{q,q})
    +
    (\hat m_{a,j}^{c,q} 
    +\hat m_{a,j}^{q,c})
    (    \hat m_{b,j}^{c,q}
    +   
    \hat m_{b,j}^{q,c})
   \Big]
   \nonumber\\
\end{eqnarray}
where 
we have introduced the classical-quantum representation,
\begin{eqnarray}
    d_{\mu,a}^c (t)
    = \frac{1}{\sqrt{2}} 
    (d_{\mu,a}^+(t) + d_{\mu,a}^-(t)),
    \qquad
    d_{\mu,a}^q (t)
    = \frac{1}{\sqrt{2}} 
    (d_{\mu,a}^+(t) - d_{\mu,a}^-(t)),
\end{eqnarray}
and ($l_1,l_2=+,-$ or $c,q$)
\begin{eqnarray}
    \hat m_{a,j}^{l_1,l_2}[d,\bar d]
    =\sum_{\mu,\nu=\uparrow,\downarrow}
    \bar d_{\mu,a}^{l_1}
    \hat \sigma_j^{\mu\nu}
    d_{\nu,a}^{l_2}.
\end{eqnarray}

In the following, we will perform a saddle point approximation to obtain a semiclassical description of spins, namely the Landau-Lifshitz-Gilbert (LLG) equation modified to exhibit non-reciprocal interactions (Eq.~(6) in the main text).
For this purpose, we introduce a set of auxiliary fields $m$ and Lagrange multipliers $\lambda$ as
\begin{eqnarray}
    Z&=&\int {\mathcal D}[d,\bar d] e^{i S_{\rm eff}[d,\bar d]}
    =\int {\mathcal D}[d,\bar d] e^{i( S_d^0[d,\bar d] + S_\gamma[d,\bar d] +S_M[\hat m[d,\bar d]])}
    \nonumber\\
    &=&\int {\mathcal D}[m] \int {\mathcal D}[d,\bar d] 
    \delta(\hat m[d,\bar d]-m)
    e^{i(S_d^0[d,\bar d] + S_\gamma[d,\bar d] +S_M[m])}
    \nonumber\\
    &=&    
    \label{SIeq: SM+S0+Slambda}
    \int {\mathcal D}[m]
    e^{i S_M[m]}
    \int {\mathcal D}[\lambda]
    \int {\mathcal D}[d,\bar d] 
    e^{i S_d^0[d,\bar d] + i S_\gamma[d,\bar d] }
    e^{iS_\lambda[\lambda,m,\hat m[d,\bar d]]}
    \\
    \label{SIeq: SM+SB}
    &\equiv&    
    \int {\mathcal D}[m]
    e^{i S_M[m]}
    e^{i S_B[m]}
\end{eqnarray}
where 
\begin{eqnarray}
    &&S_\lambda[\lambda,m,\hat m[d,\bar d]]
    =
    \int dt    
    \sum_a
    \sum_{l_1,l_2=q,c}
    \sum_{j=0}^3
    \lambda_{a,j}^{l_1,l_2}(t)
    \big[
    m_{a,j}^{l_1,l_2}(t)
    -\hat m_{a,j}^{l_1,l_2}[d(t),\bar d(t)]
    \big]
\end{eqnarray}
with
\begin{eqnarray}
    \bm\lambda_a(t) = 
    \begin{pmatrix}
        \bm\lambda_a^{q,q}(t)& \bm\lambda_a^{q,c}(t) \\
        \bm\lambda_a^{c,q}(t) & \bm\lambda_a^{c,c}(t) 
    \end{pmatrix}, 
    \qquad
    \bm m_a(t) = 
    \begin{pmatrix}
        \bm m_a^{q,q}(t)& \bm m_a^{q,c}(t) \\
        \bm m_a^{c,q}(t) & \bm m_a^{c,c}(t) 
    \end{pmatrix},
    \qquad
    \bm {\hat m}_a(t) = 
    \begin{pmatrix}
        \hat {\bm m}_a^{q,q}[d(t),\bar d(t)] & \hat {\bm m}_a^{q,c}[d(t),\bar d(t)] \\
        \hat {\bm m}_a^{c,q}[d(t),\bar d(t)] & \hat {\bm m}_a^{c,c}[d(t),\bar d(t)]
    \end{pmatrix}.
\nonumber\\
\end{eqnarray}
Note that $\bm m_a^{c,q}(t)=[\bm m_a^{c,q}(t)]^*$ and $\bm \lambda_a^{c,q}(t)=[\bm \lambda_a^{q,c}(t)]^*$, 
where 
$\bm m_a^{l_1,l_2}=( m_{a,1}^{l_1,l_2},m_{a,2}^{l_1,l_2},m_{a,3}^{l_1,l_2})$
and
$\bm\lambda_a^{l_1,l_2}=(\lambda_{a,1}^{l_1,l_2},\lambda_{a,2}^{l_1,l_2},\lambda_{a,3}^{l_1,l_2})$.
Here, the idea is to replace all $\hat m$'s composed of a product of Grassmann variables with real numbers $m$.

The physical meaning of $\bm m_a^{l_1,l_2}(t)$ become clear by
taking the saddle point of Eq.~\eqref{SIeq: SM+S0+Slambda} in terms of $\lambda$ as $0=\delta S_\lambda[\lambda,m,\hat m[d,\bar d]]/\delta\lambda_{a,j}^{l_1,l_2}$, or
\begin{eqnarray}
    \label{SIeq: m saddle point}
    &&
    \begin{pmatrix}
        m_{a,j}^{q,q}(t) &
        m_{a,j}^{q,c}(t) \\
        m_{a,j}^{c,q}(t) & 
        m_{a,j}^{c,c}(t) 
    \end{pmatrix}
    =
    \begin{pmatrix}
        \avg{
        \hat m_{a,j}^{q,q}[d(t),\bar d(t)]
        }& 
        \avg{\hat m_{a,j}^{q,c}[d(t),\bar d(t)]} \\
        \avg{
        \hat m_{a,j}^{c,q}[d(t),\bar d(t)]
        }& 
        \avg{\hat m_{a,j}^{c,c}[d(t),\bar d(t)]}
    \end{pmatrix}
    \nonumber\\
    &&=
    \begin{pmatrix}
        \avg{
        \sum_{\mu,\nu}
        \bar d_{\mu,a}^{q}(t)
        \sigma_j^{\mu,\nu}
        d_{\nu,a}^q(t)}
        & 
        \avg{
        \sum_{\mu,\nu}
        \bar d_{\mu,a}^{q}(t)
        \sigma_j^{\mu,\nu}
        d_{\nu,a}^c(t)}
        \\
        \avg{
        \sum_{\mu,\nu}
        \bar d_{\mu,a}^{c}(t)
        \sigma_j^{\mu,\nu}
        d_{\nu,a}^q(t)}
        & 
        \avg{
        \sum_{\mu,\nu}
        \bar d_{\mu,a}^{c}(t)
        \sigma_j^{\mu,\nu}
        d_{\nu,a}^c(t)}
    \end{pmatrix}
    \nonumber\\
    &&=
    \begin{pmatrix}
        0
        & 
        \theta(t-t')
        \avg{
        \sum_{\mu,\nu}
        \sigma_j^{\mu,\nu}
        \{
        \hat d_{\mu,a}^{\dagger}(t),
        \hat d_{\nu,a}(t')
        \}
        }
        \big|_{t'\rightarrow t + \delta}
        \\
        -\theta(t'-t)
        \avg{
        \sum_{\mu,\nu}
        \sigma_j^{\mu,\nu}
        \{
        \hat d_{\mu,a}^{\dagger}(t),
        \hat d_{\nu,a}(t')
        \}
        }
        \big|_{t'\rightarrow t -\delta}
        & 
        2
        \avg{
        \sum_{\mu,\nu}
        \hat d_{\mu,a}^{\dagger}(t)
        \sigma_j^{\mu,\nu}
        \hat d_{\nu,a}(t)}
    \end{pmatrix}
    \nonumber\\
    &&=
    \begin{pmatrix}
        0 &  0
        \\
        0 & 
        2 m_{a,j}
    \end{pmatrix}.
\end{eqnarray}
Here, $\bm m_a=(m_{a,1},m_{a,2},m_{a,3})
=\avg{
        \sum_{\mu,\nu}
        \hat d_{\mu,a}^{\dagger}(t)
        \bm \sigma^{\mu,\nu}
        \hat d_{\nu,a}(t)}$ is the average magnetization of $a$-site.
The off-diagonal component vanishes because the equal-time response function vanishes.

The Berry phase contribution is given by,
\begin{eqnarray}
    &&
    e^{iS_B[m]}
    \equiv
    \int {\mathcal D}[\lambda]
    \int {\mathcal D}[d,\bar d] 
    e^{i S_d^0[d,\bar d] + i S_\gamma[d,\bar d] }
    e^{i S_\lambda[\lambda,m,\hat m[d,\bar d]]}
    \nonumber\\
    &&=
    \int {\mathcal D}[\lambda]
    e^{i
    \int dt \sum_{a}\sum_{j=0}^3 
    \sum_{l_1,l_2}
    \lambda_{a,j}^{l_1,l_2} m_{a,j}^{l_1,l_2}}
    \int {\mathcal D}[d,\bar d] 
    \exp\bigg[
    i \int dt \sum_a 
    \Psi_{d,a}^\dagger(t) 
    \big(
    \bm G^{-1}_{0a}
    -\bm \Sigma_a^\lambda[\lambda_a(t)]
    \big)
    \Psi_{d,a}(t)
    \bigg]
    \nonumber\\
    &&=
    \int {\mathcal D}[\lambda]
    e^{i\int dt
    \sum_{l_1,l_2}
    \lambda_{a,j}^{l_1,l_2} m_{a,j}^{l_1,l_2}}
    \exp\bigg[
    i\cdot (-i)
    {\sum_a
    \ln\det\Big[    
    (-i)
    \big(
    \bm G^{-1}_{0a}
    -\bm \Sigma_a^\lambda[\lambda_a(t)]
    \big)
    \Big]}
    \bigg]    
    \nonumber\\
    &&\equiv    
    \int {\mathcal D}[\lambda]
    e^{iS_B^\lambda[\lambda,m]},
\end{eqnarray}
where
\begin{eqnarray}
    \Psi_{d,a}(t)=
    (d_{a,\uparrow}^q,
     d_{a,\downarrow}^q,  
     d_{a,\uparrow}^c,
     d_{a,\downarrow}^c)^{\mathrm{T}}, 
    \qquad
    \Psi_{d,a}^\dagger(t) = 
    (\bar d_{a,\uparrow}^q,
     \bar d_{a,\downarrow}^q,  
     \bar d_{a,\uparrow}^c,
     \bar d_{a,\downarrow}^c). 
\end{eqnarray}
and
\begin{eqnarray}
    \bm G^{-1}_{0a}
    = \begin{pmatrix}
        -i \gamma_a n
        \hat 1 
        & (i\partial_t + i\gamma_a n/2 )\hat 1
        \\
        (
        i\partial_t - i\gamma_a n/2 )\hat 1
        & 0
    \end{pmatrix},
    \qquad
    \bm \Sigma_a^\lambda[\lambda_a(t)]
    =\sum_{j=0}^3 
    \begin{pmatrix}
        \lambda_{a,j}^{qq}
        \hat\sigma_j
        &
        \lambda_{a,j}^{qc}
        \hat\sigma_j
        \\
        \lambda_{a,j}^{cq}
        \hat\sigma_j
        &
        \lambda_{a,j}^{cc}
        \hat\sigma_j
    \end{pmatrix}
\end{eqnarray}
and $\bm G_a^{-1}=\bm G_{0,a}^{-1}-\bm \Sigma_a$.

We perform a saddle point approximation to the $\lambda$ integral, i.e.,
\begin{eqnarray}
    e^{iS_B[m]}
    \approx e^{iS_B^\lambda[\lambda=\lambda_0,m]}
\end{eqnarray}
where $\lambda_0$ is determined from the saddle point condition,
\begin{eqnarray}
    &&0=\frac{\delta S_B^\lambda[\lambda,m]}{\delta\lambda_{a,j}^{l_1,l_2}(t)}\Bigg|_{\lambda=\lambda_0}
    = m_{a,j}^{l_1,l_2}(t) 
    -i {\rm Tr}\Bigg[
    \bm G_a 
    \frac{\delta\bm G_a^{-1}}{\delta \lambda_{a,j}^{l_1,l_2}(t)}
    \Bigg|_{\lambda=\lambda_0}
    \Bigg]
    \nonumber\\
    &&= {\rm tr}_{\rm sp}
    [
    \hat\sigma_j 
    (\bm m_{a}^{l_1,l_2}(t)
    \cdot \bm {\hat \sigma})
    ]
    \nonumber\\
    &&
    -i 
    \int dt_1 dt_2
    {\rm tr}_{s,\sigma}
    \Bigg[
        \begin{pmatrix}
        G_a^{q,q}(t_1,t_2) 
        & G_a^{q,c}(t_1,t_2) \\
        G_a^{c,q}(t_1,t_2) 
        & G_a^{c,c}(t_1,t_2) 
    \end{pmatrix}
    \frac{\delta
    \begin{pmatrix}
        (G^{-1})_a^{q,q}(t_2,t_1+0^+) 
        & (G^{-1})_a^{q,c}(t_2,t_1+0^+) \\
        (G^{-1})_a^{c,q}(t_2,t_1+0^+) 
        & (G^{-1})_a^{c,c}(t_2,t_1+0^+) 
    \end{pmatrix}
    }{\delta \lambda_{a,j}^{l_1,l_2}(t)}
    \Bigg|_{\lambda=\lambda_0}
    \Bigg]
\end{eqnarray}
Explicit components are computed as follows.
For $(l_1,l_2)=(q,q)$, 
\begin{eqnarray}
    &&0=\frac{\delta S_B^\lambda[\lambda,m]}{\delta\lambda_{a,j}^{q,q}(t)}\Bigg|_{\lambda=\lambda_0} 
    = {\rm tr}_{\rm sp}
    [
    \hat\sigma_j 
    (\bm m_{a}^{q,q}(t)
    \cdot \bm {\hat \sigma})
    ]
    \nonumber\\
    &&
    -i 
    \int dt_1 dt_2
    {\rm tr}_{s,\sigma}
    \Bigg[
        \begin{pmatrix}
        G_a^{q,q}(t_1,t_2) 
        & G_a^{q,c}(t_1,t_2) \\
        G_a^{c,q}(t_1,t_2) 
        & G_a^{c,c}(t_1,t_2) 
    \end{pmatrix}
    \frac{\delta
    \begin{pmatrix}
        (G^{-1})_a^{q,q}(t_2,t_1+0^+) 
        & (G^{-1})_a^{q,c}(t_2,t_1+0^+) \\
        (G^{-1})_a^{c,q}(t_2,t_1+0^+) 
        & (G^{-1})_a^{c,c}(t_2,t_1+0^+) 
    \end{pmatrix}
    }{\delta \lambda_{a,j}^{q,q}(t)}
    \Bigg|_{\lambda=\lambda_0}
    \Bigg]
    \nonumber\\
    &&    = {\rm tr}_{\rm sp}
    \Bigg[
    \bigg[
    (\bm m_{a}^{q,q}(t)
    \cdot \bm {\hat \sigma})
    +i 
    \int dt_1 
        \hat G_a^{q,q}(t_1,t) 
    \delta(t-t_1+0^+)
    \bigg]
    \hat\sigma_j 
    \Bigg]
\end{eqnarray}
As this holds for all $j=0,1,2,3$, this implies,
\begin{eqnarray}
    (\bm m_{a}^{q,q}(t)
    \cdot \bm {\hat \sigma})
    +i 
    \int dt_1 
        \hat G_a^{q,q}(t_1,t) 
    \delta(t-t_1+0^+)=0.
\end{eqnarray}
For $(l_1,l_2)=(q,c)$, 
\begin{eqnarray}
    &&0=\frac{\delta S_B^\lambda[\lambda,m]}{\delta\lambda_{a,j}^{q,c}(t)}\Bigg|_{\lambda=\lambda_0} 
    = {\rm tr}_{\rm sp}
    \Bigg[
    \bigg[
    (\bm m_{a}^{q,c}(t)
    \cdot \bm {\hat \sigma})
    +i 
    \int dt_1 
        \hat G_a^{c,q}(t_1,t;\lambda) 
    \delta(t-t_1+0^+)
    \bigg]
    \hat\sigma_j 
    \Bigg]
\end{eqnarray}
and so on. Therefore,
\begin{eqnarray}
    (\bm m_{a}^{q,q}(t)
    \cdot \bm {\hat \sigma})
    +i 
    \int dt_1 
        \hat G_a^{q,q}(t_1,t;\lambda) 
    \delta(t-t_1+0^+)=0, \\
    (\bm m_{a}^{q,c}(t)
    \cdot \bm {\hat \sigma})
    +i 
    \int dt_1 
        \hat G_a^{c,q}(t_1,t;\lambda) 
    \delta(t-t_1+0^+)=0, \\
    (\bm m_{a}^{c,q}(t)
    \cdot \bm {\hat \sigma})
    +i 
    \int dt_1 
        \hat G_a^{q,c}(t_1,t;\lambda) 
    \delta(t-t_1+0^+)=0, \\
    (\bm m_{a}^{c,c}(t)
    \cdot \bm {\hat \sigma})
    +i 
    \int dt_1 
        \hat G_a^{c,c}(t_1,t;\lambda) 
    \delta(t-t_1+0^+)=0,
\end{eqnarray}
or
\begin{eqnarray}
    \begin{pmatrix}
        \bm m_{a}^{q,q}(t)
        \cdot \bm {\hat \sigma} &
        \bm m_{a}^{c,q}(t)
        \cdot \bm {\hat \sigma}
        \\
        \bm m_{a}^{q,c}(t)
        \cdot \bm {\hat \sigma} &
        \bm m_{a}^{c,c}(t)
        \cdot \bm {\hat \sigma}   
    \end{pmatrix}
    =-i\int dt_1 
    \begin{pmatrix}
        \hat G_a^{q,q}(t_1,t;\lambda) &
        \hat G_a^{q,c}(t_1,t;\lambda) \\
        \hat G_a^{c,q}(t_1,t;\lambda) &
        \hat G_a^{c,c}(t_1,t;\lambda)
    \end{pmatrix}
    \delta(t-t_1+0^+).
\end{eqnarray}
We will use Eq.~\eqref{SIeq: m saddle point}, i.e., $\bm m_a^{c,c}=
2\bm m_a$ and
$\bm m_a^{q,q}=\bm m_a^{c,q}=\bm m_a^{q,c}=0$ 
from below.
Applying $\hat G^{-1}$ to both sides from the left, one obtains,
\begin{eqnarray}
    \Bigg[
    \begin{pmatrix}
        -i \gamma_a n 
        \hat 1 
        & (i\partial_t/2 + i\gamma_a n/2 )\hat 1
        \\
        (
        i\partial_t/2 - i\gamma_a n /2 )\hat 1
        & 0
    \end{pmatrix}
    -
    \begin{pmatrix}
        \bm\lambda_{a}^{qq}\cdot\bm{\hat\sigma} & \bm\lambda_{a}^{qc}\cdot\bm{\hat\sigma} \\
        \bm\lambda_{a}^{cq}\cdot\bm{\hat\sigma} & \bm\lambda_{a}^{cc}\cdot\bm{\hat\sigma}    
        \end{pmatrix}
    \Bigg]
        \begin{pmatrix}
        0 &
        0
        \\
        0 &
        \bm m_{a}(t)
        \cdot \bm {\hat \sigma}   
    \end{pmatrix}
    =
    \frac{1}{2}
    (-i)\delta(t_1-t)|_{t_1\rightarrow t+0^+} 
\end{eqnarray}
where the left-hand side can be computed as,
\begin{eqnarray}
    {\rm LHS}&=&
    \begin{pmatrix}
        0
        &
        (i\partial_t/2 + i\gamma_a n/2 )
        (\bm m_a \cdot\bm{\hat\sigma})
        \\
        0
        &
        0
    \end{pmatrix}
    -
    \begin{pmatrix}
        0
        &
        (\bm\lambda_{a}^{q,c}\cdot\bm{\hat\sigma})
        (\bm m_a\cdot\bm{\hat\sigma})
        \\
        0
        &
        (\bm\lambda_{a}^{c,c}\cdot\bm{\hat\sigma})
        (\bm m_a\cdot\bm{\hat\sigma})
    \end{pmatrix}
\end{eqnarray}
Similarly, when applying $\hat G^{-1}$ from the right, one obtains 
\begin{eqnarray}
        \begin{pmatrix}
        0 & 0
        \\
        0 &
        \bm m_{a}(t)
        \cdot \bm {\hat \sigma}   
    \end{pmatrix}
    \Bigg[
    \begin{pmatrix}
        -i \gamma_a n
        \hat 1 
        & (
        -
        i\overleftarrow{\partial}_t/2 
        + i\gamma_a n /2 )\hat 1
        \\
        (
        -
        i\overleftarrow{\partial}_t/2
        - i\gamma_a n /2 )\hat 1
        & 0
    \end{pmatrix}
    -
    \begin{pmatrix}
        \bm\lambda_{a}^{qq}\cdot\bm{\hat\sigma} & \bm\lambda_{a}^{qc}\cdot\bm{\hat\sigma} \\
        \bm\lambda_{a}^{cq}\cdot\bm{\hat\sigma} & \bm\lambda_{a}^{cc}\cdot\bm{\hat\sigma}    
        \end{pmatrix}
    \Bigg]
    =
    \frac{1}{2}
    (-i)\delta(t_1-t)|_{t_1\rightarrow t+0^+}
    \nonumber\\
\end{eqnarray}
with
\begin{eqnarray}
    {\rm LHS}
    &=&
    \begin{pmatrix}
        0 & 0
        \\
        (-i\partial_t/2
        -i\gamma_a n/2)
        (\bm m_{a}(t)
        \cdot \bm {\hat \sigma})
        &
        0
    \end{pmatrix}
    -
    \begin{pmatrix}
        0 & 0
        \\
        (\bm m_{a}(t)
        \cdot \bm {\hat \sigma})
        (\bm\lambda_{a}^{c,q}\cdot\bm{\hat\sigma})
        &
        (\bm m_{a}(t)
        \cdot \bm {\hat \sigma})
        (\bm\lambda_{a}^{c,c}\cdot\bm{\hat\sigma})
    \end{pmatrix}.
\end{eqnarray}
Subtracting the two relations yields,
\begin{eqnarray}
    \label{SIeq: saddle point qc}
    &&\frac{i}{2}
        (
        \partial_t
        +
        \gamma_a n
        )
        (\bm m_a
        \cdot
        \bm{\hat\sigma}
        )
    =        (\bm\lambda_{a}^{q,c}\cdot\bm{\hat\sigma})
        (\bm m_a\cdot\bm{\hat\sigma})
        \\        
    \label{SIeq: saddle point cq}
        &&\frac{i}{2}
        (
        \partial_t
        +\gamma_a n
        )
        (\bm m_a
        \cdot
        \bm{\hat\sigma})
        =        
        -
        (\bm m_{a}(t)
        \cdot \bm {\hat \sigma})
        (\bm\lambda_{a}^{c,q}\cdot\bm{\hat\sigma})
\end{eqnarray}
Note that Eq.~\eqref{SIeq: saddle point qc} is the complex conjugate of Eq.~\eqref{SIeq: saddle point cq} (because $\bm \lambda_a^{q,c}(t) (= [\bm \lambda_a^{c,q}(t)]^*)$) and is therefore equivalent. 
From these two equations, one obtains,
\begin{eqnarray}
    &&
    i(\partial_t + \gamma_a n)
    (\bm m_a\cdot\bm{\hat\sigma})
    =    
        (\bm\lambda_{a}^{q,c}\cdot\bm{\hat\sigma})
        (\bm m_a\cdot\bm{\hat\sigma})
        -
        (\bm m_{a}(t)
        \cdot \bm {\hat \sigma})
        ([\bm\lambda_{a}^{q,c}]^*\cdot\bm{\hat\sigma}).
\end{eqnarray}
Using
\begin{eqnarray}
    &&\sum_{j=0}^3 f_j
    {\rm tr}_\sigma
    [\hat\sigma_j\hat\sigma_i]
    =
    2f_i, 
    \\
    &&i\sum_{l,j,k=1}^3 
    a_l b_j \epsilon_{ljk}
    {\rm tr}_\sigma
    [\hat\sigma_k \hat\sigma_i]
    =2i\sum_{l,j,k}^3 
    a_l b_j \epsilon_{ljk}
    \delta_{ik}
    =2i\sum_{l,j,k}^3 
    a_l b_j \epsilon_{lji}
    =2i(\bm a\times\bm b)_i,
\end{eqnarray}
we find
\begin{eqnarray}
    i(\partial_t
    +\gamma_a n
    )\bm m_a
    &=&
    (
    \bm\lambda_a^{qc}
    -[\bm\lambda_a^{qc}]^*
    )
    +
    (
    \lambda_{a,0}^{qc}
    -[\lambda_{a,0}^{qc}]^*
    )
    \bm m_{a}   
    +i(
    \bm\lambda_a^{qc}
    +
    [\bm\lambda_a^{qc}]^*
    )
    \times 
    \bm m_{a}.
\end{eqnarray}

Let us now determine $\lambda_{a,j}^{q,c}(=[\lambda_{a,j}^{c,q}]^*)$ by
finding the saddle point of 
$S_{\rm eff}[\lambda,m]\equiv  S_B^\lambda[\lambda,m]+S_M[m]$ as
\begin{eqnarray}
    &&0=\frac{\delta S_{\rm eff}[\lambda,m]}{\delta m_{a,j}^{q,c}(t)}
    =\lambda_{a,j}^{q,c}(t)
    +\sum_b 
    \frac{J_{a,b}(\bm R_{a,b})}{2}
    m_{b,j}(t)
    -
    \frac{\alpha_a}{2}
    \frac{d m_{a,j}^{c,c}(t)}{dt}    
    +i\sum_b 
    \frac{\Omega_{a,b}(\bm R_{a,b})}{2}
    m_{b,j}(t),
\end{eqnarray}
which gives,
\begin{eqnarray}
     \lambda_{a,j}^{q,c}(t)
    +[\lambda_{a,j}^{q,c}(t)]^*
    &=&-\sum_b 
    J_{a,b}
    m_{b,j}(t)
    +
    \alpha_a
    \frac{d m_{a,j}(t)}{dt}  
    \\
     \lambda_{a,j}^{q,c}(t)
    -[\lambda_{a,j}^{q,c}(t)]^*
    &=&
    -i\sum_b 
    \Omega_{a,b}
    m_{b,j}(t).
\end{eqnarray}
This yields the desired LLG equation modified by the light-induced dissipation
\begin{eqnarray}
    \partial_t
    \bm m_a
    =
    -\gamma_a n
    \bm m_a
    -\sum_{b
    (\ne a)
    }
    \Omega_{a,b}(\bm R_{a,b})
    \bm m_b(t)
    -
    \Big[
    \sum_b 
    J_{a,b}(\bm R_{a,b})
    \bm m_{b}(t)
    -
    \alpha_a
    \bm {\dot m}_a(t)
    \Big]
    \times 
    \bm m_{a}(t),
\end{eqnarray}
where we have ignored the dissipative self-interaction $\Omega_{a,a}$.
This completes the derivation of Eq.~(6) in the main text.

\section{
Microscopic derivation of GKSL equation (3)
}
\label{SIsec: microscopic derivation of GKSL equation}

Our starting point of this work was the GKSL equation (3) in the main text (Eq.~\eqref{SIeq: GKSL master equation dissipation engineering} in the SI),
which we reproduce below for convenience: 
\begin{eqnarray}
    \label{SIeq: GKSL master equation dissipation engineering 2}
    \partial_t \hat\rho 
    &=&
    -i [\hat H_{\rm A},\hat \rho]
    +\sum_{a,\sigma} \kappa_a
    {\mathcal D}[\hat d_{\sigma,a}
    \hat P_{\uparrow\downarrow}^a]
    \hat\rho.
\end{eqnarray}
Here, the Hamiltonian in the first term $\hat H_{\rm A}=\hat H_{c0}+\hat H_{d0}+\hat H_{cd}$ (See Eqs.~\eqref{SIeq: Hc0}, \eqref{SIeq: Hd0} and \eqref{SIeq: Hcd}) is the magnetic metal Hamiltonian described by the Anderson impurity model that we have introduced in Sec.~\ref{SIsec: model}. 
The second term describes the decay of localized electrons at site $a$ that occurs exclusively when it is in a double-occupied state, which we assumed to be introduced by the time-dependent drive from the laser that couples the double-occupied state to the higher-level state.
We have introduced this term phenomenologically by taking advantage of the fact that Markovian dynamics can always be described in a GKSL form, leaving $\kappa_a$ as a free parameter that should be determined experimentally.

In this section, to complement our results from a more microscopic point of view, we derive this GKSL equation microscopically by explicitly treating the time-dependent drive from the laser and its coupling to the higher-level state.
(See e.g., Ref.~\cite{Diehl2008, Kraus2008, Tomita2017} that uses a similar approach to derive the GKSL equation from microscopics.)
We model this system as the sum of three parts:
\begin{eqnarray}
    \hat H_{\rm tot}
    =\hat H_{\rm A} 
    + \hat H_{\rm drive}(t)
    + \hat H_f
    .
\end{eqnarray}
Here,  
the second term, given by
\begin{eqnarray}
    \hat H_{\rm drive}(t)=\sum_{a,\sigma=\uparrow,\downarrow}
    \Omega_a 
    e^{-i h\nu t}
    \hat f_{\sigma,a}^\dagger
        \hat d_{\sigma,a}
    +{\rm h.c.},
\end{eqnarray}
describes the time-dependent drive (within the rotating wave approximation, which is justified close to resonance) that 
couples the localized electrons to a higher energy level, described by the third term 
$\hat H_f
    =  \hat H_f^0 + \hat H_{fb}+\hat H_b$
given by 
\begin{eqnarray}
    \hat H_f^0 
    &=& \sum_a \varepsilon_{f,a}
    \hat f^\dagger_{a,\sigma}
    \hat f_{\sigma,a}, \\
    \hat  H_{\rm b}
    &=& 
    \sum_a 
    \sum_{\bm k}
    \varepsilon_{b,\bm k,a}\hat b_{\bm k,\sigma}^{a\dagger}\hat b_{\bm k,\sigma}^a,
    \\
    \hat H_{\rm bf}
    &=& 
    \sum_a 
    v_a^{bf}
    \sum_\sigma
    \sum_{\bm R_a}\hat b^{a\dagger}_{\bm R_a,\sigma}\hat f_{\sigma,a} 
    +{\rm h.c.}
\end{eqnarray}
Here, $\hat f_{a,\sigma}$ is the annihilation operator of the higher-energy level state of the site $a$ and spin $\sigma=\uparrow,\downarrow$ with energy $\varepsilon_{f,a}$.
$\Omega_a$ is the coupling strength between the localized electrons and their higher-energy level induced by the laser that has the frequency $\nu$.
The higher-level state is assumed to be short-lived due to the coupling to the surrounding environment. 
We model this by coupling to a free fermion bath $\hat H_{\rm b}$, where $\hat b_{\bm k,\sigma}^a$ is the fermionic annihilation operator that hops between the higher-order state and bath with the hopping amplitude $v_a^{bf}$.
Since the higher energy level $\varepsilon_{f,a}$ is assumed to be far above the Fermi energy $\varepsilon_{\rm F}$, we assume that the bath fermion is unoccupied at this energy, i.e. $f_{\rm b}(\omega\approx \varepsilon_{f,a})=0$.

It is useful to move the coordinate into a rotating frame to eliminate the time dependence in $\hat H_{\rm drive}(t)$. 
By replacing $\hat f_{\sigma,a}\rightarrow e^{ih\nu t} \hat f_{\sigma,a},
\hat b_{\bm R_a,\sigma}^a
\rightarrow 
e^{ih\nu t}\hat b_{\bm R_a,\sigma}^a$, 
the $\hat H_{\rm drive}(t),\hat H_f^0, \hat H_{\rm b}^0$ are transformed to 
\begin{eqnarray}
    \hat H_{\rm drive}'&=&
    \sum_{a,\sigma}
    \Omega_a 
    \hat f_{\sigma,a}^\dagger
        \hat d_{\sigma,a}
    +{\rm h.c.},
    \\
    \hat H_f^{0}{}' 
    &=& \sum_a (\varepsilon_{f,a} - h\nu)
    \hat f^\dagger_{a,\sigma}
    \hat f_{\sigma,a}, \\
    \hat  H_{\rm b}'
    &=& 
    \sum_a 
    \sum_{\bm k}
    (\varepsilon_{b,\bm k,a}
    -h\nu)
    \hat b_{\bm k,\sigma}^{a\dagger}\hat b_{\bm k,\sigma}^a,
\end{eqnarray}
while other parts stay the same.
For later use, we introduce a notion for the drive Hamiltonian $\hat H_{\rm drive}'=\sum_{a,\sigma} [\hat V_{f\Leftarrow d}^{\sigma,a}+\hat V_{d\Leftarrow f}^{\sigma,a}]$, where 
\begin{eqnarray}
    V_{f\Leftarrow d}^{\sigma,a} 
    = \Omega_a \hat f_{\sigma,a}^\dagger \hat d_{\sigma,a}, 
    \qquad
    V_{d\Leftarrow f}^{\sigma,a} 
    = \Omega_a^* \hat d_{\sigma,a}^\dagger \hat f_{\sigma,a}
    (=(\hat V_{f\Leftarrow d}^{\sigma,a} )^\dagger), 
\end{eqnarray}

Below, we integrate out the higher-energy degrees of freedom as well as the bath that is attached to it, to derive the GKSL equation~\eqref{SIeq: GKSL master equation dissipation engineering 2}.
We first integrate out the bath degrees of freedom attached to the higher level.
This is performed by using the Keldysh formalism reviewed in Sec.~\ref{SIsubsubsec: Keldysh-GKSL}, which results in the GKSL equation for the density operator $\hat \rho_{\rm tot}$ that consists of conduction electrons and localized electrons \textit{including} the higher level state,
\begin{eqnarray}
    \partial_t\hat\rho_{\rm tot}
    &=&
    {\mathcal L}_{\rm tot}\hat\rho_{\rm tot}
    \nonumber\\
    &=&-i[
    \hat H_{\rm A}
    +\hat H_f^0{}'
    + \hat H_{\rm drive}',
    \hat\rho_{\rm tot}]
    +
    \sum_{a,\sigma} \Gamma_{f,a}{\mathcal D}[\hat f_{\sigma,a}]\hat\rho_{\rm tot},
\end{eqnarray}
where 
$\Gamma_{f,a}=\pi |v_a^{{\rm b}f}|^2\rho_{\rm b}$
is the decay rate of the higher level state (with $\rho_{{\rm b},a}=\sum_{\bm k}\delta(\omega-\varepsilon_{{\rm b},\bm k,a}+h\nu)\approx {\rm const.}$ being the density of states of the bath that is assume to be Markovian.)

We will be concerned with the case where the laser strength is weak enough that $H_{\rm drive}'$ can be regarded as a perturbation.
Recalling that we are interested in the regime where the c-d mixing can be regarded as a perturbation, 
the total Lindbladian can be split into the non-perturbative (${\mathcal L}_{\rm tot}^{(0)}$) and perturbative part (${\mathcal L}_{\rm drive}^{(1)}$ and ${\mathcal L}_{cd}^{(1)}$),
\begin{eqnarray}
    {\mathcal L}_{\rm tot}
    = {\mathcal L}_{\rm tot}^{(0)}
    +{\mathcal L}_{\rm drive}^{(1)}
    +{\mathcal L}_{cd}^{(1)}
\end{eqnarray}
where
\begin{eqnarray}
    {\mathcal L}_{\rm tot}^{(0)}
    \hat\rho_{\rm tot}
    &=&
    -i[\hat H_{c0}+\hat H_{d0}
    +\hat H_f^0{}',
    \hat\rho_{\rm tot}]
    +
    \sum_{a,\sigma} \Gamma_{a,f}{\mathcal D}[\hat f_{\sigma,a}]\hat\rho_{\rm tot},
    \\
    {\mathcal L}_{\rm drive}^{(1)}
    \hat\rho_{\rm tot}
    &=&
    -i[
    \hat H_{\rm drive}',
    \hat\rho_{\rm tot}], 
    \qquad
    {\mathcal L}_{cd}^{(1)}
    \hat\rho_{\rm tot}
    =
    -i[
    \hat H_{cd},
    \hat\rho_{\rm tot}].
\end{eqnarray}

Having in mind that the contribution from the c-d mixing ${\mathcal L}_{cd}^{(1)}$ would be incorporated perturbatively later in Sec.~\ref{SIsec: sd}, we compute here perturbatively the contribution from the drive Lindbladian ${\mathcal L}_{\rm drive}^{(1)}=\sum_a {\mathcal L}_{{\rm drive},a}^{(1)}$, 
with
\begin{eqnarray}
    {\mathcal L}_{{\rm drive},a}^{(1)}
    \hat\rho_{\rm tot}
    =-i\sum_\sigma 
    [\hat V_{f\Leftarrow d}^{\sigma,a} 
    +\hat V_{d\Leftarrow f}^{\sigma,a} ,
    \hat\rho_{\rm tot}].
\end{eqnarray}
Particularly, we derive below the effective Lindbladian by using the projection method reviewed in Sec.~\ref{SIsubsec: projection},
\begin{eqnarray}
    \label{SIeq: effective Lindbladian matrix element drive}
    ({\mathcal L}_{\rm eff}^{\rm drive})_{n_l,n_r}
    \approx
    ({\mathcal L}_{\rm tot}^{(0)})_{n_l,n_r}
    +
    ( {\mathcal L}_{\rm drive}^{(1)}
    )_{n_l,n_r}
    -
    \sum_{m\in \mathfrak{f}}
    \frac{
    {\rm tr}[\hat l_{n_l}^{{\rm tot}(0)\dagger} 
    {\mathcal L}_{\rm drive}^{(1)}
    \hat r_m^{{\rm tot}(0)}]
    {\rm tr}[\hat l_m^{{\rm tot}(0)\dagger}
    {\mathcal L}_{\rm drive}^{(1)}
    \hat r_{n_r}^{{\rm tot}(0)}]
    }
    {\lambda_m^{{\rm tot}(0)}}.
\end{eqnarray}
within the second-order perturbation in ${\mathcal L}_{\rm drive}^{(1)}$.
Here, 
$\hat r_{n}^{{\rm tot}(0)} (\hat l_{n}^{{\rm tot}(0)})$ is the right (left) eigenstate of the non-perturbative part ${\mathcal L}_{\rm tot}^{(0)}$ with an eigenvalue $\lambda_n^{\rm tot(0)}$,
and 
${\mathcal A}_{n_l,n_r}
\equiv {\rm tr}[\hat l_{n_l}^{{\rm tot}(0)}{\mathcal A}\hat r_{n_r}^{{\rm tot}(0)}]$
is the $(n_r,n_l)$ component of the superoperator $\mathcal A$. 
The sum of the final term is taken over the fast variables.
(See Sec.~\ref{SIsubsec: projection} for details.)

The analysis below is more or less parallel to the procedure performed in Sec.~\ref{SIsec: sd}. 
Similar to the analysis done there, the first step to perform this sort of calculation is to characterize the eigenvalues and eigenstates of the non-perturbative part 
$   {\mathcal L}_{\rm tot}^{(0)}
   =\sum_a [
   {\mathcal L}_{d,a}^{(0)}
   +
   {\mathcal L}_{f,a}^{(0)}
   ]
   +{\mathcal L}_c^{(0)},$
where
\begin{eqnarray}
    {\mathcal L}_{d,a}^{(0)}
    \hat\rho_{\rm tot}
    &=&-i
    [
    \sum_\sigma
    \varepsilon_{d,a}
    \hat d_{\sigma,a}^\dagger\hat d_{\sigma,a}
    +U_a 
    \hat d^\dagger_{\uparrow,a}
    \hat d_{\uparrow,a}
    \hat d^\dagger_{\downarrow,a}
    \hat d_{\downarrow,a} ,
    \hat\rho_{\rm tot}]
    \\
    {\mathcal L}_{f,a}^{(0)}
    \hat\rho_{\rm tot}
    &=&-i
    \sum_\sigma
    [\varepsilon_{f,a}
    \hat f_{\sigma,a}^\dagger\hat f_{\sigma,a},
    \hat\rho_{\rm tot}]
    +\Gamma_{f,a}\sum_\sigma 
    {\mathcal D}[\hat f_{\sigma,a}]\hat\rho_{\rm tot},   
    \\
    {\mathcal L}_c^{(0)}
    \hat\rho_{\rm tot}
    &=&
    -i \sum_{\bm k,\sigma}
    [\varepsilon_{\bm k,\sigma}\hat c^\dagger_{\bm k,\sigma}\hat c_{\bm k,\sigma},\hat\rho_{\rm tot}].
\end{eqnarray}

As we will be primarily interested in the case where the drive is tuned to be resonantly coupled to the double-occupied state (i.e. $h\nu=\varepsilon_{d,a}+U_a - \varepsilon_{f,a}$), 
we are mainly concerned with the unperturbed state with double-occupied state,
\begin{eqnarray}
    \hat r_{[(\uparrow\downarrow,\varnothing,F)];[(\uparrow\downarrow,\varnothing,F)]}^{{\rm tot}(0)}
    =
    (\ket{\uparrow\downarrow}\bra{\uparrow\downarrow})_d
    \otimes
    (\ket{\varnothing}\bra{\varnothing})_f
    \otimes \ket{F}\bra{F},
\end{eqnarray}
which has zero eigenvalue ${\mathcal L}_{\rm tot}^{(0)}\hat r_{\uparrow\downarrow}^{{\rm tot}(0)}=0$.
Here, ${(\cdots)}_{d}$ and ${(\cdots)}_f$ labels states for the localized electrons and their higher energy state, respectively, and $\ket{F}$ is the ground state of the conduction electrons forming a Fermi surface at $\omega=\varepsilon_{\rm F}$.

Again in parallel to what we find in Sec.~\ref{SIsec: sd}, 
taking the slow modes to be the $T_1$ modes (where the number of electrons in ket and bra space is the same; see Sec.~\ref{SIsubsubsec: Ld0})
the first-order correction term is shown to vanish $({\mathcal L}_{\rm drive}^{(1)})_{n_l,n_r}=0$. 
We, therefore, focus below on the second-order process,
\begin{eqnarray}
    ({\mathcal L}_{\rm eff}^{{\rm drive}(2)})_{n_l,[(\uparrow\downarrow,\varnothing,F);(\uparrow\downarrow,\varnothing,F)]}
    =
    -
    \sum_{m\in \mathfrak{f}}
    \frac{
    {\rm tr}\big[\hat l_{n_l}^{{\rm tot}(0)\dagger} 
    {\mathcal L}_{\rm drive}^{(1)}
    \hat r_m^{{\rm tot}(0)}\big]
    {\rm tr}\big[\hat l_m^{{\rm tot}(0)\dagger}
    {\mathcal L}_{\rm drive}^{(1)}
    \hat r_{[(\uparrow\downarrow,\varnothing,F)];[(\uparrow\downarrow,\varnothing,F)]}^{{\rm tot}(0)}
    \big]
    }
    {\lambda_m^{{\rm tot}(0)}}
    \nonumber\\
\end{eqnarray}
The intermediate process contains processes such as
(where we did not write explicitly the conduction electron state below, as it is unaffected by ${\mathcal L}_{\rm drive}^{(0)}$ where it stays being in their ground state $\ket{F}\bra{F}$),
\begin{eqnarray}
    (\ket{\uparrow\downarrow}
     \bra{\uparrow\downarrow})_d
    \otimes
    (\ket{\varnothing}
    \bra{\varnothing})_f
    \xrightarrow{{\mathcal L}_{{\rm drive},a}^{(1)}}
    (\ket{\uparrow}
    \bra{\uparrow\downarrow})_d
    \otimes
    (\ket{\downarrow}
    \bra{\varnothing})_f
    \xrightarrow{{\mathcal L}_{{\rm drive},a}^{(1)}}
    (\ket{\uparrow\downarrow}
    \bra{\uparrow\downarrow}))_d
    \otimes
    (\ket{\varnothing}
    \bra{\varnothing})_f.
\end{eqnarray}
In the first operation of ${\mathcal L}_{\rm drive}^{(1)}$, one of the electrons in the double-occupied state ($\downarrow$-electron in the above case) has hopped to the higher-level orbital, which hops back to the original state by the second operation of ${\mathcal L}_{\rm drive}^{(1)}$.
As listed in Table~\ref{SItable: eigenstates localized electrons 2}, the intermediate state involved in this second-order process has an eigenvalue $\lambda_{[(\uparrow,\downarrow,F)];[(\uparrow\downarrow,\varnothing,F)]}=-i(\varepsilon_{d,a}+U_a-\varepsilon_{f,a}+h\nu) - \Gamma_{f,a}/2$. 
By performing a similar analysis to what is done in Sec.~\ref{SIsec: sd}, we obtain the effective Lindbladian that applies to the double-occupied state reads,
\begin{eqnarray}
    ({\mathcal L}_{\rm eff}^{{\rm drive}(2)})_{n_l,[(\uparrow\downarrow,\varnothing,F);(\uparrow\downarrow,\varnothing,F)]}
    =\kappa_a(\nu) \sum_\sigma 
    (
    {\mathcal D}[\hat d_{\sigma,a}]
    )_{n_l, [(\uparrow\downarrow,\varnothing,F);(\uparrow\downarrow,\varnothing,F)]},
\end{eqnarray}
where the decay rate is given by 
\begin{eqnarray}
    \kappa_a(\nu) = {\rm Re}
    \frac{|\Omega_a|^2}{(-\lambda_m)}
    =
    \frac{|\Omega_a|^2 (\Gamma_{f,a} /2)}{(\varepsilon_{d,a}+U_a-\varepsilon_{f,a}+h\nu)^2+\Gamma_{f,a}^2/4}.
\end{eqnarray}
Here, we have ignored the contribution to the coherent dynamics (i.e., the ``Lamb shift'') as it merely shifts the energy level.

We note that the laser drive ${\mathcal L}_{\rm drive}^{(1)}$ may also induce an unwanted decay in the single-occupied state, due to processes such as, 
\begin{eqnarray}
    (\ket{\uparrow}
    \bra{\uparrow})_d
    \otimes
    (\ket{\varnothing}
    \bra{\varnothing})_f
    \xrightarrow{{\mathcal L}_{{\rm drive},a}^{(1)}}
    (\ket{\varnothing}
    \bra{\uparrow})_d
    \otimes
    (\ket{\uparrow}
    \bra{\varnothing})_f
    \xrightarrow{{\mathcal L}_{{\rm drive},a}^{(1)}}
    (\ket{\uparrow}
    \bra{\uparrow})_d
    \otimes
    (\ket{\varnothing}
    \bra{\varnothing})_f,
\end{eqnarray}
which involves an intermediate state with $\lambda_{[(\uparrow,\varnothing,F);(\varnothing,\uparrow,F)]}^{(0)}=-i(\varepsilon_{d,a}-\varepsilon_{f,a}+h\nu)-\Gamma_{f,a}/2$. 
(See Table~\ref{SItable: eigenstates localized electrons 2}.)
This results in a decay of a single-occupied state
\begin{eqnarray}
    ({\mathcal L}_{\rm eff}^{{\rm drive}(2)})_{n_l,[(\sigma,\varnothing,F);(\sigma',\varnothing,F)]}
    =\kappa_a'(\nu) \sum_\sigma 
    (
    {\mathcal D}[\hat d_{\sigma,a}]
    )_{n_l, [(\sigma,\varnothing,F);(\sigma',\varnothing,F)]}
\end{eqnarray}
with a rate
\begin{eqnarray}
    \kappa_a'(\nu)={\rm Re}\frac{|\Omega_a|^2}{(-\lambda_m)}
    =\frac{|\Omega_a|^2(\Gamma_{f,a}/2)}{(\varepsilon_{d,a}-\varepsilon_{f,a}+h\nu)^2+(\Gamma_{f,a}^2/4)}.
\end{eqnarray}

However, the decay rate of this unwanted single-electron loss process $\kappa_a'(\nu)$ can be made much smaller than those of the double-occupied state $\kappa_a$,
by taking the laser frequency to be resonant with the double-occupied state $h\nu =\varepsilon_{f,a}-(\varepsilon_{d,a}+U_a)$, 
one finds
\begin{eqnarray}
    \kappa_a'
    =\frac{|\Omega_a|^2(\Gamma_{f,a}/2)}{U_a^2+(\Gamma_{f,a}^2/4)}
    \ll 
    \kappa_a = \frac{2|\Omega_a|^2}{\Gamma_{f,a}}.
\end{eqnarray}
For instance, taking $\Gamma_{f,a}=10{\rm meV}$ and $U_a=1{\rm eV}$ leads to the ratio $\kappa_a'/\kappa_a \sim \Gamma_{f,a}^2/U_a^2=  10^{-4}$.

Combining the two contributions, the effective Lindbladian is derived as,
\begin{eqnarray}
    {\mathcal L}_{\rm eff}^{\rm drive}
    \hat\rho
    =-i[\hat H_{c0}+\hat H_{d0},\hat\rho]
    +\sum_{a,\sigma}
    \Big[
    \kappa_a(\nu)
    {\mathcal D}[\hat d_{\sigma,a} 
    \hat P_{\uparrow\downarrow}^a]
    \hat\rho
    +
    \kappa_a'(\nu)
    {\mathcal D}[\hat d_{\sigma,a}
    \hat P_s^a]
    \hat\rho    \Big].
\end{eqnarray}
Neglecting the decay of the single-occupied state and putting ${\mathcal L}_{cd}^{(1)}$ back, we obtain the desired GKSL equation, governed by the effective Lindbladian,
\begin{eqnarray}
    {\mathcal L}_{\rm eff}
    \hat\rho
    =
    ({\mathcal L}_{\rm eff}^{\rm drive}
    +{\mathcal L}_{cd}^{(1)})
    \hat\rho
    =-i[\hat H_{\rm A},\hat\rho]
    +\sum_{a,\sigma}
    \kappa_a
    {\mathcal D}[\hat d_{\sigma,a} 
    \hat P_{\uparrow\downarrow}^a]
    \hat\rho.
\end{eqnarray}
This completes our microscopic derivation of Eq.~(3) in the main text.


\begin{table}[t]
    \centering
    \caption{Eigenvalues and eigenstates of ${\mathcal L}_{d,a}^{(0)}
    +{\mathcal L}_{f,a}^{(0)}$ in the fast variable space}
    \label{SItable: eigenstates localized electrons 2}
    \begin{tabular}{c|c|c|c}
        \hline
         &
        Eigenvalue $\lambda_{a,n}^{d(0)}
        +\lambda_{a,n}^{f(0)}
        $ & Right eigenstate $\hat r_{a,n}^{d(0)}\otimes
        \hat r_{a,n}^{f(0)}$
        & Left eigenstate $\hat l_{a,n}^{d(0)}\otimes
        \hat l_{a,n}^{f(0)}$
        \\
        \hline
        \multirow{4}{*}{fast} &
        $ i (\varepsilon_{d,a}+U_a)
        
        -i(\varepsilon_{f,a}-h\nu)
        -\frac{\Gamma_{f,a}}{2}$ 
        & 
        $
        (\ket{\sigma}\bra{\uparrow\downarrow})_d
        \otimes 
        (\ket{\sigma'}\bra{\varnothing})_f$ & 
        $
        (\ket{\sigma}\bra{\uparrow\downarrow})_d
        \otimes 
        (\ket{\sigma'}\bra{\varnothing})_f$
        \\ 
        &
        $ -i (\varepsilon_{d,a}+U_a)
        +i(\varepsilon_{f,a}-h\nu)
        -\frac{\Gamma_{f,a}}{2}$ 
        & 
        $
        (\ket{\uparrow\downarrow}\bra{\sigma})_d
        \otimes 
        (\ket{\varnothing}\bra{\sigma'})_f$ & 
        $
        (\ket{\uparrow\downarrow}\bra{\sigma})_d
        \otimes 
        (\ket{\varnothing}\bra{\sigma'})_f$
        \\ 
        &
        $ -i \varepsilon_{d,a}
        +i(\varepsilon_{f,a}-h\nu)
        -\frac{\Gamma_{f,a}}{2}$ 
        & 
        $
        (\ket{\sigma}\bra{\varnothing})_d
        \otimes 
        (\ket{\varnothing}\bra{\sigma'})_f$ & 
        $
        (\ket{\uparrow\downarrow}\bra{\sigma})_d
        \otimes 
        (\ket{\varnothing}\bra{\sigma})_f$
        \\ 
        &
        $ i \varepsilon_{d,a}
        -i(\varepsilon_{f,a}-h\nu)
        -\frac{\Gamma_{f,a}}{2}$ 
        & 
        $
        (\ket{\varnothing}\bra{\sigma})_d
        \otimes 
        (\ket{\sigma'}\bra{\varnothing})_f$ & 
        $
        (\ket{\varnothing}\bra{\sigma})_d
        \otimes 
        (\ket{\sigma'}\bra{\varnothing})_f$
        \\ 
        \hline
    \end{tabular}
\end{table}

\section{Condition for the emergence of the chiral phase and additional data}
\label{SIsec: chiral regime}

In the main text, we have proposed to apply our dissipation engineering protocol to layered ferromagnets, to show that a non-reciprocal phase transition \cite{Fruchart2021} to a chiral phase emerges (where magnetization exhibits a persistent many-body chase-and-runaway dynamics).
In this section, we provide an estimate for the parameter regime where the chiral phase arises.
We also provide additional data that gives more details on the properties of these phases.

\begin{figure*}[t]
\centering
\includegraphics[width=0.9\linewidth,keepaspectratio]{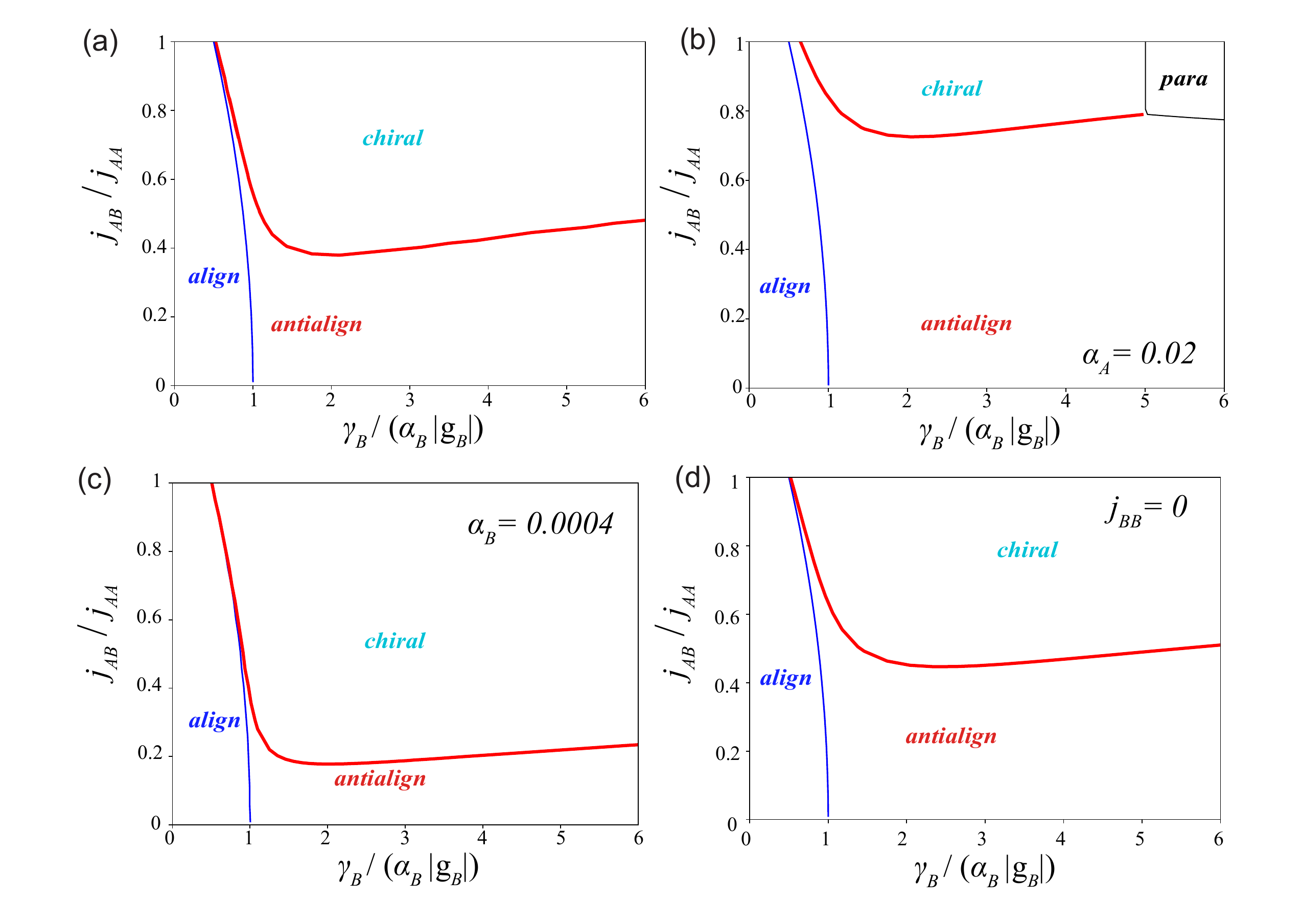}
\caption{
\textbf{Phase diagrams at various parameters.}
(a) We set $\alpha_{\rm A} = 0.1, \alpha_{\rm B}=2\times 10^{-3}, g_{\rm B}=-10{\rm meV}, n=1, j_{\rm AA}=10{\rm meV}, j_{\rm BB}=5{\rm meV}, k_{\rm B}T=5{\rm meV}$ [The same parameter as the dashed line in Fig.~1(c) in the main text]
(b) $\alpha_{\rm A}=0.02$. 
$\alpha_{\rm B}=4\times 10^{-4}$.
(d) $j_{\rm BB}=0$. 
In panels (b)-(d), other parameters are the same as panel (a). 
Here, ``aligned'' and ``anti-aligned'' correspond to a static phase where the relative angle between the direction of the magnetization $\Delta\varphi = \varphi_{\rm A}-\varphi_{\rm B}$ is $\Delta\varphi=0$ and $\Delta\varphi=\pi$, respectively, and ``chiral'' corresponds to the time-dependent phase where the two magnetization exhibits a persistent many-body chase-and-runaway motion
with $\Delta\varphi\ne 0,\pi$.
}
\label{fig: phase_SI}
\end{figure*}

As explained in the main text, we study the dynamics of the magnetization of the A ($\bm m_{\rm A}$) and B layers ($\bm m_{\rm B}$), governed by,
\begin{subequations}
\label{SIeq: non-reciprocal meanfield}
    \begin{align}
        &\dot{\bm m}_{\rm A}
        =\alpha_{\rm A}
        \bigg[
        -k_{\rm B} T \bm m_{\rm A}
        +\Big[
        1-\frac{(\bm h_{\rm eff}^{\rm A})^2}
        {3(k_{\rm B}T)^2}
        \Big]
        \bm h_{\rm eff}^{\rm A}
        \bigg]
        ,
        \\
        &\dot{\bm m}_{\rm B}
        =\alpha_{\rm B}
        \bigg[
        -k_{\rm B}T \bm m_{\rm B}
        +\Big[
        1-\frac{(\bm h_{\rm eff}^{\rm B})^2}
        {3(k_{\rm B}T)^2}
        \Big]
        \bm h_{\rm eff}^{\rm B}
        \bigg]
        -\gamma_{\rm B} n
        \bm m_{\rm B}
        -\Omega_{\rm BA}
        \bm m_{\rm A},
    \end{align}
\end{subequations}
where $\bm h_{\rm eff}^a=\sum_{b={\rm A,B}} j_{ab}\bm m_b$ is the effective field applied to $\bm m_a$ and $\Omega_{\rm BA}=(\gamma_{\rm B}/|g_{\rm B}|)j_{\rm AB}$ is the light-induced torque.
To obtain a rough estimate of where different phases emerge, it is often convenient to linearize the above equation, which gives,
\begin{eqnarray}
    \partial_t 
    \begin{pmatrix}
        \bm m_{\rm A} \\
        \bm m_{\rm B}
    \end{pmatrix}
    &=&
    \begin{pmatrix}
        \alpha_{\rm A}(j_{\rm AA} - k_{\rm B}T )
        &  \alpha_{\rm A}j_{\rm AB}
        \\
        \Big( 
        \alpha_{\rm B} - \frac{\gamma_{\rm B}}{|g_{\rm B}|} 
        \Big)
        j_{\rm AB} 
        & \alpha_{\rm B}(j_{\rm BB} - k_{\rm B}T )
    \end{pmatrix}
    \begin{pmatrix}
        \bm m_{\rm A} \\
        \bm m_{\rm B}
    \end{pmatrix}
    =
    - \hat A 
    \begin{pmatrix}
        \bm m_{\rm A} \\
        \bm m_{\rm B}
    \end{pmatrix}
\end{eqnarray}
The eigenvalues of the dynamical matrix $\hat A$, given by,
\begin{eqnarray}
    \Gamma_\pm 
    = -\frac{1}{2}
    \big[ 
    \alpha_{\rm A}(j_{\rm AA}-k_{\rm B}T)
    +\alpha_{\rm B}(j_{\rm BB}-k_{\rm B}T)
    -\gamma_{\rm B} n 
    \pm\sqrt{\Lambda}    
    \big], 
\end{eqnarray}
with
\begin{eqnarray}
    \label{SIeq: Lambda}
    \Lambda = 
    \big[ 
    \alpha_{\rm A}(j_{\rm AA}-k_{\rm B}T)
    -\alpha_{\rm B}(j_{\rm BB}-k_{\rm B}T)
    +\gamma_{\rm B} n 
    \big]^2
    + 4\alpha_{\rm A} 
    \Big(
    \alpha_{\rm B}
    -\frac{\gamma_{\rm B}}{|g_{\rm B}|}
    \Big)j_{\rm AB}^2,
\end{eqnarray}
determines the stability of the paramagnetic phase.
If the real part of (one of) the eigenvalues were negative ${\rm Re}\Gamma_\pm < 0$, it means that the paramagnetic phase is unstable, implying the emergence of an ordered phase $\bm m_a(t\rightarrow\infty)\ne 0$.
If the imaginary part of $\Gamma_+$ and $\Gamma_-$ were non-zero, which is the case for $\Lambda<0$, then the dynamics would involve oscillation.
These imply that, when ${\rm Re}\Gamma_\pm <0$ and  $\Lambda < 0$
are simultaneously satisfied, 
we can expect the chiral phase to emerge, as long as the system is sufficiently close to the order-disorder transition point.


Let us now assume that the Gilbert damping rate in the B-layer $\alpha_{\rm B}$ is small, or more precisely, $\alpha_{\rm B}\ll \alpha_{\rm A}$ and  
$\alpha_{\rm B}j_{\rm BB}\ll \alpha_{\rm A}j_{\rm AA}$. 
This is motivated by the property that it is a necessary condition for the light-generated torque generated to the B-layer ferromagnet, $\gamma_{\rm B}/|g_{\rm B|}j_{\rm AB}$, to exceed in magnitude those arising from the Gilbert damping $\alpha_{\rm B}j_{\rm AB}$, to achieve $\Lambda<0$.
Since 
\begin{eqnarray}
    \label{SIeq: alphaB<gammaB/gB}
    \alpha_{\rm B}
    <
    \frac{\gamma_{\rm B}}{g_{\rm B}},
\end{eqnarray}
is required for this to be achieved, a small Gilbert damping of the B layer $\alpha_{\rm B}$ is favorable. 

With this assumption, the condition for the magnetic order to emerge ${\rm Re}\Gamma_\pm<0$ reads,
\begin{eqnarray}
    \label{SIeq: ReGamma<0}
    \alpha_{\rm A}(j_{\rm AA}-k_{\rm B}T)
    >\gamma_{\rm B} n 
\end{eqnarray}
and the condition $\Lambda<0$ reads
\begin{eqnarray}
    \label{SIeq: Lambda<0}
    \big[ 
    \alpha_{\rm A}(j_{\rm AA}-k_{\rm B}T)
    +\gamma_{\rm B} n 
    \big]^2
    <    
    4\alpha_{\rm A} 
    \Big(
    \frac{\gamma_{\rm B}}{|g_{\rm B}|}
    -    \alpha_{\rm B}
    \Big)j_{\rm AB}^2.
\end{eqnarray}
Using 
\eqref{SIeq: ReGamma<0}, we can evaluate Eq.~\eqref{SIeq: Lambda<0} as
\begin{eqnarray}
    \gamma_{\rm B}^2 n^2
    <
    \frac{1}{4} \big[ 
    \alpha_{\rm A}(j_{\rm AA}-k_{\rm B}T)
    +\gamma_{\rm B} n 
    \big]^2
    <    
    \alpha_{\rm A} 
    \Big(
    \frac{\gamma_{\rm B}}{|g_{\rm B}|}
    -    \alpha_{\rm B}
    \Big)j_{\rm AB}^2,
\end{eqnarray}
which yields,
\begin{eqnarray}
    \gamma_{\rm B}^2 
    -    
    \frac{\alpha_{\rm A}j_{\rm AB}^2}{n^2|g_{\rm B}|}
    \gamma_{\rm B}
    +        
    \alpha_{\rm A} 
    \alpha_{\rm B}
    \frac{j_{\rm AB}^2
    }{n^2}
    < 0
\end{eqnarray}
or
\begin{eqnarray}
    \frac{1}{2}
    \Bigg[
    \frac{\alpha_{\rm A}j_{\rm AB}^2}{n^2|g_{\rm B}|}
    -\sqrt{\frac{\alpha_{\rm A}^2j_{\rm AB}^4}{n^4|g_{\rm B}|^2}
    -
    4\alpha_{\rm A} 
    \alpha_{\rm B}
    \frac{j_{\rm AB}^2
    }{n^2}}
    \Bigg]    <
    \gamma_{\rm B}
    <
    \frac{1}{2}
    \Bigg[
    \frac{\alpha_{\rm A}j_{\rm AB}^2}{n^2|g_{\rm B}|}
    +\sqrt{\frac{\alpha_{\rm A}^2j_{\rm AB}^4}{n^4|g_{\rm B}|^2}
    -
    4\alpha_{\rm A} 
    \alpha_{\rm B}
    \frac{j_{\rm AB}^2
    }{n^2}}
    \Bigg].
\end{eqnarray}
Combining Eq.~\eqref{SIeq: alphaB<gammaB/gB},
\begin{eqnarray}
    {\rm max}
    \Bigg[
    \frac{1}{2}
    \Bigg[
    \frac{\alpha_{\rm A}j_{\rm AB}^2}{n^2|g_{\rm B}|}
    -\sqrt{\frac{\alpha_{\rm A}^2j_{\rm AB}^4}{n^4|g_{\rm B}|^2}
    -
    4\alpha_{\rm A} 
    \alpha_{\rm B}
    \frac{j_{\rm AB}^2
    }{n^2}}
    \Bigg], 
    \alpha_{\rm B}|g_{\rm B}|
    \Bigg]
    <
    \gamma_{\rm B}
    <
    \frac{1}{2}
    \Bigg[
    \frac{\alpha_{\rm A}j_{\rm AB}^2}{n^2|g_{\rm B}|}
    +\sqrt{\frac{\alpha_{\rm A}^2j_{\rm AB}^4}{n^4|g_{\rm B}|^2}
    -
    4\alpha_{\rm A} 
    \alpha_{\rm B}
    \frac{j_{\rm AB}^2
    }{n^2}}
    \Bigg]
    <
    \frac{\alpha_{\rm A}j_{\rm AB}^2}{n^2|g_{\rm B}|}.
    \nonumber\\
\end{eqnarray}
This yields a rough estimate of a necessary condition for the chiral phase to emerge,
which is 
\begin{eqnarray}
    \alpha_{\rm B}|g_{\rm B}|
    < \frac{\alpha_{\rm A}j_{\rm AB}^2}{n^2|g_{\rm B}|}.
\end{eqnarray}
or
\begin{eqnarray}
    \label{SIeq: condition chiral}
    \frac{\alpha_{\rm B}}{\alpha_{\rm A}}
    \frac{g_{\rm B}^2}{j_{\rm AB}^2n^2}
    \lesssim 1. 
\end{eqnarray}

\begin{figure*}[t]
\centering
\includegraphics[width=0.7\linewidth,keepaspectratio]{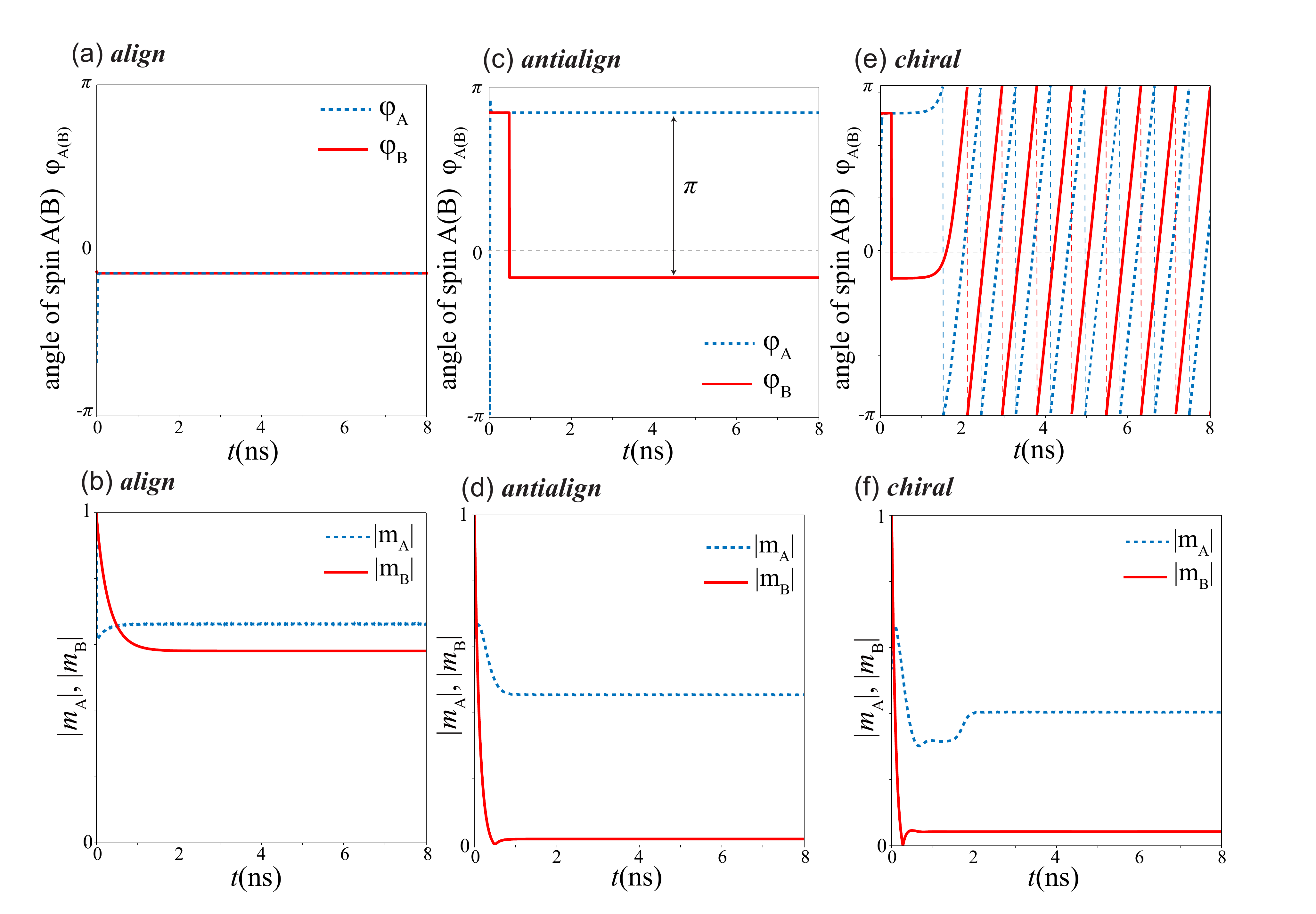}
\caption{
\textbf{Time dependence of magnetization.}
(a),(b) Aligned phase, which converges to $\Delta\varphi=\varphi_{\rm A}-\varphi_{\rm B}=0$. 
$\gamma_{\rm B}=0$.
(c),(d) Anti-aligned phase, which converges to $\Delta\varphi=\varphi_{\rm A}-\varphi_{\rm B}=\pi$. 
$\gamma_{\rm B}/(\alpha_{\rm B}|g_{\rm B})=1.1$.
(e),(f) Chiral phase, which converges to $\Delta\varphi=\varphi_{\rm A}-\varphi_{\rm B}\ne 0, \pi$.
$\gamma_{\rm B}/(\alpha_{\rm B}|g_{\rm B})=1.5$
(a),(c),(e) Phase $\varphi_{\rm A},\varphi_{\rm B}$ dynamics. 
(b),(d),(f) Amplitude $|\bm m_{\rm A}|,|\bm m_{\rm B}|$ dynamics.
We set $j_{\rm AB}=5{\rm meV}$ and $\alpha_{\rm A} = 0.1, \alpha_{\rm B}=2\times 10^{-3}, g_{\rm B}=-10{\rm meV}, n=1, j_{\rm AA}=10{\rm meV}, j_{\rm BB}=5{\rm meV}, k_{\rm B}T=9{\rm meV}$.
The initial state is taken randomly. 
}
\label{fig: timedep_SI}
\end{figure*}

The above analysis gives insights into the parameter regime where the chiral phase can be achieved. 
Figure \ref{fig: phase_SI}(a) is the phase diagram for the case $\alpha_{\rm A}=0.1, \alpha_{\rm B}=2\times 10^{-3}, g_{\rm B}=-10{\rm meV}, n=1$ (which is reproduced from the main text). 
We have also set $j_{\rm BB}=5{\rm meV}, T=5{\rm meV}$.
As one sees in the figure, the chiral phase may emerge when $j_{\rm AB} \gesim 0.4 j_{\rm AA}=4{\rm meV}$.
This is roughly consistent with the condition \eqref{SIeq: condition chiral},
\begin{eqnarray}
    j_{\rm AB}  
    \gesim
    \sqrt{\frac{\alpha_{\rm B}}{\alpha_{\rm A}}}
    \frac{g_{\rm B}}{n}
    =1.4{\rm meV}.
\end{eqnarray}
Condition \eqref{SIeq: condition chiral} implies that the smaller $\alpha_{\rm A}$ ($\alpha_{\rm B}$) shrinks (enlarges) the region of the chiral phase. This is confirmed in our numerical results shown in Figs.~\ref{fig: phase_SI}(b) and (c). 
Furthermore, the above analysis implies that the magnitude of the intra-layer ferromagnetic interaction in the B layer $j_{\rm BB}$ is irrelevant. 
This is confirmed in Fig.~\ref{fig: phase_SI}(d), where the case where the B-layer is replaced with a paramagnetic material $j_{\rm BB}=0$ is presented.

We also report here the time dependence of the magnetization in Fig.~\ref{fig: timedep_SI}.
We note that we did not see any initial state dependence on the phase difference $\Delta\varphi=\varphi_{\rm A}-\varphi_{\rm B}$ and the amplitude $|\bm m_{\rm A}|, |\bm m_{\rm B}|$ (except the sign of $\Delta\varphi$ in the chiral phase that signals a spontaneous $\mathbbm{Z}_2$ symmetry breaking~\cite{Fruchart2021}.

\end{document}